%% file: BergesLesHouches.tex
\documentclass[12pt]{article} 

\usepackage{latexsym} 	
\usepackage{amssymb}  	
\usepackage{amsbsy}
\usepackage{epsfig}     
\usepackage{graphics,color}
\usepackage{a4}
\usepackage{mathrsfs}
\usepackage{dsfont}
\usepackage{ebezier}
\usepackage{axodraw}
\usepackage{dcolumn}

\newcommand{\ep}{\varepsilon}

\newcommand{\om}{\omega}

\newcommand{\tr}{\mbox{tr}\,}
\newcommand{\Tr}{\mbox{Tr}\,}

\renewcommand{\C}{{\cal C}}

\newcommand{\sgn}{\mbox{sgn}}

\newcommand{\vecp}{{\mathbf p}}

\newcommand{\bk}{{\mathbf k}}
\newcommand{\bp}{{\mathbf p}}
\newcommand{\bq}{{\mathbf q}}

\newcommand{\bs}{{\mathbf s}}
\newcommand{\bx}{{\mathbf x}}
\newcommand{\by}{{\mathbf y}}
\newcommand{\bz}{{\mathbf z}}
\newcommand{\bl}{{\mathbf l}}
\newcommand{\br}{{\mathbf r}}

\newcommand{\cl}{{\rm cl}}
\newcommand{\rmd}{{\mathrm d}}

\newcommand{\Or}{{\mathcal O}}

\newcommand{\beq}{\begin{equation}}
\newcommand{\eeq}{\end{equation}}
\newcommand{\bea}{\begin{eqnarray}}
\newcommand{\eea}{\end{eqnarray}}
\newcommand{\bean}{\begin{eqnarray*}}
\newcommand{\eean}{\end{eqnarray*}}
\newcommand{\nn}{\nonumber \\}
\newcommand{\nnn}{\nonumber}

\newcommand{\sign}{{\rm sgn}}
\newcommand{\ol}{\overline}
\newcommand{\ul}{\underline}

\newcommand{\bit}{\begin{itemize}}
\newcommand{\eit}{\end{itemize}}
\newcommand{\bi}{\begin{itemize}\setlength{\itemsep}{0\parsep}}
\newcommand{\ei}{\end{itemize}}
\newcommand{\bc}{\begin{center}}
\newcommand{\ec}{\end{center}}

\newcommand{\gm}{\gamma}

\definecolor{DarkRed}{rgb}{0.75, 0.0, 0.0}

\definecolor{DarkGreen}{rgb}{0.20, 0.55, 0.20}

\definecolor{DarkBlue}{rgb}{0.00, 0.00, 0.70}
  
\def\db{\color{DarkBlue}}
\def\rr{\color{DarkRed}}

\newcommand{\pic}[1]{\;\parbox[c]{30pt}{\begin{picture}(30,30)(0,0)
\SetWidth{1.0}\SetScale{1.0} #1 \end{picture}}\;}
\newcommand{\pib}[1]{\;\parbox[c]{36pt}{\begin{picture}(22.5,30)(0,0)
\SetWidth{1.0}\SetScale{1.0} #1 \end{picture}}\;}
\newcommand{\picb}[1]{\;\parbox[c]{45pt}{\begin{picture}(45,30)(0,0)
\SetWidth{1.0}\SetScale{1.0} #1 \end{picture}}\;}
\newcommand{\picc}[1]{\;\parbox[c]{60pt}{\begin{picture}(60,30)(0,0)
\SetWidth{1.0}\SetScale{1.0} #1 \end{picture}}\;}

\newcommand{\spic}[1]{\;\parbox[c]{21pt}{\begin{picture}(21,21)(0,0)
\SetWidth{1.0}\SetScale{0.7} #1 \end{picture}}\;}

\newcommand{\spicb}[1]{\;\parbox[c]{32pt}{\begin{picture}(32,21)(0,0)
\SetWidth{1.0}\SetScale{0.7} #1 \end{picture}}\;}
\newcommand{\spicc}[1]{\;\parbox[c]{42pt}{\begin{picture}(42,21)(0,0)
\SetWidth{1.0}\SetScale{0.7} #1 \end{picture}}\;}

\def\Lwidth{1}

\def\Agl(#1,#2)(#3,#4,#5){\PhotonArc(#1,#2)(#3,#4,#5){\Lwidth}
{6.283 #3 mul 360 div #4 #5 sub #4 #5 sub mul sqrt mul Ldensity mul}}
\def\Lgl(#1,#2)(#3,#4){\Photon(#1,#2)(#3,#4){\Lwidth}
{#1 #3 sub #1 #3 sub mul #2 #4 sub #2 #4 sub mul add sqrt Ldensity mul}}
\def\Agh(#1,#2)(#3,#4,#5){\DashArrowArc(#1,#2)(#3,#4,#5){1}}
\def\Aagh(#1,#2)(#3,#4,#5){\DashArrowArcn(#1,#2)(#3,#5,#4){1}}
\def\Lgh(#1,#2)(#3,#4){\DashArrowLine(#1,#2)(#3,#4){1}}
\def\Lagh(#1,#2)(#3,#4){\DashArrowLine(#3,#4)(#1,#2){1}}
\def\Ahh(#1,#2)(#3,#4,#5){\DashCArc(#1,#2)(#3,#4,#5){1}}
\def\Lhh(#1,#2)(#3,#4){\DashLine(#1,#2)(#3,#4){1}}
\def\Aqu(#1,#2)(#3,#4,#5){\ArrowArc(#1,#2)(#3,#4,#5)}
\def\Aaqu(#1,#2)(#3,#4,#5){\ArrowArcn(#1,#2)(#3,#5,#4)}
\def\Lqu(#1,#2)(#3,#4){\ArrowLine(#1,#2)(#3,#4)}
\def\Laqu(#1,#2)(#3,#4){\ArrowLine(#3,#4)(#1,#2)}
\def\Aqq(#1,#2)(#3,#4,#5){\CArc(#1,#2)(#3,#4,#5)}
\def\Lqq(#1,#2)(#3,#4){\Line(#1,#2)(#3,#4)}
\def\Asc(#1,#2)(#3,#4,#5){\CArc(#1,#2)(#3,#4,#5)}
\def\Lsc(#1,#2)(#3,#4){\Line(#1,#2)(#3,#4)}
\def\DAsc(#1,#2)(#3,#4,#5){\DashCArc(#1,#2)(#3,#4,#5){3}}
\def\DLsc(#1,#2)(#3,#4){\DashLine(#1,#2)(#3,#4){3}}
\def\TAsc(#1,#2)(#3,#4,#5){\SetWidth{2.0}\CArc(#1,#2)(#3,#4,#5)\SetWidth{1.0}}
\def\TLsc(#1,#2)(#3,#4){\SetWidth{2.0}\Line(#1,#2)(#3,#4)\SetWidth{1.0}}

\def\TopoVR(#1){\pic{#1(15,15)(15,-90,270)}}
\def\TopoVRoo(#1){\;\pic{#1(15,15)(15,0,180) #1(15,15)(15,180,360)%
 \GCirc(0,15){5}{0.75} \GCirc(30,15){5}{0.75}}\;}
\def\TopoVRor(#1){\;\pic{#1(15,15)(15,0,180) #1(15,15)(15,180,360)%
 \GCirc(0,15){5}{0.75} \BBoxc(30,15)(6,6)}\;}
\def\TopoVRoi(#1){\;\pic{#1(15,15)(15,0,180) #1(15,15)(15,180,360)%
 \GCirc(0,15){5}{0.75} \GBoxc(30,15)(6,6){0}}\;}
\def\TopoVRooo(#1){\pic{#1(15,15)(15,-30,90) #1(15,15)(15,90,210)%
 #1(15,15)(15,210,330) \GCirc(15,30){3}{0} \GCirc(2,7.5){3}{0}%
 \GCirc(28,7.5){3}{0}}}
\def\TTopoVRoo(#1){\;\pic{#1(15,15)(15,0,180) #1(15,15)(15,180,360)%
 \GCirc(0,15){5}{0.75} \Text(0,15)[c]{$\scriptstyle 1$}
 \GCirc(30,15){5}{0.75} \Text(30,15)[c]{$\scriptstyle 1$}}\;}
\def\TTopoVRor(#1){\;\pic{#1(15,15)(15,0,180) #1(15,15)(15,180,360)%
 \GCirc(0,15){5}{0.75} \Text(0,15)[c]{$\scriptstyle 1$}
 \GBoxc(30,15)(9,9){0.75} \Text(30,15)[c]{$\scriptstyle 2$} }\;}
\def\TTopoVRoi(#1){\;\pic{#1(15,15)(15,0,180) #1(15,15)(15,180,360)%
 \GCirc(0,15){5}{0.75} \Text(0,15)[c]{$\scriptstyle 1$} 
 \GCirc(30,15){5}{0.75} \Text(30,15)[c]{$\scriptstyle 2$} }\;}
\def\TTopoVRooo(#1){\pic{#1(15,15)(15,-30,90) #1(15,15)(15,90,210)%
 #1(15,15)(15,210,330) 
 \GCirc(15,30){5}{0.75}\Text(15,30)[c]{$\scriptstyle 1$}%
 \GCirc(2,7.5){5}{0.75}\Text(2,7.5)[c]{$\scriptstyle 1$}%
 \GCirc(28,7.5){5}{0.75}\Text(28,7.5)[c]{$\scriptstyle 1$}}}
\def\TTopoVRoooo(#1){\pic{#1(15,15)(15,-45,45) #1(15,15)(15,45,135)%
 #1(15,15)(15,135,225) #1(15,15)(15,225,315) 
 \GCirc(25,25){5}{0.75}\Text(25,25)[c]{$\scriptstyle 1$}%
 \GCirc(25,5){5}{0.75}\Text(25,5)[c]{$\scriptstyle 1$}%
 \GCirc(5,25){5}{0.75}\Text(5,25)[c]{$\scriptstyle 1$}%
 \GCirc(5,5){5}{0.75}\Text(5,5)[c]{$\scriptstyle 1$}}}
\def\TTopoVRooi(#1){\pic{#1(15,15)(15,-30,90) #1(15,15)(15,90,210)%
 #1(15,15)(15,210,330) 
 \GCirc(15,30){5}{0.75}\Text(15,30)[c]{$\scriptstyle 2$}%
 \GCirc(2,7.5){5}{0.75}\Text(2,7.5)[c]{$\scriptstyle 1$}%
 \GCirc(28,7.5){5}{0.75}\Text(28,7.5)[c]{$\scriptstyle 1$}}}
\def\TTopoVRoor(#1){\pic{#1(15,15)(15,-30,90) #1(15,15)(15,90,210)%
 #1(15,15)(15,210,330) 
 \GBoxc(15,30)(9,9){0.75}\Text(15,30)[c]{$\scriptstyle 2$}%
 \GCirc(2,7.5){5}{0.75}\Text(2,7.5)[c]{$\scriptstyle 1$}%
 \GCirc(28,7.5){5}{0.75}\Text(28,7.5)[c]{$\scriptstyle 1$}}}
\def\TTopoVRiie(#1){\;\pic{#1(15,15)(15,0,180) #1(15,15)(15,180,360)%
 \GCirc(0,15){5}{0.75} \Text(0,15)[c]{$\scriptstyle 2$} 
 \GCirc(30,15){5}{0.75} \Text(30,15)[c]{$\scriptstyle 2$} }\;}
\def\TTopoVRire(#1){\;\pic{#1(15,15)(15,0,180) #1(15,15)(15,180,360)%
 \GCirc(0,15){5}{0.75} \Text(0,15)[c]{$\scriptstyle 2$}
 \GBoxc(30,15)(9,9){0.75} \Text(30,15)[c]{$\scriptstyle 2$} }\;}
\def\TTopoVRiif(#1){\;\pic{#1(15,15)(15,0,180) #1(15,15)(15,180,360)%
 \GCirc(0,15){5}{0.75} \Text(0,15)[c]{$\scriptstyle 1$} 
 \GCirc(30,15){5}{0.75} \Text(30,15)[c]{$\scriptstyle 3$} }\;}
\def\TTopoVRirf(#1){\;\pic{#1(15,15)(15,0,180) #1(15,15)(15,180,360)%
 \GCirc(0,15){5}{0.75} \Text(0,15)[c]{$\scriptstyle 1$}
 \GBoxc(30,15)(9,9){0.75} \Text(30,15)[c]{$\scriptstyle 3$} }\;}
\def\TTopoVRirrf(#1){\;\pic{#1(15,15)(15,0,180) #1(15,15)(15,180,360)%
 \GCirc(0,15){5}{0.75} \Text(0,15)[c]{$\scriptstyle 1$}
 \GBoxc(30,15)(12,12){0.75}
 \GBoxc(30,15)(8,8){0.75} \Text(30,15)[c]{$\scriptstyle 3$} }\;}
\def\STTopoVRoo(#1){\;\spic{#1(15,15)(15,0,180) #1(15,15)(15,180,360)%
 \GCirc(0,15){5}{0.75} \Text(0,10.5)[c]{$\scriptscriptstyle 1$}
 \GCirc(30,15){5}{0.75} \Text(21,10.5)[c]{$\scriptscriptstyle 1$}}\;}
\def\STTopoVRor(#1){\;\spic{#1(15,15)(15,0,180) #1(15,15)(15,180,360)%
 \GCirc(0,15){5}{0.75} \Text(0,10.5)[c]{$\scriptscriptstyle 1$}
 \GBoxc(30,15)(9,9){0.75} \Text(21,10.5)[c]{$\scriptscriptstyle 2$} }\;}
\def\STTopoVRoi(#1){\;\spic{#1(15,15)(15,0,180) #1(15,15)(15,180,360)%
 \GCirc(0,15){5}{0.75} \Text(0,10.5)[c]{$\scriptscriptstyle 1$} 
 \GCirc(30,15){5}{0.75} \Text(21,10.5)[c]{$\scriptscriptstyle 2$} }\;}
\def\STTopoVRooo(#1){\spic{#1(15,15)(15,-30,90) #1(15,15)(15,90,210)%
 #1(15,15)(15,210,330) 
 \GCirc(15,30){5}{0.75}\Text(10.5,21)[c]{$\scriptscriptstyle 1$}%
 \GCirc(2,7.5){5}{0.75}\Text(1.4,5.25)[c]{$\scriptscriptstyle 1$}%
 \GCirc(28,7.5){5}{0.75}\Text(19.6,5.25)[c]{$\scriptscriptstyle 1$}}}
\def\STTopoVRoooo(#1){\spic{#1(15,15)(15,-45,45) #1(15,15)(15,45,135)%
 #1(15,15)(15,135,225) #1(15,15)(15,225,315) 
 \GCirc(25,25){5}{0.75}\Text(17.5,17.5)[c]{$\scriptscriptstyle 1$}%
 \GCirc(25,5){5}{0.75}\Text(17.5,3.5)[c]{$\scriptscriptstyle 1$}%
 \GCirc(5,25){5}{0.75}\Text(3.5,17.5)[c]{$\scriptscriptstyle 1$}%
 \GCirc(5,5){5}{0.75}\Text(3.5,3.5)[c]{$\scriptscriptstyle 1$}}}
\def\STTopoVRooi(#1){\spic{#1(15,15)(15,-30,90) #1(15,15)(15,90,210)%
 #1(15,15)(15,210,330) 
 \GCirc(15,30){5}{0.75}\Text(10.5,21)[c]{$\scriptscriptstyle 2$}%
 \GCirc(2,7.5){5}{0.75}\Text(1.4,5.25)[c]{$\scriptscriptstyle 1$}%
 \GCirc(28,7.5){5}{0.75}\Text(19.6,5.25)[c]{$\scriptscriptstyle 1$}}}
\def\STTopoVRoor(#1){\spic{#1(15,15)(15,-30,90) #1(15,15)(15,90,210)%
 #1(15,15)(15,210,330) 
 \GBoxc(15,30)(9,9){0.75}\Text(10.5,21)[c]{$\scriptscriptstyle 2$}%
 \GCirc(2,7.5){5}{0.75}\Text(1.4,5.25)[c]{$\scriptscriptstyle 1$}%
 \GCirc(28,7.5){5}{0.75}\Text(19.6,5.25)[c]{$\scriptscriptstyle 1$}}}
\def\STTopoVRiie(#1){\;\spic{#1(15,15)(15,0,180) #1(15,15)(15,180,360)%
 \GCirc(0,15){5}{0.75} \Text(0,10.5)[c]{$\scriptscriptstyle 2$} 
 \GCirc(30,15){5}{0.75} \Text(21,10.5)[c]{$\scriptscriptstyle 2$} }\;}
\def\STTopoVRire(#1){\;\spic{#1(15,15)(15,0,180) #1(15,15)(15,180,360)%
 \GCirc(0,15){5}{0.75} \Text(0,10.5)[c]{$\scriptscriptstyle 2$}
 \GBoxc(30,15)(9,9){0.75} \Text(21,10.5)[c]{$\scriptscriptstyle 2$} }\;}
\def\STTopoVRiif(#1){\;\spic{#1(15,15)(15,0,180) #1(15,15)(15,180,360)%
 \GCirc(0,15){5}{0.75} \Text(0,10.5)[c]{$\scriptscriptstyle 1$} 
 \GCirc(30,15){5}{0.75} \Text(21,10.5)[c]{$\scriptscriptstyle 3$} }\;}
\def\STTopoVRirf(#1){\;\spic{#1(15,15)(15,0,180) #1(15,15)(15,180,360)%
 \GCirc(0,15){5}{0.75} \Text(0,10.5)[c]{$\scriptscriptstyle 1$}
 \GBoxc(30,15)(9,9){0.75} \Text(21,10.5)[c]{$\scriptscriptstyle 3$} }\;}
\def\STTopoVRirrf(#1){\;\spic{#1(15,15)(15,0,180) #1(15,15)(15,180,360)%
 \GCirc(0,15){5}{0.75} \Text(0,10.5)[c]{$\scriptscriptstyle 1$}
 \GBoxc(30,15)(12,12){0.75}
 \GBoxc(30,15)(8,8){0.75} \Text(21,10.5)[c]{$\scriptscriptstyle 3$} }\;}
\def\TopoVRooT(#1,#2,#3){\;\pic{#1(15,15)(15,0,180) #1(15,15)(15,180,360)%
 \GCirc(0,15){3}{0} \GCirc(30,15){3}{0} \Text(5,15)[l]{{$\scriptstyle #2$}}%
 \Text(25,15)[r]{{$\scriptstyle #3$}}}\;}
\def\TopoVRorT(#1,#2,#3){\;\pic{#1(15,15)(15,0,180) #1(15,15)(15,180,360)%
 \GCirc(0,15){3}{0} \BBoxc(30,15)(6,6) \Text(5,15)[l]{{$\scriptstyle #2$}}%
 \Text(25,15)[r]{{$\scriptstyle #3$}}}\;}
\def\TopoVRoiT(#1,#2,#3){\;\pic{#1(15,15)(15,0,180) #1(15,15)(15,180,360)%
 \GCirc(0,15){3}{0} \GBoxc(30,15)(6,6){0} \Text(5,15)[l]{{$\scriptstyle #2$}}%
 \Text(25,15)[r]{{$\scriptstyle #3$}}}\;}
\def\TopoVRoooT(#1,#2,#3,#4){\pic{#1(15,15)(15,-30,90) #1(15,15)(15,90,210)%
 #1(15,15)(15,210,330) \GCirc(15,30){3}{0} \GCirc(2,7.5){3}{0}%
 \GCirc(28,7.5){3}{0} \Text(15,25)[t]{{$\scriptstyle #2$}}%
 \Text(6,9)[bl]{{$\scriptstyle #3$}} \Text(24,9)[br]{{$\scriptstyle #4$}}}}
\def\ToptVS(#1,#2,#3){\pic{#1(15,15)(15,0,180) #2(15,15)(15,180,360)%
 #3(30,15)(0,15)}}
\def\SToptVS(#1,#2,#3){\spic{#1(15,15)(15,0,180) #2(15,15)(15,180,360)%
 #3(30,15)(0,15)}}
\def\ToptVSTxt(#1,#2,#3,#4,#5){\;\pic{#1(15,15)(15,0,180)%
 #2(15,15)(15,180,360) #3(30,15)(0,15) \GCirc(0,15){3}{0} \GCirc(30,15){3}{0}%
 \Text(4,17)[bl]{{$\scriptstyle #4$}} \Text(26,13)[tr]{{$\scriptstyle #5$}}}\;}
\def\ToptVE(#1,#2){\picc{#1(15,15)(15,0,360) #2(45,15)(15,-180,180)}}
\def\SToptVE(#1,#2){\spicc{#1(15,15)(15,0,360) #2(45,15)(15,-180,180)}}
\def\ToprVM(#1,#2,#3,#4,#5,#6){\pic{#3(15,15)(15,-30,90) #1(15,15)(15,90,210)%
 #2(15,15)(15,210,330) #5(2,7.5)(15,15) #6(15,15)(15,30) #4(28,7.5)(15,15)}}
\def\SToprVM(#1,#2,#3,#4,#5,#6){\spic{#3(15,15)(15,-30,90)%
 #1(15,15)(15,90,210)%
 #2(15,15)(15,210,330) #5(2,7.5)(15,15) #6(15,15)(15,30) #4(28,7.5)(15,15)}}
\def\ToprVV(#1,#2,#3,#4,#5){\!\!\picb{#2(26.25,15)(15,256,76)%
 #3(30,30)(15,30) #1(18.75,15)(15,104,284) #4(15,30)(22.5,0)%
 #5(30,30)(22.5,0)}\!\!}
\def\SToprVV(#1,#2,#3,#4,#5){\!\!\spicb{#2(26.25,15)(15,256,76)%
 #3(30,30)(15,30) #1(18.75,15)(15,104,284) #4(15,30)(22.5,0)%
 #5(30,30)(22.5,0)}\!\!}
\def\ToprVB(#1,#2,#3,#4){\picb{#1(30,15)(15,-120,120) #2(30,15)(15,120,240)%
 #3(15,15)(15,60,300) #4(15,15)(15,-60,60)}}
\def\SToprVB(#1,#2,#3,#4){\spicb{#1(30,15)(15,-120,120) #2(30,15)(15,120,240)%
 #3(15,15)(15,60,300) #4(15,15)(15,-60,60)}}
\def\TopfVX(#1,#2,#3,#4,#5,#6,#7,#8,#9){\picb{#1(15,15)(15,90,270)%
 #2(30,15)(15,-90,90) #4(30,30)(15,30) #3(15,0)(30,0) #6(15,0)(15,15)%
 #5(15,15)(30,30) #8(15,30)(20,25) #8(25,20)(30,15) #7(30,15)(30,0)%
 #9(15,15)(30,15)}}
\def\STopfVX(#1,#2,#3,#4,#5,#6,#7,#8,#9){\spicb{#1(15,15)(15,90,270)%
 #2(30,15)(15,-90,90) #4(30,30)(15,30) #3(15,0)(30,0) #6(15,0)(15,15)%
 #5(15,15)(30,30) #8(15,30)(20,25) #8(25,20)(30,15) #7(30,15)(30,0)%
 #9(15,15)(30,15)}}
\def\TopfVH(#1,#2,#3,#4,#5,#6,#7,#8,#9){\picb{#1(15,15)(15,90,270)%
 #2(30,15)(15,-90,90) #4(30,30)(15,30) #3(15,0)(30,0) #6(15,0)(15,15)%
 #5(15,15)(15,30) #8(30,30)(30,15) #7(30,15)(30,0) #9(15,15)(30,15)}}
\def\STopfVH(#1,#2,#3,#4,#5,#6,#7,#8,#9){\spicb{#1(15,15)(15,90,270)%
 #2(30,15)(15,-90,90) #4(30,30)(15,30) #3(15,0)(30,0) #6(15,0)(15,15)%
 #5(15,15)(15,30) #8(30,30)(30,15) #7(30,15)(30,0) #9(15,15)(30,15)}}
\def\TopfVW(#1,#2,#3,#4,#5,#6,#7,#8){\pic{#1(15,15)(15,90,180)%
 #3(15,15)(15,180,270) #2(15,15)(15,270,360) #4(15,15)(15,0,90)%
 #5(15,15)(15,30) #7(15,15)(15,0) #6(0,15)(15,15) #8(30,15)(15,15)}}
\def\STopfVW(#1,#2,#3,#4,#5,#6,#7,#8){\spic{#1(15,15)(15,90,180)%
 #3(15,15)(15,180,270) #2(15,15)(15,270,360) #4(15,15)(15,0,90)%
 #5(15,15)(15,30) #7(15,15)(15,0) #6(0,15)(15,15) #8(30,15)(15,15)}}
\def\TopfVV(#1,#2,#3,#4,#5,#6,#7,#8){\!\!\picb{#2(26.25,15)(15,256,346)%
 #3(26.25,15)(15,-14,76) #4(30,30)(15,30) #1(18.75,15)(15,104,284)%
 #7(22.5,0)(15,30) #6(30,30)(26.25,15) #8(26.25,15)(22.5,0)%
 #5(25.25,15)(39.8,11.4)}\!\!}
\def\STopfVV(#1,#2,#3,#4,#5,#6,#7,#8){\!\!\spicb{#2(26.25,15)(15,256,346)%
 #3(26.25,15)(15,-14,76) #4(30,30)(15,30) #1(18.75,15)(15,104,284)%
 #7(22.5,0)(15,30) #6(30,30)(26.25,15) #8(26.25,15)(22.5,0)%
 #5(25.25,15)(39.8,11.4)}\!\!}
\def\TopfVB(#1,#2,#3,#4,#5,#6,#7){\picb{#2(30,15)(15,-120,120)%
 #6(30,15)(15,120,180) #5(30,15)(15,180,240) #1(15,15)(15,60,300)%
 #4(15,15)(15,-60,0) #3(15,15)(15,0,60) #7(30,15)(15,15)}}
\def\STopfVB(#1,#2,#3,#4,#5,#6,#7){\spicb{#2(30,15)(15,-120,120)%
 #6(30,15)(15,120,180) #5(30,15)(15,180,240) #1(15,15)(15,60,300)%
 #4(15,15)(15,-60,0) #3(15,15)(15,0,60) #7(30,15)(15,15)}}
\def\TopfVN(#1,#2,#3,#4,#5,#6,#7){\picb{#1(15,15)(15,90,270)%
 #2(30,15)(15,-90,90) #4(30,30)(15,30) #3(15,0)(30,0)%
 #5(15,0)(15,30) #6(30,30)(30,0) #7(15,30)(30,0)}} 
\def\STopfVN(#1,#2,#3,#4,#5,#6,#7){\spicb{#1(15,15)(15,90,270)%
 #2(30,15)(15,-90,90) #4(30,30)(15,30) #3(15,0)(30,0)%
 #5(15,0)(15,30) #6(30,30)(30,0) #7(15,30)(30,0)}} 
\def\TopfVU(#1,#2,#3,#4,#5,#6,#7){\pic{#3(15,15)(15,0,90)%
 #2(15,15)(15,90,180) #4(15,15)(15,180,270) #1(15,15)(15,270,360)%
 #6(0,15)(15,30) #7(15,0)(0,15) #5(30,15)(15,0)}}
\def\STopfVU(#1,#2,#3,#4,#5,#6,#7){\spic{#3(15,15)(15,0,90)%
 #2(15,15)(15,90,180) #4(15,15)(15,180,270) #1(15,15)(15,270,360)%
 #6(0,15)(15,30) #7(15,0)(0,15) #5(30,15)(15,0)}}
\def\TopfVT(#1,#2,#3,#4,#5,#6){\pic{#1(15,15)(15,90,210)%
 #2(15,15)(15,210,330) #3(15,15)(15,-30,90) #4(2,7.5)(15,30)%
 #6(28,7.5)(2,7.5) #5(15,30)(28,7.5)}}
\def\STopfVT(#1,#2,#3,#4,#5,#6){\spic{#1(15,15)(15,90,210)%
 #2(15,15)(15,210,330) #3(15,15)(15,-30,90) #4(2,7.5)(15,30)%
 #6(28,7.5)(2,7.5) #5(15,30)(28,7.5)}}
\def\STopEVa(#1,#2){\spicb{#1(15,15)(15,90,270)%
 #1(30,15)(15,-90,90) #2(30,30)(15,30) #2(15,0)(30,0) #2(15,0)(15,30)%
 #2(30,30)(30,0) #2(15,15)(30,15) #2(22.5,15)(22.5,0)}}
\def\STopEVb(#1,#2){\spicb{#1(15,15)(15,90,270)%
 #1(30,15)(15,-90,90) #2(30,30)(15,30) #2(15,0)(30,0) #2(15,0)(15,30)%
 #2(30,30)(30,0) #2(15,10)(30,10) #2(15,20)(30,20)}}
\def\STopEVc(#1,#2){\spicb{#1(15,15)(15,90,270)%
 #1(30,15)(15,-90,90) #2(30,30)(15,30) #2(15,0)(30,0) #2(15,0)(15,30)%
 #2(30,30)(30,0) #2(15,10)(30,20) #2(15,20)(19.5,17) #2(25.5,13)(30,10)}}
\def\STopEVd(#1,#2){\spicb{#1(15,15)(15,90,270)%
 #1(30,15)(15,-90,90) #2(30,30)(15,30) #2(15,0)(30,0) #2(15,0)(15,30)%
 #2(30,30)(30,0) #2(15,11.25)(30,11.25) #2(22.5,9)(22.5,0)%
 #1(15,11.25)(7.5,15,90)}}
\def\STopEVe(#1,#2){\spic{#1(15,15)(15,0,90)%
 #1(15,15)(15,90,180) #1(15,15)(15,180,270) #1(15,15)(15,270,360)%
 #2(0,15)(15,30) #2(10,15)(10,1) #2(30,15)(0,15) #2(20,15)(20,1)}}
\def\STopEVf(#1,#2){\spic{#1(15,15)(15,0,90)%
 #1(15,15)(15,90,180) #1(15,15)(15,180,270) #1(15,15)(15,270,360)%
 #2(0,15)(15,30) #2(15,6)(24,15) #2(30,15)(0,15) #2(15,15)(15,0)}}
\def\STopEVg(#1,#2){\spicb{#1(15,15)(15,90,270)%
 #1(30,15)(15,-90,90) #2(30,30)(15,30) #2(15,0)(30,0) #2(15,0)(15,30)%
 #2(30,30)(30,0) #2(15,15)(45,15)}}
\def\STopEVh(#1,#2){\!\!\spicb{#1(26.25,15)(15,256,346)%
 #1(26.25,15)(15,-14,76) #2(30,30)(15,30) #1(18.75,15)(15,104,284)%
 #2(22.5,0)(15,30) #2(30,30)(26.25,15) #2(26.25,15)(22.5,0)%
 #2(25.25,15)(39.8,11.4) #2(19.75,15)(5.2,11.4)}\!\!}
\def\STopEVi(#1,#2){\spic{#1(15,15)(15,0,90)%
 #1(15,15)(15,90,180) #1(15,15)(15,180,270) #1(15,15)(15,270,360)%
 #2(15,15)(15,30) #2(0,15)(30,15) #2(15,0)(10,15) #2(15,0)(20,15)}}
\def\STopEVj(#1,#2){\spicb{#1(15,15)(15,90,270)%
 #1(30,15)(15,-90,90) #2(30,30)(15,30) #2(15,0)(30,0) #2(15,0)(15,20)%
 #2(15,20)(22.5,30) #2(22.5,30)(30,20) #2(30,20)(30,0)%
 #2(15,10)(30,20) #2(15,20)(19.5,17) #2(25.5,13)(30,10)}}
\def\STopEVk(#1,#2){\spicb{#1(15,15)(15,90,270)%
 #1(30,15)(15,-90,90) #2(30,30)(15,30) #2(15,0)(30,0) #2(15,0)(15,30)%
 #2(30,0)(30,30) #2(15,0)(30,24) #2(20,30)(24.5,19.2) #2(26.5,14.4)(30,6)}}
\def\STopEVl(#1,#2){\!\!\spicb{#1(26.25,15)(15,256,346)%
 #1(26.25,15)(15,-14,76) #2(30,30)(15,30) #1(18.75,15)(15,104,284)%
 #2(22.5,0)(5.2,11.4) #2(30,30)(26.25,15) #2(26.25,15)(22.5,0)%
 #2(26.25,15)(39.8,11.4) #2(15,30)(5.2,11.4) }\!\!}
\def\STopEVm(#1,#2){\spicb{#1(30,15)(15,-120,120) #1(30,15)(15,120,240)%
 #1(15,15)(15,60,300) #1(15,15)(15,-60,60) #2(22.5,15)(30,15)%
 #2(22.5,15)(17,22.5) #2(22.5,15)(17,7.5) }}
\def\STopEVn(#1,#2){\spicb{#1(15,15)(15,90,270)%
 #1(30,15)(15,-90,90) #2(30,30)(15,30) #2(15,0)(30,0)%
 #2(15,0)(15,30) #2(30,30)(30,0) #2(15,30)(30,0) #2(30,15)(45,15)}}
\def\STopEVo(#1,#2){\spicb{#1(30,15)(15,-120,120) #1(30,15)(15,120,240)%
 #1(15,15)(15,60,300) #1(15,15)(15,-60,60)%
 #2(16,20.5)(29,20.5) #2(16,9.5)(29,9.5) }}
\def\STopEVp(#1,#2){\spic{#1(15,15)(15,0,90)%
 #1(15,15)(15,90,180) #1(15,15)(15,180,270) #1(15,15)(15,270,360)%
 #2(0,15)(15,30) #2(0,15)(30,15) #2(15,0)(10,15) #2(15,0)(20,15)}}
\def\STopEVq(#1,#2){\spic{#1(15,15)(15,0,90)%
 #1(15,15)(15,90,180) #1(15,15)(15,180,270) #1(15,15)(15,270,360)%
 #2(0,15)(15,30) #1(19,15)(4,0,360) #2(30,15)(23,15) #2(15,15)(0,15)%
 #2(15,15)(15,0)}}
\def\STopEVr(#1,#2){\spic{#1(15,15)(15,0,90)%
 #1(15,15)(15,90,180) #1(15,15)(15,180,270) #1(15,15)(15,270,360)%
 #2(0,15)(15,30) #1(26,15)(4,0,360) #2(22,15)(0,15)%
 #2(15,15)(15,0)}}
\def\STopEVs(#1,#2){\spic{#1(15,15)(15,0,90)%
 #1(15,15)(15,90,180) #1(15,15)(15,180,270) #1(15,15)(15,270,360)%
 #2(0.75,10.35)(29.25,10.35) #2(15,0)(15,10.35)%
 #2(0.75,10.35)(6.2,27.1) #2(29.25,10.35)(23.8,27.1)}}
\def\STopEVt(#1,#2){\spic{#1(15,15)(15,0,90)%
 #1(15,15)(15,90,180) #1(15,15)(15,180,270) #1(15,15)(15,270,360)%
 #2(0,15)(15,30) #1(15,4)(4,0,360) #2(30,15)(0,15) #2(15,15)(15,8)}}
\def\STopEVu(#1,#2){\spic{#1(15,15)(15,0,90)%
 #1(15,15)(15,90,180) #1(15,15)(15,180,270) #1(15,15)(15,270,360)%
 #2(0,15)(15,30) #2(15,7.5)(30,15) #2(30,15)(0,15) #2(15,15)(15,0)}}
\def\STopEVv(#1,#2){\spic{#1(15,15)(15,0,90)%
 #1(15,15)(15,90,180) #1(15,15)(15,180,270) #1(15,15)(15,270,360)%
 #2(0,15)(30,15) #2(0,15)(15,7.5) #2(15,0)(15,15) #2(15,0)(22.5,15)}}
\def\STopEVw(#1,#2){\spicb{#1(30,15)(15,-120,120) #1(30,15)(15,120,240)%
 #1(15,15)(15,60,300) #1(15,15)(15,-60,60) #2(0,15)(15,15) #2(30,15)(45,15)}}
\def\STopEVx(#1,#2){\spicb{#1(15,15)(15,90,270)%
 #1(30,15)(15,-90,90) #2(30,30)(15,30) #2(15,0)(30,0) #2(15,0)(15,15)%
 #2(15,0)(30,15) #2(30,0)(30,15) #2(15,15)(20.5,9.5) #2(24.5,5.5)(30,0)%
 #2(15,15)(22.5,22.5) #2(30,15)(22.5,22.5) #2(22.5,30)(22.5,22.5)}}
\def\STopEVy(#1,#2){\spicb{#1(15,15)(15,90,270)%
 #1(30,15)(15,-90,90) #2(30,30)(15,30) #2(15,0)(30,0) #2(15,0)(15,30)%
 #2(30,0)(30,30) #2(15,0)(30,24) #2(15,30)(22.375,18.2) #2(26.375,11.8)(30,6)}}
\def\STopEVz(#1,#2){\spic{#1(15,15)(15,0,90)%
 #1(15,15)(15,90,180) #1(15,15)(15,180,270) #1(15,15)(15,270,360)%
 #2(15,0)(0.75,10.35) #2(15,0)(29.25,10.35)%
 #2(0.75,10.35)(6.2,27.1) #2(29.25,10.35)(23.8,27.1)}}
\def\STopEVaa(#1,#2){\spic{#1(15,15)(15,0,90)%
 #1(15,15)(15,90,180) #1(15,15)(15,180,270) #1(15,15)(15,270,360)%
 #2(0,15)(15,30) #2(15,30)(30,15) #2(30,15)(0,15) #2(15,15)(15,0)}}
\def\STopEVab(#1,#2){\spic{#1(15,15)(15,0,90)%
 #1(15,15)(15,90,180) #1(15,15)(15,180,270) #1(15,15)(15,270,360)%
 #2(15,0)(0.75,10.35) #2(15,0)(23.8,27.1)%
 #2(0.75,10.35)(6.2,27.1) #2(29.25,10.35)(23.8,27.1)}}
\def\STopEVac(#1,#2){\spic{#1(15,15)(15,0,90)%
 #1(15,15)(15,90,180) #1(15,15)(15,180,270) #1(15,15)(15,270,360)%
 #2(0,15)(15,30) #2(15,0)(30,15) #2(30,15)(0,15) #2(15,15)(15,0)}}
\def\STopEVad(#1,#2){\!\!\spicb{#1(26.25,15)(15,256,346)%
 #1(26.25,15)(15,-14,76) #2(30,30)(15,30) #1(18.75,15)(15,104,284)%
 #2(22.5,0)(15,30) #2(30,30)(26.25,15) #2(26.25,15)(22.5,0)%
 #1(33,13.2)(7,0,360)}\!\!}
\def\STopEVae(#1,#2){\spic{#1(15,15)(15,90,180)%
 #1(15,15)(15,180,270) #1(15,15)(15,270,360) #1(15,15)(15,0,90)%
 #2(15,15)(15,30) #2(15,15)(15,0) #2(0,15)(15,15) #2(30,15)(15,15)%
 #2(15,0)(30,15)}}
\def\STopEVaf(#1,#2){\spicb{#1(15,15)(15,90,270)%
 #1(30,15)(15,-90,90) #2(30,30)(15,30) #2(15,0)(30,0)%
 #2(15,0)(15,30) #2(30,30)(30,0) #2(15,30)(22.5,0) #2(22.5,0)(30,30)}}
\def\STopEVag(#1,#2){\spicb{#1(15,15)(15,90,270)%
 #1(30,15)(15,-90,90) #2(30,30)(15,30) #2(15,0)(30,0) #2(15,0)(15,15)%
 #2(15,15)(22.5,30) #2(22.5,30)(30,15) #2(30,15)(30,0) #2(15,15)(30,15)%
 #2(15,0)(30,15)}}
\def\STopEVah(#1,#2){\spic{#1(15,15)(15,0,90)%
 #1(15,15)(15,90,180) #1(15,15)(15,180,270) #1(15,15)(15,270,360)%
 #2(0,15)(15,30) #2(15,0)(0,15) #2(30,15)(15,0) #2(30,15)(15,30)}}
\def\STopEVai(#1,#2){\spicb{#1(30,15)(15,-120,120) #1(30,15)(15,120,240)%
 #1(15,15)(15,60,300) #1(15,15)(15,-60,60) #1(22.5,15)(7.5,0,360) }}
\def\TopoSTxt(#1,#2){\picb{#1(0,15)(20,15) #1(25,15)(45,15)%
 \GCirc(22.5,15){3}{0} \Text(22.5,20)[b]{{$\scriptstyle #2$}}}}
\def\TopoSB(#1,#2,#3){\picb{#1(0,15)(7.5,15) #2(22.5,15)(15,0,180)%
 #3(22.5,15)(15,180,360) #1(37.5,15)(45,15)}}
\def\STopoSB(#1,#2,#3){\spicb{#1(0,15)(7.5,15) #2(22.5,15)(15,0,180)%
 #3(22.5,15)(15,180,360) #1(37.5,15)(45,15)}}
\def\TopoST(#1,#2){\picb{#1(0,0)(22.5,0) #1(22.5,0)(45,0)%
 #2(22.5,15)(15,-90,270)}} 
\def\STopoST(#1,#2){\spicb{#1(0,0)(22.5,0) #1(22.5,0)(45,0)%
 #2(22.5,15)(15,-90,270)}} 
\def\ToptSiTxt(#1,#2){\picb{#1(0,15)(20,15) #1(25,15)(45,15)%
 \GBoxc(22.5,15)(6,6){0} \Text(22.5,20)[b]{{$\scriptstyle #2$}}}}
\def\ToptSM(#1,#2,#3,#4,#5,#6){\picb{#1(0,15)(7.5,15) #1(37.5,15)(45,15)%
 #2(22.5,15)(15,0,90) #3(22.5,15)(15,90,180) #4(22.5,15)(15,180,270)%
 #5(22.5,15)(15,270,360) #6(22.5,30)(22.5,0)}}
\def\ToptSAl(#1,#2,#3,#4,#5){\picb{#1(0,15)(7.5,15) #1(37.5,15)(45,15)%
 #2(22.5,15)(15,0,90) #3(22.5,15)(15,90,180) #4(22.5,15)(15,180,360)%
 #5(7.5,30)(15,270,360)}}
\def\ToptSAr(#1,#2,#3,#4,#5){\picb{#1(0,15)(7.5,15) #1(37.5,15)(45,15)%
 #2(22.5,15)(15,0,90) #3(22.5,15)(15,90,180) #4(22.5,15)(15,180,360)%
 #5(37.5,30)(15,180,270)}}
\def\ToptSE(#1,#2,#3,#4,#5){\picb{#1(0,15)(7.5,15) #1(37.5,15)(45,15)%
 #3(15,15)(7.5,0,180) #4(15,15)(7.5,180,360)%
 #2(30,15)(7.5,0,180) #5(30,15)(7.5,180,360)}} 
\def\ToptSS(#1,#2,#3,#4){\picb{#1(0,15)(7.5,15) #1(37.5,15)(45,15)%
 #4(7.5,15)(37.5,15) #2(22.5,15)(15,0,180) #3(22.5,15)(15,180,360)}}
\def\SToptSM(#1,#2,#3,#4,#5,#6){\spicb{#1(0,15)(7.5,15) #1(37.5,15)(45,15)%
 #2(22.5,15)(15,0,90) #3(22.5,15)(15,90,180) #4(22.5,15)(15,180,270)%
 #5(22.5,15)(15,270,360) #6(22.5,30)(22.5,0)}}
\def\SToptSAl(#1,#2,#3,#4,#5){\spicb{#1(0,15)(7.5,15) #1(37.5,15)(45,15)%
 #2(22.5,15)(15,0,90) #3(22.5,15)(15,90,180) #4(22.5,15)(15,180,360)%
 #5(7.5,30)(15,270,360)}}
\def\SToptSAr(#1,#2,#3,#4,#5){\spicb{#1(0,15)(7.5,15) #1(37.5,15)(45,15)%
 #2(22.5,15)(15,0,90) #3(22.5,15)(15,90,180) #4(22.5,15)(15,180,360)%
 #5(37.5,30)(15,180,270)}}
\def\SToptSE(#1,#2,#3,#4,#5){\spicb{#1(0,15)(7.5,15) #1(37.5,15)(45,15)%
 #3(15,15)(7.5,0,180) #4(15,15)(7.5,180,360)%
 #2(30,15)(7.5,0,180) #5(30,15)(7.5,180,360)}} 
\def\SToptSS(#1,#2,#3,#4){\spicb{#1(0,15)(7.5,15) #1(37.5,15)(45,15)%
 #4(7.5,15)(37.5,15) #2(22.5,15)(15,0,180) #3(22.5,15)(15,180,360)}}
\def\ToptSrTxt(#1,#2){\picb{#1(0,15)(20,15) #1(25,15)(45,15)%
 \GBoxc(22.5,15)(6,6){1} \Text(22.5,20)[b]{{$\scriptstyle #2$}}}}
\def\ToptSBB(#1,#2,#3){\picb{#1(0,15)(7.5,15)  #1(37.5,15)(45,15)%
 #2(22.5,15)(15,0,90) #2(22.5,15)(15,90,180) #3(22.5,15)(15,180,360)%
 \GCirc(22.5,30){5}{0.75} \Text(22.5,30)[c]{$\scriptstyle 1$}}}
\def\SToptSBB(#1,#2,#3){\spicb{#1(0,15)(7.5,15)  #1(37.5,15)(45,15)%
 #2(22.5,15)(15,0,90) #2(22.5,15)(15,90,180) #3(22.5,15)(15,180,360)%
 \GCirc(22.5,30){5}{0.75} \Text(16,21)[c]{$\scriptscriptstyle 1$}}}
\def\ToptSBBTxt(#1,#2,#3,#4){\picb{#1(0,15)(7.5,15)  #1(37.5,15)(45,15)%
 #2(22.5,15)(15,0,90) #2(22.5,15)(15,90,180) #3(22.5,15)(15,180,360)%
 \GCirc(22.5,30){3}{0} \Text(22.5,25)[t]{{$\scriptstyle #4$}}}}
\def\ToptSTB(#1,#2){\picb{#1(0,0)(22.5,0) #1(22.5,0)(45,0)%
 #2(22.5,15)(15,-90,90) #2(22.5,15)(15,90,270) \GCirc(22.5,30){5}{0.75}%
 \Text(22.5,30)[c]{$\scriptstyle 1$}}} 
\def\SToptSTB(#1,#2){\spicb{#1(0,0)(22.5,0) #1(22.5,0)(45,0)%
 #2(22.5,15)(15,-90,90) #2(22.5,15)(15,90,270) \GCirc(22.5,30){5}{0.75}%
 \Text(16,21)[c]{$\scriptscriptstyle 1$}}} 
\def\ToptSTBTxt(#1,#2,#3){\picb{#1(0,0)(22.5,0) #1(22.5,0)(45,0)%
 #2(22.5,15)(15,-90,90) #2(22.5,15)(15,90,270) \GCirc(22.5,30){3}{0}%
 \Text(22.5,25)[t]{{$\scriptstyle #3$}}}} 
\def\TopoT(#1,#2,#3){\!\pic{#1(15,30)(15,15) #2(2,7.5)(15,15)%
 #3(28,7.5)(15,15) \GCirc(15,15){3}{0}}\!} 
\def\TopoTTxt(#1,#2,#3,#4){\!\pic{#1(15,30)(15,15) #2(2,7.5)(15,15)%
 #3(28,7.5)(15,15) \GCirc(15,15){3}{0}%
 \Text(19,17)[bl]{{$\scriptstyle #4$}}}\!}
\def\TopoTS(#1,#2,#3,#4,#5,#6){\!\pic{#1(15,30)(15,23) #2(2,7.5)(8,11)%
 #3(28,7.5)(22,11) #6(15,15)(8,210,330) #5(15,15)(8,90,210)%
 #4(15,15)(8,-30,90)}\!}
\def\TopoTAo(#1,#2,#3,#4,#5){\!\pic{#1(15,30)(15,23) #2(2,7.5)(15,7)%
 #3(28,7.5)(15,7) #4(15,15)(8,-90,90) #5(15,15)(8,90,270)}\!}
\def\TopoTAr(#1,#2,#3,#4,#5){\!\pic{#1(15,30)(8,19) #2(2,7.5)(8,19)%
 #3(28,7.5)(22,11) #4(15,15)(8,-45,135) #5(15,15)(8,135,315)}\!}
\def\TopoTAl(#1,#2,#3,#4,#5){\!\pic{#1(15,30)(23,19) #2(2,7.5)(8,11)%
 #3(28,7.5)(23,19) #4(15,15)(8,45,225) #5(15,15)(8,-135,45)}\!}

\def\ToptrFpt(#1,#2,#3,#4){\picb{#1(0,0)(15,15) #2(15,15)(30,0)%
 #3(15,15)(30,30) #4(0,30)(15,15)}}
\def\ToptrFpta(#1,#2,#3,#4,#5){\picb{#1(0,0)(15,0) #2(15,0)(30,0)%
 #3(15,30)(30,30) #4(0,30)(15,30) #5(15,30)(15,0)}}
\def\ToptrFptb(#1,#2,#3,#4,#5){\picb{#1(0,0)(7,15) #2(23,15)(30,0)%
 #3(23,15)(30,30) #4(0,30)(7,15) #5(7,15)(23,15)}}
\def\ToptrFptc(#1,#2,#3,#4,#5){\picb{#1(0,0)(15,0) #2(15,30)(30,0)%
 #3(15,0)(20,10) #3(25,20)(30,30) #4(0,30)(15,30) #5(15,30)(15,0)}}
\def\ToptSfull(#1,#2){\picb{#1(0,15)(20,15) #1(25,15)(45,15)%
 \GCirc(22.5,15){3}{0.5} \Text(22.5,20)[b]{{$\scriptstyle #2$}}}}
\def\TopoSBfull(#1,#2,#3,#4){\picb{#1(0,15)(7.5,15) #2(22.5,15)(15,0,180)%
 #3(22.5,15)(15,180,360) #1(37.5,15)(45,15) \GCirc(37.5,15){3}{0.5}%
 \Text(32.5,15)[r]{{$\scriptstyle #4$}}}}
\def\ToptSSamp(#1,#2,#3,#4,#5){\picb{#1(0,15)(7.5,15) #1(37.5,15)(45,15)%
 #4(7.5,15)(37.5,15) #2(22.5,15)(15,0,180) #3(22.5,15)(15,180,360)%
 \GBoxc(37.5,15)(6,6){0.5} \Text(33.5,13)[tr]{{$\scriptstyle #5$}}}}
\def\ToptSSfull(#1,#2,#3,#4,#5){\picb{#1(0,15)(7.5,15) #1(37.5,15)(45,15)%
 #4(7.5,15)(37.5,15) #2(22.5,15)(15,0,180) #3(22.5,15)(15,180,360)%
 \GCirc(37.5,15){3}{0.5} \Text(33.5,13)[tr]{{$\scriptstyle #5$}}}}
\def\ToptSAlfull(#1,#2,#3,#4,#5,#6,#7){\picb{#1(0,15)(7.5,15)%
 #1(37.5,15)(45,15) #2(22.5,15)(15,0,90) #3(22.5,15)(15,90,180)%
 #4(22.5,15)(15,180,360) #5(7.5,30)(15,270,360)%
 \GCirc(37.5,15){3}{0.5} \GCirc(22.5,30){3}{0.5}%
 \Text(33.5,13)[tr]{{$\scriptstyle #6$}}%
 \Text(24.5,26)[tl]{{$\scriptstyle #7$}}}}
\def\ToptrFptamp(#1,#2,#3,#4){\picb{#1(4,4)(15,15) #2(15,15)(26,4)%
 #3(15,15)(26,26) #4(4,26)(15,15) \GBoxc(15,15)(6,6){0.5}}}
\def\ToptrFptfull(#1,#2,#3,#4){\picb{#1(4,4)(15,15) #2(15,15)(26,4)%
 #3(15,15)(26,26) #4(4,26)(15,15) \GCirc(15,15){3}{0.5}}}
\def\ToptrFptafull(#1,#2,#3,#4,#5){\picb{#1(4,4)(15,8) #2(15,8)(26,4)%
 #3(15,22)(26,26) #4(4,26)(15,22) #5(15,22)(15,8)%
 \GCirc(15,8){3}{0.5} \GCirc(15,22){3}{0.5}}}
\def\ToptrFptbfull(#1,#2,#3,#4,#5){\picb{#1(4,4)(8,15) #2(22,15)(26,4)%
 #3(22,15)(26,26) #4(4,26)(8,15) #5(8,15)(22,15)%
 \GCirc(8,15){3}{0.5} \GCirc(22,15){3}{0.5}}}
\def\ToptrFptcfull(#1,#2,#3,#4,#5){\picb{#1(4,4)(15,8) #2(15,22)(26,4)%
 #3(15,8)(18,12.9) #3(21,17.8)(26,26) #4(4,26)(15,22) #5(15,22)(15,8)%
 \GCirc(15,8){3}{0.5} \GCirc(15,22){3}{0.5}}}

\def\ToptVSblob(#1,#2,#3){\pic{#1(15,15)(15,0,180) #2(15,15)(15,180,360)%
 #3(30,15)(0,15) \GCirc(30,15){3}{0}}\;}
\def\ToptVSblobTxt(#1,#2,#3,#4){\pic{#1(15,15)(15,0,180) #2(15,15)(15,180,360)%
 #3(30,15)(0,15) \GCirc(30,15){3}{0} \Text(26,13)[tr]{{$\scriptstyle #4$}}}\;}
\def\ToprVBamp(#1,#2,#3,#4){\picb{#1(30,15)(15,-120,120) #2(30,15)(15,120,240)%
 #3(15,15)(15,60,300) #4(15,15)(15,-60,60) \GBoxc(22.5,28)(6,6){0}}}

\def\ToptVSblobsh(#1,#2,#3){\pic{#1(15,15)(15,0,180) #2(15,15)(15,180,360)%
 #3(30,15)(0,15) \GCirc(30,15){3}{0.5}}\;}
\def\SToptVSblobsh(#1,#2,#3){\spic{#1(15,15)(15,0,180) #2(15,15)(15,180,360)%
 #3(30,15)(0,15) \GCirc(30,15){3}{0.5}}\;}
\def\ToprVBampsh(#1,#2,#3,#4){\picb{#1(30,15)(15,-120,120)%
 #2(30,15)(15,120,240)%
 #3(15,15)(15,60,300) #4(15,15)(15,-60,60) \GBoxc(22.5,28)(6,6){0.5}}}
\def\ToprVVblob(#1,#2,#3,#4,#5){\!\!\picb{#2(26.25,15)(15,256,76)%
 #3(30,30)(15,30) #1(18.75,15)(15,104,284) #4(15,30)(22.5,0)%
 #5(30,30)(22.5,0) \GCirc(29,28){3}{0}}\!\!}
\def\ToprVVblobbsh(#1,#2,#3,#4,#5){\!\!\picb{#2(26.25,15)(15,256,76)%
 #3(30,30)(15,30) #1(18.75,15)(15,104,284) #4(15,30)(22.5,0)%
 #5(30,30)(22.5,0) \GCirc(29,28){3}{0.5} \GCirc(16,28){3}{0.5}}\!\!}
\def\SToprVVblobbsh(#1,#2,#3,#4,#5){\!\!\spicb{#2(26.25,15)(15,256,76)%
 #3(30,30)(15,30) #1(18.75,15)(15,104,284) #4(15,30)(22.5,0)%
 #5(30,30)(22.5,0) \GCirc(29,28){3}{0.5} \GCirc(16,28){3}{0.5}}\!\!}
\def\ToprVBblob(#1,#2,#3,#4){\picb{#1(30,15)(15,-120,120)%
 #2(30,15)(15,120,240)%
 #3(15,15)(15,60,300) #4(15,15)(15,-60,60) \GCirc(22.5,28){3}{0}}}
\def\ToprVBblobsh(#1,#2,#3,#4){\picb{#1(30,15)(15,-120,120)%
 #2(30,15)(15,120,240)%
 #3(15,15)(15,60,300) #4(15,15)(15,-60,60) \GCirc(22.5,28){3}{0.5}}}
\def\SToprVBblobsh(#1,#2,#3,#4){\spicb{#1(30,15)(15,-120,120)%
 #2(30,15)(15,120,240)%
 #3(15,15)(15,60,300) #4(15,15)(15,-60,60) \GCirc(22.5,28){3}{0.5}}}
\def\ToptrFptblob(#1,#2,#3,#4){\picb{#1(4,4)(15,15) #2(15,15)(26,4)%
 #3(15,15)(26,26) #4(4,26)(15,15) \GCirc(15,15){3}{0} }}

\def\SDgeneric(#1,#2,#3){\!\pic{#1(0,15)(15,15) #2(2,7.5)(15,15)%
 #3(2,22.5)(15,15) \GCirc(15,15){5}{0.75}%
 \Text(12,8)[c]{$\scriptstyle \ddots$}%
 \Text(8,2)[t]{{$\scriptstyle n$}}}\!} 
\def\SDoneLHS(#1){\pic{#1(0,15)(10,15) \GCirc(15,15){5}{0.5}}}
\def\SDoneA(#1,#2,#3){\picb{#1(0,15)(7.5,15) #2(22.5,15)(15,0,180)%
 #3(22.5,15)(15,180,360)}}
\def\SDoneB(#1,#2,#3,#4){\picb{#1(0,15)(7.5,15) #4(7.5,15)(37.5,15)%
 #2(22.5,15)(15,0,180) #3(22.5,15)(15,180,360) \GCirc(37.5,15){3}{0.5}}}
\def\SSDoneA(#1,#2,#3){\spicb{#1(0,15)(7.5,15) #2(22.5,15)(15,0,180)%
 #3(22.5,15)(15,180,360)}}
\def\SSDoneB(#1,#2,#3,#4){\spicb{#1(0,15)(7.5,15) #4(7.5,15)(37.5,15)%
 #2(22.5,15)(15,0,180) #3(22.5,15)(15,180,360) \GCirc(37.5,15){3}{0.5}}}
\def\SDtwoLHS(#1){\pic{#1(0,15)(10,15) #1(20,15)(30,15)%
 \GCirc(15,15){5}{0.5}}}
\def\SDtwoA(#1,#2,#3){\picb{#1(0,15)(7.5,15) #2(22.5,15)(15,0,180)%
 #3(22.5,15)(15,180,360) #1(37.5,15)(45,15) \GCirc(37.5,15){3}{0.5}}}
\def\SDtwoB(#1,#2,#3,#4){\!\picb{#1(0,0.5)(22.5,0)%
 #2(45,0.5)(22.5,0) #3(22.5,15)(15,-90,90) #4(22.5,15)(15,90,270)}}
\def\SDtwoC(#1,#2,#3,#4,#5){\picb{#1(0,15)(7.5,15) #1(37.5,15)(45,15)%
 #2(22.5,15)(15,0,90) #3(22.5,15)(15,90,180) #4(22.5,15)(15,180,360)%
 #5(7.5,30)(15,270,360) \GCirc(22.5,30){3}{0.5} \GCirc(37.5,15){3}{0.5}}}
\def\SDtwoD(#1,#2,#3,#4){\picb{#1(0,15)(7.5,15) #1(37.5,15)(45,15)%
 #4(7.5,15)(37.5,15) #2(22.5,15)(15,0,180) #3(22.5,15)(15,180,360)%
 \GCirc(37.5,15){3}{0.5}}}
\def\SSDtwoA(#1,#2,#3){\spicb{#1(0,15)(7.5,15) #2(22.5,15)(15,0,180)%
 #3(22.5,15)(15,180,360) #1(37.5,15)(45,15) \GCirc(37.5,15){3}{0.5}}}
\def\SSDtwoB(#1,#2,#3,#4){\!\spicb{#1(0,0.5)(22.5,0)%
 #2(45,0.5)(22.5,0) #3(22.5,15)(15,-90,90) #4(22.5,15)(15,90,270)}}
\def\SSDtwoC(#1,#2,#3,#4,#5){\spicb{#1(0,15)(7.5,15) #1(37.5,15)(45,15)%
 #2(22.5,15)(15,0,90) #3(22.5,15)(15,90,180) #4(22.5,15)(15,180,360)%
 #5(7.5,30)(15,270,360) \GCirc(22.5,30){3}{0.5} \GCirc(37.5,15){3}{0.5}}}
\def\SSDtwoD(#1,#2,#3,#4){\spicb{#1(0,15)(7.5,15) #1(37.5,15)(45,15)%
 #4(7.5,15)(37.5,15) #2(22.5,15)(15,0,180) #3(22.5,15)(15,180,360)%
 \GCirc(37.5,15){3}{0.5}}}
\def\SDthreeLHS(#1,#2,#3){\!\pic{#1(15,30)(15,15) #2(2,7.5)(15,15)%
 #3(28,7.5)(15,15) \GCirc(15,15){5}{0.5}%
 \Text(0,10)[b]{{$\scriptstyle 1$}} \Text(30,10)[b]{{$\scriptstyle 2$}}%
\Text(19,34)[t]{{$\scriptstyle 3$}}}\!} 
\def\SDthree(#1,#2,#3){\!\pic{#1(12,27)(12,12) #2(-1,4.5)(12,12)%
 #3(25,4.5)(12,12)}\!}
\def\SSDthree(#1,#2,#3){\!\spic{#1(15,30)(15,15) #2(2,7.5)(15,15)%
 #3(28,7.5)(15,15)}\!}
\def\SSDthreeA(#1,#2,#3,#4,#5,#6){\!\spicb{#1(22.5,30)(22.5,40)%
 #2(-4.5,0)(9.5,7.5)%
 #3(48.5,0)(35.5,7.5) #6(22.5,15)(15,210,330) #5(22.5,15)(15,90,210)%
 #4(22.5,15)(15,-30,90) \GCirc(22.5,30){3}{0.5} \GCirc(35.5,7.5){3}{0.5}}\!}
\def\SSDthreeB(#1,#2,#3,#4,#5){\!\spicb{#1(22.5,40)(35.5,22.5)%
 #2(-4.5,0)(9.5,7.5)%
 #3(48.5,0)(35.5,22.5) #4(22.5,15)(15,45,225) #5(22.5,15)(15,-135,45)%
 \GCirc(35.5,22.5){3}{0.5}}\!}
\def\SSDthreeC(#1,#2,#3,#4,#5){\!\spicb{#1(22.5,30)(22.5,40)%
 #2(-4.5,0.5)(22.5,0)%
 #3(48.5,0.5)(22.5,0) #4(22.5,15)(15,-90,90) #5(22.5,15)(15,90,270)%
 \GCirc(22.5,30){3}{0.5}}\!}
\def\SSDthreeD(#1,#2,#3,#4,#5,#6,#7){\!\spicb{#1(22.5,30)(22.5,40)%
 #2(-4.5,0)(9.5,7.5)%
 #3(48.5,0)(35.5,7.5) #6(22.5,15)(15,210,330) #5(22.5,15)(15,90,210)%
 #4(22.5,15)(15,-30,90) #7(22.5,-5)(18,52,131) \GCirc(22.5,30){3}{0.5}%
 \GCirc(35.5,7.5){3}{0.5}}\!}
\def\SSDthreeE(#1,#2,#3,#4,#5,#6,#7,#8){\!\spicb{#1(22.5,30)(22.5,40)%
 #2(-4.5,0)(9.5,7.5)%
 #3(48.5,0)(35.5,7.5) #7(22.5,15)(15,210,330) #5(22.5,15)(15,90,150)%
 #4(22.5,15)(15,-30,90)  #6(22.5,15)(15,150,210) #8(7,15)(8,-80,80)%
 \GCirc(22.5,30){3}{0.5} \GCirc(35.5,7.5){3}{0.5} \GCirc(9.5,22.5){3}{0.5}}\!}
\def\SSDthreeF(#1,#2,#3,#4,#5,#6,#7,#8){\!\spicb{#1(22.5,30)(22.5,40)%
 #2(-4.5,0)(9.5,7.5) #3(48.5,0)(35.5,7.5)%
 #7(22.5,15)(15,210,330) #6(22.5,15)(15,90,210)%
 #4(22.5,15)(15,-30,30) #5(22.5,15)(15,30,90) #8(9.5,7.5)(35.5,22.5)%
 \GCirc(22.5,30){3}{0.5} \GCirc(35.5,22.5){3}{0.5} \GCirc(35.5,7.5){3}{0.5}}\!}
\def\SSDthreeG(#1,#2,#3,#4,#5,#6,#7){\!\spicb{#1(22.5,40)(35.5,22.5)%
 #2(-4.5,0)(9.5,7.5)%
 #3(48.5,0)(35.5,22.5) #4(22.5,15)(15,45,225) #5(22.5,15)(15,-135,-90)%
 #6(22.5,15)(15,-90,45)%
 #7(7,15)(8,-80,80) \GCirc(35.5,22.5){3}{0.5} \GCirc(9.5,22.5){3}{0.5}}\!}
\def\SSDthreeH(#1,#2,#3,#4,#5,#6){\!\spicb{#1(22.5,40)(35.5,22.5)%
 #2(-4.5,0)(9.5,7.5)%
 #3(48.5,0)(35.5,22.5) #4(22.5,15)(15,45,225) #5(22.5,15)(15,-135,45)%
 #6(9.5,7.5)(35.5,22.5) \GCirc(35.5,22.5){3}{0.5}}\!}
\def\SDfourLHS(#1,#2,#3,#4){\!\pic{#1(4,4)(15,15) #2(15,15)(26,4)%
 #3(15,15)(26,26) #4(4,26)(15,15) \GCirc(15,15){5}{0.5}%
 \Text(0,2)[b]{{$\scriptstyle 1$}}\Text(30,2)[b]{{$\scriptstyle 2$}}%
 \Text(30,28)[t]{{$\scriptstyle 3$}}\Text(0,28)[t]{{$\scriptstyle 4$}}%
}\!}
\def\SDfour(#1,#2,#3,#4){\!\pic{#1(4,4)(15,15) #2(15,15)(26,4)%
 #3(15,15)(26,26) #4(4,26)(15,15)}\!}
\def\SSDfour(#1,#2,#3,#4){\!\spic{#1(4,4)(15,15) #2(15,15)(26,4)%
 #3(15,15)(26,26) #4(4,26)(15,15)}\!}
\def\SSDfourA(#1,#2,#3,#4,#5,#6,#7,#8){\!\spicb{#1(0,-5)(12,4.5)%
 #2(45,-5)(33,4.5) #3(45,35)(33,25.5) #4(0,35)(12,25.5)%
 #5(22.5,15)(15,-135,-45) #6(22.5,15)(15,-45,45) #7(22.5,15)(15,45,135)%
 #8(22.5,15)(15,135,-135) \GCirc(33,4.5){3}{0.5}%
 \GCirc(33,25.5){3}{0.5} \GCirc(12,25.5){3}{0.5}}\!}
\def\SSDfourB(#1,#2,#3,#4,#5,#6,#7){\!\spicb{#1(0,-5)(12,4.5)%
 #2(45,-5)(33,4.5) #3(45,35)(22.5,30) #4(0,35)(22.5,30)%
 #5(22.5,15)(15,-135,-45) #6(22.5,15)(15,-45,90) #7(22.5,15)(15,90,-135)%
 \GCirc(33,4.5){3}{0.5} \GCirc(22.5,30){3}{0.5}}\!}
\def\SSDfourC(#1,#2,#3,#4,#5,#6){\!\spicb{#1(0,-5)(12,4.5)%
 #2(45,15)(33,25.5) #3(45,35)(33,25.5) #4(22.5,35)(33,25.5)%
 #5(22.5,15)(15,-135,45) #6(22.5,15)(15,45,-135) \GCirc(33,25.5){3}{0.5}}\!}
\def\SSDfourD(#1,#2,#3,#4,#5,#6,#7){\!\spicb{#1(0,-5)(22.5,0)%
 #2(45,-5)(22.5,0) #3(45,35)(33,25.5) #4(0,35)(12,25.5)%
 #5(22.5,15)(15,-90,45) #6(22.5,15)(15,45,135) #7(22.5,15)(15,135,-90)%
 \GCirc(33,25.5){3}{0.5} \GCirc(12,25.5){3}{0.5}}\!}
\def\SSDfourE(#1,#2,#3,#4,#5,#6){\!\spicb{#1(0,-5)(22.5,0)%
 #2(45,-5)(22.5,0) #3(45,35)(22.5,30) #4(0,35)(22.5,30)%
 #5(22.5,15)(15,-90,90) #6(22.5,15)(15,90,-90) \GCirc(22.5,30){3}{0.5}}\!}
\def\SDfiveLHS(#1,#2,#3,#4,#5){\!\pic{#1(15,15)(15,29) #2(15,15)(2.5,21)%
 #3(15,15)(7.5,3) #4(15,15)(22.5,3)  #5(15,15)(27.5,21)%
 \GCirc(15,15){5}{0.5}}\!}

\def\FullProp(#1){\pic{#1(5,15)(25,15) \Text(15,18)[b]{{\tiny (full)}}}}
\def\Prop(#1){\pic{#1(5,15)(25,15)}}
\def\PropPiProp(#1){\picb{#1(0,15)(15,15) #1(25,15)(40,15)
 \GCirc(20,15){5}{0.75} \Text(20,15)[c]{$\textstyle \pi$}}}
\def\PropFreeProp(#1){\picb{#1(0,15)(17.5,15) #1(22.5,15)(40,15)
 \Line(17.5,10)(22.5,20) \GCirc(20,15){2}{0} }}
\def\SPropPiProp(#1){\!\pic{#1(5,15)(15,15) #1(25,15)(35,15)
 \GCirc(20,15){5}{0.75} \Text(20,15)[c]{$\textstyle \pi$}}}
\def\SPropFreeProp(#1){\!\pic{#1(5,15)(17.5,15) #1(22.5,15)(35,15)
 \Line(17.5,10)(22.5,20) \GCirc(20,15){2}{0} }}
\def\FProp(#1){\picb{#1(0,15)(17.5,15) #1(27.5,15)(45,15)
 \GCirc(22.5,15){5}{0.75} }}
\def\FFProp(#1){\picb{#1(0,15)(17.5,15)}}
\def\PropPiPropPiProp(#1){\picc{#1(0,15)(15,15) #1(25,15)(35,15)%
 #1(45,15)(60,15) \GCirc(20,15){5}{0.75} \Text(20,15)[c]{$\textstyle \pi$}%
 \GCirc(40,15){5}{0.75} \Text(40,15)[c]{$\textstyle \pi$}}}
\def\SDvacO(#1,#2){\pic{#1(15,15)(15,-90,270)%
 \Line(13,-5)(17,5) \GCirc(15,0){2}{0} \Text(15,26)[t]{{\tiny #2}}}}
\def\SSDvacO(#1,#2){\spic{#1(15,15)(15,-90,270)%
 \Line(13,-5)(17,5) \GCirc(15,0){2}{0} \Text(15,26)[t]{{\tiny #2}}}}
\def\SDvacOpi(#1){\pic{#1(15,15)(15,91,89)%
 \GCirc(15,30){5}{0.75} \Text(15,30)[c]{$\textstyle \pi$}}}
\def\SSDvacOpi(#1){\spic{#1(15,15)(15,91,89)%
 \GCirc(15,30){5}{0.75} \Text(10.5,21)[c]{$\scriptstyle \pi$}}}
\def\ToprVMa(#1,#2,#3,#4,#5,#6){\spic{#3(15,15)(15,-30,90)%
 #1(15,15)(15,90,210)%
 #2(15,15)(15,210,330) #5(2,7.5)(15,15) #6(15,15)(15,30) #4(28,7.5)(15,15)%
 \GCirc(15,30){3}{0.5} \GCirc(28,7.5){3}{0.5}}}
\def\SToprVMa(#1,#2,#3,#4,#5,#6){\spic{#3(15,15)(15,-30,90)%
 #1(15,15)(15,90,210)%
 #2(15,15)(15,210,330) #5(2,7.5)(15,15) #6(15,15)(15,30) #4(28,7.5)(15,15)%
 \GCirc(15,30){3}{0.5} \GCirc(28,7.5){3}{0.5}}}
\def\ToprVMb(#1,#2,#3,#4,#5,#6,#7){\pic{#3(15,15)(15,-30,90)%
 #1(15,15)(15,90,210) #2(15,15)(15,210,330) #5(2,7.5)(15,15)%
 #6(15,15)(15,30) #4(28,7.5)(15,15)%
 \GCirc(15,30){3}{0.5} \Text(18,29)[bl]{#7}}}
\def\SToprVMb(#1,#2,#3,#4,#5,#6,#7){\spic{#3(15,15)(15,-30,90)%
 #1(15,15)(15,90,210) #2(15,15)(15,210,330) #5(2,7.5)(15,15)%
 #6(15,15)(15,30) #4(28,7.5)(15,15)%
 \GCirc(15,30){3}{0.5} \Text(18,29)[bl]{#7}}}
\def\ToprVVb(#1,#2,#3,#4,#5){\!\!\picb{#2(26.25,15)(15,256,76)%
 #3(30,30)(15,30) #1(18.75,15)(15,104,284) #4(15,30)(22.5,0)%
 #5(30,30)(22.5,0) \GCirc(22.5,0){3}{0.5}}\!\!}
\def\SToprVVb(#1,#2,#3,#4,#5){\!\!\spicb{#2(26.25,15)(15,256,76)%
 #3(30,30)(15,30) #1(18.75,15)(15,104,284) #4(15,30)(22.5,0)%
 #5(30,30)(22.5,0) \GCirc(22.5,0){3}{0.5}}\!\!}
\def\ToprVVc(#1,#2,#3,#4,#5,#6){\!\!\picb{#2(26.25,15)(15,256,76)%
 #3(30,30)(15,30) #1(18.75,15)(15,104,284) #4(15,30)(22.5,0)%
 #5(30,30)(22.5,0) \GCirc(22.5,0){3}{0.5} \Text(26,-1)[tl]{#6}}\!\!}
\def\TopfVXa(#1,#2,#3,#4,#5,#6,#7,#8,#9){\picb{#1(15,15)(15,90,270)%
 #2(30,15)(15,-90,90) #4(30,30)(15,30) #3(15,0)(30,0) #6(15,0)(15,15)%
 #5(15,15)(30,30) #8(15,30)(20,25) #8(25,20)(30,15) #7(30,15)(30,0)%
 #9(15,15)(30,15) \GCirc(15,0){3}{0.5} \GCirc(30,0){3}{0.5}%
 \GCirc(30,30){3}{0.5}}}
\def\TopfVHa(#1,#2,#3,#4,#5,#6,#7,#8,#9){\picb{#1(15,15)(15,90,270)%
 #2(30,15)(15,-90,90) #4(30,30)(15,30) #3(15,0)(30,0) #6(15,0)(15,15)%
 #5(15,15)(15,30) #8(30,30)(30,15) #7(30,15)(30,0) #9(15,15)(30,15)%
 \GCirc(30,0){3}{0.5} \GCirc(30,30){3}{0.5}}}
\def\TopfVHb(#1,#2,#3,#4,#5,#6,#7,#8,#9){\picb{#1(15,15)(15,90,270)%
 #2(30,15)(15,-90,90) #4(30,30)(15,30) #3(15,0)(30,0) #6(15,0)(15,15)%
 #5(15,15)(15,30) #8(30,30)(30,15) #7(30,15)(30,0) #9(15,15)(30,15)%
 \GCirc(15,0){3}{0.5} \GCirc(30,0){3}{0.5} \GCirc(30,30){3}{0.5}}}
\def\TopfVWa(#1,#2,#3,#4,#5,#6,#7,#8){\pic{#1(15,15)(15,90,180)%
 #3(15,15)(15,180,270) #2(15,15)(15,270,360) #4(15,15)(15,0,90)%
 #5(15,15)(15,30) #7(15,15)(15,0) #6(0,15)(15,15) #8(30,15)(15,15)
 \GCirc(15,0){3}{0.5} \GCirc(30,15){3}{0.5} \GCirc(15,30){3}{0.5}}\,}
\def\TopfVWb(#1,#2,#3,#4,#5,#6,#7,#8){\pic{#1(15,15)(15,90,180)%
 #3(15,15)(15,180,270) #2(15,15)(15,270,360) #4(15,15)(15,0,90)%
 #5(15,15)(15,30) #7(15,15)(15,0) #6(0,15)(15,15) #8(30,15)(15,15)
 \GCirc(15,15){3}{0.5} \GCirc(30,15){3}{0.5}}\,}
\def\TopfVWc(#1,#2,#3,#4,#5,#6,#7,#8){\pic{#1(15,15)(15,90,180)%
 #3(15,15)(15,180,270) #2(15,15)(15,270,360) #4(15,15)(15,0,90)%
 #5(15,15)(15,30) #7(15,15)(15,0) #6(0,15)(15,15) #8(30,15)(15,15)
 \GCirc(30,15){3}{0.5} \GCirc(15,30){3}{0.5}}\,}
\def\TopfVVa(#1,#2,#3,#4,#5,#6,#7,#8){\!\!\picb{#2(26.25,15)(15,256,346)%
 #3(26.25,15)(15,-14,76) #4(30,30)(15,30) #1(18.75,15)(15,104,284)%
 #7(22.5,0)(15,30) #6(30,30)(26.25,15) #8(26.25,15)(22.5,0)%
 #5(26.25,15)(39.8,11.4) \GCirc(26.25,15){3}{0.5}%
 \GCirc(39.8,11.4){3}{0.5}}\!}
\def\TopfVVb(#1,#2,#3,#4,#5,#6,#7,#8){\!\!\picb{#2(26.25,15)(15,256,346)%
 #3(26.25,15)(15,-14,76) #4(30,30)(15,30) #1(18.75,15)(15,104,284)%
 #7(22.5,0)(15,30) #6(30,30)(26.25,15) #8(26.25,15)(22.5,0)%
 #5(26.25,15)(39.8,11.4) \GCirc(15,30){3}{0.5} \GCirc(30,30){3}{0.5}%
 \GCirc(39.8,11.4){3}{0.5}}\!}
\def\TopfVVc(#1,#2,#3,#4,#5,#6,#7,#8){\!\!\picb{#2(26.25,15)(15,256,346)%
 #3(26.25,15)(15,-14,76) #4(30,30)(15,30) #1(18.75,15)(15,104,284)%
 #7(22.5,0)(15,30) #6(30,30)(26.25,15) #8(26.25,15)(22.5,0)%
 #5(26.25,15)(39.8,11.4) \GCirc(22.5,0){3}{0.5}}\!}
\def\TopfVVd(#1,#2,#3,#4,#5,#6,#7,#8){\!\!\picb{#2(26.25,15)(15,256,346)%
 #3(26.25,15)(15,-14,76) #4(30,30)(15,30) #1(18.75,15)(15,104,284)%
 #7(22.5,0)(15,30) #6(30,30)(26.25,15) #8(26.25,15)(22.5,0)%
 #5(26.25,15)(39.8,11.4) \GCirc(15,30){3}{0.5}}\!}
\def\TopfVVe(#1,#2,#3,#4,#5,#6,#7,#8){\!\!\picb{#2(26.25,15)(15,256,346)%
 #3(26.25,15)(15,-14,76) #4(30,30)(15,30) #1(18.75,15)(15,104,284)%
 #7(22.5,0)(15,30) #6(30,30)(26.25,15) #8(26.25,15)(22.5,0)%
 #5(26.25,15)(39.8,11.4) \GCirc(22.5,0){3}{0.5} \GCirc(39.8,11.4){3}{0.5}}\!}
\def\TopfVBa(#1,#2,#3,#4,#5,#6,#7){\picb{#2(30,15)(15,-120,120)%
 #6(30,15)(15,120,180) #5(30,15)(15,180,240) #1(15,15)(15,60,300)%
 #4(15,15)(15,-60,0) #3(15,15)(15,0,60) #7(30,15)(15,15)%
 \GCirc(22.5,2.5){3}{0.5} \GCirc(30,15){3}{0.5}}}
\def\TopfVBb(#1,#2,#3,#4,#5,#6,#7){\picb{#2(30,15)(15,-120,120)%
 #6(30,15)(15,120,180) #5(30,15)(15,180,240) #1(15,15)(15,60,300)%
 #4(15,15)(15,-60,0) #3(15,15)(15,0,60) #7(30,15)(15,15)%
 \GCirc(15,15){3}{0.5} \GCirc(30,15){3}{0.5}}}
\def\TopfVBc(#1,#2,#3,#4,#5,#6,#7){\picb{#2(30,15)(15,-120,120)%
 #6(30,15)(15,120,180) #5(30,15)(15,180,240) #1(15,15)(15,60,300)%
 #4(15,15)(15,-60,0) #3(15,15)(15,0,60) #7(30,15)(15,15)%
 \GCirc(22.5,2.5){3}{0.5}}}
\def\TopfVNa(#1,#2,#3,#4,#5,#6,#7){\picb{#1(15,15)(15,90,270)%
 #2(30,15)(15,-90,90) #4(30,30)(15,30) #3(15,0)(30,0)%
 #5(15,0)(15,30) #6(30,30)(30,0) #7(15,30)(30,0) \GCirc(30,30){3}{0.5}}} 
\def\TopfVNb(#1,#2,#3,#4,#5,#6,#7){\picb{#1(15,15)(15,90,270)%
 #2(30,15)(15,-90,90) #4(30,30)(15,30) #3(15,0)(30,0)%
 #5(15,0)(15,30) #6(30,30)(30,0) #7(15,30)(30,0)%
 \GCirc(15,30){3}{0.5} \GCirc(30,30){3}{0.5}}} 
\def\TopfVUa(#1,#2,#3,#4,#5,#6,#7){\pic{#3(15,15)(15,0,90)%
 #2(15,15)(15,90,180) #4(15,15)(15,180,270) #1(15,15)(15,270,360)%
 #6(0,15)(15,30) #7(15,0)(0,15) #5(30,15)(15,0) \GCirc(15,0){3}{0.5}}}
\def\TopfVTa(#1,#2,#3,#4,#5,#6){\pic{#1(15,15)(15,90,210)%
 #2(15,15)(15,210,330) #3(15,15)(15,-30,90) #4(2,7.5)(15,30)%
 #6(28,7.5)(2,7.5) #5(15,30)(28,7.5) \GCirc(15,30){3}{0.5}}}
\def\SDlatA(#1,#2,#3,#4,#5,#6){\!\!\picb{#2(26.25,15)(15.5,256,76)%
 #4(30,30)(15,30) #1(18.75,15)(15.5,104,284) #5(15,30)(22.5,0)%
 #6(30,30)(22.5,0) #3(14.3,18)(20,-70,33) \GCirc(22.5,0){3}{0.5}}\!\!}
\def\SDlatB(#1,#2,#3,#4,#5,#6,#7){\!\!\picb{#2(26.25,15)(15.5,256,76)%
 #3(22.5,30)(15,30) #4(30,30)(22.5,30) #1(18.75,15)(15.5,104,284)%
 #5(15,30)(22.5,0) #6(30,30)(22.5,0) #7(22.5,0)(22.5,30)%
 \GCirc(22.5,0){3}{0.5}}\!\!}
\def\ToprVVblobshTxt(#1,#2,#3,#4,#5,#6){\!\!\picb{#2(26.25,15)(15,256,76)%
 #3(30,30)(15,30) #1(18.75,15)(15,104,284) #4(15,30)(22.5,0)%
 #5(30,30)(22.5,0) \GCirc(29,28){3}{0.5}%
 \Text(31,28)[bl]{#6}}\!\!}
\def\SToprVVblobshTxt(#1,#2,#3,#4,#5,#6){\!\!\spicb{#2(26.25,15)(15,256,76)%
 #3(30,30)(15,30) #1(18.75,15)(15,104,284) #4(15,30)(22.5,0)%
 #5(30,30)(22.5,0) \GCirc(29,28){3}{0.5}%
 \Text(31,28)[bl]{#6}}\!\!}

\def\STopLA(#1,#2){\spicc{#1(15,15)(15,0,360) #1(45,15)(15,0,360)%
 #2(30,15)(60,15)}}
\def\STopLB(#1,#2){\;\;\spic{#2(0,15)(30,15) #2(7.5,28)(22.5,2)%
 #2(22.5,28)(7.5,2)%
 #1(15,30)(7.5,330,210) #1(2,7.5)(7.5,90,330) #1(28,7.5)(7.5,210,90)}}
\def\STopLa(#1,#2){\spicb{#1(15,15)(15,90,270)%
 #1(30,15)(15,-90,90) #2(30,30)(15,30) #2(15,0)(30,0) #2(22.5,0)(22.5,30)%
 #2(22.5,0)(15,30) #2(22.5,0)(30,30)}}
\def\STopLb(#1,#2){\!\!\spicb{#1(26.25,15)(15.5,256,76)%
 #2(30,30)(15,30) #1(18.75,15)(15.5,104,284) #2(15,30)(22.5,0)%
 #2(30,30)(22.5,0) #1(15,17.8)(19.3,292.8,39.1)}\!\!}
\def\STopLc(#1,#2){\spicb{#1(30,15)(15,-120,120)%
 #1(30,15)(15,120,240) #1(15,15)(15,60,300) #1(15,15)(15,-60,60)%
 #2(22.5,3)(22.5,27)}}

\def\TopoSsh(#1){\pib{#1(0,15)(10,15) #1(12.5,15)(22.5,15)%
 \GCirc(11.25,15){2.0}{0}}}
\def\SPropCircProp(#1,#2){\!\pib{#1(5,15)(15,15) #1(25,15)(35,15)
 \GCirc(20,15){5}{0.75} \Text(20,15)[c]{$\scriptstyle #2$}}}
\def\SPropBoxProp(#1,#2){\!\pib{#1(5,15)(15,15) #1(25,15)(35,15)
 \GBoxc(20,15)(9,9){0.75} \Text(20,15)[c]{$\scriptstyle #2$}}}
\def\TopoS(#1){\SPropCircProp(#1,1)}
\def\ToptSi(#1){\SPropCircProp(#1,2)}
\def\ToptSr(#1){\SPropBoxProp(#1,2)}

\begin{document}

\title{Nonequilibrium Quantum Fields:\\ From Cold Atoms to Cosmology}

\author{J{\"u}rgen Berges\\
Institute for Theoretical Physics\\
Heidelberg University}

\date{}

\begin{titlepage}
\maketitle
\def\thepage{}  

\begin{abstract}
\centerline{Lecture notes of the Les Houches Summer School on}

\centerline{\it Strongly interacting quantum systems out of equilibrium.}
\end{abstract}

\end{titlepage}

\renewcommand{\thepage}{\arabic{page}}

\tableofcontents

\include{ch_introLH}

\include{ch_neqqftLH}

\include{ch_approx2PILH}

\include{ch_clstatLH}

\include{ch_instabLH}

\include{ch_transport_fixedpointsLH}

\end{document}

%% file: ch_introLH.tex
\section{Introduction}
\label{sec:intro}
\setcounter{equation}{0}

\subsection{Units}

We will mostly work with natural units where 
Planck's constant divided by $2\pi$, the speed of light and Boltzmann's constant are set to one, i.e.\ 
\begin{eqnarray}
\db \hbar = c = k_B = 1 \, . 
\end{eqnarray}
As a consequence, for instance, the mass of a particle ($m$) is equal to its rest energy ($mc^2$) and also to its inverse Compton wavelength ($mc/\hbar$). More generally, we have that the units of
\begin{eqnarray}
\db \left[ \mathrm{energy}\right] \, = \, \left[ \mathrm{mass}\right] \, = \, \left[ \mathrm{temperature}\right] \, = \, \left[ \mathrm{length}\right]^{-1} \, = \, \left[ \mathrm{time}\right]^{-1}
\nonumber
\end{eqnarray}
may all be expressed in terms of the basic unit of energy where we take the electronvolt (eV), i.e.~the amount of energy gained by the charge of a single electron moved across an electric potential difference of one volt. Recall that
\begin{eqnarray}
\db \mathrm{GeV} \, = \, 10^3 \, \mathrm{MeV} \, = \, 10^6 \, \mathrm{KeV} \, = \, 10^9 \, \mathrm{eV} \, . 
\nonumber 
\end{eqnarray}
The conversion to other units, such as length, can be obtained from $\hbar c/ \mathrm{MeV} \simeq 197.33 \cdot 10^{-15}\, m \equiv 197.33 \, \mathrm{fm}$ with $c=2.9979 \cdot 10^8 \, m/s$ and $\hbar = 6.5822 \cdot 10^{-22} \, \mathrm{MeV}s$. For later use we have for the conversion to length, time, temperature and mass:
\begin{eqnarray}
 \db \mathrm{MeV}^{-1} & \db = & \db 197.33\, \mathrm{fm} \, ,
 \nonumber\\
 \db \mathrm{MeV}^{-1} & \db = & \db 6.5822 \cdot 10^{-22}\, s \, ,
 \nonumber\\
 \db \mathrm{MeV} & \db = & \db 1.1605 \cdot 10^{10}\, K \, ,
 \nonumber\\
 \db \mathrm{MeV} & \db = & \db 1.7827 \cdot 10^{-27}\, g \, .
 \nonumber
\end{eqnarray}

\subsection{Isolated quantum systems in extreme conditions}

In recent years we have witnessed a dramatic convergence of research on nonequilibrium quantum systems in extreme conditions across traditional lines of specialization. Prominent examples include the evolution of the early universe shortly after a period of strongly accelerated expansion called inflation, the initial stages in collisions of ultrarelativistic nuclei at giant laboratory facilities, as well as table-top experiments with degenerate quantum gases far from equilibrium. Even though the typical energy scales of these systems vastly differ, they can show very similar dynamical properties. Certain characteristic numbers can even be quantitatively the same and one may use this {\it\rr universality} to learn from table-top experiments with ultracold atoms something about the dynamics during the very early stages of our universe.

Here the focus lies on isolated quantum systems, whose dynamics is governed by {\it\rr unitary time evolution}. This sets them apart from other condensed matter systems where the influence of the environment is largely unavoidable. Isolated systems offer the possibility to study fundamental aspects of quantum statistical mechanics, such as {\it\rr nonequilibrium instabilities} at early times and late-time {\it\rr thermalization} from first principles. A quantum many-body system in thermal equilibrium is independent of its history in time and characterized by a few conserved quantities only. Therefore, any thermalization
process starting from a nonequilibrium initial state requires an effective loss of details of the initial conditions at sufficiently long times. 

An effective {\it\rr partial memory loss} of the initial state can already be observed at earlier characteristic stages of the nonequilibrium unitary time evolution. The corresponding {\it\rr prethermalization} is characterized in terms of approximately conserved quantities, which evolve quasi-stationary even though the system is still far from equilibrium. 
An extreme case occurs if the nonequilibrium dynamics becomes {\it\rr self-similar}. This amounts to an enormous reduction of the sensitivity to details of the underlying theory and initial conditions. The time evolution in this self-similar regime is described in terms of universal scaling exponents and scaling functions associated with {\it\rr nonthermal fixed points}, which is similar in spirit to the description of critical phenomena in thermal equilibrium. While the notion of thermal fixed points describing different universality classes is well established for systems in thermal equilibrium, the classification of far-from-equilibrium universality classes is a rapidly progressing research topic.

\begin{figure}[t]
\centerline{
\hspace*{2cm}\epsfig{file=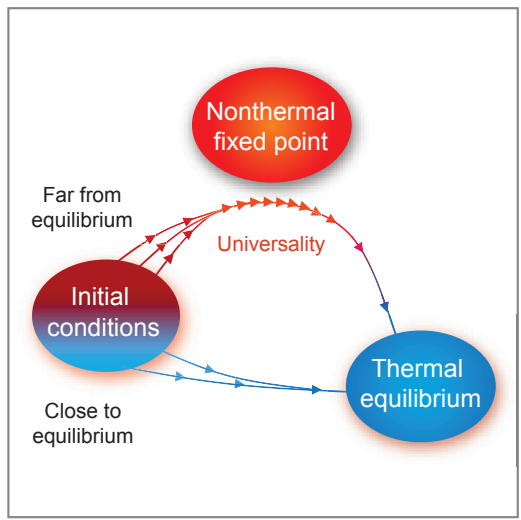,width=10.cm}}
\caption{\label{fig:detour} Schematic evolution towards thermal equilibrium as exemplified in sections~\ref{sec:Earlyuniverseinflation} and~\ref{sec:Bosegas}. Starting far from equilibrium, the system is attracted towards a nonthermal fixed point, thus acquiring its universal properties before relaxing towards thermal equilibrium. (K.~Boguslavski)}
\end{figure}

In the following, we will outline these concepts for the examples of relativistic inflaton dynamics in the early universe in section~\ref{sec:Earlyuniverseinflation}, and for the nonrelativistic dynamics of an ultracold Bose gas in section~\ref{sec:Bosegas}. These sections are meant to provide an overview of the different stages, starting far from equilibrium such that the system approaches a nonthermal fixed point before eventual thermalization. A schematic view is given in figure~\ref{fig:detour}. 

Introducing the basic ingredients for nonequilibrium quantum field theory in more detail in section \ref{sec:funcmethods}, we analyze the notion of thermalization in isolated quantum systems using functional integral techniques in section~\ref{sec:2PI}. Section~\ref{sec:clstat} explains to what extent nonequilibrium quantum field theory can be mapped onto a classical-statistical field theory problem, which provides powerful simulation techniques. Section~\ref{sec:instab} is devoted to nonequilibrium instabilities and the subsequent evolution towards nonthermal fixed points is explained in section~\ref{sec:transport}.

\subsection{Heating the universe after inflation}
\label{sec:Earlyuniverseinflation}

{\bf\emph{Inflation:}} It is one of the most striking developments in our understanding of the universe, that its properties at smallest length scales described by quantum field theory are tightly connected to its properties at largest scales up to the size of the observable universe today. In order to discuss some aspects of this, which will be relevant for these lectures, we recall that the universe is expanding with cosmological time~$t$. The change is described by the Hubble parameter
\begin{equation}
\db H \, = \, \frac{\dot{a}(t)}{a(t)} \, ,
\end{equation}   
which is defined in terms of the scale factor $a(t)$ for a homogeneous and isotropic universe and the dot denotes the time derivative. The scale factor relates the proper distance between a pair of objects such that space itself is expanding with time. At the present moment $t_0$ we take $a(t_0) = 1$, where roughly $t_0 \simeq 1/H_0 \simeq 14\, \mathrm{Gyr} \simeq 4.4 \cdot 10^{17}s$. 

The expansion has not been the same at all times.
Inflation denotes a period at very early times, where the expansion is strongly accelerated with $\ddot{a} > 0$. During this period a tiny region of space expands to a huge size. As a consequence, spatial curvature is decreased and the universe acquires a practically {\it\rr flat geometry}. Furthermore, the accelerated expansion becomes so fast that points on the sky which are not in causal contact today were indeed correlated before inflation. This explains why we can see cosmic microwave radiation with practically the same temperature from all directions of the sky, thus providing a solution of this so-called {\it\rr horizon problem}. Most strikingly, quantum fluctuations stretched to macroscopic sizes by the expansion can explain the {\it\rr origin of initial perturbations} in the matter density. The gravitational instability of these initial perturbations then leads to the observed large-scale structure formation into clusters and superclusters of galaxies along with the fluctuations of the cosmic microwave background radiation, all in remarkable agreement with observations. 

Knowledge about the amplitude of the perturbations can give important information about the energy scale of inflation. A typical characteristic energy density may be around 
$\epsilon_{\rm inflation} \sim ( 10^{15} - 10^{16} \mathrm{GeV} )^4$, which is reminiscent of the grand unification scale above which the electroweak and strong forces may become similar in strength. However, many different scenarios still exist and in the following we will concentrate on some generic aspects. The energy density at a given time, $\epsilon$, determines the Hubble parameter by the Friedmann equation
\begin{equation}
\db H^2 \, = \, \frac{8 \pi G}{3}\, \epsilon \, .
\label{eq:Friedmann}
\end{equation}
Here $G$ denotes Newton's gravitational constant and we define the reduced Planck mass $M_P \equiv (8 \pi G)^{-1/2} \simeq 2.435 \cdot 10^{18} \mathrm{GeV}$. 
 
In a simplest model of inflation, the dominant contribution to the energy density at the time is described in terms of a spatially homogeneous scalar field $\phi(t)$, which can be decomposed as
\begin{equation}
\db \epsilon_\phi \, = \, \frac{1}{2}\, \dot{\phi}^2 + V(\phi) 
\end{equation}
into a kinetic energy part $\dot{\phi}^2/2$ and a potential energy part $V(\phi)$. 
Likewise, the pressure is given by the difference
\begin{equation}
\db P_\phi \, = \, \frac{1}{2}\, \dot{\phi}^2 - V(\phi) \, .
\end{equation}
As a consequence, a time-independent field has {\it\rr negative pressure}: $P_\phi = -\epsilon_\phi$ with {\it\rr constant energy density}. According to (\ref{eq:Friedmann}), in this case $H = \mathrm{const}$ such that
\begin{equation}
{\db \frac{\dot{a}}{a} \, = \, \mathrm{const}} \quad, \mbox{i.e.} \quad 
{\db a(t) \, \sim \, e^{H t} }\, .
\label{eq:inflation}
\end{equation} 
To be consistent with the observed properties of the cosmic microwave radiation, one may need about 60 e-folds of growth before the end of inflation. This requires the specification of a suitable potential $V(\phi)$ along with initial conditions that allow for such dynamics. It should be stressed, however, that there is a large amount of models of inflation with different degrees of freedom and levels of sophistication. Since this will not be relevant for our purposes, we continue for the moment with our simple single-field model.\\

\noindent
{\bf\emph{Coherent field evolution:}} Generically, the previous matter (with pressure $P_m = 0$) and radiation ($P_r = \epsilon_r/3$ describing relativistic particles) content of the universe is diluted away during inflation according to
\begin{equation}
\db \epsilon_m \, \sim \, \frac{1}{a^3} \qquad , \qquad \epsilon_r \, \sim \, \frac{1}{a^4} \, . 
\end{equation}
The suppression factor for the radiation can be understood from 
the fact that its energy density $\epsilon_r \sim n_r \omega_r$ is given by the number density scaling with inverse volume as $n_r \sim 1/a^3$ times the mean energy per particle being red-shifted as $\omega_r \sim 1/a$, since the wavelength is stretched by the expansion.  As a consequence of the dramatic dilution, after inflation all the energy is stored in the large coherent field amplitude of the inflaton. Inflation ends once the time-dependence of the initially slowly moving field $\phi(t)$ can no longer be neglected. In this case, there is no constant energy density dominating and the solution (\ref{eq:inflation}) is no longer valid.   

To discuss the subsequent time evolution, we first note that the 
{\it\rr classical equation of motion} for the scalar field is given by 
\begin{equation}
\db \ddot{\phi} + 3 H \dot{\phi} + \frac{\rmd V}{\rmd \phi} \, = \, 0 \, .
\label{eq:classicalH}
\end{equation}
This may be seen as the equation of motion for a relativistic scalar field with a ``friction'' term $3 H \dot{\phi}$ due to the expansion of the universe. For the following, we also specify to
a quartic potential of the form
\begin{equation}
\db V(\phi) \, = \, \frac{1}{2}\, m^2 \phi^2 + \frac{\lambda}{4!}\, \phi^4 \, .
\label{eq:potential}
\end{equation}
Furthermore, we note that for relativistic particles dominating the energy density the scale factor evolves as
\begin{equation}
\db a(t) \, \sim \, t^{1/2} \quad , \qquad H \sim \frac{1}{t} \, , 
\label{eq:radiationdom}
\end{equation}
in agreement with $H^2 \sim \epsilon_r \sim 1/a^4$ according to (\ref{eq:Friedmann}). In this case the equation of motion for the field can be rewritten in a form that is very useful for later analysis: We introduce {\it\rr conformal time} $t_c$ and a {\it\rr field rescaling} as
\begin{equation}
\db t_c \, = \, \int \frac{\rmd t}{a} \quad , \qquad \phi_c \, = \, a \, \phi \, ,
\label{eq:arad}
\end{equation}
where the integration boundaries run from some initial time to the final time (today).
Then the field equation (\ref{eq:classicalH}) for the potential (\ref{eq:potential}) becomes
\begin{equation}
\db \phi_c^{\prime\prime} + \left(m^2 a^2-\frac{a^{\prime\prime}}{a}\right) \phi_c + \frac{\lambda}{6} \, \phi_c^3 \, = \, 0 \, ,
\label{eq:eomtau}
\end{equation}
where primes denote $t_c$-derivatives. 
For radiation domination we have $a(t) \sim t^{1/2}$ and with (\ref{eq:arad}) that $t_c \sim 2\, t^{1/2}$. Consequently, $a \sim t_c$
leads to $a^{\prime\prime}=0$. From (\ref{eq:eomtau}) we observe that for massless fields, where $m^2 = 0$, the dynamics in expanding space is conformally equivalent to dynamics in Minkowski space-time with no expansion in this case, i.e.\
\begin{equation}
\mbox{\framebox{\rr 
$\displaystyle \,\, \phi_c^{\prime\prime} + \frac{\lambda}{6} \, \phi_c^3 \, = \, 0 \,\,$}}
\label{eq:eomtau2}
\end{equation}
For a given non-zero initial field amplitude, this equation describes an oscillating field in a quartic potential. For notational simplicity, in the following we will drop index labels and employ $\phi(t)$ to denote the rescaled field in conformal time.\\

\noindent
{\bf\emph{Nonequilibrium instabilities and preheating:}} As a classical homogeneous field, the solution of (\ref{eq:eomtau2}) would continue oscillating. However, in a {\it\rr quantum field theory} the macroscopic inflaton field amplitude $\phi(t)$ corresponds to the expectation value of a (real) scalar Heisenberg field operator $\Phi (t, \mathbf{x})$: 
\begin{equation}
\db \phi(t) = \langle \Phi (t, \mathbf{x}) \rangle \equiv \Tr\{ \varrho_0\, \Phi (t, \mathbf{x})\} \, ,
\end{equation}
which involves the density operator $\varrho_0$ as will be explained in more detail in section~\ref{sec:funcmethods}. After inflation ends at some time, which we may set to $t = 0$, the initial state for the subsequent evolution is approximately described by a vacuum-like, pure-state density operator $\varrho_0 \equiv \varrho(t = 0)$ with trace $\Tr \varrho_0 = \Tr \varrho_0^2 = 1$. Typical physical normalizations of the inflaton model employ an initially large field amplitude $\phi(t=0) \sim 0.3 M_P$ with a very weak coupling of about $\lambda \sim 10^{-13}$. We note that the fluctuating quantum field $\Phi (t, \mathbf{x})$ explicitly depends on the spatial variable $\mathbf x$, while its expectation value $\phi(t)$ is homogeneous in our case.

It is a central topic of these lectures to show how the transition from a pure state described by a coherent macroscopic field amplitude proceeds towards radiation, thus connecting to the thermal history of the universe at later times. This can involve dramatic far-from-equilibrium phenomena at intermediate times. Indeed, the situation of the classically oscillating macroscopic field $\phi(t)$ turns out to be unstable in a quantum world, which will be shown in detail in section~\ref{sec:instab}. Here we will outline main aspects of the dynamics in order to introduce some important notions of nonequilibrium quantum field theory. Seeded by quantum fluctuations, a {\it\rr parametric resonance instability} occurs. This instability is signaled by an exponential growth of correlation functions, which corresponds to very fast particle production called {\it\rr preheating}. 

While the macroscopic field is described by a one-point function $\langle \Phi (t, \mathbf{x}) \rangle$, particle production can be extracted from {\it\rr two-point correlation functions}. More precisely, one considers the symmetrized correlator $\langle \Phi(t,\mathbf x) \Phi(t^\prime,\mathbf x^\prime) + \Phi(t^\prime,\mathbf x^\prime) \Phi(t,\mathbf x)\rangle$, which is the anti-commutator expectation value of two fields evaluated at equal times $t=t^\prime$. The spatially homogeneous correlation function may be used to define a time-dependent occupation number distribution $f(t,\mathbf p)$ of particle modes with momentum $\mathbf p$ after Fourier transformation:
\begin{equation}
\db \frac{{\rr f(t,\mathbf p)} + 1/2}{\omega(t,\mathbf p)} + (2 \pi)^3 \delta({\mathbf p}) \phi^2(t)
\equiv \frac{1}{2} \int\! \rmd^3x\, e^{-i {\mathbf p} {\mathbf x}}\, \langle \Phi(t,\mathbf x) \Phi(t,0) + \Phi(t,0) \Phi(t,\mathbf x) \rangle 
\label{eq:fphi}
\end{equation}
for a given (in general time-dependent) dispersion relation $\omega(t,\mathbf p)$.
Because of spatial isotropy, the functions $f(t,\mathbf p)$ and $\omega(t,\mathbf p)$ depend on the modulus of momentum, and we will frequently write $f(t,|\mathbf p|)$ etc. Here the term $\sim \phi^2(t)$ coming together with the Dirac \mbox{$\delta$-function} denotes the coherent field part of the correlator at zero momentum. We note that in a finite volume $V$, we have $(2\pi)^3\delta(\mathbf 0) \rightarrow V$ and the scaling of the field term with volume just reflects the presence of a coherent zero mode spreading over the entire volume. For the following, all integrals will be considered to be suitably regularized, say, by some high-momentum cutoff if necessary.

After inflation, all previous particle content is diluted away and the absence of particles corresponds to an initial state with $f(t=0,\mathbf p) = 0$ for the subsequent evolution. Apart from the coherent field zero-mode, in this case only the ``quantum-half'' coming from the vacuum in (\ref{eq:fphi}) contributes. These quantum fluctuations seed the parametric resonance instability, which leads to a decay of the macroscopic field $\phi$ together with an exponentially growing mode occupancy after inflation:
\begin{equation}
\mbox{\framebox{\rr 
$\displaystyle \,\, f(t,Q) \, \sim \, \exp\left(\gamma_Q \, t \right) \,\,$}}
\label{eq:instability}
\end{equation} 
Here, $\gamma_Q$ denotes a real and positive growth rate for some range of momenta around a fast growing characteristic mode $Q$. The far-from-equilibrium dynamics of an instability with exponentially growing modes may be seen as the opposite of what happens during the relaxation of some close-to-equilibrium modes, which decay exponentially. It is important to note that for transient phenomena the presence of instabilities is as common as relaxation is for the late-time approach to thermal equilibrium. 

The exponential growth (\ref{eq:instability}) leads to an unusually large occupancy of modes at the characteristic momentum scale $Q$. Most importantly, this over-occupied system becomes strongly correlated despite the presence of a very weak coupling $\lambda \ll 1$. Qualitatively, this may be understood from a ``mean-field'' approximation for quantum corrections. In this approximation, the quartic interaction term in the potential (\ref{eq:potential}) is replaced in the quantum theory by 
\begin{equation}
\db \frac{\lambda}{4!}\, \Phi^4(t, \mathbf x) \, \rightarrow \, \frac{\lambda}{4!}\, 6 \langle \Phi^2(t, \mathbf x) \rangle\, \Phi^2(t, \mathbf x) \, .
\label{eq:potentialshift}
\end{equation}
Here, $\langle \Phi^2(t, \mathbf x) \rangle$ means the anti-commutator expectation value appearing also in (\ref{eq:fphi}), now evaluated at equal space-time points. Applying proper renormalization conditions as in standard vacuum quantum field theory, we concentrate on the finite part of the term $\lambda \langle \Phi^2(t, \mathbf x) \rangle$ given by
\begin{equation}
\db \lambda \left( \int \frac{\rmd^3p}{(2 \pi)^3} \frac{f(t,\mathbf p)}{\omega(t,\mathbf p)} + \phi^2(t) \right)  \, 
\label{eq:MFshift}
\end{equation}
according to (\ref{eq:fphi}).

Strong correlations may be expected when interaction effects become comparable in size to the contributions from the typical kinetic energies of particles. For our case of particles with characteristic momentum $Q$, this happens at the time $t_*$ when their occupation number $f(t_*,Q)$ grows so high that their contribution to the mean-field shift (\ref{eq:MFshift}) is of the same order than their kinetic term $\sim Q^2$. The latter is characteristic for the relativistic theory, which is second-order in space-time derivatives. Altogether we thus have
\begin{equation}
\db \lambda \int \frac{\rmd^3p}{(2 \pi)^3} \frac{f(t_*,\mathbf p)}{\omega(t_*,\mathbf p)} \, \stackrel{!}{\sim} \, Q^2 \, .  
\end{equation}
Parametrically, we can estimate the integral for a relativistic dispersion $\omega \simeq |\mathbf p|$ as
\begin{equation}
\db \lambda \int^Q \rmd p\, |\mathbf p|^2 \frac{f(t_*,\mathbf p)}{|\mathbf p|} \, \sim \,  \lambda f(t_*,Q)\, Q^2 \, . 
\end{equation}
This is of order $\sim Q^2$ if the occupancy is as large as 
\begin{equation}
\mbox{\framebox{\rr 
$\displaystyle \,\, f(t_*,Q) \sim \frac{1}{\lambda} \,\,$}}
\label{eq:overocc}
\end{equation}
Taking into account (\ref{eq:instability}), such an extreme nonequilibrium condition is expected to occur at the time
\begin{equation}
\db t_* \simeq \frac{1}{\gamma_Q}\, \ln \left(\frac{1}{\lambda}\right) \, 
\end{equation}
after inflation.

The maximally amplified mode turns out to be given by $Q \sim \sqrt{\lambda}\, \phi(t=0)$ and approximately $Q/\gamma_Q \sim \mathcal{O}(10)$. It is striking that the quantum correction in (\ref{eq:potentialshift}) cannot be neglected no matter how small the coupling $\lambda$ is. Though this is a robust feature of the preheating dynamics, we will see in section~\ref{sec:instab} that for a quantitative analysis one has to go beyond a mean-field analysis. In particular, strongly nonlinear quantum corrections have to be taken into account even before $t_*$, which are not included in the mean-field approach. Apart from the primary growth (\ref{eq:instability}) starting early, nonlinear effects lead to an important \mbox{\it\rr secondary amplification} period of fluctuations with enhanced growth rates for higher momentum modes. As a
consequence, a wide range of growing modes lead to a {\it\rr prethermalization} of the equation of state, given by an almost constant ratio of pressure over energy density, at the end of
this early stage while the distribution function itself is still far from equilibrium.\\

\noindent  
{\bf\emph{Nonthermal fixed points and turbulence:}} The exponentially fast transfer of the zero-mode energy into fluctuations during preheating stops around $t_*$, when the energy densities stored in the coherent field and in fluctuations become of the same order. Quantum-statistical corrections far beyond the mean-field approximation govern the subsequent dynamics of the strongly correlated system. If everything is strongly correlated, one might naively expect a fast evolution. However, the opposite turns out to be true and the evolution slows down considerably after $t_*$. In contrast to the earlier instability regime, which exhibits a characteristic scale for the growth of fluctuations, the subsequent evolution is governed by a {\it\rr nonthermal fixed point}. This is an attractor solution with {\it\rr self-similar scaling behavior} of correlation functions.  

A somewhat similar phenomenon is well-known close to thermal equilibrium, where strong correlations near continuous phase transitions lead to scaling behavior and a corresponding critical slowing down of the dynamics. However, the far-from-equilibrium situation after preheating is rather different. The typical momentum of a relativistic system in thermal equilibrium is given by the temperature $T$, and the Bose-Einstein distribution is of order one for momenta $|\mathbf p| \sim T$. Moreover, thermal equilibrium respects {\it\rr detailed balance}, where each process is equilibrated by its reverse process as will be discussed in section~\ref{sec:2PI}. In contrast, the strong overoccupation (\ref{eq:overocc}) of typical momenta $\sim Q$ represents an extreme nonequilibrium distribution of modes, which leads to a net flux of energy and particles across momentum scales, thus violating detailed balance. 

To get a first understanding of this point, we may compare the characteristic energy density $\epsilon \sim \int^Q \rmd^3p/(2\pi)^3\,  |\mathbf p| f(\mathbf p) \sim Q^4/\lambda$ of the far-from-equilibrium state to a Stefan-Boltzmann law $\sim T^4$ in thermal equilibrium. One observes that the nonequilibrium distribution exhibits a smaller characteristic momentum $Q \sim \lambda^{1/4} T$
than the corresponding thermal system for the same energy density. Therefore, we expect a transport of energy from lower to higher momentum scales on the way towards thermalization. Indeed, the underlying mechanism for the slow dynamics is related to the transport of conserved quantities, which leads to power-law behavior as will be discussed in detail in section~\ref{sec:transport}. 

The situation is further complicated in the presence of additional conserved quantities, apart from energy conservation. Though total particle number is a priori not conserved for the relativistic system, an effectively conserved particle number emerges during the transient nonequilibrium evolution. To get a simple illustration of the consequences of both energy and number conservation, one may think of an idealized three-scales problem: We consider particles that are ``injected'' at the characteristic momentum scale $Q$ as a consequence of the decay of the coherent field as described above. One then asks for the particles and the energy transported to a characteristic lower momentum scale $K$ and to a higher scale $\Lambda$ with $K \ll Q \ll \Lambda$. We let $\dot{n}_Q$ denote the injection rate for the number density at the scale $Q$, while $\dot{n}_K$ and $\dot{n}_\lambda$ characterize the corresponding ``ejection'' rates at $K$ and $\Lambda$. Then number and energy conservation in this simple setup correspond to
\begin{equation}
\db \dot{n}_Q = \dot{n}_K + \dot{n}_\lambda \quad , \quad Q \dot{n}_Q = K\dot{n}_K + \Lambda\dot{n}_\Lambda \, .
\end{equation}
These can be solved for $\dot{n}_K$ and $\dot{n}_\lambda$ as
\begin{equation}
\db \dot{n}_K = \frac{\Lambda - Q}{\Lambda - K}\, \dot{n}_Q \simeq \dot{n}_Q \quad , \quad
\db \dot{n}_\Lambda = \frac{Q - K}{\Lambda - K}\, \dot{n}_Q \simeq \frac{Q}{\Lambda}\, \dot{n}_Q \, ,
\label{eq:eKQL}
\end{equation}
where the latter approximate equalities exploit the separation of scales. According to the first equation the particles are transported to lower scales, while the second says $\Lambda \dot{n}_\Lambda \simeq Q \dot{n}_Q$ such that the energy transport occurs towards higher scales.   

\begin{figure}[t]
\centerline{
\epsfig{file=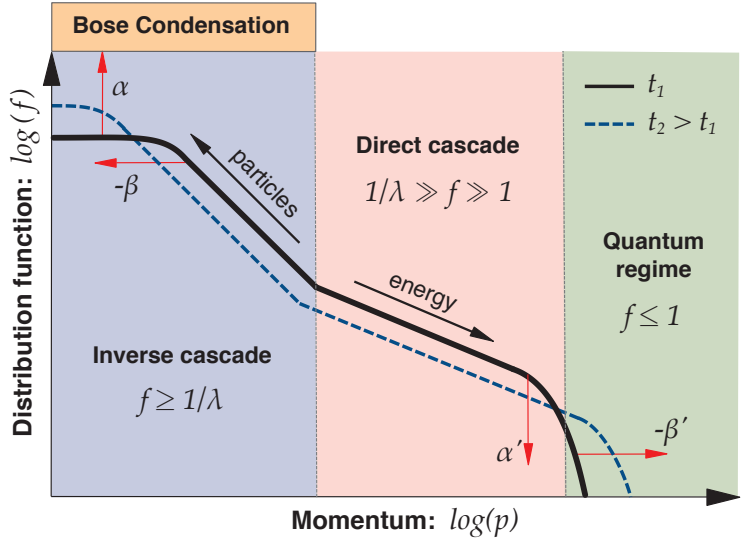,width=10.cm}}
\caption{\label{fig:dual_cascade} Schematic behavior of the occupation number distribution near the nonthermal fixed point as a function of momentum $|\mathbf p|$ for two times $t=t_1$ with $t_2 > t_1$. The scaling exponents $\alpha$ and $\beta$ characterize the rate and direction of the inverse particle cascade, while $\alpha^\prime$ and $\beta^\prime$ describe the direct energy cascade during the self-similar evolution. (A.~Pi\~neiro Orioli)}
\end{figure}

To illustrate this further, figure~\ref{fig:dual_cascade} shows the schematic behavior of the distribution function $f(t,\mathbf p)$ in the vicinity of the nonthermal fixed point. The distribution is given as a function of momentum for two subsequent times $t=t_1$ and $t_2 > t_1$. One observes that the particle transport towards low momenta is part of a {\it\rr dual cascade}, in which energy is also transfered towards higher momenta. It turns out that there is no single scaling solution conserving both energy and particle number in this case. Instead, a dual cascade emerges such that in a given inertial range of momenta only one conservation law
governs the scaling behavior. 

The inverse particle cascade occurs in a highly occupied infrared range of momenta with $|\mathbf p| \ll Q$. In this nonperturbative regime with $f(t,\mathbf p) \gtrsim 1/\lambda$, the effectively conserved particle number receives a dominant contribution at the time-dependent scale $K(t)$, where the integrand for particle number $\sim |\mathbf p|^2 f(t,\mathbf p)$ has its maximum. The cascade exhibits an approximate power-law behavior in an inertial range of momenta. This power-law is similar to the one known from stationary wave turbulence in the presence of external sources or sinks. However, we are dealing with an isolated system. Instead of simple power laws, the transport in isolated systems is described in terms of the more general notion of a {\it\rr self-similar evolution}, where the distribution function acquires a scaling form: 
\begin{equation}
\mbox{\framebox{\rr 
$\displaystyle \,\, f(t,\mathbf p)=t^\alpha\,f_S(t^\beta \mathbf p) \, . \,\,$}}
\label{eq:selfsim}
\end{equation}
Here, all quantities are considered to be dimensionless by use of some suitable momentum scale. The values of the scaling exponents $\alpha$ and $\beta$, as well as the form of the time-independent {\it\rr fixed point distribution} $f_S(t^\beta \mathbf p)$ are universal. More precisely, all models in the same universality class can be related by a multiplicative rescaling of $t$ and $|\mathbf p|$. Quantities which are invariant under this rescaling are universal.  

The exponents for the infrared particle cascade are well described by    
\begin{equation}
\db	\alpha = \beta\,d \quad , \quad \beta=\,\frac{1}{2} \,,
\label{eq:exponentsIPC}
\end{equation}
where $d$ denotes the spatial dimension and here $d=3$. The positive values for $\alpha$ and $\beta$ determine the rate and direction of the particle number transport towards low momenta, since a given characteristic momentum scale $K(t_1)=K_1$ evolves as $K(t)=K_1(t/t_1)^{-\beta}$ with amplitude $f(t,K(t))\sim t^\alpha$ according to (\ref{eq:selfsim}). The fixed relation between $\alpha$ and $\beta$ reflects the conservation of particle number density $n=\int\rmd^d p\,f(t,\mathbf p) \sim t^{\alpha - d \beta}$ in this inertial range by using the self-similarity (\ref{eq:selfsim}). 

The emergence of an effectively conserved particle number for infrared modes can be explained by the dynamical generation of a mass gap for the (massless) relativistic theory, which arises from an intriguing interplay of condensation and medium effects. Indeed, the inverse particle cascade towards the infrared continuously (re-)populates the zero-mode, which leads to a power-law behavior also for the condensate field $\phi(t)$. In particular, if we would start the evolution from overoccupation (\ref{eq:overocc}) without an initial coherent field part ($\phi(t=0) = 0)$, the inverse cascade would lead to the formation of a Bose condensate in this far-from-equilibrium state, which is shown in section~\ref{sec:transport}. 

The particle transport towards low momenta is accompanied by energy transport towards higher momenta. In figure~\ref{fig:dual_cascade} we indicate that in the direct energy cascade regime other scaling exponents $\alpha^\prime$ and $\beta^\prime$ with a different scaling function $f_S^\prime$ are found than for the inverse particle cascade. In the perturbative higher momentum range we have $1/\lambda \gg f(\mathbf p) \gg 1$ and the energy is dominated by the scale $\Lambda(t)$.  For the hard modes with $|\mathbf p| \sim \Lambda$ the negative exponent $\beta= -1/5$ determines the evolution of characteristic momenta and $\alpha=-4/5$ of their occupancy. The latter exponent determines the parametric time $t \sim t_* \lambda^{-5/4}$ at which the occupancies of the hard modes $f(t,\Lambda(t))$, which are order $1/\lambda$ at $t_*$, dropped to become comparable to the ``quantum-half''. At this stage a subsequent approach to a thermal equilibrium distribution with the help of elastic and inelastic scattering processes sets in.

\subsection{Ultracold quantum gases far from equilibrium}
\label{sec:Bosegas}

Experimentally, the investigation of quantum systems in extreme conditions is boosted by the physics of ultracold quantum gases. Because these gases are very cold and dilute, they can be used to obtain clean realizations of microscopic Hamiltonians built from low-energy elementary scattering and interactions with applied fields.  By using optical or atom chip traps, they provide a flexible testbed with tuneable interactions, symmetries and dimensionality, with connections to a wide variety of systems. Setups employing ultracold quantum gases can be largely isolated in contrast to many other condensed matter systems, where significant couplings to the environment are mostly inevitable. This offers the possibility to study fundamental aspects of far-from-equilibrium dynamics and the thermalization process in a highly isolated environment.

We have seen in section \ref{sec:Earlyuniverseinflation} that the dynamics of the early universe after inflation can have different characteristic stages. After a very rapid evolution caused by nonequilibrium instabilities, the dynamics becomes self-similar. The time evolution in this self-similar regime is described in terms of universal scaling exponents and scaling functions. The universality observed opens the exciting possibility to find other systems in the same universality class, which may be easier accessible experimentally. Ultracold quantum gases at nanokelvins can differ in their characteristic energy scale by more than $38$ orders of magnitude from inflationary physics. Nevertheless, they can show very similar dynamical behavior and universal properties can even be quantitatively the same. Therefore, one may learn from experiments with cold atoms aspects about the dynamics during the early stages of our universe.   

We will consider here interacting bosons with s-wave scattering length $a$, and leave the inclusion of fermions to later sections. For a gas of density $n$ the average interatomic distance is $n^{-1/3}$. Together with the scattering length $a$, this can be used to define a dimensionless ``diluteness parameter''
\begin{equation}
\db \zeta = \sqrt{n a^3} \, .
\label{eq:dilute}
\end{equation}
For a typical scattering length of, e.g., $a \simeq 5 \,{\rm nm}$ and bulk density $n \simeq 10^{14} \,{\rm cm}^{-3}$ the diluteness parameter $\zeta \simeq 10^{-3}$ is very small, and in the following we will always assume $\zeta \ll 1$. The density and scattering length can also be used to define a characteristic ``coherence length'', whose inverse is described by the momentum scale 
\begin{equation}
\db Q = \sqrt{16\pi a n} \, .
\label{eq:Q}
\end{equation}

To observe the dynamics near nonthermal fixed points for the interacting Bose gas, an unusually large occupancy of modes at the inverse coherence length scale $Q$ has to be prepared. More precisely, for a weakly coupled gas of average density $n = \int \rmd^3p/(2\pi)^3 f_{\rm nr}(\mathbf p)$ this requires a characteristic mode occupancy as large as
\begin{equation}
\mbox{\framebox{\rr 
$\displaystyle \,\, f_{\rm nr}(Q) \sim \frac{1}{\zeta} \,\,$}} 
\label{eq:overpopulation}
\end{equation}
This represents a far-from-equilibrium distribution of modes, and we note that the diluteness parameter $\zeta$ plays the corresponding role of the coupling in (\ref{eq:overocc}).
Such an extreme condition may be obtained, for instance, from a quench or nonequilibrium instabilities similar to the previously discussed relativistic case.

Most importantly, the system in this overoccupied regime is strongly correlated. These properties may be understood from a Gross-Pitaevskii equation for a nonrelativistic complex Bose field $\chi$: 
\begin{equation}
\db	i\partial_t\chi(t,\mathbf x)= \left( -\frac{\nabla^2}{2m} + g|\chi(t,\mathbf x)|^2 \right)\chi(t,\mathbf x).
\label{eq:gpe}
\end{equation}
Here the coupling $g$ is not dimensionless and determined from the mass $m$ and scattering length as $g = 4\pi a/m$. The total number of particles is given by $\int \rmd^3 x |\chi(t, \mathbf x)|^2$ and is conserved. 

In the mean-field approximation the effect of the interaction term in the Gross-Pitaevskii
equation is a constant energy shift for each particle,
\begin{equation}
\db \Delta E = 2 g \langle |\chi|^2 \rangle = 2 g n = 2 g \int \frac{\rmd^3p}{(2\pi)^3} f_{\rm nr}(\mathbf p) \, , 
\end{equation}
which can be absorbed in a redefinition of the chemical potential. However, we note that for the very high occupancy (\ref{eq:overpopulation}) of the typical momentum $Q$ the shift in energy is not small compared to the relevant kinetic energy $Q^2/2m$, i.e.\ $2 g n \sim Q^2/2m$. Parametrically, this can be directly verified using (\ref{eq:overpopulation}):
\begin{equation}
\db g \int \rmd^3p\, f_{\rm nr}(\mathbf p) \sim g\, Q^3 f_{\rm nr}(Q) \sim g \frac{Q^3}{\zeta} \sim g \frac{Q^3}{m g Q} \sim \frac{Q^2}{m} \, .
\end{equation}
Here we have used that with $a = m g/(4\pi)$ equation (\ref{eq:Q}) implies $Q = 2 \sqrt{m g n}$ and (\ref{eq:dilute}) gives $\zeta = m g Q/(16 \pi^{3/2})$. Most importantly, the energy shift $2gn$ is of the order of the kinetic energy $Q^2/2m$ irrespective of the coupling strength $g$. This already hints at a strongly correlated system, where the dependence on the details of the underlying model parameters is lost.

To study the nonequilibrium evolution of such a system, one may define a time-dependent occupation number distribution $f_{\rm nr}(t,\mathbf p)$ from the equal-time two-point correlation function:
\begin{equation}
\db {\rr f_{\rm nr}(t,\mathbf p)} + \frac{1}{2} + (2 \pi)^3 \delta({\mathbf p}) |\chi_0|^2(t) \equiv \frac{1}{2} \int \rmd^3x e^{-i {\mathbf p} {\mathbf x}} \langle \chi(t,\mathbf x)\chi^*(t,0) + \chi(t,0)\chi^*(t,\mathbf x)\rangle . 
\label{eq:n_def_stat_nonrel}
\end{equation} 
The term $|\chi_0|^2(t)$ coming together with the Dirac $\delta$-function determines the condensate fraction at zero momentum. These definitions are in complete analogy to equation (\ref{eq:fphi}) for the relativistic theory. The only major difference is the appearance of the dispersion in the definition for the relativistic case, which is a consequence of the second-order differential equation (\ref{eq:eomtau2}) while the nonrelativistic equation (\ref{eq:gpe}) is first-order in time.

Since the far-from-equilibrium dynamics of the strongly correlated system is expected to become insensitive to the details of the initial conditions, we may choose the initial condensate fraction to be zero. Moreover, it turns out that in the overoccupied regime, where $f_{\rm nr}(t,Q)$ is much larger than the ``quantum-half'', the quantum-statistical dynamics is essentially classical-statistical such that the quantum and the corresponding classical field theories belong to the same universality class. The classical aspects of nonequilibrium quantum theories will be addressed in section~\ref{sec:clstat}.

\begin{figure}[t]
\centerline{
\epsfig{file=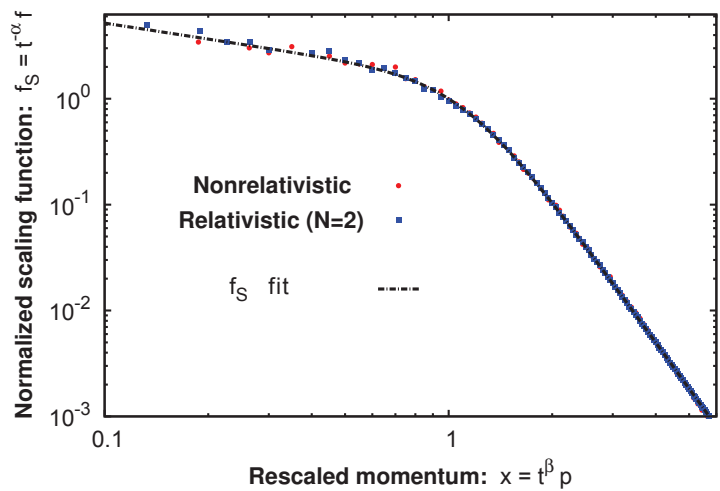,width=10.cm}}
\caption{\label{fig:compare_rel_nonrel} Shown is the normalized fixed point distribution $f_S(t^\beta \mathbf p) = t^{-\alpha} f(t,\mathbf p)$ in the infrared regime of the inverse particle cascade. The simulation results for the relativistic (circles) inflaton model with quartic interaction and the nonrelativistic (squares) Gross-Pitaevskii field theory agree to very good accuracy.}
\end{figure}

Figure~\ref{fig:compare_rel_nonrel} shows a striking example of universality far from equilibrium: Displayed is the normalized fixed point distribution $f_{S,{\rm nr}}(t^\beta \mathbf p) = t^{-\alpha} f_{\rm nr}(t,\mathbf p)$ in the infrared, after evolving from overoccupied initial conditions for the nonrelativistic theory (circles) using the simulation techniques of section \ref{sec:clstat}. The corresponding fixed point distribution can be extracted for relativistic theories, such as discussed in section \ref{sec:Earlyuniverseinflation}, which we can generalize to include $N$-component inflaton models as done in section~\ref{sec:instab}. If the relativistic and the nonrelativistic theories belong to the same universality class, then the normalized fixed point distribution has to agree. That this is indeed the case in the infrared scaling regime is exemplified in figure~\ref{fig:compare_rel_nonrel} for the example of the relativistic $N=2$ component field theory (squares). 

A crucial ingredient for the universality observed is the presence of strong correlations, which can occur even for weakly coupled systems if they are overoccupied. These extreme occupancies of modes can be found in a variety of isolated systems in extreme conditions and the identification of nonequilibrium universality classes represents a crucial step for their understanding. This is of particular relevance also for nonequilibrium gauge theories, such as employed in the description of the initial stages of ultrarelativistic heavy-ion collision experiments. Though we cannot cover this in depth in these lectures, we suggest some further reading at the end of each chapter.

\subsection{Bibliography}

\begin{itemize}
\item Part of the presentation about nonthermal universality classes follows J.~Berges, K.~Boguslavski, S.~Schlichting and R.~Venugopalan, {\it Universality far from equilibrium: From superfluid Bose gases to heavy-ion collisions}, Phys.\ Rev.\ Lett.\ {\bf 114} (2015) 061601, where a version of figure~\ref{fig:detour} appeared on https://journals.aps.org/prl/issues/114/6 to highlight Editors' Suggestions. Figures~\ref{fig:dual_cascade} and \ref{fig:compare_rel_nonrel} as well as part of the discussion about self-similar evolution are taken from A.~Pi\~neiro Orioli, K.~Boguslavski and J.~Berges, {\it Universal self-similar dynamics of relativistic and nonrelativistic field theories near nonthermal fixed points}, http://arxiv.org/abs/1503.02498.
\item Parts of these lectures are based on J.~Berges, {\it Introduction to nonequilibrium quantum field theory,}  AIP Conf.\ Proc.\  {\bf 739} (2005) 3,  http://arxiv.org/abs/hep-ph/0409233. The latter covers the phenomenon of (pre)thermalization and aspects of renormalization in detail, while here we discuss in particular the important notion of self-similar evolution near nonthermal fixed points.  
\item  For overviews in cosmology about aspects of nonequilibrium inflaton dynamics see L.~Kofman, {\it Preheating after inflation,}  Lect.\ Notes Phys.\  {\bf 738} (2008) 55. 
\item The preheating phenomenon was first developed in L.~Kofman, A.~D.~Linde and A.~A.~Starobinsky, {\it Reheating after inflation}, Phys.\ Rev.\ Lett.\  {\bf 73} (1994) 3195. The subsequent direct energy cascade in the perturbative regime was shown in this context in R.~Micha and I.~Tkachev, {\it Relativistic turbulence: A Long way from preheating to equilibrium}, Phys.\ Rev.\ Lett.\  {\bf 90} (2003) 121301. The inverse particle cascade in the nonperturbative regime and the underlying nonthermal fixed point has been found in J.~Berges, A.~Rothkopf and J.~Schmidt, {\it Non-thermal fixed points: Effective weak-coupling for strongly correlated systems far from equilibrium}, Phys.\ Rev.\ Lett.\  {\bf 101} (2008) 041603.   
\item For an overview about nonthermal fixed points in Bose gases, see the contribution by B.~Nowak et al., {\it Non-thermal fixed points: universality, topology, \& turbulence in Bose gases}, http://arxiv.org/abs/1302.1448.
\item For an overview about nonthermal fixed points in heavy-ion collisions, see J.~Berges, K.~Boguslavski, S.~Schlichting and R.~Venugopalan, {\it Universal attractor in a highly occupied non-Abelian plasma}, Phys.\ Rev.\ D {\bf 89} (2014) 114007.
\item A perspective on cold atoms and cosmology with related references may be found in J.~Schmiedmayer and J.~Berges,
  {\it Cold Atom Cosmology},
  Science {\bf 341} (2013) 6151,  1188.
\end{itemize}

%% file: ch_neqqftLH.tex
\section{Nonequilibrium quantum field theory}
\label{sec:funcmethods}
\setcounter{equation}{0}

\subsection{How to describe nonequilibrium quantum fields} 
\label{sec:thermalization}

There are very few ingredients for the description of nonequilibrium quantum fields. {\rr\it Nonequilibrium dynamics typically
requires the specification of an initial state} at some given time $t_0$. 
This may include a density operator  
$\varrho_0 \equiv \varrho(t_0)$ in a mixed ($\Tr \varrho_0^2 < 1$) or pure state
($\Tr \varrho_0^2 = 1$). Nonequilibrium means that $\varrho_0$ does not correspond to a thermal
equilibrium density operator:
$\varrho_0 \not =  \varrho^{\rm (eq)}$ 
with, for instance, $\varrho^{\rm (eq)} \sim e^{-\beta H}$ for the
case of a canonical thermal ensemble with inverse temperature $\beta$ for
given Hamilton operator $H$. 

Completely equivalent to the specification of the initial density 
operator $\varrho_0$ is the knowledge of all initial correlation functions: 
the initial one-point function
$\Tr\{\varrho_0 \Phi(t_0,\bx)\}$, two-point function
$\Tr\{\varrho_0 \Phi(t_0,\bx)\Phi(t_0,\by)\}$,
three-point function etc. To be specific, here
$\Phi(x)$ may denote a scalar Heisenberg field operator depending on time and space, $x=(x^0,\bx)$. Typically, the ``experimental setup'' 
requires only knowledge about a few lowest correlation functions
at the time $t_0$, whereas complicated higher correlation functions
may build up at later times. From the 
behavior of the correlation functions for times $x^0 > t_0$
one can extract the time evolution of all other quantities.

Once the nonequilibrium initial state is 
specified, for {\rr\it closed systems} the time-evolution is 
completely determined by  
the Hamiltonian. Equivalently, for a given classical action $S$
the dynamics can be described in terms of a functional integral. From the latter one obtains 
the {\rr\it effective action $\Gamma$}, which is the generating
functional for all correlation functions of the quantum theory. 
{\rr\it There are no further ingredients involved concerning the dynamics 
than what is known from standard vacuum quantum field theory.}
It should be stressed that during the nonequilibrium time evolution 
there is no loss of information in any strict sense.
The important process of thermalization is a nontrivial question in
a calculation from first principles. Thermal equilibrium keeps no memory about 
the time history except for the values of a few conserved charges.
Equilibrium is time-translation invariant and cannot be reached from
a nonequilibrium evolution on a fundamental level. It is striking 
to observes that the evolution can go very closely 
towards thermal equilibrium without
ever starting again to deviate from it for accessible times.
The observed {\rr\it effective loss of details about
the initial conditions} can mimic very accurately
the irreversible dynamics obtained from effective
descriptions in their range of applicability.

For out-of-equilibrium calculations there are additional 
complications which do not appear in vacuum or thermal 
equilibrium. A major new aspect for approximate descriptions concerns {\rr\it secularity:}
The standard perturbative time evolution suffers from
the presence of spurious, so-called secular terms, which grow with 
time and invalidate the expansion even in the presence of a weak 
coupling. Here it is important  
to note that the very same problem can appear as well for nonperturbative
approximation schemes such as standard $1/N$ expansions beyond leading order, where $N$ denotes the number of field components. To obtain a uniform approximation in time, where an expansion parameter controlling the error at early times is valid also at late times, requires selective summation of an infinite series of contributions. 

If a nonequilibrium initial state is evolved from early to very late times, it is in general crucial that energy is conserved exactly for the given approximation. If thermal equilibrium is approached, then the late-time result is insensitive to the details of the initial conditions and becomes uniquely determined by the energy density and further conserved charges.
This property can be conveniently implemented, if the dynamics is obtained from an effective action by a variational principle. The analogue in classical mechanics is well known: if the 
equations of motion can be derived from the principle of least 
action, then they will not admit any friction term without further approximations. 

Many of these requirements can indeed be achieved using efficient functional integral techniques based on so-called $n$-particle irreducible ($n$PI) effective actions. They often provide a practical means to describe far-from-equilibrium dynamics as well as thermalization from first principles. Closely related are approximations based on identities relating different $n$-point functions such as Dyson-Schwinger equations, or identities between derivatives of correlators described by the functional renormalization group.

Local symmetries as encountered in gauge theories pose even stronger restrictions on possible approximation schemes. Here, the derivation of a sequence of effective theories, which disentangle dynamics on different time and distance scales can be very efficient. In particular, for weakly-coupled theories close to equilibrium at high temperature a wealth of information can be obtained from effective descriptions such as kinetic theory. 

Some aspects of nonequilibrium systems are often 
successfully described using classical-statistical field theory methods. Important examples involve the dynamics of nonequilibrium instabilities such as parametric resonance, or universal properties of dynamic critical phenomena near second-order phase transitions. Classical-statistical field theory can be implemented on a space-time lattice in Minkowski space and simulated on computers. Classical Rayleigh-Jeans divergences and the lack of 
genuine quantum effects limit their use. It is, therefore, important to understand for which time and distance scales classical simulations can adequately reproduce the behavior of the underlying quantum theory of interest. 

Of course, this list of methods is not complete and these notes are restricted to only some of the major field-theoretic techniques and applications. In the following we will introduce the functional integral formulation of nonequilibrium quantum field theory, which provides a very efficient starting point to derive the different approaches that will be discussed in later chapters.

\subsection{Nonequilibrium generating functional}
\label{sec:nonequgenfunc}

We consider first a scalar quantum field theory  
with Heisenberg field operator $\Phi(x)$. Though most of the developments are completely equivalent for relativistic as well as nonrelativistic theories, in the following we will always assume a relativistic setup for definiteness and point out the relevant differences where necessary.

Knowing all correlation functions, which are represented as the trace over time-dependent Heisenberg field operators for given initial density operator $\varrho_0$, completely specifies the quantum system at any time. Therefore, 
all information about nonequilibrium quantum field
theory is contained in the {\rr\it generating functional}
for correlation functions
\bea\db 
Z[J,R;{\rr \varrho_0}] &\db =& {\db 
\Tr\left\{{\rr \varrho_0}\, \textrm{T}_{\C}\, e^{i \left( \int_{x,\C}\!
J(x) \Phi(x)+\frac{1}{2}
\int_{x y,\C}\! \Phi(x) R(x,y)\Phi(y)\right)} \right\} } \, 
\label{eq:definingZneq} 
\eea
for given $\varrho_0$. It is the generalization of the {\it \rr partition function} for nonequilibrium systems in the presence of sources $J(x)$ and $R(x,y)$. Since this is a generating functional for the trace over Heisenberg field operators, the time argument $x^0$ of the field operator $\Phi(x^0,\bx)$ is evaluated along a {\rr\it closed real-time contour 
$\mathcal{C}$} appearing in the integrals, where we write $\int_{x,\C} \equiv \int_{\C} \rmd x^0 \int {\rmd}^d x$ in $d$ spacial dimensions. This will be explained in more detail in section~\ref{sec:functionalintegral} and should be distinguished from computations restricted to scattering matrix elements between asymptotic incoming and outgoing states, which can be formulated without a closed time contour $\mathcal{C}$. Employing the latter is more general, in the sense that vacuum quantum field theory as well as thermal and nonequilibrium systems can be described. The time-path $\mathcal{C}$ starts at time $t_0$ running forward along the real-time axis and then runs back to $t_0$. Graphically, $\mathcal{C}$ is depicted as

\begin{picture}(20,58) (-85,10)
\put(0,25){\line(1,0){200}}
\put(25,0){\line(0,1){50}}
\put(170,45){\line(0,1){15}}
\put(170,45){\line(1,0){15}}
\thicklines
\put(25,30){\vector(1,0){75}}
\put(100,30){\line(1,0){75}}
\put(175,20){\vector(-1,0){75}}
\put(25,20){\line(1,0){75}}
\put(25,30){\line(1,0){150}}
\put(25,20){\line(1,0){150}}
\cCircle[30](175,25){5}[r]
\put(25,20){\circle*{2.4}}
\put(25,30){\circle*{2.4}}
\put(188,10){\makebox(0,0){$\rightarrow \infty$}}
\put(110,40){\makebox(0,0){$\mathcal{C}^{+}$}}
\put(110,8){\makebox(0,0){$\mathcal{C}^{-}$}}
\put(179,54){\makebox(0,0){$x^{0}$}}
\put(18,15){\makebox(0,0){$t_{0}$}}
\end{picture}\\

\noindent 
where we shifted parts of the curve slightly away from the real axis purely for visualization purposes. Contour time ordering along this real-time path is denoted by $T_{\C}$. It corresponds to usual time ordering along the forward piece $\C^+$ and reversed 
ordering on the backward piece $\C^-$. In particular, any time on 
$\C^-$ is considered later than any time on $\C^+$. 

We have introduced the 
generating functional with two source terms on the contour, 
$J(x)$ and $R(x,y)$, which can be extended straightforwardly to take into account
further source terms if necessary. Standard functional differentiation with respect to sources is extended to include time arguments along the closed time path. In particular,
\beq \db
\frac{\delta J(x)}{\delta J(y)} \, = \, \delta_\C (x-y) \, \equiv \, \delta_\C(x^0-y^0)\,
\delta(\bx -\by) \, ,
\eeq 
where the Dirac $\delta_\C(x^0-y^0)$ is defined on the closed time contour to be zero everywhere except at $x^0=y^0$ if either $x^0$ and $y^0$ are both on $\C^+$ or both on $\C^-$, where it is infinite with
\beq \db
\int_\C \rmd x^0 \, \delta_\C (x^0) \, = \, 1 \, .
\eeq

Setting all source terms to zero in (\ref{eq:definingZneq}) we obtain the normalized partition sum
\begin{equation}
\db \left. Z[J,R;{\rr \varrho_0}]\right|_{J,R=0} \, = \, \Tr\left\{{\rr \varrho_0}\right\} \, = \, 1 \, .
\end{equation}
{\rr\it Nonequilibrium correlation functions} correspond to expectation values of products of Heisenberg field operators. These can be obtained
by functional differentiation 
of (\ref{eq:definingZneq}). For instance, the one-point
function or {\rr\it macroscopic field} $\rr \phi$ reads
\begin{equation}\db
\frac{\delta Z[J,R;{\rr \varrho_0}]}{i \delta J(x)}\Big|_{J,R=0} \, = \, \db \Tr\{{\rr \varrho_0}\, \Phi(x)\}  \, \db \equiv \, \db {\rr\langle} \Phi(x) {\rr\rangle} \, \equiv \, {\rr \phi(x)} \, .
\label{eq:onepointneq}
\end{equation}
Two-point functions are obtained from second functional derivatives and $n$-point functions involve $n$ derivatives. Together with the time-ordered two-point function
\begin{equation}\db
\frac{\delta^2 Z[J,R;{\rr \varrho_0}]}{i \delta J(x) i \delta J(y)}\Big|_{J,R=0} \, = \, \Tr\{{\rr \varrho_0}\, \textrm{T}_{\C} \Phi(x)\Phi(y)\} 
\, \equiv \, {\rr\langle} \textrm{T}_{\C} \Phi(x)\Phi(y) {\rr\rangle}
\label{eq:twopointneq}
\end{equation}
we can introduce the {\rr\it connected two-point function} or {\rr\it propagator} as 
\begin{equation}\rr
G(x,y) \,\, {\db\equiv} \,\,  {\db {\rr\langle} \textrm{T}_{\C} \Phi(x)\Phi(y)} {\rr\rangle} - {\rr \phi(x) \phi(y) \,} .
\label{eq:connectedG}
\end{equation}
Here $G(x,y)$ is defined on the contour $\C$ and the time-ordering of $\textrm{T}_{\C} \Phi(x)\Phi(y)$ has to be evaluated according to the positions of the time arguments of the fields. In terms of the Heaviside step function $\theta(x^0 -y^0)$ this reads explicitly
\bea\db
\textrm{T}_{\C} \Phi(x)\Phi(y)
&\db \!\!\! = \!\!\! & \left\{
\begin{array}{l}
\!{\db \Phi(x) \Phi(y) \, \theta(x^{0} - y^{0}) + \Phi(y) \Phi(x) \, \theta(y^{0} - x^{0})}
\,\,\,\, \textrm{for $x^0$, $y^0$ on $\mathcal{C}^+$}\nonumber\\
\!{\db \Phi(x) \Phi(y) \, \theta(y^{0} - x^{0}) + \Phi(y) \Phi(x) \, \theta(x^{0} - y^{0})}
\,\,\,\, \textrm{for $x^0$, $y^0$ on $\mathcal{C}^-$}\nonumber\\
\!{\db \Phi(y) \Phi(x)} 
\,\,\,\, \textrm{for $x^0$ on $\mathcal{C}^+$, $y^0$ on $\mathcal{C}^-$}\nonumber\\
\!{\db \Phi(x) \Phi(y)} 
\,\,\,\, \textrm{for $x^0$ on $\mathcal{C}^-$, $y^0$ on $\mathcal{C}^+$}
\end{array}
\right. 
\nonumber\\
&\db \!\!\! \equiv \!\!\! & {\db \Phi(x) \Phi(y) \, \theta_\C(x^{0} - y^{0}) + \Phi(y) \Phi(x) \, \theta_\C(y^{0} - x^{0})} \, ,
\label{eq:contourstep}
\eea
where the last equation defines the contour step function $\theta_\C(x^0 -y^0)$.

For an efficient notation, the field $\Phi(x)$ may be written 
as $\Phi^{\pm}(x^{0},\bf{x})$ where the $\pm$-index denotes on which part of the contour $\mathcal{C}^{\pm}$ the time argument $x^0$ is located. Then the contour integration for the linear source term in (\ref{eq:definingZneq}) takes the form
\begin{eqnarray}
\db \int_{x,\mathcal{C}} \Phi(x) J(x) & \db \equiv& \db \int_{t_{0}}^{\infty} \rmd x^{0} \int \rmd^{d}x \left( \Phi^{+}(x) J^{+}(x) - \Phi^{-}(x) J^{-}(x) \right) ~,
\end{eqnarray} 
where the minus sign comes from the reversed time integration along $\mathcal{C}^{-}$. Similarly, the bilinear source term in (\ref{eq:definingZneq}) can be decomposed into four terms introducing $R^{++}(x,y)$, $R^{--}(x,y)$, $R^{+-}(x,y)$ and $R^{-+}(x,y)$ according to the different possibilities to locate $x^0$ and $y^0$ on $\mathcal{C}^{\pm}$. For instance, the one-point
function for vanishing sources is obtained as
\begin{equation}\db
\frac{\delta Z[J,R;{\rr\varrho_0}]}
{i \delta J^+(x)}\Big|_{J,R=0} \, = \, \rr
\phi(x) \, .
\end{equation}
We emphasize that the field expectation value is the same if obtained from a derivative with respect to either $J^+$ or $J^-$ in the absence of sources, since time-ordering plays no role for a one-point function. Taking the second functional derivative of the generating functional (\ref{eq:definingZneq}) with respect to the source $J^{+}$ and setting $J$ and $R$ to zero afterwards, we obtain
\bea\db
\frac{\delta^{2} Z[J , R ; {\rr \varrho_0}]}{i\delta J^{+}(x) \, i \delta J^{+}(y)} \Big|_{J,R=0} 
&\db = & {\rr \left\langle {\db \Phi(x) \Phi(y) \, \theta(x^{0} - y^{0})} 
{\db \, + \, \Phi(y) \Phi(x) \, \theta(y^{0} - x^{0})} \right\rangle} \nonumber\\
&\db \equiv &  {\rr G^{++}(x,y)} + {\rr \phi(x) \phi(y)} \, .
\label{eq:Gpp}
\eea
Here we used that contour ordering corresponds to standard time ordering if all time arguments are on $\mathcal{C}^+$. 
We also introduced the notation $G^{++}(x,y)$ in order to distinguish this correlator from the other possible second functional derivatives with respect to the sources $J^{+}$ and $J^{-}$ setting $J,R = 0$ afterwards. These can be written as:
\begin{eqnarray}\db
\frac{\delta^{2} Z[J , R ; {\rr \varrho_0}]}{i\delta J^{-}(x) \, i \delta J^{-}(y)} \Big|_{J,R=0} &\db = & {\rr \left\langle {\db \Phi(x) \Phi(y) \, \theta(y^{0} - x^{0})} 
{\db \, + \, \Phi(y) \Phi(x) \, \theta(x^{0} - y^{0})} \right\rangle} \nonumber\\
&\db \equiv &\db {\rr G^{--} (x,y)} + {\rr \phi(x) \phi(y)} \, , \nonumber\\
\db
\frac{\delta^{2} Z[J , R ; {\rr \varrho_0}]}{i\delta J^{+}(x) \, i \delta J^{-}(y)} \Big|_{J,R=0}
&\db = & \db {\rr \left\langle {\db \Phi(y) \Phi(x)}  \right\rangle} \, \equiv \, {\rr G^{+-} (x,y)} + {\rr \phi(x) \phi(y)} \, , \nonumber\\
\db
\frac{\delta^{2} Z[J , R ; {\rr \varrho_0}]}{i\delta J^{-}(x) \, i \delta J^{+}(y)} \Big|_{J,R=0}
&\db = & \db {\rr \left\langle {\db \Phi(x) \Phi(y)}  \right\rangle} \, \equiv \, {\rr G^{-+} (x,y)} + {\rr \phi(x) \phi(y)} \, .
\label{eq:Gpm}
\end{eqnarray}
While any nonequilibrium two-point correlation function can be written in terms of the components (\ref{eq:Gpp}) -- (\ref{eq:Gpm}), not all of them are independent. In particular, using the property
$\theta(x^{0}-y^{0}) + \theta(y^{0}-x^{0}) = 1$ of the Heaviside step function one obtains
the algebraic identity for $J,R=0$:
\begin{equation}
\db {\rr G^{++} (x,y)} + {\rr G^{--}(x,y)} \, = \, {\rr G^{+-}(x,y)} + {\rr G^{-+}(x,y)}~.
\label{eq:AlgebraicIdentity}
\end{equation}
Later we will observe that a further, so-called fluctuation-dissipation relation exists for the special case of vacuum or thermal equilibrium density matrices. However, for general out-of-equilibrium situations this is not the case.

\subsection{Functional integral representation}
\label{sec:functionalintegral}

Above the time-ordering along the closed contour ${\mathcal{C}}$ appears as a bookkeeping device that allows one to conveniently describe different components of nonequilibrium correlation functions. To simplify the evaluation of correlation functions, in the following we write the generating functional (\ref{eq:definingZneq}) in terms of a functional integral representation. 
In this construction the closed time-path appears because we want to compute correlation functions which are given as the {\rr\it trace} over the density operator with time-ordered products of Heisenberg field operators. Representing the trace as a path integral will require a time path where the initial and final times are identified.  

We evaluate the trace using eigenstates of the field operator $\Phi^{\pm}$ at time $t_0$,
\begin{equation}
\db \Phi^{\pm}(t_{0},{\bf{x}}) \, |\varphi^{\pm}\rangle \, = \, \varphi_0^{\pm}(\bf{x}) \,|\varphi^{\pm}\rangle ~,
\end{equation}
such that (\ref{eq:definingZneq}) may be written as
\begin{eqnarray}
\db Z[J,R; {\rr \varrho_0}] 
&\db =& \db \int [\rmd \varphi_{0}^{+}] \, \langle \varphi^{+}| \: {\rr \varrho_0} \, \textrm{T}_{\mathcal{C}} \,\exp i \left\{ \int_{x, \mathcal{C}} \Phi(x) J(x) \right. 
\nonumber\\
&& \db +\: \left. \frac{1}{2} \int_{x,y, \mathcal{C}} \Phi(x) R(x,y) \Phi(y) \right\} |\varphi^{+}\rangle ~. 
\label{GeneratingFunctional1}
\end{eqnarray}
The integration measure is 
\begin{equation}
\db \int [\rmd\varphi_{0}^{\pm}] \, \equiv \int \prod_{\bf{x}} \rmd\varphi_{0}^{\pm}({\bf{x}}) ~.
\label{FunctionalMeasure}
\end{equation}
With the insertion 
\begin{equation}
\db \int [\rmd \varphi_{0}^{-}] \: |\varphi^{-}\rangle \langle \varphi^{-}| = \mathds{1}~,
\label{UnitOperator}
\end{equation}
we may bring (\ref{GeneratingFunctional1}) to a form 
\begin{eqnarray}
\db Z[J,R; {\rr \varrho_0}] &\db =& \db \int [\rmd \varphi_{0}^{+}] [\rmd \varphi_{0}^{-}] \: \langle\varphi^{+}| \, {\rr \varrho_0} \, |\varphi^{-}\rangle 
\left( \varphi^{-},t_{0} \,|\, \varphi^{+},t_{0} \right)_{J,R} ~.  
\label{GeneratingFunctional2}
\end{eqnarray}
Here the transition amplitude in the presence of the sources is given by
\begin{eqnarray}
\db \left( \varphi^{-},t_{0} \,|\, \varphi^{+},t_{0} \right)_{J,R}  
&\db \equiv&\db \langle \varphi^{-}| \, \textrm{T}_{\mathcal{C}} \,\exp i \left\{ \int_{x, \mathcal{C}} \Phi(x) J(x) \right. 
\nonumber\\
&& \db +\: \left. \frac{1}{2} \int_{x,y, \mathcal{C}} \Phi(x) R(x,y) \Phi(y) \right\} \,  |\varphi^{+}\rangle ~. 
\label{MatrixElement}
\end{eqnarray} 
This matrix element can be written as a functional integral over ``classical'' fields $\varphi(x)$ on the closed time path $\mathcal{C}$, whose derivation is given below. For a quantum theory with classical action $S[\varphi]$ it reads
\begin{eqnarray}
\db \left( \varphi^{-},t_{0} \,|\, \varphi^{+},t_{0} \right)_{J,R} &\db =& \db \int\limits_{\varphi_0^+}^{\varphi_0^-} \mathscr{D}'\varphi  \exp i \left\{ S[\varphi] + \int_{x,\mathcal{C}} \varphi(x) J(x) \right.
\nonumber\\
&& \db +\: \left. \frac{1}{2} \int_{x,y,\mathcal{C}} \varphi(x) R(x,y) \varphi(y) \right\}~.
\label{eq:transition}
\end{eqnarray}
Here the classical action includes the time contour integral over the respective Lagrangian density ${\mathcal L}(x)$, i.e.\ $S = \int_{\C} \rmd x^0 \int {\rmd}^{d} x\, {\mathcal L}(x)$. The functional integration goes over the field configurations $\varphi(x)$ depending on space and times $x^0 > t_0$ that satisfy the boundary conditions $\rr \varphi^{\pm}(x^{0} = t_{0},{\bf{x}}) = \varphi_{0}^{\pm}({\bf{x}})$ for $x^0$ on ${\mathcal{C}}^\pm$. The prime on the functional measure is indicating that the integration over the fields at $x^0 = t_0$ is excluded.

Putting everything together, we may write the generating functional (\ref{eq:definingZneq}) as
\beq \db
Z [J,R;{\rr \varrho_0}] = {\rr\underbrace{
\int_{_{_{_{_{_{_{_{_{_{_{_{_{}}}}}}}}}}}}} [\rmd \varphi_{0}^{+}] [\rmd \varphi_{0}^{-}]  \langle\varphi^{+}|  \varrho_0  |\varphi^{-}\rangle}} 
\underbrace{\int\limits_{{\db \varphi_0^{+}}}^{{\db \varphi_0^{-}}}
\! \mathscr{D}'\varphi  e^{i  \left\{S[\varphi] 
+ \int_{x,\C}\! J(x)\varphi(x)  + \frac{1}{2}
\int_{x y,\C}\! \varphi(x) R(x,y)\varphi(y)\right\}}} 
\label{eq:neqgen}
\eeq
\hspace*{3.2cm}{\rr initial conditions}\hspace*{3.2cm}{\db quantum
dynamics} 

\vspace*{0.3cm}

\noindent
The above expression displays two important ingredients entering nonequilibrium quantum field theory: the quantum fluctuations described by the functional integral with action $S$, and the statistical fluctuations encoded in the averaging procedure with the matrix elements of the initial density operator~$\varrho_0$.  

Formulations of closed time path generating
functionals typically employ a time interval starting at some $t_0$ and extending to the far future as mentioned above. Causality implies that for any $n$-point function with finite time
arguments the contributions of an infinite time path cancel for times 
exceeding the largest time argument of the $n$-point function. We will see this explicitly when we derive time evolution equations for correlation functions in section \ref{sec:exactevoleq}.
To avoid unnecessary cancellations of infinite time path contributions we will often
consider finite time paths. The largest time of the path  
can be kept as a parameter and is evolved
in the time evolution equations.\\

\noindent
{\bf \large Details: Construction of the path integral}\\

\noindent
To simplify the evaluation of correlation functions we wrote the above generating functional in terms of a functional integral representation using (\ref{eq:transition}). Here we describe the construction of the corresponding transition amplitude following standard presentations and refer to the literature for more information.

We start from a Schr{\"o}dinger picture where the field operators are time-independent.
For the scalar field theory the Hamiltonian $H[\Pi,\Phi]$   
is expressed in terms of the field $\Phi(\bx)$ and the conjugate momentum field operator $\Pi(\bx)$. We work with basis sets of eigenstates for which $\Phi(\bx)$ or $\Pi(\bx)$ are multiplicative operators, respectively, 
\begin{equation}\db
\Phi(\bx)|\varphi\rangle = \varphi(\bx)|\varphi\rangle \, , \qquad
\Pi(\bx)|\pi\rangle = \pi(\bx)|\pi\rangle \, .
\nonumber
\end{equation}
The completeness and orthogonality conditions read
\begin{equation}
\db \int [\rmd \varphi]\: |\varphi \rangle \langle \varphi| = \mathds{1}\quad,
\qquad \langle \varphi_j|\varphi_i \rangle 
= \delta\left[\varphi_j -\varphi_i \right] 
\label{eq:cophi}
\end{equation}
and
\begin{equation}
\db \int \left[\frac{\rmd \pi}{2\pi}\right] |\pi \rangle \langle \pi| = \mathds{1}\quad,
\qquad \langle \pi_j|\pi_i \rangle 
= \delta[\pi_j -\pi_i] \, .
\label{eq:copi}
\end{equation}
To become familiar with the notation we recall that the $\delta$-functional may be represented as a product of individual $\delta$-functions, one for each point $\bx$ in space,
\begin{equation}\db
\delta[\varphi_j -\varphi_i] \, = \,
\prod\limits_{\bx} \, 
\delta\left( \varphi_j(\bx)-\varphi_i(\bx) \right) \, .
\end{equation}
If we write each $\delta$-function in the momentum representation
\begin{equation}\db
\delta\left( \varphi_j(\bx)-\varphi_i(\bx) \right)\, = \, 
\int \frac{\rmd \pi_i(\bx)}{2\pi}\, 
\exp \Big\{ i\, \pi_i(\bx) \left(\varphi_j(\bx) - \varphi_i(\bx)\right)\! \Big\}
\end{equation}
we obtain
\begin{eqnarray}\db
\delta[\varphi_j -\varphi_i] &\db = &\db  
\prod\limits_{\bx}\, \int \frac{\rmd \pi_i(\bx)}{2\pi}\, 
\exp \Big\{ i\, \pi_i(\bx) \left(\varphi_j(\bx) - \varphi_i(\bx)\right)\! \Big\}
\nonumber\\
&\db = &\db \int \left[\frac{\rmd \pi_i}{2\pi}\right]
\exp \left\{ i \int \rmd^d x\, \pi_i(\bx) \left(\varphi_j(\bx) - \varphi_i(\bx)\right) \right\}
\, .
\label{eq:Fourierdelta}
\end{eqnarray}
Writing the same $\delta$-functional using orthogonality and completeness conditions as
$\delta[\varphi_j -\varphi_i] \, = \, \int \left[{\rmd \pi_i}/(2\pi)\right]
\langle \varphi_j|\pi_i\rangle\langle \pi_i|\varphi_i \rangle$, 
we recover by comparison with (\ref{eq:Fourierdelta}) the familiar overlap 
\begin{equation}\db
\langle\varphi_j|\pi_i\rangle \, = \,  \exp \left\{ i \int \rmd^d x\, \pi_i(\bx) \varphi_j(\bx) \right\} \, .
\label{eq:overlap}
\end{equation}

If the dynamics is described by a Hamiltonian $H$ then the state $|\varphi_0\rangle$ at time $t_0$ evolves into $e^{-i H (t_f-t_0)} |\varphi_0\rangle$ at time $t_f$. The transition amplitude for the field in configuration $\varphi_0(\bx)$ at time $t_0$ to evolve to $\varphi_f(\bx)$ at time $t_f$ is thus
\begin{equation}
\db \left( \varphi_f,t_f \,|\, \varphi_0,t_0 \right)  
\, \equiv \, \langle \varphi_f|\, e^{-i H (t_f-t_0)} \, |\varphi_0\rangle \, .
\end{equation}
To conveniently evaluate this amplitude we write it as a path integral. For this we divide the interval $t_f-t_0$ into $N+1$ equal periods. Next, we insert a complete set of states at each division and consider the limit $N \to \infty$:
\begin{eqnarray}\db
\left( \varphi_f,t_f \,|\, \varphi_0,t_0 \right) &\db = &\db
\lim\limits_{N \to \infty} \int \prod\limits_{i=1}^{N}\, [\rmd \varphi_i]\, 
\langle \varphi_f |  e^{-i H (t_f - t_N)} | \varphi_N \rangle
\langle \varphi_N |  e^{-i H (t_N - t_{N-1})} | \varphi_{N-1} \rangle
\nonumber\\
&& \db \cdots \,
\langle \varphi_1 |  e^{-i H (t_1 - t_0)} | \varphi_0 \rangle \, .
\end{eqnarray} 
As $N$ gets large, $\Delta t = (t_{i+1} - t_i) \to 0$ and we may expand the exponentials,
\begin{eqnarray}
\db \langle \varphi_{i+1} |\,  e^{-i H \Delta t} \, | \varphi_i \rangle & \db \simeq &
\db \langle \varphi_{i+1} |\,  1 - i H \Delta t \, | \varphi_i \rangle 
\nonumber\\
&\db = &\db \delta\left[\varphi_{i+1} - \varphi_i \right] - i \Delta t \, 
\langle \varphi_{i+1} |\, H \, | \varphi_i \rangle \, .
\label{eq:expand}
\end{eqnarray}
The first term in the above sum is a $\delta$-functional which can be written in terms of its momentum representation (\ref{eq:Fourierdelta}). 
It remains to evaluate the operators in the matrix element 
$\langle \varphi_{i+1} |\, H[\Pi,\Phi] \, | \varphi_i \rangle$. We consider first the case of a matrix element for a contribution $h_\pi[\Pi]$ that only depends on the conjugate field momentum. We write, inserting a complete set of states,  
\begin{eqnarray}\db
\lefteqn{\langle \varphi_{i+1} |\, h_\pi[\Pi] \, | \varphi_i \rangle
\, = \,
\int \left[\frac{\rmd \pi_i}{2\pi}\right]  
\langle \varphi_{i+1} |\, h_\pi[\Pi] \, |\pi_i\rangle \langle \pi_i | \varphi_i \rangle}
\nonumber\\
&\db = &\db \int \left[\frac{\rmd \pi_i}{2\pi}\right]  
h_\pi[\pi_i] \, \exp \left\{ i \int \rmd^d x\, \pi_i(\bx) \left( \varphi_{i+1}(\bx) - \varphi_i(\bx) \right) \right\} \, ,
\end{eqnarray}
where (\ref{eq:overlap}) is employed for the second equality. Typical Hamiltonians can be written as a sum of a field and a conjugate momentum term, i.e.\ $H[\Pi,\Phi] = h_\varphi[\Phi] + h_\pi[\Pi]$. For the field contribution $h_\varphi[\Phi]$, we employ a symmetric operator ordering. In general, matrix elements including also products of $\Phi$ and $\Pi$ operators can be evaluated by a symmetric operator ordering. For instance, the symmetrization of a term like $\Phi \Pi \rightarrow \{\Phi,\Pi\}/2 \equiv (\Phi \Pi + \Pi \Phi)/2$ leads to the matrix element
\begin{eqnarray}\db
\lefteqn{
\langle \varphi_{i+1} |\, \frac{\Phi \Pi + \Pi \Phi}{2} \, | \varphi_i \rangle \, = \,
\frac{\varphi_{i+1} + \varphi_i}{2}\, \langle \varphi_{i+1} |\, \Pi \, | \varphi_i \rangle }
\nonumber\\
&\db = &\db \bar{\varphi_i} \int \left[\frac{\rmd \pi_i}{2\pi}\right]  
\pi_i \, \exp \left\{ i \int \rmd^d x\, \pi_i(\bx) \left( \varphi_{i+1}(\bx) - \varphi_i(\bx) \right) \right\}
\, ,
\end{eqnarray} 
where
\begin{equation}\db
\bar{\varphi}_i \, \equiv \, \frac{\varphi_{i+1} + \varphi_i}{2} \,\, . 
\end{equation}
In general, we can write
\begin{equation} \db
\langle \varphi_{i+1} |\, H[\Pi,\Phi] \, | \varphi_i \rangle \, = \, 
\int \left[\frac{\rmd \pi_i}{2\pi}\right] H[\pi_i,\bar{\varphi}_i]  
 \, \exp \left\{ i \int \rmd^d x\, \pi_i(\bx) \left( \varphi_{i+1}(\bx) - \varphi_i(\bx) \right) \right\}  .
\label{eq:hPhi}
\end{equation}
Therefore, (\ref{eq:expand}) may be represented as
\begin{eqnarray} \db
\lefteqn{\langle \varphi_{i+1} |\,  e^{-i H \Delta t} \, | \varphi_i \rangle 
\, = \, \int \left[\frac{\rmd \pi_i}{2\pi}\right] \left( 1 - i\, \Delta t\, 
H[\pi_i,\bar{\varphi}_i] + {\mathcal O}(\Delta t^2) \right) } 
\nonumber\\
& \db \times & \db \exp \left\{ i \int \rmd^d x\, \pi_i(\bx) \left( \varphi_{i+1}(\bx) - \varphi_i(\bx) \right) \right\}
\nonumber\\
& \db \simeq & \db \int \left[\frac{\rmd \pi_i}{2\pi}\right]
\exp \left\{ i \int \rmd^d x\, \pi_i(\bx) \left( \varphi_{i+1}(\bx) - \varphi_i(\bx) \right) - i \, \Delta t \, H[\pi_i,\bar{\varphi}_i] \right\} \, . \quad 
\end{eqnarray}
Putting everything together gives
\begin{eqnarray}\db
\lefteqn{\left( \varphi_f,t_f \,|\, \varphi_0,t_0 \right) = 
\lim\limits_{N \to \infty} \int \prod\limits_{i=1}^{N}\, [\rmd \varphi_i]\,
\prod\limits_{j=0}^{N} \left[\frac{\rmd \pi_i}{2\pi}\right]}
\nonumber\\
&\db \times &\db  \exp \left\{ i \Delta t \sum\limits_{j=0}^{N} \left(\int \rmd^d x\, \pi_j(\bx)
\frac{\varphi_{j+1}(\bx) - \varphi_j(\bx)}{\Delta t} - H[\pi_j,\bar{\varphi}_j]\right) 
\right\}
\nonumber\\
&\db \equiv& \db \!\!\!\!
\int\limits_{\varphi(t_0,\bx) = \varphi_0(\bx)}^{\varphi(t_f,\bx) = \varphi_f(\bx)}
\!\!\!\!\!\!
\mathscr{D}' \varphi \mathscr{D} \pi \exp \left\{ i \int_{t_0}^{t_f} \rmd x^0 \left( \int \rmd^d x\,
\pi(x) \frac{\partial \varphi(x)}{\partial x^0} - H[\pi, \varphi] \right) \right\}. 
\label{eq:pathintegral}
\end{eqnarray}
Above we have defined $\varphi_{N+1}(\bx) = \varphi_f(\bx)$ and identify 
$(\varphi_i(\bx),t_i) = \varphi(t_i,\bx)$ employed for the continuum notation. We emphasize that all references to operators are gone in (\ref{eq:pathintegral}).

If the Hamiltonian is quadratic in $\pi(x)$, as for a real scalar field theory with
\begin{equation} \db
H[\pi, \varphi] \, = \, \int \rmd^d x \left( \frac{1}{2}\,\pi^2 + \frac{1}{2}\,\left(\nabla \varphi\right)^2 + \frac{m^2}{2}\, \varphi^2 + V(\varphi) \right)
\label{eq:Hscalar}
\end{equation}
with mass parameter $m$ and interaction part $V(\varphi)$, then one can carry out the functional integration over $\pi$ in (\ref{eq:pathintegral}). Completing the squares one finds from the Gaussian integral
\begin{equation}
\db  \int \frac{\rmd\pi_j(\bx)}{2\pi} 
\, \exp \left\{ i\, \pi_j (\varphi_{j+1}-\varphi_j) - \frac{1}{2}\,\pi_j^2 \Delta t \right\}
\, = \, (2\pi i \Delta t)^{-\frac{1}{2}}\, 
\exp \left\{i \frac{\left(\varphi_{j+1}-\varphi_{j}\right)^2}{2 \Delta t}\right\} 
\end{equation}
that the net effect is to replace $\pi$ by the time derivative of $\varphi$ in the exponential of 
(\ref{eq:pathintegral}). Applying the continuum notation the matrix element then becomes
\begin{equation}\db
\left( \varphi_f,t_f \,|\, \varphi_0,t_0 \right) \,\, = \,
\int\limits_{\varphi(t_0,\bx) = \varphi_0(\bx)}^{\varphi(t_f,\bx) = \varphi_f(\bx)}
\mathscr{D}' \varphi\, e^{i S[\varphi]} \, ,
\label{eq:transitionpath}
\end{equation}
where the classical action reads
\begin{equation} \db
S[\varphi] = \int_{t_0}^{t_f} \rmd x^0 \int \rmd^d x\, \left\{ \frac{1}{2} 
\left(\frac{\partial\varphi}{\partial x^0}\right)^2 
- \frac{1}{2}(\nabla\varphi)^2 - \frac{m^2}{2} \varphi^2 - V(\varphi)\right\} \, .
\label{eq:classicalactionC}
\end{equation}
Here the measure means 
\begin{equation} \db
\int \mathscr{D}' \varphi\, \, = \, 
\lim\limits_{N \to \infty} \int \prod\limits_{i=1}^{N}\, \prod\limits_\bx \rmd \varphi(t_i,\bx)\,
(2 \pi i \Delta t)^{-\frac{1}{2}} \quad, \quad N \Delta t = t_f - t_0 \, . 
\end{equation}

The final time $t_f$ in the above matrix elements is, of course, arbitrary. Applying the same construction to the closed time path of section \ref{sec:nonequgenfunc}, first to the forward piece ${\mathcal{C}}^+$ and subsequently to the backward piece ${\mathcal{C}}^-$, is straightforward. Adding to the Hamiltonian appropriate source terms as employed in section \ref{sec:nonequgenfunc} leads to the generating functional for correlation functions.

\subsection{Initial conditions}
\label{sec:initialconditions}

To understand in more detail how the initial density matrix 
enters calculations, we consider first the example of a 
{\rr \it Gaussian density matrix} for a real scalar field theory. 
For simplicity, we neglect for 
a moment the spatial dependencies setting $d=0$, i.e.\ we consider quantum mechanics. The generalization to higher $d$ will typically be straightforward.\footnote{We note that for homogeneous field 
expectation values taking into account spatial
dependencies essentially amounts to adding a 
momentum label in Fourier space (see also below).} In this case the most general form of a Gaussian density matrix can be parametrized in terms of {\rr \it five real parameters} $\rr \phi_0$, $\rr \dot{\phi}_0$, $\rr \xi$, $\rr \eta$ and $\rr \sigma$: 
\bea
&&\db \!\!\!\!\!\!  
\langle \varphi^+| \varrho_0 | \varphi^- \rangle = \nn
&&\db \!\! \frac{1}{\sqrt{2 \pi {\rr \xi}^2}}
\exp \Big\{ i {\rr \dot{\phi}_0} (\varphi_0^+ - \varphi_0^-)
\!-\! \frac{{\rr \sigma}^2+1}{8 {\rr \xi}^2} 
\left[(\varphi_0^+ - {\rr \phi_0})^2 
\!+\! (\varphi_0^- - {\rr \phi_0})^2 \right] \nn
&&\db \!\! + i\, \frac{\rr \eta}{2 {\rr \xi}} 
\left[(\varphi_0^+ - {\rr \phi_0})^2 
- (\varphi_0^- - {\rr \phi_0})^2 \right]
+ \frac{{\rr \sigma}^2-1}{4 {\rr \xi}^2} 
(\varphi_0^+ - {\rr \phi_0}) 
(\varphi_0^- - {\rr \phi_0}) \Big\}  , \quad 
\label{eq:GaussianrhoD}
\eea
whose values will be further discussed below. Gaussianity refers here to the fact that the highest power of $\varphi_0^\pm$ appearing in the exponential of (\ref{eq:GaussianrhoD}) is two. Therefore, density matrix averages of field operators at initial time only involve Gaussian integrals.

In order to interpret the above parameters and to see that we indeed consider the most general
Gaussian density matrix, we first note that (\ref{eq:GaussianrhoD}) 
is equivalent to the following set of initial conditions for one- and 
two-point functions:
\begin{eqnarray}
{\rr \phi_0} &\db \!=\!&\db 
\Tr\left\{\varrho_0 \Phi(t) \right\}_{|t=t_0} \quad , \quad
{\rr \dot{\phi}_0} \,=\, 
\Tr\left\{\varrho_0 \partial_t\Phi(t) \right\}_{|t=t_0} \, ,
\label{eq:equirhoa}
\\[0.3cm]
{\rr \xi^2} &\db\!=\!&\db \Tr\left\{\varrho_0
\Phi(t)\Phi(t') \right\}_{|t=t'=t_0} - {\rr \phi_0 \phi_0} 
 \, ,\\ 
{\rr \xi \eta} &\db\!=\!&\db 
\frac{1}{2} \Tr\left\{\varrho_0
\left( \partial_t\Phi(t)\Phi(t') + \Phi(t) \partial_{t'}\Phi(t') 
\right) \right\}_{|t=t'=t_0} - {\rr \dot{\phi}_0 \phi_0}
\, ,\label{eq:equirhoc}\\
{\rr \eta^2 + \frac{\sigma^2}{4 \xi^2}}
&\db\!=\!&\db 
  \Tr\left\{\varrho_0
\partial_t\Phi(t)\partial_{t'}\Phi(t') \right\}_{|t=t'=t_0} 
- {\rr \dot{\phi}_0 \dot{\phi}_0} \, 
\label{eq:sigini}. 
\end{eqnarray} 
In contrast to the symmetrized combination (\ref{eq:equirhoc}), 
we note that the anti-symmetrized initial correlator involving the 
commutator of $\Phi$ and $\partial_t{\Phi}$ at $t_0$ is not independent 
because of the operator commutation relation
\begin{equation}\db
\left[ \Phi(t), \partial_t \Phi(t) \right] \, = \, i \, .
\end{equation} 
The equivalence between the initial density 
matrix and the initial conditions for the correlators can be verified explicitly.
For instance
\begin{eqnarray}\db
\Tr \varrho_0 &\db =&\db \int_{-\infty}^{\infty} {\rm d} \varphi_0^+\, 
\langle \varphi^+| \varrho_0 | \varphi^+ \rangle \nn
&\db =&\db \frac{1}{\sqrt{2 \pi {\rr \xi}^2}} \int_{-\infty}^{\infty}
{\rm d} \varphi_0^+ \exp \Big\{- \frac{1}{2 {\rr\xi}^2}(\varphi_0^+ - {\rr \phi_0})^2
\Big\} \, = \, 1 \, , \label{eq:initialdis}
\end{eqnarray}
where one notes that the RHS of (\ref{eq:GaussianrhoD}) does not depend on $\dot{\phi}_0$, $\sigma$ and $\eta$ for $\varphi_0^- = \varphi_0^+$. Similarly, 
\begin{eqnarray}\db
\db \Tr \left\{\varrho_0 \Phi(t_0)\right\} &\db =&\db
\frac{1}{\sqrt{2 \pi {\rr \xi}^2}} \int_{-\infty}^{\infty}
{\rm d} \varphi_0^+\, \varphi_0^+ 
\exp \Big\{- \frac{1}{2 {\rr \xi}^2}(\varphi_0^+ - {\rr \phi_0})^2
\Big\} \, = \, {\rr \phi_0} \, , \quad 
\end{eqnarray}
which is simply obtained by the shift $\varphi_0^+ \to \varphi_0^+ + \phi_0$, etc.
Similarly, since only Gaussian integrations 
appear one finds that all
initial $n$-point functions with $n > 2$ can be expressed 
in terms of products of the one- and two-point functions.

Of course, {\rr\it higher initial time 
derivatives are not independent} as can be observed
from the field equation of motion. For instance, for the above scalar theory Hamiltonian (\ref{eq:Hscalar}) with mass $m$ and the interaction part $V(\Phi) = \lambda \Phi^4/4!$ the corresponding field equation reads:
\bea \db
\langle \partial^2_t{\Phi} \rangle =  - m^2 \langle \Phi \rangle 
- \frac{\lambda}{6}\, \langle \Phi^3 \rangle \, .
\eea
Since $\langle \Phi^3 \rangle$ is given at initial time in terms
of one- and two-point functions for Gaussian 
$\varrho_0$, also second and higher time derivatives
are not independent. We conclude that for our case the most general
Gaussian density matrix is indeed described by the five parameters
appearing in (\ref{eq:GaussianrhoD}).
In particular, all observable information contained in the density
matrix can be conveniently expressed in terms of the correlation
functions (\ref{eq:equirhoa})--(\ref{eq:sigini}). 

For further interpretation of the initial conditions 
we note that
\begin{eqnarray}\db 
\Tr\, \varrho_0^2 &\db = &\db \int_{-\infty}^{\infty} {\rm d} \varphi_0^+
{\rm d} \varphi_0^-\,
\langle \varphi^+| \varrho_0 | \varphi^- \rangle
\langle \varphi^-| \varrho_0 | \varphi^+ \rangle
\nonumber\\
& \db = &\db \frac{1}{2\pi {\rr \xi}^2} \int_{-\infty}^{\infty} {\rm d} \varphi_0^+
{\rm d} \varphi_0^-\, 
\exp \Big\{- \frac{{\rr \sigma}^2 + 1}{4 {\rr \xi}^2}\left[(\varphi_0^+)^2 +  (\varphi_0^-)^2\right]
+ \frac{{\rr \sigma}^2 - 1}{2 {\rr \xi}^2} \varphi_0^+ \varphi_0^-
\Big\}
\nonumber\\
& \db = & \db \frac{1}{\rr \sigma} \: , 
\end{eqnarray}
where again we shifted the integration variables by $\phi_0$ to arrive at the second equality. 
The latter shows that for $\rr \sigma > 1$ the density
matrix describes a mixed state, which may be stated in terms
of a non-zero occupation number $f$ with $\sigma = 1 + 2 f$.\footnote{For the special case of
thermal equilibrium the occupation number is given by the Bose-Einstein distribution.} 
For $\sigma = 1$ the ``mixing term'' in (\ref{eq:GaussianrhoD}) is absent and one obtains
a pure-state density matrix of the product form,
\beq \db 
\varrho_0 =
| \Psi \rangle \langle \Psi | \, ,
\eeq
with Schr{\"o}dinger wave function
\beq \db 
\langle \varphi^+ | \Psi \rangle
= \frac{1}{(2 \pi {\rr \xi}^2)^{1/4}}
\exp \Big\{ i {\rr \dot{\phi}_0} \varphi_0^+
-\Big(\frac{1}{4 {\rr \xi}^2} + i\, \frac{\rr \eta}{2 {\rr \xi}} \Big)
(\varphi_0^+ - {\rr \phi_0})^2 \Big\} \, .
\eeq

In order to go beyond Gaussian initial density matrices and to field theory in $d$ spatial dimensions, one may generalize the above example and parametrize the most 
general density matrix as 
\beq\db
\langle \varphi^+ | \varrho_0 | \varphi^- \rangle 
\, = \, \mathcal{N}\, e^{i {\db h_\C[\varphi]}} \, ,
\label{eq:denpara}
\eeq
with normalization $\mathcal{N}$ and 
$h_\C[\varphi]$ expanded in powers of the fields:  
\bea\db
h_\C[\varphi] &\db = &\db {\rr \alpha_0} + \int_{x,\C} {\rr \alpha_1(x)} \varphi(x)
+ \frac{1}{2} \int_{x y,\C} {\rr \alpha_2(x,y)} \varphi(x)\varphi(y)
\nonumber\\ 
&& \db + \frac{1}{3!} \int_{x y z,\C} {\rr \alpha_3(x,y,z)} 
\varphi(x)\varphi(y)\varphi(z)
\nn &&\db
+ \frac{1}{4!} \int_{x y z w,\C} {\rr \alpha_4(x,y,z,w)} 
\varphi(x)\varphi(y)\varphi(z)\varphi(w) 
+ \ldots 
\label{eq:expansiondensity}
\eea
Here the integrals have to be evaluated along the forward piece, $\C^+$, and the backward piece, $\C^-$, of the closed time path with $\rr \varphi^+(t_0,\bx) = \varphi_0^+(\bx)$ and 
$\rr \varphi^-(t_0,\bx)=\varphi_0^-(\bx)$ as described in sections \ref{sec:nonequgenfunc} and \ref{sec:functionalintegral}. Since the density matrix $\varrho_0$ is specified at time
$t_0$ only, all time integrals in (\ref{eq:expansiondensity}) can contribute only at the endpoints of the closed time contour. As a consequence, the coefficients $\alpha_1(x)$,
$\alpha_2(x,y)$, $\alpha_3(x,y,z)$, \ldots vanish identically
for times different than~$t_0$. More precisely, one has 
\bea \db
\int_{x,\C} {\rr \alpha_1(x)} \varphi(x) &\db \equiv&\db \int \rmd^d x
\left\{{\rr \alpha_1^+(\bx)}\, \varphi_0^+(\bx)
+ {\rr \alpha_1^-(\bx)}\, \varphi_0^-(\bx) \right\}\, ,
\nonumber\\ \db
\int_{x y,\C} {\rr \alpha_2(x,y)} 
\varphi(x) \varphi(y) &\db \equiv&\db \int \rmd^d x \rmd^d y \bigg\{
{\rr \alpha_2^{++}}(\bx,\by) \varphi_0^{+}(\bx)\varphi_0^{+}(\by)
\nonumber\\
&& \db 
\qquad\qquad + \, {\rr \alpha_2^{+-}}(\bx,\by)\, \varphi_0^{+}(\bx)\varphi_0^{-}(\by)
\nonumber\\
&& \db 
\qquad\qquad + \, {\rr \alpha_2^{-+}}(\bx,\by)\, \varphi_0^{-}(\bx)\varphi_0^{+}(\by)
\nonumber\\
&& \db 
\qquad\qquad + \, {\rr \alpha_2^{--}}(\bx,\by)\, \varphi_0^{-}(\bx)\varphi_0^{-}(\by)\bigg\}
\eea
and so on. We note that $\alpha_0$ appearing in (\ref{eq:expansiondensity}) is an irrelevant constant that can be taken into account by a rescaling of the overall normalization $\mathcal{N}$ in (\ref{eq:denpara}).
For a physical density matrix the other coefficients
are, of course, not arbitrary. Hermiticity of the density matrix, $\varrho_0 = \varrho_0^\dagger$, implies 
$i\alpha_1^+ = (i \alpha_1^{-})^*$,
$i \alpha_2^{++} = (i \alpha_2^{--})^*$, 
$i\alpha_2^{+-} = (i \alpha_2^{-+})^*$, etc. 
If the initial state is invariant under symmetries,
there are further constraints. For example, for an initial
state which is invariant under $\Phi \rightarrow -\Phi$, all $\alpha_n(x_1, \ldots, x_n)$ with odd $n$ vanish. If the initial state is homogeneous in space, the $\alpha_n(x_1, \ldots, x_n)$ are invariant under space translations $\bx_i \rightarrow \bx_i + {\bf a}$ for $i=1,\ldots,n$ and arbitrary spatial vector ${\bf a}$. In this case, they can be conveniently described in spatial momentum space,
\begin{eqnarray} \rr
\alpha_n^{\pm \cdots}(\bx_1, \ldots, \bx_n) &\db =& \db \int \frac{\rmd^d p_1}{(2\pi)^d} \cdots \frac{\rmd^d p_n}{(2\pi)^d}
\, \exp \left\{ i \sum\limits_{j=1}^{n} \bp_j \bx_j \right\}
(2\pi)^d \, \delta (\bp_1 + \ldots + \bp_n) 
\nonumber \\
&\db \times &  {\rr \alpha_n^{\pm \cdots}(\bp_1, \ldots, \bp_n)} \, .
\end{eqnarray}
 
Using the parametrization (\ref{eq:denpara}) and 
(\ref{eq:expansiondensity}) for the most general initial density matrix,
one observes that the generating functional (\ref{eq:neqgen}) introduced
above may be written as
\beq\db
Z [{\rr J,R;\varrho_0}] = \!
\int \!\! \mathscr{D}\! \varphi e^{i  \left(S[\varphi] 
+ \int_{x,\C} \! {\rr J(x)} \varphi(x)  + \frac{1}{2}
\int_{x y,\C}\! {\rr R(x,y)} 
\varphi(x)\varphi(y) 
+ \frac{1}{3!} \int_{x y z,\C}\! {\rr \alpha_3(x,y,z)} 
\varphi(x)\varphi(y)\varphi(z) + \ldots\right)} .
\label{eq:Zneqgen}
\eeq
Here we have neglected an irrelevant normalization constant and  
rescaled the sources in (\ref{eq:neqgen})
as $J(x) + {\rr \alpha_1(x)} \rightarrow {\rr J(x)}$ and 
$R(x,y) + {\rr \alpha_2(x,y)} \rightarrow {\rr R(x,y)}$.
The sources $J$ and $R$ may, therefore, be conveniently used to absorb the 
lower linear and quadratic contributions
coming from the density matrix specifying the initial state.
This exploits the fact that {\rr\it the initial density matrix 
is completely encoded in terms of initial-time sources} 
for the functional integral.  

The generating functional (\ref{eq:Zneqgen}) can be
used to describe situations involving initial
density matrices for arbitrarily complex physical situations. 
However, often the initial conditions of an experiment 
may be described by only a few lowest $n$--point functions.
For many practical purposes the initial
density matrix is well described by a Gaussian one.
For instance, the initial 
conditions for the reheating dynamics in the early universe at the
end of inflation are described by a Gaussian density matrix to rather good 
accuracy. Also many effective theories for the description of nonequilibrium 
dynamics, such as Boltzmann equations, exploit an assumption about Gaussianity of
initial conditions.
 
From (\ref{eq:Zneqgen}) one observes that for Gaussian initial
density matrices, for which $\alpha_i \equiv 0$ for $i \ge 3$, one has 
\beq \db
Z\left[{\rr J,R; \varrho_0^{\rm (Gauss)}}\right] \,\,\to \,\, Z[{\rr J, R}] \, .
\eeq
As a consequence, in this case
the nonequilibrium generating functional corresponds to the 
generating functional with linear and bilinear source terms introduced in (\ref{eq:definingZneq})
for a closed time path. 

In field theory non-Gaussian initial density matrices pose no problems in principle
but require taking into account additional initial-time sources. An alternative to initial-time 
sources is to represent a non-Gaussian initial density matrix with the help of an
additional imaginary piece of the time contour preceding the closed time path $\C$. 
We refer to the literature at the end of this chapter for further details
concerning this alternative representation.

\subsection{Effective actions}
\label{sec:effectiveactions}

The functional $Z[J,R;\varrho_0]$ given by (\ref{eq:Zneqgen}) is the nonequilibrium
quantum field theoretical generalization of the thermodynamic partition function
in the presence of source terms. In thermodynamics, Legendre transforms of the
logarithm of the partition function lead to equivalent descriptions of the physics.
Similarly, in nonequilibrium quantum field theory Legendre transforms with
respect to the various source terms lead to different representations of a free
energy functional $\Gamma$. The Legendre transform with
respect to the linear source term, $\sim J$, will lead to the so-called
one-particle irreducible (1PI) effective action $\Gamma[\phi]$ parametrized by
the one-point function $\phi$. An additional Legendre transform with respect to the
bilinear source term, $\sim R$, gives the two-particle irreducible (2PI)
effective action $\Gamma[\phi,G]$, which is parametrized in addition by the
two-point function or propagator $G$. In the presence of higher source terms this procedure
can, in principle, be continued to $n$-particle irreducible ($n$PI) effective actions up to
arbitrarily high~$n$. 

In the absence of approximations, all these generating functionals give 
equivalent descriptions of the physics and any choice is only a 
matter of convenience or efficiency. For nonequilibrium initial-value problems it is often particularly helpful to employ effective actions, since they are described in terms of correlation functions whose initial values are typically much better accessible than those of source terms. Furthermore, the ability to find suitable 
approximation schemes in practice can strongly depend on the appropriate
choice of the functional representation. In a following chapter 
it will be demonstrated how 2PI effective actions can be applied to describe important processes such as thermalization in quantum field theory. 
Later chapters include also further applications, such as nonequilibrium instabilities or an efficient derivation of kinetic theory in its range of validity.

In this section we discuss the construction of 1PI and 2PI effective actions. The latter
provides a powerful starting point for
approximations in nonequilibrium quantum field theory with Gaussian initial density matrix as specified in section \ref{sec:initialconditions}. It should be emphasized again that the use
of a 2PI effective action for Gaussian initial density matrices represents a priori
no approximation for the dynamics --- in an interacting quantum field theory higher irreducible 
correlations build up in general corresponding to non-Gaussian 
density matrices for times $t > t_0$. It only restricts 
the possible setup described by the initial conditions
for correlation functions.

To be able to explain a variety of approximation schemes in a following chapter, we consider here the example a quantum field theory for a real, $N$--component scalar field 
$\varphi_a$ ($a=1,\ldots,N$) with $O(N)$-symmetric classical action 
\beq\db
S[\varphi] \, = \, \int_{x,\C}
	\left\{\frac{1}{2}\partial^\mu\varphi_a(x)\partial_\mu\varphi_a(x)  
	-\frac{m^2}{2}\varphi_a\!(x)\varphi_a\!(x) 
	-\frac{\lambda}{4!N}\left(\varphi_a\!(x)\varphi_a\!(x)\right)^2
 \right\}  \, .
\label{eq:classical}
\eeq
Here summation over repeated indices is implied as well as the integration
over the closed time path $\C$ as introduced in section \ref{sec:nonequgenfunc}.
Starting from $Z[J,R]$, given by (\ref{eq:Zneqgen}) with $\alpha_i \equiv 0$ for $i \ge 3$,
we write 
\bea \db
\lefteqn{
Z[J,R] \, \equiv \, \exp\left(i W[J,R] \right) } \nonumber\\
	&\db = & \db \int\! \mathscr{D}
\varphi\, \exp i \left\{ S[\varphi]+ \int_{x,\C}\!
J_a(x)\varphi_a(x)+\frac{1}{2}
\int_{x y,\C}\!\! R_{ab}(x,y) \varphi_a(x)\varphi_b(y)
\right\} \qquad
\label{modZ}
\eea
with real $J_a(x)$, and $R_{ab}(x,y) = R_{ba}(y,x)$. 

The {\rr\it macroscopic field $\phi_a$} and the {\rr\it connected two-point function $G_{ab}$}
in the presence of the source terms $\sim J_a(x)$ and $\sim R_{ab}(x,y)$ can be obtained by variation of $W[J,R]$,
\bea \db
\frac{\delta W[J,R]}{\delta J_a(x)}
& \db = & \db \frac{-i \delta \ln Z[J,R]}{\delta J_a(x)} \,\, = \,\, \frac{1}{Z[J,R]}
\frac{\delta Z[J,R]}{i \delta J_a(x)}
\,\, \equiv \, \,
{\rr \phi_a(x)}\, , 
\label{eq:fielddef} \\
\db \frac{\delta W[J,R]}{\delta R_{ab}(x,y)}
& \db \equiv & \db
\frac{1}{2}\Big(\phi_a(x)\phi_b(y)+ {\rr G_{ab}(x,y)} 
\Big) \, .
\label{deltaWdeltaR}
\eea
These definitions are in accordance with (\ref{eq:onepointneq}) and (\ref{eq:connectedG}) for vanishing sources, i.e.\ $J,R = 0$. However, we keep $J$ and $R$ nonvanishing in (\ref{deltaWdeltaR}) and to ease the notation
the dependence of $\phi = \phi(J,R)$ and $G = G(J,R)$ on the sources is suppressed. 

Before constructing the 2PI effective action, we 
consider first the {\rr\it 1PI effective action}. It is
constructed via a Legendre transform
with respect to the source term linear in the field, 
\bea \db
\Gamma^{R}[\phi] \, = \, W[J,R]-\int_{x,\C} \frac{\delta W[J,R]}{\delta J_a(x)}\, J_a(x) \,=\, W[J,R]-\int_{x,\C} J_a(x) \phi_a(x)  \, .
\label{GammaR}
\eea
To complete this construction one has to note that the relation between $\phi$ and $J$ is $R$-dependent, i.e.\ inverting $\phi = \delta W[J,R]/\delta J$ yields $J= J^{R}(\phi)$ on the RHS of (\ref{GammaR}). Taking the functional derivative with respect to the field one finds
\bea\db
\frac{\delta \Gamma^{R}[\phi]}{\delta \phi_a(x)} & \! \db = \! & \db \int_{y,\C}
{\rr \frac{\delta W[J,R]}{\delta J_b(y)}} \frac{\delta J_b(y)}{\delta \phi_a(x)} 
- \int_{y,\C} {\rr \phi_b(y)} \frac{\delta J_b(y)}{\delta \phi_a(x)} - J_a(x)
\, = \, - J_a(x)  . \qquad
\label{eq:phieom}
\eea
and the source derivative yields
\bea \db
\frac{\delta \Gamma^{R}[\phi]}{\delta R_{ab}(x,y)}&\db =&\db
\frac{\delta W[J,R]}{\delta R_{ab}(x,y)}
+ \int_{z,\C} {\rr \frac{\delta W[J,R]}{\delta J_c(z)} } 
\frac{\delta J_c(z)}{\delta R_{ab}(x,y)}
-\int_{z,\C} {\rr \phi_c(z)}\frac{\delta J_c(z)}{\delta R_{ab}(x,y)} \nonumber\\
&\db =&\db
\frac{\delta W[J,R]}{\delta R_{ab}(x,y)} \, .
\label{eq:gaw}
\eea
The second equalities both in (\ref{eq:phieom}) and (\ref{eq:gaw}) arise from cancellations using (\ref{eq:fielddef}). 

The standard 1PI effective action, $\Gamma[\phi]$, is obtained from (\ref{GammaR}) for $R = 0$, i.e.\ $\Gamma[\phi] = \Gamma^{R = 0}[\phi]$. For non-zero $R$ the functional $\Gamma^{R}[\phi]$ may be viewed as the 1PI effective action for a theory governed by a modified classical action, 
$S^{R}[\varphi]$, with
\bea \db
S^{R}[\varphi] \, \equiv \, S[\varphi] + \frac{1}{2} \int_{xy,\C} R_{ab}(x,y) 
\varphi_a(x) \varphi_b(y) \, . 
\label{eq:SR}  
\eea
As a consequence, it is straightforward to recover for $\Gamma^{R}[\phi]$
``textbook'' relations for the 1PI effective action taking into account
$R$. For instance, evaluating $\Gamma^R[\phi]$ in a saddle-point approximation (``one-loop'') around the macroscopic field configuration $\phi_a(x)$ one finds up to irrelevant constants 
\bea\db 
\Gamma^{R {\rr ({\rm 1loop})}}[\phi] &\db = &\db 
S^R[\phi] + {\rr \frac{i}{2}\, \Tr_\C\ln [G_0^{-1}(\phi) - i R]} \, , 
\label{G1Roneloop}
\eea
where the trace involves the integration over space-time coordinates on the closed time path $\C$ as well as summation over field indices. We derive (\ref{G1Roneloop}) below, which is indeed the standard one-loop result for the 1PI effective action with $S[\phi] \to S^R[\phi]$ and $G_0^{-1}(\phi) \to G_0^{-1}(\phi) - i R$. Here the classical inverse propagator 
$i G_{0,ab}^{-1}(x,y;\phi) \equiv
\delta^2 S[\phi]/\delta \phi_a(x) \delta \phi_b(y)$ for the action (\ref{eq:classical}) reads
\bea \db
i G^{-1}_{0,ab}(x,y;\phi) &\db =&\db - \left( \square_x + m^2 
+ \frac{\lambda}{6 N}\, \phi_c(x)\phi_c(x) \right) \delta_{ab}
\delta_\C(x-y) \nonumber\\ 
&&\db - \frac{\lambda}{3 N}\, \phi_a(x) \phi_b(x) \delta_\C(x-y) \, . 
\label{classprop}
\eea
It is instructive to compare this to what is obtained 
from the approximation (\ref{G1Roneloop}) for the propagator by taking the functional derivative  
with respect to $R$: $\delta \Gamma^{R ({\rm 1loop})}[\phi]/\delta R = (\phi \phi + [G_0^{-1}(\phi) - i R]^{-1})/2$. Since this has to agree to $(\phi \phi + G)/2$ using (\ref{eq:gaw}) and the definition (\ref{GammaR}), we obtain $G^{-1} = G_0^{-1}(\phi) - i R$ in this approximation. 

We now perform a further Legendre transform of $\Gamma^{R}[\phi]$ with
respect to the source~$R$ in order to arrive at the {\rr\it 2PI effective
action}, $\Gamma[\phi,G]$, which is a functional of the field $\phi_a(x)$ and the propagator $G_{ab}(x,y)$:
\bea \db
\Gamma[\phi, G] & \db =& \db \Gamma^{R}[\phi] 
- \int_{xy,\C} 
\underbrace{\frac{\delta \Gamma^{R}[\phi]}{\delta R_{ab}(x,y)}} R_{ab}(x,y)
\nonumber\\
&& \db \qquad\quad \qquad \,\,\,\,\, \frac{\delta W[J,R]}{\delta R_{ab}(x,y)}
=\frac{1}{2}\left[\phi_a(x)\phi_b(y) + G_{ab}(x,y)\right] \nn
&\db =& \db \Gamma^{R}[\phi] - \frac{1}{2} \int_{xy,\C} 
 \left[ \phi_a(x)\phi_b(y) + G_{ab}(x,y) \right] R_{ab}(x,y) \, .
\label{twostepLT}
\eea
Here we have used first (\ref{eq:gaw}) and then the definition (\ref{deltaWdeltaR}).
Of course, the two subsequent Legendre transforms, which have
been used to arrive at
(\ref{twostepLT}), agree with a simultaneous Legendre 
transform of $W[J,R]$ with respect to both source terms:
\bea \db
\lefteqn{\Gamma[\phi,G] \, = \, \db W[J,R]-\int_{x,\C} \frac{\delta W[J,R]}{\delta J_a(x)}\, 
J_a(x) - \int_{xy,\C} \frac{\delta W[J,R]}{\delta R_{ab}(x,y)}\, R_{ab}(x,y) }
\nonumber\\
&\db = \! & \db  W[J,R]-\int_{x,\C} \! \phi_a(x) J_a(x) - \frac{1}{2} \int_{xy,\C} \!
 \left[ \phi_a(x) \phi_b(y) + G_{ab}(x,y) \right] R_{ab}(x,y) \, .\qquad
\label{2PIGdirect}
\eea
Taking the field and propagator dependence of the sources into account, one obtains from (\ref{2PIGdirect}) after cancellations the stationarity conditions
\bea \db
\frac{\delta \Gamma[\phi,G]}
{\delta \phi_a(x)}& \db = &\db - J_a(x) - \int_{y,\C} R_{ab}(x,y) \phi_b(y)  \, ,
\label{stationphi} \\
\db \frac{\delta \Gamma[\phi,G]}{\delta G_{ab}(x,y)}
&\db =& \db -\frac{1}{2} R_{ab}(x,y) \, , 
\label{eq:station}
\eea 
which are the {\rr\it quantum equations of motion for $\phi$ and $G$}.  

To get familiar with the 2PI effective action, we note that the following approximate expression
\begin{eqnarray} \db
\Gamma^{\rr ({\rm 1loop})}[\phi,G] &\db = & \db S[\phi] + {\rr \frac{i}{2} \Tr_\C \ln G^{-1} }
          + \frac{i}{2} \Tr_\C \left\{ \left(G_0^{-1}\! (\phi) - G^{-1}\right)\, G\right\}  
          \nonumber\\
 &\db =& \db \Gamma^{R {\rr ({\rm 1loop})}}[\phi] - \frac{1}{2} \int_{xy,\C} 
 \left[ \phi_a(x)\phi_b(y) + G_{ab}(x,y) \right] R_{ab}(x,y) \qquad 
\label{eq:2PIoneloop} 
\end{eqnarray}
indeed corresponds to the Legendre transform of the one-loop result (\ref{G1Roneloop}) for the 1PI effective action $\Gamma^{R}[\phi]$ as written in the second line of (\ref{eq:2PIoneloop}). To see this, we first observe that the stationarity equation (\ref{eq:station}) yields in this case
\begin{equation}
\db \frac{\delta \Gamma^{\rr ({\rm 1loop})}[\phi,G]}{\delta G_{ab}(x,y)}
\, = \, {\rr -\frac{i}{2} G^{-1}_{ab}(x,y) + \frac{i}{2} G_{0,ab}^{-1}(x,y;\phi) \, = \, -\frac{1}{2} R_{ab}(x,y) }
\end{equation}
or $G^{-1} = G_0^{-1}(\phi) - i R$. Using this to replace $\left(G_0^{-1}(\phi)-G^{-1}\right)$ in the first line of (\ref{eq:2PIoneloop}) yields with the definitions (\ref{eq:SR}) and (\ref{twostepLT}) the result.

To go beyond this approximation it is convenient to write the exact 
$\Gamma[\phi,G]$ as the one-loop type expression in the first line of (\ref{eq:2PIoneloop})
and a {\rr\it ``rest''}, $\Gamma_2[\phi,G]$: 
\bea
\mbox{\framebox{$\displaystyle 
{\db \,\, \Gamma[\phi,G] = S[\phi] + \frac{i}{2} \Tr_\C \ln G^{-1} 
          + \frac{i}{2} \Tr_\C \, G_0^{-1}\! (\phi)\, G } 
{\rr          \, + \, \Gamma_2[\phi,G]}\db + {\rm const} \,\, $}}
\label{2PIaction}
\eea
Here we have written $\Tr_\C G^{-1} G$ as an irrelevant constant which can be adjusted 
for normalization. 
To get an understanding of {\rr\it $\Gamma_2[\phi,G]$} 
we vary this expression with respect to $G$, which yields  
\bea \db
G_{ab}^{-1}(x,y) &\db =& \db G_{0,ab}^{-1}(x,y;\phi)  
- i R_{ab}(x,y) 
- {\rr \Sigma_{ab}(x,y;\phi,G)} \, ,
\label{SchwingerDysonR}
\eea
where we have defined: 
\bea
\mbox{\framebox{\rr
$\displaystyle \,\, \Sigma_{ab}(x,y;\phi,G) \equiv   
2 i\, \frac{\delta \Gamma_2[\phi,G]}{\delta G_{ab}(x,y)}\,\,$}}
\label{exactsigma}
\eea
This gives the {\rr\it proper self-energy}, $\Sigma_{ab}(x,y;\phi,G)$, in terms of a functional derivative of $\Gamma_2[\phi,G]$ with respect to $G_{ab}(x,y)$. 

Inverting (\ref{SchwingerDysonR}), we may express for further interpretation the full propagator $G$ as an infinite series using a compact matrix notation: 
\bea\db
G &\db =&\db (G_0^{-1} - i R)^{-1} + (G_0^{-1} - i R)^{-1}\, \Sigma\,
(G_0^{-1} - i R)^{-1} \nonumber\\
&&\db + \, (G_0^{-1} - i R)^{-1}\, \Sigma\, (G_0^{-1} - i R)^{-1}
\, \Sigma\, (G_0^{-1} - i R)^{-1} + \ldots 
\label{expandG}
\eea
A term is called {\it \rr one-particle irreducible} (1PI) if it remains connected when an arbitrary internal propagator $(G_0^{-1} - iR)^{-1}$ is removed. For instance, removing in the expression $(G_0^{-1} - iR)^{-1} \, \Sigma \, (G_0^{-1} - iR)^{-1}\, \Sigma \, (G_0^{-1} - iR)^{-1}$ the inner propagator, we end up with two disconnected pieces $(G_0^{-1} - iR)^{-1} \, \Sigma $ and $\Sigma \, (G_0^{-1} - iR)^{-1}$. Therefore, that contribution is {\rr \it one-particle  reducible}. If we would start classifying all one-particle reducible and irreducible structures contributing to $G$ accordingly, we would end up with the series (\ref{expandG}) with the self-energy $\Sigma$ containing all 1PI contributions. If this would not be the case, then we could always write $G$ as the series in (\ref{expandG}) plus a correction $\Delta G$. However, since the sum of the series is known to be given by (\ref{SchwingerDysonR}) this correction must be zero, i.e.\ $\Delta G = 0$. 

Most importantly, from the fact that $\Sigma(\phi,G)$
contains only 1PI contributions one can conclude the important property of $\Gamma_2[\phi,G]$ that it only contains {\it \rr two--particle irreducible} (2PI) contributions with respect to the full propagator $G$. A contribution is said to be two-particle
irreducible if it does not become disconnected by removing two inner propagators. 
Suppose $\Gamma_2[\phi,G]$ 
had a {\rr\it two--particle reducible} contribution. The latter
could be written as $\tilde{\Gamma} G G \tilde{\Gamma}'$, where $GG$ denotes 
in a matrix notation two propagators connecting two parts 
$\tilde{\Gamma}$ and $\tilde{\Gamma}'$ of a contribution.
Then $\Sigma(\phi,G)$ would contain a contribution of the form 
$\tilde{\Gamma} G \tilde{\Gamma}'$ since it is given by a derivative
of $\Gamma_2$ with respect to $G$. Such a structure is {\rr\it 
one-particle reducible} and cannot occur for the proper self-energy. Therefore, 
two-particle reducible contributions to $\Gamma_2[\phi,G]$ have to be absent.

This can be conveniently illustrated using a diagrammatic language, where propagators are associated to lines that can meet in the presence of interactions. For instance, for a quartic interaction this is exemplified
for a two- and a three-loop graph contributing to $\Gamma_2$ below. 
Diagrammatically, the graphs contributing to $\Sigma(\phi,G)$
are obtained by opening one propagator line in graphs 
contributing to $\Gamma_2[\phi,G]$, where the corresponding contributions are schematically given on the right:

\vspace*{0.3cm}

$\qquad$ \parbox{1.cm}{$\db \Gamma_2$:}
\parbox{2cm}
{\centerline{\epsfig{file=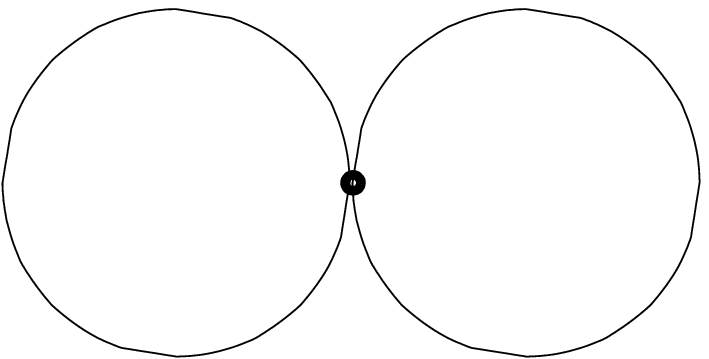,width=2.25cm}}

\vspace*{0.1cm}

\centerline{\epsfig{file=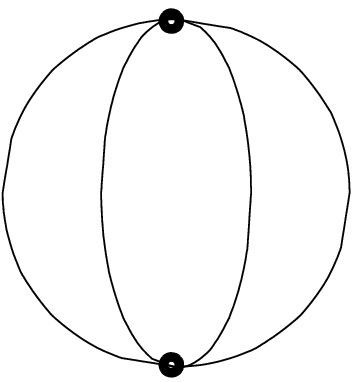,width=1.2cm}}
} 
\parbox{2.3cm}{

\vspace*{0.4cm}

$\qquad\longrightarrow\qquad\quad$}
\parbox{2.8cm}{$\db \Sigma \sim \delta \Gamma_2/\delta G$:}
\parbox{2cm}
{\centerline{\epsfig{file=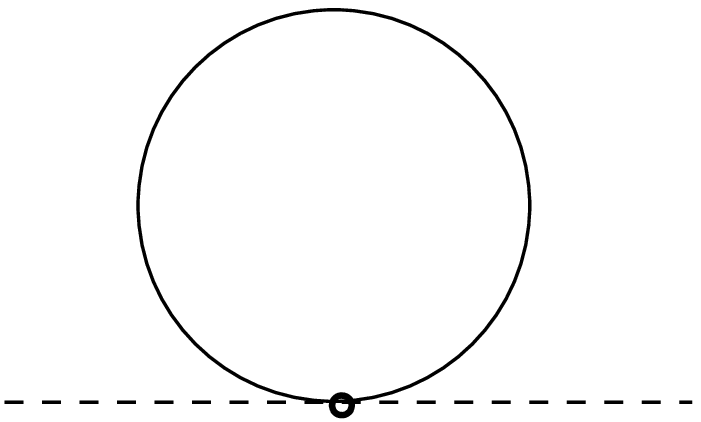,width=2.cm}}
\vspace*{0.1cm}

\centerline{\epsfig{file=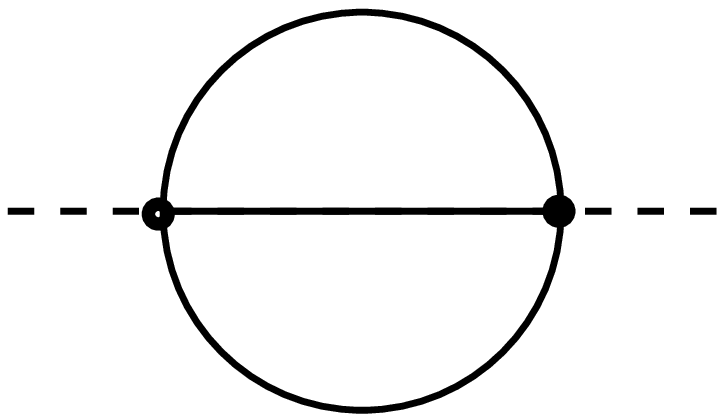,width=2.1cm}}
}

\vspace*{0.3cm}

\noindent
In these diagrams all lines are associated to the propagator $G$.
Using the expansion (\ref{expandG}) we see that each diagram corresponds to an infinite 
series of diagrams with lines associated to the ``classical'' propagator $(G_0^{-1} - i R)^{-1}$, and the above two- and three-loop diagrams
contain, e.g., so-called ``daisies'' and
``ladder'' resummations: 
\begin{figure}[h]
\centerline{\epsfig{file=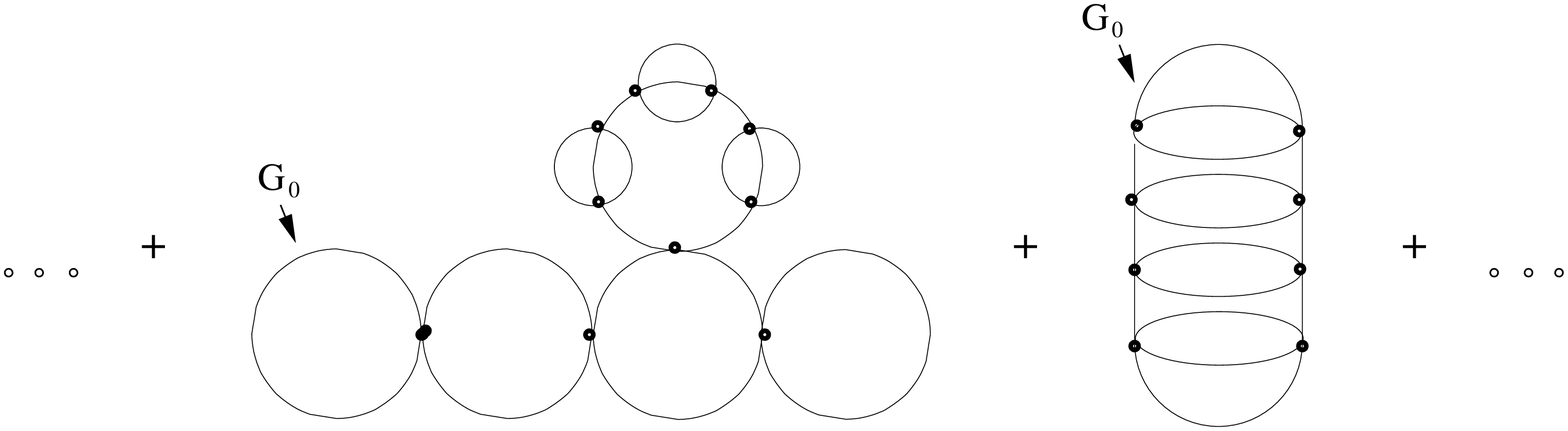,width=11.cm}}
\end{figure}

\noindent
The diagrammatic loop expansion of the 2PI effective action will be discussed in detail in a subsequent chapter.

\pagebreak

\noindent
{\bf \large Details: One-loop effective action}\\

\noindent
We derive the approximate result (\ref{G1Roneloop}) for the effective action $\Gamma^{R}[\phi]$ using a saddle-point approximation. Starting from the Legendre transform (\ref{GammaR}), we have 
\begin{equation} \db
\Gamma^{R}[\phi] \, = \, - i \ln Z[J,R] - \int_{x} J_a(x) \phi_a(x) \, .
\end{equation}
Using the defining functional integral (\ref{modZ}) this can be written as
\begin{eqnarray} \db
e^{i \Gamma^{R}[\phi]} &\db = & \db \int \mathscr{D} \varphi \, \exp i \left\{ S^R[\varphi] +
\int_{x} J_a(x) (\varphi_a(x) - \phi_a(x)) \right\}
\nonumber\\
&\db = & \db \int \mathscr{D} \varphi \, \exp i \left\{ S^R[\phi + \varphi] +
\int_{x} J_a(x) \varphi_a(x) \right\} 
\nonumber\\
&\db = & \db e^{i S^R[\phi]} \, \int \mathscr{D} \varphi \, \exp i \left\{ S^R[\phi + \varphi] - S^R[\phi] + \int_{x} J_a(x) \varphi_a(x) \right\} 
, \qquad
\end{eqnarray}
where for the second equality above we employed a field shift 
\begin{equation}
\db \varphi_a(x) \, \rightarrow \, \varphi_a(x) + \phi_a(x) \, .
\label{eq:fieldshift}
\end{equation} 
Separating the classical part, we may write
\begin{equation} \db
\Gamma^{R}[\phi] \, = \, S^{R}[\varphi] + \Gamma^{R}_1[\phi]
\end{equation} 
where the fluctuation part reads 
\begin{equation}\db
\Gamma^{R}_1[\phi] \, = \, -i \ln \int \mathscr{D} \varphi \,
\exp i \left\{
S^{R}[\phi + \varphi] - S^{R}[\phi] + \int_{x} J_a(x) \varphi_a(x)  
\right\} \, .
\label{eq:fluc}
\end{equation}
Furthermore, using the definitions (\ref{eq:SR}) and (\ref{classprop}) one can verify that
\begin{eqnarray}\db
S^{R}[\phi + \varphi] - S^{R}[\phi] & \db = & \db \frac{1}{2} \int_{xy}
\varphi_a(x) \left[
i G_{0,ab}^{-1}(x,y;\phi) + R_{ab}(x,y) \right] \varphi_b(y)
\nonumber\\
&&\db 
+ \,  S_{\rm int}[\varphi,\phi] + \int_{x} \varphi_a(x)\, \frac{\delta S^{R}[\phi]}{\delta \phi_a(x)} \, .
\label{eq:powers}
\end{eqnarray}
Here we introduced the interaction term 
\begin{equation} \db
S_{\rm int}[\varphi,{\rr \phi}] \, = \, {\rr -\frac{\lambda}{6 N} \int_{x}
\phi_a(x) \varphi_a(x) \varphi_b(x)\varphi_b(x)}
- \frac{\lambda}{4! N} \int_{x} \left( \varphi_a(x)\varphi_a(x) \right)^2 \, . 
\label{eq:Sint}
\end{equation}
Starting from a classical action (\ref{eq:classical}) with a quartic self-interaction, one observes from (\ref{eq:Sint}) that an effective cubic interaction term appears in the functional integral for the effective action in the presence of a non-zero field $\phi$. 

From plugging (\ref{eq:powers}) into (\ref{eq:fluc}) we obtain with the help of the field equation (\ref{eq:phieom}), which reads
\begin{equation} \db
\frac{\delta S^R[\phi]}{\delta \phi_a(x)} + \frac{\delta \Gamma^R_1[\phi]}{\delta \phi_a(x)}
\, = \, - J_a(x) \, ,
\end{equation}
an exact functional integro-differential equation for $\Gamma^{R}_1[\phi]$:
\begin{eqnarray}\db
\Gamma^{R}_1[\phi] & \db = & \db -i \ln \int \mathscr{D} \varphi \,
\exp i \Bigg\{ \frac{1}{2} \int_{xy}
\varphi_a(x) \left[i G_{0,ab}^{-1}(x,y;\phi) + R_{ab}(x,y) \right] \varphi_b(y)
\nonumber\\
&&\db 
+ \, S_{\rm int}[\varphi,\phi] - \int_{x} \varphi_a(x) \, \frac{\delta \Gamma^R_1[\phi]}{\delta \phi_a(x)} \Bigg\} \, .
\label{eq:intdiffG1}
\end{eqnarray}
The exponential in the integrand contains a term that is quadratic in the fields, a contribution $S_{\rm int}$ with cubic and quartic field powers as well as a linear source-like term $\sim \delta \Gamma^R_1[\phi]/\delta \phi$. The linear term represents a so-called tadpole which  guarantees a vanishing expectation value of the fluctuating field, $\langle \varphi \rangle_{J,R} = 0$, after the shift (\ref{eq:fieldshift}). Its role is taken into account by taking $\phi$ to describe the macroscopic field. If we neglect the cubic and quartic field powers contained in  $S_{\rm int}$, we arrive at the approximate saddle-point or one-loop expression     
\begin{eqnarray}\db
\Gamma^{R {\rr ({\rm 1loop})}}_1[\phi]
 & \db \!  = \! & \rr -i \ln \int \!\! \mathscr{D} \varphi \,
\exp \Bigg\{\! - \frac{1}{2} \int_{xy} \!\!\!\!
\varphi_a(x) \left[G_{0,ab}^{-1}(x,y;\phi) - i R_{ab}(x,y) \right] \varphi_b(y) \Bigg\} 
\nonumber\\
& \db \!  = \! & \db { \rr -i \ln \left( {\det} \left[ G_0^{-1}(\phi) -i R \right] \right)^{-\frac{1}{2}} }
+ \, {\rm const}
\nonumber\\
& \db \!  = \! & \db {\rr \frac{i}{2}\, \Tr \ln \left[ G_0^{-1}(\phi) - i R \right]} + \, {\rm const} \, ,
\end{eqnarray}
which leads to (\ref{G1Roneloop}) up to an irrelevant constant.

\subsection{Nonequilibrium evolution equations}
\label{sec:exactevoleq}

\subsubsection{Propagator evolution equations}
\label{sec:specstat}

The functional representation of the nonequilibrium 2PI effective action,
$\Gamma[\phi,G]$, employs a one-point, $\phi$, and a 
two-point function, $G$, whose physical values may be computed 
for all times of interest starting from the initial time $t_0$.
The equations of motion are given by the 
stationarity conditions (\ref{stationphi}) and (\ref{eq:station}). 
We will consider first the scalar field theory in the  
symmetric regime, where $\mbox{$\Gamma[\phi=0,G]$} \equiv \Gamma[G]$ 
is sufficient. We come back to the case of a nonvanishing field expectation
value and consider also additional field degrees of freedom including 
fermionic and gauge fields below. 

The stationarity condition (\ref{eq:station}) for the propagator equation leads to (\ref{SchwingerDysonR}), which reads 
$G_{ab}^{-1}(x,y) = G_{0,ab}^{-1}(x,y) - \Sigma_{ab} (x,y) -i R_{ab}(x,y)$. 
Alternatively to (\ref{expandG}), we may invert this equation by writing
\begin{equation} \db
G_{ab}(x,y) \, = \, G_{0,ab}(x,y) + \int_{zw,\C} G_{0,ac}(x,z) \left[ 
\Sigma_{cd}(z,w) + i R_{cd}(z,w) \right] G_{db}(w,y) \, .
\label{eq:invSD} 
\end{equation}
This can be directly verified with the help of (\ref{SchwingerDysonR}) by acting with $G^{-1}$ on the left of (\ref{eq:invSD}) to get $G G^{-1} = \mathds{1}$ and with $G_0^{-1} - \Sigma - i R$ on the right of (\ref{eq:invSD}) to get in matrix notation:
\bea
\db 
 \Big(G_{0} + G_{0} \left[ 
\Sigma + i R \right] G \Big) 
\left( G_0^{-1} - \Sigma - i R \right)
&\db = &\db G_0 G_0^{-1} - G_0 \left[ 
\Sigma + i R \right] \nonumber\\ 
&\db + &\db  G_0 \left[ 
\Sigma + i R \right] G \underbrace{\left( G_0^{-1} - \Sigma - i R \right)}
\nonumber\\
&\db = & \db \mathds{1} \qquad\qquad\qquad\qquad\,\, G^{-1}
\eea
 
Following section \ref{sec:initialconditions}, the initial Gaussian density matrix leads to a non-zero source term $R(x,y)$ at initial times $x^0,y^0=t_0$ only. Therefore, (\ref{eq:invSD}) may be used to relate source terms to initial values of the propagator and its first derivatives for the relativistic theory (\ref{eq:classical}). However, typically the initial values for correlation functions are known and it would be favorable to prescribe them directly without recourse to sources. In particular, we may 
rewrite the equation of motion as a {\rr\it partial differential equation}. For instance, starting from (\ref{SchwingerDysonR}) convolution with $G$ using
$\int_{z,\C} G_{ac}^{-1}(x,z)G_{cb}(z,y) = \delta_{ab} \delta_\C (x-y)$ leads to 
\beq \db
\int_{z,\C} G_{0,ac}^{-1}(x,z)G_{cb}(z,y) - \int_{z,\C} 
\left[\Sigma_{ac}(x,z) + i R_{ac}(x,z)\right] G_{cb}(z,y) \,=\, \delta_{ab} \delta_\C(x-y) \, .
\label{eq:exactG}
\eeq
In the absence of a macroscopic field we have
\begin{equation} \db
i G_{0,ab}^{-1}(x,y) \, = \, - \left[\square_x + m^2\right] \delta_{ab} \delta_\C(x-y) 
\label{eq:freeprop}
\end{equation}
and one obtains the evolution equation for the 
propagator for $\phi=0$:
\bea
\mbox{\framebox{\rr
$\,\, \displaystyle
\left( \square_x + m^2 \right) G_{ab}(x,y) 
+\, i \int_{z,\C} \! \left[ \Sigma_{ac}(x,z) + i R_{ac}(x,z) \right] G_{cb}(z,y)
 = - i \delta_{ab} \delta_\C(x-y) $
}}
\!\!\!\!\!\!\!\!\nonumber\\
&& 
\label{eq:exactevolG}
\eea
which is an exact equation for known self-energy $\Sigma$. Since a source term representing the initial density matrix has support at initial time only, it gives a vanishing contribution to $\int_{z,\C} R_{ac}(x,z) G_{cb}(z,y)$ for $x^0 > t_0$. To solve this equation for some given approximation of $\Sigma_{ab}(x,y)$, the values for $G_{ab}(x,y)$ and first derivatives have to be prescribed at $x^0,y^0=t_0$. Then (\ref{eq:exactevolG}) is used to evolve $G_{ab}(x,y)$ for times larger than $t_0$.  We use this to describe in the following the evolution equation for physical propagators setting $R = 0$ in (\ref{eq:exactevolG}).    

It is convenient to introduce a decomposition of the
two-point function $G$ into spectral and statistical components.
The corresponding evolution
equations for the spectral function and statistical 
propagator are fully equivalent to the evolution equation
for $G$, but have a simple physical interpretation.
While the spectral function encodes the
spectrum of the theory, the statistical 
propagator gives information about occupation numbers.
Loosely speaking, the decomposition makes explicit
what states are available and how often they are
occupied. This interpretation will become more clear as we proceed. 

For the neutral scalar field theory 
there are two linearly independent real--valued two--point functions, which 
may be associated to the real and the imaginary part of $G$. More precisely, we consider the expectation value of the commutator and 
the anti-commutator of two fields,
\bea
\mbox{\rr commutator:} 
&&\rr \rho_{ab}(x,y) \, = \, i \langle [\Phi_a(x),\Phi_b(y)] \rangle \, ,
\label{eq:comrho}\\
\mbox{\db anti-commutator ($\phi=0$):} 
&&\db F_{ab}(x,y) \, = \, \frac{1}{2} \langle \{\Phi_a(x),\Phi_b(y)\} \rangle \, .
\label{eq:anticomF}
\eea
Here $\rho_{ab}(x,y)$ denotes the {\it \rr spectral function} and 
$F_{ab}(x,y)$ the {\it \rr statistical two-point function}. For neutral scalar fields the real functions obey 
$F_{ab}(x,y)=F_{ba}(y,x)$ and $\rho_{ab}(x,y)=-\rho_{ba}(y,x)$. 
We note from (\ref{eq:comrho}) that the spectral function $\rho$ 
encodes the equal-time commutation relations
\beq\rr
\rho_{ab}(x,y)|_{x^0=y^0} = 0 \quad, \quad 
\partial_{x^0}\rho_{ab}(x,y)|_{x^0=y^0} = \delta_{ab} \delta(\bx-\by) \, .
\label{eq:bosecomrel}
\eeq 
The spectral function is also directly related to the retarded or advanced propagator, $G^R_{ab}(x,y) = \rho_{ab} (x,y) \Theta (x^0 - y^0) = G^A_{ab}(y,x)$, respectively.

The decomposition identity for spectral and statistical 
components of the propagator in the absence of sources reads:
\beq
\mbox{\framebox{\db
$\,\, \displaystyle
G_{ab}(x,y) \, = \, F_{ab}(x,y) - \frac{i}{2}\, {\rr\rho_{ab}(x,y)}\, \sgn_{\C}(x^0 - y^0)$
}}\label{eq:decompid}
\eeq
Here $\sgn_\C (x^0 - y^0) = \theta_\C (x^0 - y^0) - \theta_\C (y^0 - x^0)$
using the contour step function introduced in (\ref{eq:contourstep}).
With this the above decomposition is directly understood as 
\bea \db 
G_{ab}(x,y) &\db \stackrel{(\phi=0)}{=}&\db \langle \Phi_a(x) \Phi_b(y) \rangle \theta_{\C}(x^0 - y^0) 
+ \langle \Phi_b(y) \Phi_a(x) \rangle \theta_{\C}(y^0 - x^0) \nn
&\db =&\db \frac{1}{2} \langle \{\Phi_a(x), \Phi_b(y)\} \rangle
\left( \theta_{\C}(x^0 - y^0) + \theta_{\C}(y^0 - x^0) \right) \nn
&\db -& \db   \frac{i}{2}\, {\rr i \langle [\Phi_a(x), \Phi_b(y)] \rangle}
\underbrace{\left(\theta_{\C}(x^0 - y^0) - \theta_{\C}(y^0 - x^0)\right)}  \nn
&&\db  \qquad\qquad \qquad\qquad \qquad  \sign_{\C}(x^0 - y^0)  \, . \nnn
\eea
Making the contour ordering explicit, we distinguish the four propagators 
\bea
\db G^{++}_{ab}(x,y) &\db = & \db F_{ab}(x,y) - \frac{i}{2}\, {\rr \rho_{ab}(x,y)}\, \sgn(x^0-y^0) \, ,
\nonumber\\
\db G^{--}_{ab}(x,y) &\db = & \db F_{ab}(x,y) + \frac{i}{2}\, {\rr \rho_{ab}(x,y)}\, \sgn(x^0-y^0) \, ,
\nonumber\\
\db G^{+-}_{ab}(x,y) &\db = & \db F_{ab}(x,y) + \frac{i}{2}\, {\rr \rho_{ab}(x,y)} \, , 
\nonumber\\
\db G^{-+}_{ab}(x,y) & \db = & \db F_{ab}(x,y) - \frac{i}{2}\, {\rr \rho_{ab}(x,y)}\, , \qquad
\eea
which correspond to the definitions (\ref{eq:Gpp}) -- (\ref{eq:Gpm}) with the standard sign function, $\sgn(x^0-y^0) = \theta(x^0-y^0) - \theta(y^0-x^0)$. 
Since $F_{ab}(x,y)$ and $\rho_{ab}(x,y)$ are not time-ordered correlation functions, no distinction between $\C^+$ and $\C^-$ for the location of their time arguments has to be taken into account here.

To obtain a similar decomposition for the self-energy,
we separate $\Sigma$ in a ``local'' 
and ``nonlocal'' part according to
\beq\db
\Sigma_{ab}(x,y) \, = \, - i \Sigma^{(0)}_{ab}(x)\, \delta(x-y)
+ \ol{\Sigma}_{ab}(x,y) \, . 
\label{eq:sighominh}
\eeq
Since $\Sigma^{(0)}$ just corresponds to a space-time dependent 
mass-shift it is convenient for the following to introduce the notation
\beq\db 
M^2_{ab}(x) \, \stackrel{(\phi=0)} {=} \, m^2 \delta_{ab} + \Sigma^{(0)}_{ab}(x) \, .
\label{eq:localself}
\eeq
To make the time-ordering for the non-local part of the self-energy, 
$\ol{\Sigma}_{ab}(x,y)$, explicit we can use the same identity as for
the propagator (\ref{eq:decompid}) to decompose:
\bea\db 
\ol{\Sigma}_{ab} (x,y) \, = \, \Sigma^F_{ab}(x,y) - \frac{i}{2} {\rr \Sigma^{\rho}_{ab}(x,y)}\, 
\sign_{\C}(x^0 - y^0) \, .
\label{eq:decompself}
\eea

Out of equilibrium we have to follow the
time-evolution both for the statistical propagator,
$F$, as well as for the spectral function, $\rho$.
The evolution equations are obtained from (\ref{eq:exactevolG})
with the help of the identities (\ref{eq:decompid}) and
(\ref{eq:decompself}). Most importantly, once expressed in terms
of $F$ and $\rho$, the time-ordering is explicit and the
respective sign-functions appearing in the time-ordered 
propagator can be conveniently 
evaluated along the closed real-time contour $\C$.

With the notation (\ref{eq:sighominh}) the time evolution equation for the 
time-ordered propagator (\ref{eq:exactevolG}) reads 
\beq\db
\left[ \square_x \delta_{ac} + M^2_{ab}(x) \right] G_{cb}(x,y) 
+\, i \int_{z,\C} \ol{\Sigma}_{ac}(x,z) G_{cb}(z,y)
\,=\, - i \delta_{ab} \delta_\C(x-y) \, ,
\label{eq:evoleqGM}
\eeq
where we have set $R = 0$. 
For the evaluation along the time contour $\C$ 
involved in the integration  
$\int_z \equiv \int_{\C} \rmd z^0 \int {\rmd}^d z$
we employ  (\ref{eq:decompid}) and
(\ref{eq:decompself}):
\bea 
&& \db {\db i \int_{z,\C} \ol{\Sigma}_{ac}(x,z) G_{cb}(z,y) 
\,=\, i \int_{z,\C} } \Big\{ {\db \Sigma^F_{ac}(x,z) F_{cb}(z,y) }\nn
&&  
{\db - \frac{i}{2} \Sigma^F_{ac}(x,z) \rho_{cb}(z,y)\, \sign_{\C}(z^0-y^0)} 
{\db - \frac{i}{2} \Sigma^{\rho}_{ac}(x,z) F_{cb}(z,y)\, \sign_{\C}(x^0-z^0)}
\nonumber\\
&& \db
{\rr -\frac{1}{4} \Sigma^{\rho}_{ac}(x,z) \rho_{cb}(z,y)\, 
\sign_{\C}(x^0-z^0)\, \sign_{\C}(z^0-y^0)} 
\Big\} \label{eq:evalC} \, .
\eea
The first term on the RHS~vanishes because of integration 
along the {\rr\it closed time contour $\C$}. 
To proceed for the second term one splits the 
contour integral such that the sign-functions have a definite 
value, for instance
\beq\db
\int_{\C}\! {\rmd} z^0\, \sign_{\C}(z^0-y^0) \, = \, \int_{t_0}^{y^0}\! {\rmd} z^0 
(-1) + \int_{y^0}^{t_0}\! {\rmd} z^0 \, = \, - 2 \int_{t_0}^{y^0}\! {\rmd} z^0 \, 
\eeq 
for the closed contour with initial time $t_0$. We emphasize that the contributions from times later than $y^0$ simply cancel. To evaluate the last term on the RHS~of (\ref{eq:evalC}) it is
convenient to distinguish the cases\\

\noindent
(a) \ul{$\theta_{\C}(x^0 - y^0) = 1$}:
\beq\rr
\int_{\C}\! {\rmd} z^0\,\sign_{\C}(x^0-z^0) \sign_{\C}(z^0-y^0)
\, = \, \int_{t_0}^{y^0}\! {\rmd} z^0 (-1) + \int_{y^0}^{x^0}\! {\rmd} z^0  
+ \int_{x^0}^{t_0}\! {\rmd} z^0 (-1) \, , 
\eeq
(b) \ul{$\theta_{\C}(y^0 - x^0) =1$}:
\beq\rr
\int_{\C}\! {\rmd} z^0\,\sign_{\C}(x^0-z^0) \sign_{\C}(z^0-y^0)
\, = \, \int_{t_0}^{x^0}\! {\rmd} z^0 (-1) + \int_{x^0}^{y^0}\! {\rmd} z^0  
+ \int_{y^0}^{t_0}\! {\rmd} z^0 (-1) \, .
\eeq
One observes that (a) and (b) differ only by an overall sign factor
$\sim \sign_{\C}(x^0-y^0)$. Combining the integrals therefore
gives:
\bea
&& \!\!\!\!\!\!\!\! \db i \int_{z,\C} \ol{\Sigma}_{ac}(x,z) G_{cb}(z,y)
\, = \, \int {\rmd}^d z\, \Bigg\{
\int_{t_0}^{x^0} {\rmd} z^0\, \Sigma^{\rho}_{ac}(x,z) F_{cb}(z,y)
\nonumber\\
&& \!\!\!\!\!\!\!\! \db
- \int_{t_0}^{y^0} {\rmd} z^0\, \Sigma^F_{ac}(x,z) \rho_{cb}(z,y) 
 {\rr - \frac{i}{2} \sign_{\C}(x^0-y^0) \int_{y^0}^{x^0}{\rmd} z^0\,
\Sigma^{\rho}_{ac}(x,z) \rho_{cb}(z,y)} \Bigg\} . \qquad
\label{eq:memdec}
\eea
One finally employs 
\beq\db
\square_x G_{ab}(x,y) = \square_x F_{ab}(x,y) 
- \frac{i}{2}\, \sign_{\C}(x^0 - y^0) \square_x {\rr \rho_{ab}(x,y)} 
- i \delta_{ab} \delta_\C(x-y)
\label{eq:canceld}
\eeq
such that the $\delta$-term cancels with the respective one on the RHS~of
the evolution equation (\ref{eq:evoleqGM}). Here we have used
\bea\db
-\frac{i}{2} \partial_{x^0}^2 \left[ {\rr \rho_{ab}(x,y)}\, 
\sign_{\C}(x^0 - y^0) \right]
&\db =&\db 
-\frac{i}{2} \sign_{\C}(x^0 - y^0) \partial_{x^0}^2 {\rr \rho_{ab}(x,y)} \nn
&& \db {\color{black}
\underbrace{\db - i \delta_{\C}(x^0-y^0) \partial_{x^0} {\rr \rho_{ab}(x,y)}}} \nn
&& \db \quad \quad \, - i \delta_{ab} \delta_\C(x-y) 
\,\, {\color{black},}
\nonumber
\eea
where (\ref{eq:bosecomrel}) is employed for the last line and 
the term $\sim \rho_{ab}(x,y) \delta_{\C}(x^0-y^0)$ is observed to vanish since  $\sim \rho_{ab}(x,y)|_{x^0=y^0}=0$.
Comparing coefficients, which here corresponds to
separating real and imaginary parts, one finds from (\ref{eq:memdec})
and (\ref{eq:canceld}) the equations for $F_{ab}(x,y)$ and $\rho_{ab}(x,y)$.

Using the abbreviated notation $\int_{t_1}^{t_2}
{\rm d}z \equiv \int_{t_1}^{t_2} {\rm d}z^0 
\int_{-\infty}^{\infty} {\rm d}^d z$ we arrive at the 
{\rr\it coupled evolution equations for the statistical propagator
and the spectral function:} 
\beq
\framebox{
\begin{minipage}{11.0cm} \vspace*{-0.3cm}
\bea \db
\db \left[ \square_x \delta_{ac}
+ M^2_{ac}(x) \right] F_{cb}(x,y) &\db\! =\!&\db 
- \int_{t_0}^{x^0}\!\! {\rm d} z\,
{\rr \Sigma^{\rho}_{ac}(x,z)} F_{cb}(z,y)
\nonumber\\
&&\db + \int_{t_0}^{y^0}\!\! {\rm d} z\, 
\Sigma^F_{ac}(x,z) {\rr \rho_{cb}(z,y)} \, , \nonumber\\[0.1cm]
\left[\square_x \delta_{ac} + M^2_{ac}(x) 
\right] {\rr \rho_{cb}(x,y)} &\db\! =\!&\db 
- \int_{y^0}^{x^0}\!\! {\rm d} z\, 
{\rr \Sigma^{\rho}_{ac}(x,z)} {\rr \rho_{cb}(z,y)}\, . \nonumber
\eea
\end{minipage}}
\label{eq:exactrhoF}
\eeq
For the considered case with $\phi=0$, the self-energies dependent only on $F$ and $\rho$, i.e.\ $\db M^2 = M^2(F)$, $\db \Sigma^F= \Sigma^F({\rr \rho},F)$ and 
$\rr \Sigma^{\rho} = \Sigma^{\rho}(\rho,{\db F})$.
We note that the local self-energy correction (\ref{eq:localself}) 
encoded in $M^2$ does not depend on the spectral function because
the latter vanishes for equal-time arguments. 

One observes that (\ref{eq:exactrhoF}) are {\rr\it causal equations} with characteristic 
{\rr\it ``memory'' integrals,} which integrate over the time history of the
evolution. We emphasize that the presence of
memory integrals is a property of the exact theory and in
accordance with all symmetries. In particular, the equations describe a unitary time evolution without further approximations.
The equations themselves do not single out a direction of time and
they should be clearly distinguished from phenomenological
nonequilibrium equations, where irreversibility is typically put in by hand. 
Since the nonequilibrium evolution equations are exact for known self-energies, they are fully equivalent to any kind of identity for the two-point functions and are often called Kadanoff-Baym or Schwinger-Dyson equations. 

Note that the initial-time properties of the spectral function
have to comply with the equal-time commutation relations 
(\ref{eq:bosecomrel}). In contrast, for
$F_{ab}(x,y)$ as well as its first derivatives the initial conditions
at $t_0$ need to be supplied in order to solve these equations.
To make contact with the discussion of initial conditions 
in section~\ref{sec:initialconditions}, we consider for a moment the spatially
homogeneous case for which $F_{ab}(x,y) = F_{ab}(x^0,y^0;\bx - \by) =
\int {\rm d}^d p/(2 \pi)^d \, \exp[i \bp (\bx-\by)] F_{ab}(x^0,y^0; \bp)$
and equivalently for $\rho_{ab}(x,y)$. 
In terms of the Fourier components $F_{ab}(t,t'; \bp)$  
the solution of the integro-differential equations
(\ref{eq:exactrhoF}) can be solved with the following initial 
conditions respecting the $O(N)$ symmetry of the theory:
\bea
&&\db F_{ab}(t,t';\bp)_{|t=t'=t_0} \, = \,  {\rr \xi^2_\bp\, \delta_{ab}}  \, , 
\nonumber\\ 
&& \db \frac{1}{2}\left( \partial_t F_{ab}(t,t';\bp) + \partial_{t'} F_{ab}(t,t';\bp) 
\right)_{|t=t'=t_0} \,=\, {\rr \xi_\bp \eta_\bp \, \delta_{ab}}\, , \nonumber\\
&&\db \partial_t \partial_{t'} F_{ab}(t,t';\bp)_{|t=t'=t_0} \,=\,   
{\rr \left(\eta^2_\bp + \frac{\sigma^2_\bp}{4 \xi^2_\bp} \right)  \delta_{ab}}  \,\, .
\label{eq:initialcondFp}
\eea
Here we have used that the required correlators at initial time
are identical to those given in (\ref{eq:equirhoc}) for the 
considered case $\phi \equiv 0$ if the momentum labels are attached for $d > 0$.
Accordingly, these are the very same parameters that have to be specified 
for the corresponding Gaussian initial density matrix 
(\ref{eq:GaussianrhoD}). We emphasize that the initial conditions for the
spectral function equation are completely
fixed by the properties of the theory itself:
the equal-time commutation relations (\ref{eq:bosecomrel})
specify $\rho_{ab}(t,t';\bp)|_{t=t'=t_0} = 0$, 
$\partial_{t}\rho_{ab}(t,t';\bp)|_{t=t'=t_0} = \delta_{ab}$ and
$\partial_{t}\partial_{t'}\rho_{ab}(t,t';\bp)|_{t=t'=t_0} = 0$
for the anti-symmetric spectral function.

\subsubsection{Non-vanishing field expectation value}

In the presence of a non-zero field expectation value, 
$\phi \not = 0$, the same decompositions 
(\ref{eq:decompid}) and (\ref{eq:decompself}) apply.
Equivalently, one can always view (\ref{eq:decompid}) as defining $F$ and $\rho$
from the {\it \rr connected}$\,$ propagator $G$.
This means that the symmetric part, which is described by statistical two-point function, becomes in the presence of a macroscopic field 
\beq
\db F_{ab}(x,y) \, = \, \frac{1}{2} \langle \{\Phi_a(x),\Phi_b(y)\} \rangle - \phi_a(x) \phi_b(y)\, ,
\label{eq:anticomFphi}
\eeq
whereas the anti-symmetric spectral function is still given by 
$\rho_{ab}(x,y) = i \langle [\Phi_a(x),\Phi_b(y)] \rangle$.
The general form of the scalar evolution equations for the spectral 
and statistical function (\ref{eq:exactrhoF}) remains the same for $\phi \not = 0$. The only change compared to the symmetric regime is that the
functional dependence now includes field dependent terms $\db M^2 = M^2(\phi,F)$, 
$\db \Sigma_F= \Sigma_F(\phi,{\rr \rho},F)$ and 
$\rr \Sigma_{\rho} = \Sigma_{\rho}({\db \phi},\rho,{\db F})$.
    
For the $N$-component scalar field theory (\ref{eq:classical}) with the field dependent inverse classical propagator (\ref{classprop})
acting on the l.h.s.\ of the evolution equations for $F_{ab}(x,y)$ and $\rho_{ab}(x,y)$, one has ($\phi^2 \equiv \phi_a\phi_a$):
\beq\db
M_{ab}^2(x) \, = \, \left( m^2 + \frac{\lambda}{6N}\, \phi^2(x) 
\right) \delta_{ab} + \frac{\lambda}{3N}\, \phi_a(x)\phi_b(x) + \Sigma^{(0)}_{ab}(x)  \, .
\label{Meff}
\eeq
In this case the evolution equations for the
spectral function and statistical two-point function (\ref{eq:exactrhoF}) 
are supplemented by a differential equation for $\phi$ given by 
the stationarity condition (\ref{stationphi}). For this we have to compute the 
functional derivative of (\ref{2PIaction}), i.e.\
\beq \db
\frac{\delta\Gamma}{\delta \phi_a(x)} \, = \, \frac{\delta S}{\delta \phi_a(x)} + \frac{i}{2} \frac{\delta \{ \Tr\, G_0^{-1}(\phi) G\}}{\delta \phi_a(x)}
+ \frac{\delta\Gamma_2}{\delta \phi_a(x)} \, = \, - J_a(x) \,
\eeq
according to (\ref{stationphi}) for $R = 0$. Again, since a source term representing the initial density matrix has support only at time $t_0$, we have $J_a(x) = 0$ for $x^0 > t_0$ in the absence of external sources. 
With 
\beq\db
\frac{i}{2} \Tr\, G_0^{-1}(\phi) G
= - \frac{1}{2} \int_x \left(\left[\square_x + m^2+ \frac{\lambda}{6N}
\phi^2(x)\right] \delta_{ab} + \frac{\lambda}{3N}\phi_a(x)\phi_b(x)\right)
F_{ba}(x,x) 
\eeq
for the $N$-component scalar field theory this yields the {\rr\it field evolution equation:}
\beq
\mbox{\framebox{\db$\,\displaystyle
\left\{\left(\square_x + m^2 + \frac{\lambda}{6N}
\left[\phi^2(x) + F_{cc}(x,x) \right] \right) \delta_{ab} 
+ \frac{\lambda}{3N}\, F_{ab}(x,x)  
\right\} \phi_b(x) 
\, = \, \frac{\delta \Gamma_2}{\delta \phi_a(x)}\,$}}
\label{eq:exactphi}
\eeq
The solution of this equation requires specifying the field 
and its first derivative at initial time. In the context of the above 
discussion for spatially homogeneous fields this just corresponds to specifying
$\phi_a(t_0)$ and $\partial_t \phi_a(x^0)|_{x^0 = t_0}$.

\subsubsection{Evolution equations for fermions and gauge fields}

The different dynamical roles of spectral and 
statistical components is a generic property of nonequilibrium
field theory and 
not specific to scalar field degrees of freedom. 
In terms of spectral and statistical components the equations 
for {\rr\it fermionic fields} or {\rr\it gauge fields} 
have very similar structures as well. To be specific we
consider Dirac fermions $\Psi(x)$ and $\bar{\Psi} = \Psi^\dagger \gamma^0$, in matrix 
notation with Dirac matrices $\gamma^\mu$ for $\mu =0,\ldots,3$.\footnote{Dirac matrices obey $\left\{\gamma^\mu, \gamma^\nu\right\} = 2 g^{\mu\nu}$ with $g^{\mu\nu} = {\rm diag} (1,-1,-1,-1)$.} The time-ordered fermion propagator is
$\Delta (x,y) = \langle \Psi(x) \bar{\Psi}(y) \rangle \theta_{\C}(x^0 - y^0) 
- \langle \bar{\Psi}(y) \Psi(x) \rangle \theta_{\C}(y^0 - x^0)$, where the minus sign is a consequence of the anti-commuting property of fermionic fields. Correspondingly, in contrast to bosons, for fermions the 
field anti-commutator is associated to the spectral function,\footnote{We use $\langle \Psi \rangle = \langle \bar{\Psi} \rangle = 0$ in the absence of sources.}
\bea
\mbox{\rr anti-commutator:} 
&&\rr \rho^{(f)}(x,y) = i \langle \{ \Psi(x), \bar{\Psi}(y) \} 
\rangle \, ,
\label{eq:fanticomrho}\\
\mbox{\db commutator:} 
&&\db F^{(f)}(x,y) =   \frac{1}{2} \langle
[\Psi(x), \bar{\Psi}(y)]\rangle \, .
\label{eq:fcomF}
\eea
In terms of spectral 
and statistical components the propagator reads 
\beq\db
\Delta(x,y) = F^{(f)}(x,y) - \frac{i}{2} {\rr \rho^{(f)}(x,y)}\, 
\sign_{\C}(x^0 - y^0) \, .
\label{eq:fermdec}
\eeq
The equal-time anti-commutation relations for the fields are again 
encoded in $\rho^{(f)}(x,y)$. For instance, for Dirac fermions one has
\beq\rr
\gamma^0 \rho^{(f)}(x,y)|_{x^0=y^0} = i \delta(\bx-\by) \, ,
\label{eq:rhoinitial}
\eeq
which will uniquely specify the initial conditions for the evolution equation for $\rho^{(f)}$ similar to what is observed for bosons.

For Dirac fermions with 
mass $m^{(f)}$ the free inverse propagator reads
\bea \db
i \Delta_0^{-1} (x,y) 
= [i \slash \!\!\! \partial_{x}\, - m^{(f)} ]\, \delta (x-y)\, ,
\label{eq:freedirac}
\eea
where $\slash \!\!\! \partial \equiv \gamma^\mu \partial_\mu$. 
Similarly to the bosonic case, we may write in the absence of sources
\beq\db
\Delta^{-1}(x,y)\, =\, 
\Delta_{0}^{-1}(x,y) - {\rr \Sigma^{(f)}(x,y)} \, ,
\label{eq:fermSD}
\eeq
which defines the proper fermion self-energy $\Sigma^{(f)}(x,y)$.  
The corresponding decomposition for the fermion self-energy 
reads
\beq
\db \Sigma^{(f)}(x,y) \,=\, 
\Sigma_{F}^{(f)} (x,y) - \frac{i}{2}\, {\rr \Sigma_{\rho}^{(f)}(x,y)}\, 
\sign_\C(x^0-y^0)  \, .
\label{eq:selffermdec}
\eeq
If there is a local contribution to the proper self-energy,
this is to be separated in complete analogy to the scalar equation (\ref{eq:sighominh}) and
the decomposition (\ref{eq:selffermdec}) is taken for the non-local part
of the self-energy.

From the equation of motion for
the time-ordered fermion propagator (\ref{eq:fermSD}) we obtain a suitable time evolution equation by convoluting with~$\Delta$. For a free inverse propagator as in 
(\ref{eq:freedirac}) for Dirac fermions this yields 
\beq\db
\left[i \slash \!\!\! \partial_{x} - m^{(f)} \right] 
\Delta(x,y) - i \int_{z,\C}\, \Sigma^{(f)}(x,z) \Delta(z,y)
\, = \, i \delta_{\C}(x-y) \, .
\label{eq:evolDel}
\eeq 
Following the lines of the above discussion for scalars one finds
for the fermion case the coupled evolution equations
\bea\db  
\left[i \slash \!\!\! \partial_{x} - m^{(f)} \right]  
F^{(f)}(x,y) &\!\!\db =\!\!&\db  
 \int_{t_0}^{x^0}\!\! {\rm d}z\, {\rr \Sigma^{(f)}_{\rho}(x,z)} F^{(f)}(z,y)
- \int_{t_0}^{y^0}\!\! {\rm d}z\, \Sigma^{(f)}_{F}(x,z) {\rr \rho^{(f)}(z,y)} \, , 
\nonumber\\[0.2cm] \db 
\left[i \slash \!\!\! \partial_{x} - m^{(f)} \right]  
{\rr \rho^{(f)}(x,y)} &\!\!\db =\!\!&\db  
 \int_{y^0}^{x^0}\!\! {\rm d}z\,  {\rr \Sigma^{(f)}_{\rho} (x,z)
\rho^{(f)}(z,y)} \, . 
\qquad
\label{eq:Fexact}
\eea
Similarly, the nonequilibrium evolution equations 
for gauge fields can be obtained as well. For instance, for a gauge field $A^\mu(x)$ with Lorentz indices $\mu = 0, \ldots, 3$ the
full time-ordered propagator $D^{\mu\nu}(x,y) = \langle T_\C A^\mu(x) A^\nu(y) \rangle$ may be written as 
\beq\db
D^{\mu\nu}(x,y) = 
F_D^{\mu\nu}(x,y) - \frac{i}{2} {\rr \rho_D^{\mu\nu}(x,y)}\,
{\rm sgn}_{\C} (x^0-y^0) \, .
\label{eq:decompgauge}
\eeq
For a theory with
free inverse gauge field propagator given by
\beq \db
i D^{-1}_{0,\mu\nu}(x,y) 
= \left[g_{\mu\nu}\, \square - \left( 1 - \xi^{-1}\right) 
\partial_\mu \partial_\nu \right]_x \delta(x-y) 
\eeq
for covariant gauges with gauge-fixing parameter $\xi$ 
and vanishing macroscopic field, $\langle A^\mu(x) \rangle = 0$, one 
finds the respective equations from (\ref{eq:exactrhoF}) by 
\beq \db
\left[ \square_x + {M^2} \right] {\rr \rho(x,y)} \,\,\, \longrightarrow\,\,\, 
- \left[{g^\mu}_\gamma \square - (1-\xi^{-1}) \partial^\mu 
\partial_{\gamma} \right]_x {\rr \rho_D^{\gamma\nu}(x,y)} \, 
\eeq
and equivalently for $F_D^{\gamma\nu}(x,y)$. Of course, 
the respective Lorentz and internal indices have to be attached to the corresponding
self-energies on the RHS~of the equations.\\ 

\pagebreak

\noindent
{\bf \large Details: Functional integral for fermions}\\

\noindent
In section \ref{sec:functionalintegral} we derived a bosonic functional integral in terms of eigenvalues of the bosonic field operators. Fermionic operators anti-commute, which is intimately connected with the Pauli exclusion principle. In order to represent fermionic integrals, we will have to deal with anti-commuting eigenvalues of fermionic field operators. Anti-commuting numbers and functions are called Grassmann variables. Her we recall some basic aspects which we need for our purposes and refer to the literature for more extensive discussions.

For a single anti-commuting variable $\eta$ we have  
\beq \db
\{\eta,\eta\} \, = \, 0 \, ,
\label{eq:grass}
\eeq 
such that $\eta \eta = - \eta \eta$, thus $\eta^2 = 0$. The derivative operator $\rmd / \rmd \eta$ can be defined in an analogous way as for ordinary numbers and we write
\beq \db
\left\{ \frac{\rmd}{\rmd\eta}\, ,\eta \right\} \, = \, 1 \,.
\label{eq:etadiff}
\eeq
As a consequence of (\ref{eq:grass}), the power series expansion of any suitable function $f(\eta)$ may be written as
\beq \db
f(\eta) \, = \, a + b\, \eta
\eeq
with ordinary commuting numbers $a$ and $b$. Therefore, $\rmd^2 f(\eta)/\rmd \eta^2 = 0$ and we have
$\{\rmd/\rmd\eta,\rmd/\rmd\eta \} = 0$.

In order to define integration for Grassmann variables, we start by asking what could be the analogue of the bosonic Gaussian integral, $\int \rmd x \exp \{- a x^2\} = \sqrt{\pi/a}$~? An important property is certainly the invariance under the shift of the integration variable $x \rightarrow x + b$. To carry over such a property to the fermionic integral, we define
\beq \db
\int \rmd \eta \, \left(a + b\, \eta \right) \, \stackrel{!}{=} \, b \, .
\eeq
In fact, since $\eta^2 = 0$ the integral must be linear in $\eta$ and the only linear function which respects the property of shift invariance is a constant. Consequently, integration for Grassmann variables acts just like differentiation. To write a Gaussian integral, we consider a complex anti-commuting variable $\eta = (\eta_R + i \eta_I)/\sqrt{2}$ where $\eta_R$ and $\eta_I$ are real-valued anti-commuting variables. With $\eta^* = (\eta_R - i \eta_I)/\sqrt{2}$ we find, for instance, $\eta \eta^* = (\eta_R^2 + i \eta_I \eta_R - i \eta_R \eta_I + \eta_I^2)/2 = - i \eta_R \eta_I = - \eta^* \eta$. The Gaussian integral is
\bea \db
\int \rmd \eta^* \rmd \eta\, e^{- b\, \eta^* \eta} & \db = & 
\db \int \rmd \eta^* \rmd \eta \left( 1- b\, \eta^* \eta \right)
\nonumber\\
& \db = & 
\db b \int \rmd \eta^* \rmd \eta \,\eta \eta^* \,\, = \, \, b \, ,
\eea 
where the sign change in the second equality comes from changing the order of the variables to be able to use first $\int \rmd \eta\, \eta = 1$ and then $\int \rmd \eta^* \, \eta^* = 1$.

Next we want to consider the space of anti-commuting functions, $\eta(x)$. Generalizing (\ref{eq:etadiff}), functional differentiation is defined as
\beq \db
\left\{ \frac{\delta}{\delta\eta(x)}\, ,\eta(y) \right\} \, = \, \delta(x-y) \,.
\eeq 
The corresponding Gaussian functional is
\beq \db
\int \mathscr{D}\eta^* \mathscr{D}\eta\, \exp\left\{ 
- \int_{xy} \eta^*(x) B(x,y) \eta(y) \right\} \, = \, \det B \; , 
\eeq
where $\int \mathscr{D}\eta^* \mathscr{D}\eta = \Pi_x \rmd \eta^*(x) \rmd \eta(x)$. 

The construction of the 2PI effective action for fermionic fields
proceeds along very similar lines as for bosons. However, 
one has to take into account the anti-commuting behavior of
the fermion fields. A main differences compared to the bosonic case
can be observed from the one-loop part ($\Gamma_2\equiv 0$) of the 
2PI effective action. Using again the Dirac fermions introduced in the previous section and, for comparison, a real scalar field $\varphi$ with vanishing field expectation values
it involves the integrals:  
\bea
\!\!\!\!\! \mbox{Fermions:}\,\,\,\,\,
{\db -i \ln  \int \mathscr{D} \bar{\psi}\mathscr{D} \psi\,
e^{iS_0^{(f)}} = -i \ln \left( \det \Delta_0^{-1} \right)} 
&\db =& \db {\rr -i}\, \Tr \ln \Delta_0^{-1} \, ,\nn
\!\!\mbox{Bosons:} \qquad 
{\db -i \ln  \int \mathscr{D} \varphi\,
e^{iS_0} = -i \ln \left(\det G_0^{-1}\right)^{-\frac{1}{2}} } 
&\db =& \db {\rr \frac{i}{2}} \Tr \ln G_0^{-1}  \, .
\eea
Here $S_0^{(f)} = \int {\rm d}^4 x {\rm d}^4 y\, 
\bar{\psi}(x) i \Delta_{0}^{-1}(x,y)
\psi(y)$ denotes a fermion action that is bilinear in the
Grassmann fields. For Dirac fermions the free inverse 
propagator $\Delta_{0}^{-1}(x,y)$ is given by (\ref{eq:freedirac}) 
and $\bar{\psi} = \psi^\dagger \gamma^0$.
For the bosons $S_0$ is given by the quadratic part of (\ref{eq:classical}).
Comparing the two integrals one observes that the factor $\rr 1/2$
for the bosonic case is replaced by $\rr -1$ for the fermion fields
because of their anti-commuting property. With this difference,
following along the lines as for the bosonic case, one finds
that the 2PI effective action for fermions can be written in
complete analogy to (\ref{2PIaction}). Accordingly,
for the case of vanishing fermion field expectation
values, $\langle \Psi \rangle = \langle \bar{\Psi} \rangle = 0$, 
one has:
\beq
\mbox{\framebox{\db$\,\displaystyle \Gamma[\Delta] \, = \, {\rr -i}\, 
\Tr\ln \Delta^{-1} {\rr -i}\, \Tr\, \Delta_0^{-1} \Delta
+ {\rr \Gamma_2[\Delta]} + {\rm const}\,$}}
\label{eq:fermion2PI}
\eeq
Here {\rr\it $\Gamma_2[\Delta]$ contains all 2PI
diagrams} with lines associated to the contour time ordered 
propagator $\Delta(x,y) = \langle {\rm T}_\C \Psi(x) 
\bar{\Psi}(y) \rangle$. The trace ``$\Tr$'' includes integration over time and
spatial coordinates, as well as summation over field indices.

As for the bosonic case of equation (\ref{eq:station}), the 
equation of motion for $\Delta$ in absence of external sources is 
obtained by extremizing the effective action: 
\beq\db
\frac{\delta\Gamma[\Delta]}{\delta \Delta(x,y)} = 0 \, .
\label{eq:fermstat}
\eeq 
Using (\ref{eq:fermion2PI}) this stationarity condition
can be written as
\beq\db
\Delta^{-1}(x,y)\, =\, 
\Delta_{0}^{-1}(x,y) - {\rr \Sigma^{(f)}(x,y;\Delta)} \, ,
\eeq
with the proper fermion self-energy: 
\beq 
\mbox{\framebox{\rr$\,\displaystyle
\Sigma^{(f)}(x,y;\Delta) \equiv -i 
\frac{\delta \Gamma_2[\Delta]}{\delta \Delta(y,x)}\,$}}
\label{eq:fermselfen}
\eeq

\subsection{The special case of thermal equilibrium}
\label{sec:detourthermal}

For the neutral scalar field theory the canonical thermal equilibrium density matrix is
\beq\rr
\varrho_\beta \, = \, \frac{1}{Z_\beta}\, e^{-\beta H}
\label{eq:thermaldm}
\eeq
with the partition sum $Z_\beta = \Tr e^{-\beta H}$. For some time $t_0$ we take $\varrho(t_0) = \varrho_\beta$ as the density matrix in the generating functional (\ref{eq:definingZneq}). The construction of a path integral representation follows then closely section \ref{sec:nonequgenfunc}. The main difference is that $e^{-\beta H} = e^{-i (-i\beta) H}$ may be interpreted as an evolution operator in imaginary time. This simplifies taking into account a thermal density matrix which is not Gaussian in general.
Following the steps of section \ref{sec:nonequgenfunc}, we evaluate the trace using eigenstates of the Heisenberg field operator $\Phi(x)$ at time $x^0 = t_0$, $\Phi(t_0,\bx)|\varphi\rangle = \varphi_0(\bx)|\varphi\rangle$. The partition function may then be written as
\beq
\db Z_{\rr \beta} \, = \, \int [\rmd \varphi_{0}] \, \langle \varphi| \: {\rr \varrho_\beta} \: |\varphi\rangle ~ .
\label{eq:partitionfunction}
\eeq
The interpretation of the canonical equilibrium density matrix as an evolution operator in imaginary time allows us to consider  
\beq \db
\langle \varphi | {\rr e^{-\beta H}} | \varphi \rangle \, \equiv \, \left(\varphi,t_0 {\rr -i \beta}| \varphi,t_0 \right) 
\eeq
as a transition amplitude. As a consequence, we can use (\ref{eq:transitionpath}) applied to imaginary times in order to represent it as a path integral. Since we are calculating a trace, we have to identify 
\beq \db
\varphi(t_0{\rr -i\beta},\bx) \, = \, \varphi(t_0,\bx) \, = \, \varphi_0(\bx) \, .
\label{eq:periodic}  
\eeq
This periodicity of the field in thermal equilibrium represents an important constraint which is, in general, not present out of equilibrium. 

In order to write down a generating functional also for real-time correlation functions in thermal equilibrium, we extend with $\textrm{T}_{\mathcal{C}_\beta}$ the time ordering to include the imaginary time branch from $t_0$ to $t_0-i\beta$. Here any time on the imaginary branch is considered to be later than any time on the previously discussed real-time contour $\C$. Graphically, $\mathcal{C}_\beta$ is given below:

\begin{picture}(20,120) (-80,-54)
\put(0,25){\line(1,0){200}}
\put(25,0){\line(0,1){50}}
\put(25,0){\line(0,-1){62}}
\put(170,45){\line(0,1){15}}
\put(170,45){\line(1,0){15}}
\thicklines
\put(25,30){\vector(1,0){75}}
\put(100,30){\line(1,0){75}}
\put(175,20){\vector(-1,0){75}}
\put(25,20){\line(1,0){75}}
\put(25,30){\line(1,0){150}}
\put(25,20){\line(1,0){150}}
\put(25,20){\vector(0,-1){45}}
\put(25,-25){\line(0,-1){30}}
\cCircle[30](175,25){5}[r]
\put(25,30){\circle*{2.4}}
\put(25,-55){\circle*{2.4}}
\put(49,-55){\makebox(0,0){$t_0-i \beta$}}
\put(179,54){\makebox(0,0){$x^{0}$}}
\put(16,14){\makebox(0,0){$t_0$}}
\put(188,10){\makebox(0,0){$\rightarrow \infty$}}
\end{picture}\\

\noindent
We also include sources $J(x)$ and $R(x,y)$ and use the notation $\int_{x,\C_\beta} \equiv \int_{\C_\beta} \rmd x^0 \int \rmd^d x$. The generating functional for contour time ordered correlation functions is then given by 
\begin{eqnarray}
\db Z_{\rr \beta}[J,R] 
&\db =& \db \int [\rmd \varphi_{0}] \, \langle \varphi| \: {\rr \varrho_\beta} \, \textrm{T}_{\rr \mathcal{C}_\beta} \,\exp i \left\{ \int_{x, {\rr \mathcal{C}_\beta}} \Phi(x) J(x) \right. 
\nonumber\\
&& \db +\: \left. \frac{1}{2} \int_{xy, {\rr \mathcal{C}_\beta}} \Phi(x) R(x,y) \Phi(y) \right\} |\varphi\rangle ~
\label{eq:GenFunctbeta}
\end{eqnarray}
Inserting complete sets of states along the contour following section \ref{sec:nonequgenfunc}, we obtain
\beq \db
Z_{\rr \beta}[J,R] 
\, = \, \int_{\rm periodic}\!\!\!\!\! \mathscr{D}
\varphi\, \exp i \left\{ S_{\rr \beta}[\varphi]+ \int_{x,{\rr \C_\beta}}\!\!\!
J(x)\varphi(x)+\frac{1}{2}
\int_{x y,{\rr \C_\beta}}\!\!\! R(x,y) \varphi(x)\varphi(y)
\right\},
\label{eq:generatingthermal}
\eeq
where the term periodic refers to the periodicity condition for the fields (\ref{eq:periodic}). The thermal equilibrium contour action, $S_\beta$, for a theory with quartic self-interaction reads
\beq\db
S_{\rr \beta}[\varphi] \, = \, \int_{x,{\rr \C_\beta}}
	\left\{\frac{1}{2}\partial^\mu\varphi(x)\partial_\mu\varphi(x)  
	-\frac{m^2}{2}\varphi(x)\varphi(x) 
	-\frac{\lambda}{4!}\left(\varphi(x)\varphi(x)\right)^2
 \right\}  \, .
\label{eq:classicalthermal}
\eeq

For the specific case of calculating static quantities, the real-time extent of $\C_\beta$ can be taken to zero. Static quantities can then be obtained from the corresponding generating
functional with the imaginary-time interval $[0,-i\beta]$ only, which is often 
formulated in ``Euclidean'' space-time with the introduction of a Euclidean time $\tau = i t$
and Euclidean action $S_{E} = -i S$. Because of the periodicity condition (\ref{eq:periodic}), the Euclidean time dimension is compact. In Fourier
space this leads to a discrete set of so-called Matsubara frequencies, which we will not consider here. Quantities like the spectral function are, however, difficult to obtain from such a formulation since they would require analytic continuation. The above time contour including the real-time axis gives direct access to these quantities. We emphasize that in either formulation -- including real times or not -- all correlation functions are computed here with the thermal density matrix (\ref{eq:thermaldm}) such that the system is always in equilibrium.  

Contour time ordered correlation functions in thermal equilibrium can be obtained from the generating functional (\ref{eq:generatingthermal}) by functional differentiation. For instance, the two-point function is given by
\beq\db
\frac{1}{Z_\beta}\frac{\delta^2 Z_{\rr \beta}[J,R]}{\delta J(x)\delta J(y)}\Big|_{J,R=0} \, = \, 
\langle \textrm{T}_{\rr \C_\beta} \Phi(x)\Phi(y) \rangle
\, = \, G^{\rm (eq)}(x-y) + \phi \phi \, .
\label{eq:twopointeq}
\eeq
For the last equality we have used that thermal equilibrium is translation invariant, such that  the one-point function, $\phi$, is homogeneous and, accordingly, the two-point function 
depends only on relative coordinates, $G^{\rm (eq)} (x,y) = G^{\rm (eq)} (x-y)$. Similar to (\ref{eq:decompid}), we may write
\beq \db
G^{\rm (eq)}(x-y) \, = \, F^{\rm (eq)}(x-y) - \frac{i}{2}\, \rho^{\rm (eq)}(x-y)\, \sgn_{\rr \C_\beta}(x^0 - y^0)
\label{eq:decompbeta}
\eeq
with the contour sign function $\sgn_{\C_\beta}(x^0 - y^0) = \theta_{\C_\beta}(x^0 - y^0) - \theta_{\C_\beta}(y^0 - x^0)$.

We now derive an important consequence of the periodicity condition (\ref{eq:periodic}) for correlation functions. At the initial time $t_0$ of the contour $\C_\beta$, we have
\beq \db
\langle \textrm{T}_{\rr \C_\beta} \Phi(x)\, \Phi(y) \rangle|_{x^0 = t_0}
\, = \, \langle \Phi(y)\Phi(x) \rangle|_{x^0 = t_0} 
\eeq
and at the end point $t_0-i\beta$,
\beq \db
\langle \textrm{T}_{\rr \C_\beta} \Phi(x)\, \Phi(y) \rangle|_{x^0 = t_0 {\rr - i \beta}} \, = \, \langle \Phi(x)\Phi(y) \rangle|_{x^0 = t_0 {\rr - i \beta}} \, .
\eeq
The periodicity condition for the fields in thermal equilibrium then implies\footnote{The periodicity 
condition is also called Kubo-Martin-Schwinger (``KMS'') condition.} 
\beq \db
\langle \Phi(y)\Phi(x) \rangle|_{x^0 = t_0} \, = \, \langle \Phi(x)\Phi(y) \rangle|_{x^0 = t_0 {\rr - i \beta}} \; .
\label{eq:periodictycorrelation}
\eeq
Using the decomposition (\ref{eq:decompbeta}) this reads
\beq \db 
\left( F^{\rm (eq)}(x-y) + \frac{i}{2}\, \rho^{\rm (eq)}(x-y) \right)\!\Big|_{x^0 = t_0} \!\!\! = \, \left( F^{\rm (eq)}(x-y) - \frac{i}{2}\, \rho^{\rm (eq)}(x-y) \right)\!\Big|_{x^0 = t_0 {\rr - i \beta}}.
\label{eq:period2}
\eeq
Translation invariance also makes it convenient to consider the Fourier transform with {\it real}$\,$ four-momentum, $(\omega,\bp)$, 
\beq\db
F^{\rm (eq)}(x-y) \, = \int \frac{\rmd \omega \rmd^{d} p}{(2\pi)^{d+1}} 
e^{-i \omega (x^0-y^0) + i \bp (\bx - \by)} F^{\rm (eq)}(\omega,\bp)
\eeq
and equivalently for the spectral part.
Taking for a moment for granted that these integrals can be properly regularized and defined for the considered quantum field theory, (\ref{eq:period2}) reads in Fourier space
\beq\db
F^{\rm (eq)}(\omega,\bp) + \frac{i}{2} \rho^{\rm (eq)}(\omega,\bp) 
\, = \, {\rr e^{-\beta \omega}} \left(
F^{\rm (eq)}(\omega,\bp) - \frac{i}{2} \rho^{\rm (eq)}(\omega,\bp) 
\right) \, .
\label{eq:periodicityFrho}
\eeq
Solving for $F^{\rm (eq)}(\omega,\bp)$ and using that $(1+e^{-\beta \omega})/(1-e^{\beta \omega}) =(e^{\beta \omega}+1)/(e^{\beta \omega}-1) = 1 + 2/(e^{\beta \omega}-1)$, the periodicity condition for bosons can be written as:\footnote{In our conventions
the Fourier transform of the real-valued anti-symmetric function
$\rho(x,y)$ is purely imaginary, while that of the symmetric function $F(x,y)$ is real.}
\beq
\mbox{\framebox{\db 
$\,\,\displaystyle F^{\rm (eq)}(\omega,\bp) \, = \, -i
\left(\frac{1}{2} + {\rr f_\beta (\omega)}\right) \, 
\rho^{\rm (eq)}(\omega,\bp)$}}
\label{eq:flucdissbose}
\eeq
with $f_\beta(\omega)=(e^{\beta \omega}-1)^{-1}$ 
denoting the Bose-Einstein distribution function.
Equation~(\ref{eq:flucdissbose}) relates the spectral function to the 
statistical propagator. This is also called 
{\rr\it fluctuation-dissipation relation}, which will be discussed further in later chapters. 
While $\rho^{\rm (eq)}$ encodes 
the information about the spectrum of the theory, one observes from 
(\ref{eq:flucdissbose}) that the function $F^{\rm (eq)}$ 
encodes the statistical aspects in terms of the occupation number
distribution $f_\beta(\omega)$. In the same way one obtains for the Fourier
transforms of the spectral and statistical components of the
self-energy the thermal equilibrium relation
\beq\db
\Sigma^{\rm (eq)}_F(\omega,\bp) \, = \, -i
\left(\frac{1}{2} + {\rr f_\beta(\omega)} \right) 
\, \Sigma^{\rm (eq)}_{\rho}(\omega,\bp) \, .
\label{eq:sigmadecom}
\eeq
From (\ref{eq:flucdissbose}) and (\ref{eq:sigmadecom}) one obtains the identity
\beq\db
\Sigma^{\rm (eq)}_{\rho}(\omega,\bp) F^{\rm (eq)}(\omega,\bp)
- \Sigma^{\rm (eq)}_F(\omega,\bp)  \rho^{\rm (eq)}(\omega,\bp) \, =  \, 0 \, .
\eeq
Later we will see that this condition can be rewritten
as the difference of a "gain" and "loss" term in kinetic descriptions in their range
of applicability. 

For fermions the contour-ordered propagator reads in thermal equilibrium $\Delta^{\rm (eq)}(x-y) = \langle \Psi(x) \bar{\Psi}(y) \rangle
\theta_{\C_\beta}(x^0-y^0) - \langle \bar{\Psi}(y) \Psi(x) \rangle
\theta_{\C_\beta}(y^0-x^0)$ with the minus sign from the anti-commuting property of fermions. If we proceed along the lines as for bosons above, i.e.\
evaluating the contour-ordered correlation function once at the initial time and once at the final point of the contour, this minus sign leads to the 
{\it \rr anti-periodicity condition}
\beq \db
\langle \bar{\Psi}(y) \Psi(x) \rangle|_{x^0 = t_0} \, = \, {\rr - } \langle \Psi(x)\bar{\Psi}(y) \rangle|_{x^0 = t_0 {\rr - i \beta}} \; .
\label{eq:antiperiodicty}
\eeq
Using the decomposition (\ref{eq:fermdec}) for fermions, one can translate this again into a relation between the spectral and statistical two-point function.
The difference is that the Bose-Einstein distribution in 
(\ref{eq:flucdissbose}) is replaced by the Fermi-Dirac distribution
$f_{\beta}^{(f)}(\omega)=(e^{\beta \omega}+1)^{-1}$ 
according to $1/2+f_\beta(\omega)$ $\rightarrow$ $1/2-f_\beta^{(f)}(\omega)\,$
in the respective relation.

It is important to realize that for a general out of equilibrium situation the spectral and statistical components of correlation functions are 
not related by a fluctuation-dissipation relation. Such a relation   
is a manifestation of the tremendous simplification that happens if the
system can be characterized in terms of thermal equilibrium correlation functions. 
An even more stringent reduction occurs for the
vacuum, where $f_\beta(\omega) \equiv 0$ such that the spectral and statistical
function agree up to a normalization. In this respect, nonequilibrium 
quantum field theory is more complicated, since it admits the description 
of more general situations and
encompasses the thermal equilibrium or vacuum theory as special 
cases. 

\subsection{Nonrelativistic quantum field theory limit}
\label{eq:nonrelqft}

For the nonrelativistic limit of a relativistic quantum field theory, one considers processes with typical momenta of particles that are small compared to their mass $m$. Consequently, we can approximately write $\sqrt{{\mathbf p}^2 +m^2} \simeq m + {\mathbf p}^2/(2m)$. A related important aspect of the nonrelativistic limit is the absence of anti-particles, such that the typical chemical potential associated to the difference between the particle and anti-particle numbers is about $\mu \simeq m$. Therefore, in a nonrelativistic context it is often convenient to introduce the different chemical potential 
\begin{equation}
\db \mu_{\rm nr}= \mu - m \, . 
\label{eq:munr}   
\end{equation}
To be specific, we consider a relativistic quantum theory for a complex scalar field $\varphi(t,\mathbf x)$. This theory can be directly related to the field theory for $a=1,\dots,N$ real field components $\varphi_a(t,\mathbf x)$ of section~\ref{sec:effectiveactions} by taking $N=2$ and identifying
\begin{equation}
\db \varphi = \frac{1}{\sqrt{2}}\left(\varphi_1 + i \varphi_2\right) \, .
\end{equation}
In the nonrelativistic regime, this complex scalar field theory can be effectively described in terms of the Gross-Pitaevskii field theory employed already in section~\ref{sec:Bosegas} for the physics of ultracold Bose gases. 

In order to relate the relativistic and nonrelativistic descriptions, it is convenient to start from the corresponding functional integral expression (\ref{eq:pathintegral}), which represents the partition function before the conjugate momentum field is integrated out. Since the derivation in thermal equilibrium and out of equilibrium follows mostly along the same lines using the results of the previous sections -- with the main difference that the time integration 
runs along the closed time path for the nonequilibrium system and includes the interval $[0,-i\beta]$ in thermal equilibrium -- for the following we do not distinguish them in the notation. This will be sufficient to derive, in particular, the nonrelativistic version of the relativistic evolution equations (\ref{eq:exactrhoF}).

In terms of the complex field $\varphi(t,\mathbf x)$ and its conjugate momentum field $\pi(t,\mathbf x)$, the functional integral for the partition function reads
\begin{equation}
\db Z[\mu] =
\int\mathscr{D}\varphi^{*}\mathscr{D}\varphi \mathscr{D}\pi^{*}\mathscr{D}\pi \exp \left\{ i \int \rmd t \left( \int \rmd^d x\,
\left[ \pi \partial_t \varphi + \pi^{*} \partial_t \varphi^{*} \right] - \left[ H - \mu Q \right] \right) \right\}
\label{eq:Zmu}
\end{equation}
with the Hamiltonian 
\begin{equation}
\db H = \int \rmd^{3}x \left( \pi^{*}\pi +(\nabla\varphi^{*})(\nabla\varphi)
 + m^2 \varphi^{*}\varphi + \frac{\lambda}{12} (\varphi^{*}\varphi)^2 \right)
\label{eq:ham}
\end{equation}
and charge
\begin{equation}
\db Q = i\int \rmd^{3}x\; (\varphi^{*}\pi^{*} - \pi\varphi ) \, .
\end{equation}
Here we write the partition function without sources for brevity, which can always be added along the lines of section~\ref{sec:nonequgenfunc}.

The term in the exponential of (\ref{eq:Zmu}) is essentially the classical action for the interacting theory. We first consider the quadratic part of that action by setting $\lambda = 0$ in (\ref{eq:ham}), which can be written in a compact matrix notation in spatial Fourier space as
\begin{equation}
\db i \int \rmd t \int \frac{\rmd^{d}p}{(2\pi)^d}
\left( \sqrt{\omega} \varphi^*(t,\mathbf p) , \frac{-i}{\sqrt{\omega}} \pi(t,\mathbf p)  \right)
\left( \begin{array}{cc}
-\omega & i\partial_t + \mu  \\
\db i \partial_t + \mu & -\omega
\end{array} \right)
\left( \begin{array}{c} \sqrt{\omega} \varphi(t,\mathbf p) \\
\db \displaystyle{\frac{i}{\sqrt{\omega}}} \pi^*(t,\mathbf p)
\end{array} \right)
\label{eq:quadratic}
\end{equation}  
with $\omega = \sqrt{{\mathbf p}^2 + m^2}$. To proceed, we look for a canonical transformation such that the above quadratic part is diagonal in the new fields. This is achieved by introducing
\begin{equation}
{\db \chi = \sqrt{\frac{\omega}{2}}\varphi + \frac{i}{\sqrt{2
\omega}}\pi^{*} }\quad , \quad{\db
\bar{\chi} =
\sqrt{\frac{\omega}{2}}\varphi^{*} + \frac{i}{\sqrt{2 \omega}}\pi }
\label{eq:chidef}
\end{equation}
together with their complex conjugates. Inverted, these read $\varphi = (\chi+\bar{\chi}^*)/\sqrt{2\omega}$ and $\pi = i \sqrt{\omega/2}(\chi^*-\bar{\chi})$. In terms of the new fields, the quadratic part (\ref{eq:quadratic}) becomes a sum of a particle ($+\mu$) and an anti-particle ($-\mu$) contribution:
\begin{equation}
\db i \int \rmd t \int \frac{\rmd^{d}p}{(2\pi)^d} \left\{ 
\chi^*(t,\mathbf p) \left[ i\partial_t - \omega + \mu \right] \chi(t,\mathbf p) + \bar{\chi}^*(t,\mathbf p) \left[ i\partial_t - \omega - \mu \right] \bar{\chi}(t,\mathbf p) 
\right\} \, .
\end{equation}
Using that $\omega \simeq m + {\mathbf p}^2/(2m)$ and (\ref{eq:munr}), we can write for the particle contribution in configuration space 
\begin{equation}
\db i \int \rmd t \int \rmd^d x \left\{ 
\chi^*(t,\mathbf x) \left[ i\partial_t + \frac{\nabla^2}{2m} + \mu_{\rm nr} \right] \chi(t,\mathbf x) \right\} \, .
\end{equation}
Of course, taking into account the interaction in (\ref{eq:ham}) for $\lambda \neq 0$, with  
\begin{equation}
\db \frac{\lambda}{12} (\varphi^* \varphi)^2 = \frac{\lambda}{48 \omega^2}
(\chi^* \chi + \chi \bar{\chi} + \chi^* \bar{\chi}^* + \bar{\chi}^* \bar{\chi}
)^2 
\end{equation}
one observes that particles and anti-particles do not decouple in general. However, since in the nonrelativistic limit all characteristic scales are taken to be much smaller than the mass $m$ with $\mu - m \ll m \simeq \mu$, the term $\sim \bar{\chi}^* [m + \mu] \bar{\chi} \simeq 2m \bar{\chi}^* \bar{\chi}$ dominates the quadratic anti-particle part of the action. Loop corrections involving the ``heavy'' anti-particle modes are suppressed and the coupling terms involving anti-particle fields may be approximately neglected. In section~\ref{sec:transport} we show that this is an excellent approximation for the infrared scaling regime near nonthermal fixed points.

With these approximations, the final result for the nonrelativistic limit of $Z[\mu] \simeq Z_{\rm nr}[\mu_{\rm nr}]$ reads 
\begin{equation}
\db Z_{\rm nr}[\mu_{\rm nr}] =
\int\mathscr{D}\chi^{*}\mathscr{D}\chi \exp \left\{ i \int \rmd t \int \rmd^d x\,
\left( 
\chi^* \left[ i\partial_t + \frac{\nabla^2}{2m} + \mu_{\rm nr} \right] \chi - \frac{g}{2} (\chi^* \chi)^2 \right)\right\} .
\label{eq:Zmunr}
\end{equation}
Here the nonrelativistic coupling $g$ is not dimensionless and given by
\begin{equation}
\db g = \frac{\lambda}{24 m^2} \, .
\end{equation}

To study the nonequilibrium evolution we can follow the developments of previous chapters, now starting from the nonrelativistic action for the complex field $\chi(x)$ with $t \equiv x^0$ on the closed time path $\C$, which reads for $\mu_{\rm nr} = 0$:
\begin{equation}
\db S[\chi,\chi^*] =
\int_{\C} \rmd x^0 \int \rmd^d x\,
\left( 
\chi^* \left[ i\partial_{x^0} + \frac{\nabla^2}{2m} \right] \chi - \frac{g}{2} (\chi^* \chi)^2 \right) \, .
\end{equation}
Varying this classical action with respect to the field gives the Gross-Pitaevskii equation employed in (\ref{eq:gpe}). The second derivative gives the inverse classical propagator. For a complex field there are four combinations of the two derivatives with respect to $\chi$ or $\chi^*$. These can be efficiently described using the following index notation, 
\begin{equation}
{\db \chi_1(x) \equiv \chi(x)} \quad , \quad {\db \chi_2(x) \equiv \chi^*(x)} \, ,
\label{eq:indexnotation}
\end{equation}
such that the inverse classical propagator matrix reads in terms of $\chi_a$ for $a=1,2$:
\begin{equation}
\db \frac{\delta^2S[\chi,\chi^*]}{\delta \chi_a(x) \delta \chi_b^*(y)} = \left( 
i\sigma_{ab}^3\partial_{x^0} + \delta_{ab} 
\left[ -\frac{\nabla^2}{2m} + \frac{g}{2}\,\chi_c(x)\chi_c^*(x) \right] 
+ g\,\chi_a(x)\chi_b^*(x) 
\right) \delta_{\C}(x-y) .
\end{equation}
Here $\sigma^3 = {\rm diag}(1,-1)$ denotes the third Pauli matrix. 

Following the corresponding steps of section~\ref{sec:exactevoleq}, we can write down the quantum evolution equations. For instance, the equations for the nonrelativistic statistical two-point function $F_{ab}^{({\rm nr})}(x,y)$ and the spectral function $\rho_{ab}^{({\rm nr})}(x,y)$ read in the absence of a condensate:\footnote{In full analogy to (\ref{eq:comrho}) and (\ref{eq:anticomF}), we can write in terms of the Heisenberg field operators $\hat{\chi}_a$ the two-point functions  
$F_{ab}^{({\rm nr})}(x,y)=\langle\{\hat{\chi}_a(x),\hat{\chi}_b^\dagger(y)\}\rangle/2$ and
$\rho_{ab}^{({\rm nr})}(x,y)= i\langle[\hat{\chi}_a(x),\hat{\chi}_b^\dagger(y)]\rangle$.}
\begin{eqnarray}
	\db \left[i\sigma_{ac}^3\partial_{x^0}-\Omega_{ac}(x)\right] F_{cb}^{({\rm nr})}(x,y) &\db =& \db \int_{t_0}^{x^0}\mathrm{d}z\,\Sigma_{ac}^{\rho({\rm nr})}(x,z)F_{cb}^{({\rm nr})}(z,y)\nonumber\\
	&\db -& \db \int_{t_0}^{y^0}\mathrm{d}z\,\Sigma_{ac}^{F({\rm nr})}(x,z)\rho_{cb}^{({\rm nr})}(z,y),
\label{eq:evol_eq_F_nonrel}
\\
\db \left[i\sigma_{ac}^3\partial_{x^0}-\Omega_{ac}(x)\right] \rho_{cb}^{({\rm nr})}(x,y) & \db =&\db \int_{y^0}^{x^0}\mathrm{d}z\,\Sigma_{ac}^{\rho({\rm nr})}(x,z)\rho_{cb}^{({\rm nr})}(z,y),
\label{eq:evol_eq_rho_nonrel}
\end{eqnarray}
where
\begin{equation}
\db	\Omega_{ab}(x)=\, \delta_{ab}\left[ -\frac{\nabla^2}{2m} + \frac{g}{2}\, F_{cc}^{({\rm nr})}(x,x) \right] 
	+ g\,F_{ab}^{({\rm nr})}(x,x) \,  
\end{equation}
and $\Sigma_{ab}^{F({\rm nr})}(x,y)$ and $\Sigma_{ab}^{\rho({\rm nr})}(x,y)$ denote the corresponding nonrelativistic statistical and spectral parts of the self-energies, respectively.

\subsection{Bibliography}
\label{sec:litQFT}

\begin{itemize}
\item 2PI or so-called $\Phi$-derivable approximation schemes have been put forward in J.~M.~Luttinger and J.~C.~Ward, {\it Ground state energy of a many fermion system.\ 2}, Phys.\ Rev.\  {\bf 118} (1960) 1417 and in G.~Baym,
  {\it Selfconsistent approximation in many body systems,}
  Phys.\ Rev.\  {\bf 127} (1962) 1391, or 
J.~M.~Cornwall, R.~Jackiw and E.~Tomboulis, {\it Effective Action For Composite Operators},
Phys.\ Rev.\ D {\bf 10} (1974) 2428.
\item
General discussions on 2PI or higher effective actions include  
H.~Kleinert, {\it Higher effective actions for bose systems},
Fortschritte der Physik {\bf 30} (1982) 187.
A.N.~Vasiliev, {\it Functional Methods in Quantum Field Theory and Statistical
Physics}, Gordon and Breach Science Pub.~(1998) and
J.~Berges, {\it $n$PI effective action techniques for gauge theories,} Phys.\ Rev.\ D {\bf 70} (2004) 105010.
\item In the context of nonequilibrium dynamics, apart from L.P.~Kadanoff, G.~Baym, 
{\it Quantum Statistical Mechanics,}
Benjamin, New York (1962), some reviews are 
  K.~c.~Chou, Z.~b.~Su, B.~l.~Hao and L.~Yu,
  {\it Equilibrium And Nonequilibrium Formalisms Made Unified,}
  Phys.\ Rept.\  {\bf 118} (1985) 1,
  E.~Calzetta and B.~L.~Hu,
  {\it Nonequilibrium Quantum Fields: Closed Time Path Effective Action, Wigner
  Function and Boltzmann Equation,}
  Phys.\ Rev.\  D {\bf 37} (1988) 2878, P.\ Danielewicz,
  {\it Quantum Theory of Nonequilibrium Processes, I}, Ann.\ Phys.\ {\bf 152} (1984) 239 and J.~Berges,
  {\it Introduction to nonequilibrium quantum field theory,}
  AIP Conf.\ Proc.\  {\bf 739} (2005) 3; arXiv:hep-ph/0409233.
\item An introduction to the close relation of these techniques with the nonequilibrium functional renormalization group is given in J.~Berges and D.~Mesterhazy, {\it Introduction to the nonequilibrium functional renormalization group,} Schladming lectures, arXiv:1204.1489 [hep-ph]. 
\item Aspects of renormalization are discussed in 
  H.~van Hees and J.~Knoll,
  {\it Renormalization in self-consistent approximations schemes at finite
  temperature. I: Theory,}
  Phys.\ Rev.\ D {\bf 65}, 025010 (2002).
  J.~P.~Blaizot, E.~Iancu and U.~Reinosa,
  {\it Renormalization of phi-derivable approximations in scalar field
  theories,}
  Nucl.\ Phys.\ A {\bf 736}, 149 (2004).
 J.~Berges, S.~Borsanyi, U.~Reinosa and J.~Serreau,
 {\it Nonperturbative renormalization for 2PI effective action techniques,}
 Annals Phys.\  {\bf 320} (2005) 344. U.~Reinosa and J.~Serreau,
  {\it 2PI effective action for gauge theories: Renormalization,}
  JHEP {\bf 0607} (2006) 028. The renormalization of initial-value problems is specifically addressed in M.~Garny and M.~M.~Muller,
  {\it Kadanoff-Baym Equations with Non-Gaussian Initial Conditions: The Equilibrium Limit,}  Phys.\ Rev.\ D {\bf 80} (2009) 085011.	
\item For a discussion of the relation between relativistic and nonrelativistic theories, which is available on the arXiv, see e.g.\ T.S.\ Evans, {\it The condensed matter limit of relativistic QFT}, in: Proc.\ 4th Workshop on Thermal Field Theories and their Applications, Dalian, China, 1995, Y.X.\ Guiet.\ et.\ al.\ (eds.), World Scientific, 1996; http://arxiv.org/abs/hep-ph/9510298.	
\end{itemize}

%% file: ch_approx2PILH.tex
\section{Thermalization}
\label{sec:2PI}
\setcounter{equation}{0}

\subsection{Two-particle irreducible loop or coupling expansion} 
\label{sec:loopexp}

\subsubsection{Expansion of the effective action}

Loop or coupling expansions of the 2PI effective action proceed 
along the same lines as the corresponding expansions for the standard 
1PI effective action, with the only difference that 
\bi
\item contributions are parametrized in terms of ``dressed'' propagators, 
which are obtained from the stationarity condition as in (\ref{eq:station}), 
instead of classical propagators 
\item and only 2PI contributions are kept.
\ei 
Using the decomposition (\ref{2PIaction}), the 2PI effective action contains the contributions from the classical action supplemented by a saddle-point (``one-loop'') correction and a 2PI contribution $\Gamma_2$. For the example of the $O(N)$-symmetric classical action (\ref{eq:classical}) for the scalar field $\varphi_a(x)$, the contributions to $\Gamma_2[\phi,G]$ are constructed from the effective interaction 
\beq \db
i S_{\rm int}[\phi,\varphi] =  
{\rr 
- \int_{x,\C} i \frac{\lambda}{6 N} \phi_a(x)\varphi_a(x)\varphi_b(x)\varphi_b(x)}
- \int_{x,\C} i \frac{\lambda}{4! N} \Big(\varphi_a(x)\varphi_a(x)\Big)^2 
\label{interactionS} \, .
\eeq
As discussed in section \ref{sec:effectiveactions}, this is obtained from the classical action 
by shifting $\varphi_a(x) \to \phi_a(x) + \varphi_a(x)$
and collecting all terms cubic and quartic in the fluctuating field
$\varphi_a(x)$. 
As for the case of the 1PI effective action, in 
addition to the quartic interaction there is thus
an effective cubic interaction for non-vanishing field expectation
value. To obtain loop or coupling expansions, one Taylor expands
(\ref{interactionS}) using $e^x = \sum_n x^n/n!$ such that only Gaussian functional integrals have to be performed to compute the corrections.

To be specific, we consider first the scalar field theory for a one-component field, i.e.\ $N=1$ in (\ref{interactionS}). 
To lowest order one has $\Gamma_2[\phi,G] = 0$ and we recover 
the one-loop result given in (\ref{eq:2PIoneloop}). Further corrections can be very efficiently classified with a standard graphical notation. The propagator $G$ is associated to a line and 
the interactions are represented as points where four (three) lines meet for a four- (three-)vertex: 
\beq
G(x,y) \,\, = \, {\scriptstyle x}\!\Prop(\Lsc)\!{\scriptstyle y} \, , \qquad
\lambda \int_{x,\C} \,\, = \SDfour(\Lsc,\Lsc,\Lsc,\Lsc)\!\!\!\!\!{\scriptstyle x}\,\, , \qquad
\lambda \int_{x,\C} \phi(x) \,\, =  \SDthree(\Lsc,\Lsc,\Lsc)\!\!\!\!\!\!\!{\scriptstyle x}\;\,
 \,\, .
\eeq
Since $\Gamma_2[\phi,G]$ is a functional, which associates a number
to the fields $\phi$ and $G$, only diagrams with closed loops of propagator lines and vertices can contribute. We may classify the contributions to $\Gamma_2[\phi,G]$ according to their number of closed loops, i.e.\
\beq \db
\Gamma_2[\phi,G] \, = \, \Gamma_2^{\rm (2loop)}[\phi,G] + \Gamma_2^{\rm (3loop)}[\phi,G] + \ldots 
\eeq
At two-loop order there are two contributions, $\Gamma_2^{\rm (2loop)}[\phi,G] = \Gamma_2^{(2{\rm a})}[G] + {\rr \Gamma_2^{(2{\rm b})}[\phi,G]}$. The field independent contribution is 
\beq { \db
\Gamma_2^{(2{\rm a})}[G] \, = \,  
-i\, 3\, \Big(-i \frac{\lambda}{4!}\Big) \int_{x,\C} G^2(x,x) } \, = \, 
-\frac{\lambda}{8} \, \SToptVE(\Asc,\Asc) \, ,
\label{eq:2a}
\eeq
where we have made explicit the 
different factors coming from the overall $-i$
in the defining functional integral for $\Gamma[\phi,G]$ 
(see (\ref{modZ})), the factor $3$ from the three different ways to form closed loops with a single four-vertex and the factor from the vertex itself. 
The field-dependent two-loop contribution is given by
\beq
{\rr \Gamma_2^{(2{\rm b})}[\phi,G] \, = \, -i\, 6\, \frac{1}{2}\, 
\int_{xy,\C} \Big(-i \frac{\lambda}{6}\phi(x) \Big) 
\Big(-i \frac{\lambda}{6}\phi(y) \Big)
G^3(x,y)}  \, = \, i\, \frac{\lambda^2}{12} \, \SToptVS(\Asc,\Asc,\Lsc) \, , 
\label{eq:2b} 
\eeq
where the number of different ways to form closed loops from two three-vertices is $6$, and the factor $1/2$ comes from the Taylor expansion of the exponential to second order.
At three-loop order there are three contributions, $\Gamma_2^{\rm (3loop)}[\phi,G] = \Gamma_2^{(3{\rm a})}[G] + {\rr \Gamma_2^{(3{\rm b})}[\phi,G]} + {\rr \Gamma_2^{(3{\rm c})}[\phi,G]}$, with
\bea \db
\Gamma_2^{(3{\rm a})}[G] &\db \! =\!& {\db i\, \frac{\lambda^2}{48} \int_{xy,\C}\!\! G^4(x,y) } \, = \, i\, \frac{\lambda^2}{48} \, \SToprVB(\Asc,\Asc,\Asc,\Asc) \; ,
\nonumber\\
\rr
\Gamma_2^{(3{\rm b})}[\phi,G] &\rr\! = \! & {\rr \frac{\lambda^3}{8} \int_{xyz,\C}\!\! G^2(x,y)G^2(y,z)G(z,x)
\phi(x) \phi(z) } \, = \, \frac{\lambda^3}{8} \, \SToprVV(\Asc,\Asc,\Lsc,\Lsc,\Lsc) \; ,
\nonumber\\
\rr
\Gamma_2^{(3{\rm c})}[\phi,G] &\rr \!=\!& \rr -i\, \frac{\lambda^4}{24} \int_{xyzw,\C}\!\! G(x,y)G(y,z)G(z,x) G(x,w)G(w,y)G(w,z) 
\nonumber\\
&& {\rr \times \, \phi(x) \phi(y) \phi(z) \phi(w)} \, = \, -i\, \frac{\lambda^4}{24} \, \SToprVM(\Asc,\Asc,\Asc,\Lsc,\Lsc,\Lsc) \; .
\eea

Of course, care has to be taken when estimating the order of the coupling for a given contribution. For instance, one observes already at the classical level from $\delta S/\delta \phi(x) = 0$ for the simple case of a space-time independent field 
$\phi$ that $m^2 \phi + \lambda \phi^3/6 = 0$. As a consequence, for the case with spontaneous symmetry breaking with $\phi \not = 0$ and $m^2 < 0$ one finds $\phi = \pm \sqrt{-6 m^2/\lambda}$. Since parametrically 
$\phi \sim 1/\sqrt{\lambda}$, both of the above two-loop contributions should be taken into account to first order in the coupling $\lambda$. Similarly, all three-loop contributions should be included at the next order. For general out-of-equilibrium situations the power counting can be much more involved even in the presence of a small coupling. We will consider explicit examples of time-dependent or ``dynamical'' power counting schemes
for far-from-equilibrium dynamics in later chapters.

\begin{figure}[t]
\centerline{\epsfig{file=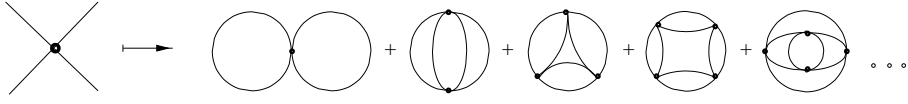,width=12cm}}
\caption{Topologically distinct diagrams in the 2PI loop expansion
shown to five-loop order for $\phi = 0$. The suppressed prefactors are given in 
(\ref{eq:2PIGfiveloop}).}
\label{fig:fiveloop}
\end{figure}

The 2PI loop expansion exhibits of course
much less topologically distinct diagrams than the
respective 1PI expansion. For instance, in the symmetric
phase ($\phi = 0$) one finds that up to four 
loops only a single diagram 
contributes at each order. 
At fifth order there are two distinct diagrams which
are shown along with the lower-loop graphs in Fig.\ \ref{fig:fiveloop} suppressing prefactors.
For later discussion, we give here
the results up to five loops for the $O(N)$-symmetric field theory with classical action (\ref{eq:classical}) for general $N$. For $\phi = 0$ by 
virtue of $O(N)$ rotations the propagator can be taken to be diagonal:
\beq\db
G_{ab}(x,y) \,=\, G(x,y) \delta_{ab} \, . 
\label{eq:symmetricG}
\eeq
This represents the most general form of the propagator for the symmetric regime if the initial conditions respect the $O(N)$ symmetry. 
One finds to five-loop order with $\Gamma_2[G] \equiv \Gamma_2[\phi=0,G]$:
\bea\db
\db\Gamma_2^{(2{\rm loop})}[G]\big|_{G_{ab}=G\delta_{ab}} &\db \! = \! &\db 
-\frac{\lambda}{8}\frac{(N+2)}{3}\int_{x,\C}\! G^2(x,x)\, ,
\nonumber\\ \db
\Gamma_2^{(3{\rm loop})}[G]\big|_{G_{ab}=G\delta_{ab}} &\db \! = \! & \db \frac{i\lambda^2}{48}\frac{(N+2)}{3N}\int_{xy,\C} \!\! 
G^4(x,y)\,,\nonumber\\
\db\Gamma_2^{(4{\rm loop})}[G]\big|_{G_{ab}=G\delta_{ab}} &\db \! = \! &\db 
\frac{\lambda^3}{48}\frac{(N+2)(N+8)}{27N^2}\int_{xyz,\C} \!\!\! 
G^2(x,y)G^2(x,z)G^2(z,y) \, ,\nonumber\\
\db\Gamma_2^{(5{\rm loop})}[G]\big|_{G_{ab}=G\delta_{ab}} &\db \! = \! &\db 
-\frac{i\lambda^4}{128}\frac{(N+2)(N^2+6N+20)}{81N^3} \int_{xyzw,\C} 
\!\!\! G^2(x,y)G^2(y,z)
\nonumber\\ 
&\db \times &\db  G^2(z,w)G^2(w,x) 
-\frac{i\lambda^4}{32}\frac{(N+2)(5N+22)}{81N^3} \int_{xyzw,\C} \!\!\! 
G^2(x,y)\nonumber\\ 
&\db \times &\db 
G(x,z)G(x,w)G^2(z,w)G(y,z)G(y,w) \, .
\label{eq:2PIGfiveloop}
\eea

\subsubsection{Self-energies}
\label{sec:sigmaloops}

In section~\ref{sec:exactevoleq} we have derived the coupled evolution
equations (\ref{eq:exactrhoF}) for the statistical 
propagator $F$ and the spectral function $\rho$ as well as the field equation (\ref{eq:exactphi}). A systematic 
approximation to the exact equations can be obtained from the
above loop or coupling expansion of the 2PI effective action. This determines all the required self-energies to a given order in the expansion from the variation of the 2PI effective action using (\ref{eq:station}). 
We emphasize that {\rr\it all classifications of contributions
are done for the effective action}. Once an approximation order
is specified on the level of the effective action, one obtains a {\rr\it closed set of equations} and there are no further approximations on the level of the evolution
equations required. This ensures the ``conserving'' 
properties of 2PI expansions such as energy conservation, since all approximate equations of motion are obtained from a variational principle.

For the computation of self-energies with
\beq \db
\Sigma(x,y) \, \equiv \, - i \Sigma^{(0)}(x)\, \delta_\C(x-y)
+ \ol{\Sigma}(x,y) \, = \,  2i\, \frac{\delta \Gamma_2[\phi,G]}{\delta G(x,y)} 
\eeq
we again start with the one-component scalar field theory using the self-energy decomposition (\ref{eq:sighominh}). The field-independent two-loop contribution (\ref{eq:2a}) gives
\beq \db
2i \frac{\delta \Gamma_2^{(2{\rm a})}[G]}{\delta G(x,y)} \, = \, 
-i \,\frac{\lambda}{2}\, G(x,x)\, \delta_\C(x-y) \, .
\eeq
With the above self-energy decomposition we obtain
\beq \db
\Sigma^{(0)} \, = \, \frac{\lambda}{2}\, F(x,x)\, ,
\label{eq:sigma0F}
\eeq
where we have used that $\rho(x,x)=0$. This contribution
corresponds to a space-time dependent mass shift 
in the evolution equations. 
Similarly, for the field-dependent contribution one finds
\beq \rr 
2i \frac{\delta \Gamma_2^{(2{\rm b})}[\phi,G]}{\delta G(x,y)} \, = \,
- \frac{\lambda^2}{2}\, G^2(x,y) \phi(x) \phi(y) \, .
\eeq
To write this in terms of statistical and spectral two-point functions, we use (\ref{eq:decompid}) and consider
\bea\db
G^2(x,y) & \db = & \db \left(
F(x,y) - \frac{i}{2}\, \sgn_\C(x^0-y^0) \rho(x,y) \right)^2
\nonumber \\
& \db = & \db
F^2(x,y) -\frac{1}{4} \rho^2(x,y) - i\, \sgn_\C(x^0-y^0) F(x,y) \rho(x,y) \, ,
\eea
where $\sgn_\C^2(x^0-y^0)=1$ has been employed.
With the self-energy decomposition (\ref{eq:decompself}) this leads to
\bea \rr
\Sigma^F(x,y) & \rr = & \rr - \frac{\lambda^2}{2} \left( F^2(x,y) -\frac{1}{4} \rho^2(x,y) \right) \phi(x) \phi(y) \, ,
\nonumber\\ \rr
\Sigma^\rho(x,y) & \rr = & \rr -\lambda^2 F(x,y) \rho(x,y) \phi(x) \phi(y) \, .
\label{eq:selfphi}
\eea
It remains to compute the loop contributions to the field evolution equation (\ref{eq:exactphi}), i.e.\
\beq\rr
\frac{\delta \Gamma_2^{(2{\rm b})}[\phi,G]}{\delta \phi(x)}
\,=\, i\, \frac{\lambda^2}{6} \int_{y,\C} G^3(x,y) \phi(y) \, .
\eeq
Similar to the above steps, we consider
\bea\db
G^3(x,y) &\db = & \db \left( 
F(x,y) - \frac{i}{2}\, \sgn_\C(x^0-y^0) \rho(x,y)
\right)^3 
\nonumber\\
&\db = & \db F^3(x,y) - \frac{3i}{2} \, \sgn_\C(x^0-y^0) F^2(x,y) \rho(x,y)
\nonumber\\
&& \db -\frac{3}{4} F(x,y) \rho^2(x,y)
+\frac{i}{8}\, \sgn_\C(x^0-y^0) \rho^3(x,y)
\eea
using, in particular, $\sgn_\C^3(x^0-y^0)=\sgn_\C(x^0-y^0)$. Following the discussion in section \ref{sec:exactevoleq}, we split the contour integration as 
\bea\db
\int_{\C} \rmd y^0\, G^3(x,y) &\db = &\db \int_{t_0}^{x^0}\rmd y^0
\left(
-\frac{3i}{2}\, F^2(x,y) \rho(x,y) + \frac{i}{8} \rho^3(x,y) 
\right) 
\nonumber\\
& \db + & \db \int_{x_0}^{t^0}\rmd y^0
\left(
\frac{3i}{2}\, F^2(x,y) \rho(x,y) - \frac{i}{8} \rho^3(x,y) 
\right)
\nonumber\\
&\db = &\db - 3i \int_{t_0}^{x^0}\rmd y^0 \rho(x,y) \left(
F^2(x,y) - \frac{1}{12} \rho^2(x,y)
\right) \, .
\eea  
Putting everything together gives
\beq\rr
\frac{\delta \Gamma_2^{(2{\rm b})}[\phi,G]}{\delta \phi(x)}
\,=\, \frac{\lambda^2}{2} \int_{t_0}^{x^0}\rmd y\, \rho(x,y) \left(
F^2(x,y) - \frac{1}{12} \rho^2(x,y)
\right) \phi(y) \, ,
\label{eq:field1}
\eeq
where we use again the abbreviated notation $\int_{t_1}^{t_2}
{\rm d}z \equiv \int_{t_1}^{t_2} {\rm d}z^0 
\int {\rm d}^d z$.
The evolution equations (\ref{eq:exactrhoF}) and (\ref{eq:exactphi}) together with the loop contributions (\ref{eq:sigma0F}), (\ref{eq:selfphi}) and (\ref{eq:field1}) represent a closed set of equations for $F(x,y)$, $\rho(x,y)$ and $\phi(x)$. A corresponding discussion can be done starting from the above three-loop expressions for $\Gamma_2[\phi,G]$ and, similarly, at higher loop orders.

For later discussions, we present here also the case of the scalar $O(N)$ symmetric 
field theory with a vanishing field expectation value starting from the three-loop $\Gamma_2$. The propagator in the symmetric regime can be written as (\ref{eq:symmetricG}). From  $\Gamma_2[G]|_{G_{ab}=G\delta_{ab}}$, we obtain the
self-energy as
\bea \db
\Sigma_{ab}(x,y) &\db = &\db 2i \frac{\delta \Gamma_2[G]|_{G_{ab}=G\delta_{ab}}}{\delta G_{ab}(x,y)}
\, = \,  2i \frac{\delta_{ab}}{N} \frac{\delta \Gamma_2[G]|_{G_{ab}=G\delta_{ab}}}{\delta G(x,y)}  \, ,
\eea 
where we have employed $G(x,y) = G_{ab}(x,y) \delta_{ab}/N$ for the last equality. Writing $\Sigma_{ab}(x,y) = \Sigma(x,y) \delta_{ab}$ and using the decomposition (\ref{eq:sighominh}) we thus have 
\beq \db
\Sigma(x,y) \, = \,  \frac{2i}{N} \frac{\delta \Gamma_2[G]|_{G_{ab}=G\delta_{ab}}}{\delta G(x,y)} \, .
\eeq
From the two-loop contribution to $\Gamma_2[G]$ given in (\ref{eq:2PIGfiveloop}) one finds 
\beq\db
\frac{2i}{N}\frac{\delta \Gamma_2^{(2)}[G]|_{G_{ab} \, = \, G\delta_{ab}}}{\delta G(x,y)} 
\, = \,
-i \lambda \frac{(N+2)}{6N}\, G(x,x)\, \delta_\C(x-y) \, .
\eeq
Using the propagator decomposition (\ref{eq:decompid}) in spectral and statistical components, this leads to the one-loop self-energy
\beq\db
\Sigma^{(0)}(x) \, = \, \lambda\frac{N+2}{6N}\,F(x,x) \, ,
\eeq
which corresponds to (\ref{eq:sigma0F}) for $N=1$.  
Similarly, from $\Gamma_2[G]$ at three-loop, one finds
\beq \db
\ol{\Sigma}^{\rm (2loop)}(x,y)
\, = \, -\lambda^2 \frac{N+2}{18 N^2} \, G^3(x,y) \, . 
\eeq
Using the decomposition (\ref{eq:decompself}) for the self-energy, we have
\bea 
\db \Sigma_F(x,y) &\db =&\db
-\lambda^2\,\frac{N+2}{18N^2}\,F(x,y)\left[F^2(x,y) 
-\frac{3}{4}\rho^2(x,y)\right] \, ,
\label{eqsigmaF3}\\
\db\Sigma_{\rho}(x,y) &\db =&\db
-\lambda^2\,\frac{N+2}{6N^2}\,\rho(x,y)\left[F^2(x,y) 
-\frac{1}{12}\rho^2(x,y)\right] \, ,
\label{eqsigmarho3}
\eea
which enter (\ref{eq:localself}) and (\ref{eq:exactrhoF}).

\subsubsection{Solving nonequilibrium evolution equations}
\label{sec:solneqeq}

The nonequilibrium evolution equations obtained from the loop expansion of the 2PI effective action can be used as a starting point for a wealth of further approximations that give analytical insight into the dynamics. For instance, we will later show that already two-loop self-energies as in (\ref{eqsigmaF3}) and (\ref{eqsigmarho3}) include and go beyond standard kinetic or Boltzmann equations describing two-to-two scattering of particles. However, it is very instructive to consider their numerical solution without further approximations. In particular, this will allow us to observe the striking process of thermalization in quantum field theory from first principles. 

Despite the apparent complexity of the evolution equations (\ref{eq:exactrhoF}) and (\ref{eq:exactphi})  
they can be very efficiently implemented and solved on a computer.
Here it is important to note that all equations are explicit
in time, i.e.\ all quantities at some time $t$ can be obtained
by integration over the explicitly known functions for earlier times $t_{\rm past}<t$
for given initial conditions. The time-evolution
equations (\ref{eq:exactrhoF}) for $\rho(x,y)|_{x^0 = t_1,y^0=t_2}$ 
and $F(x,y)|_{x^0 = t_1,y^0=t_2}$
do not depend on the RHS~on $\rho(x,y)$ and $F(x,y)$ for
times $x^0 \ge t_1$ and $y^0 \ge t_2$ and similarly for the field evolution equation 
(\ref{eq:exactphi}). To see this, we note that
the integrands vanish identically for the upper time-limits
of the memory integrals because of the anti-symmetry of the
spectral components, with $\rho(x,y)|_{x^0 = y^0} \equiv 0$ and 
$\Sigma_{\rho}(x,y)|_{x^0 = y^0} \equiv 0$. As a consequence, only
explicitly known quantities at earlier times determine the
time evolution of the unknowns at later times. {\rr\it The numerical
implementation, therefore, only involves sums over
known functions.}

Solving the evolution equations requires the specification of (Gaussian) initial conditions as discussed in section \ref{sec:initialconditions}. We consider first the one-component scalar field theory for spatially homogeneous systems and Fourier transform with respect to the spatial coordinates. The equal-time commutation relations fix the Fourier modes of the spectral function $\rho(x^0,x^0,\bp) = 0$, $\partial_{x^0}\rho(x^0,y^0;\bp) = 1$ and $\partial_{x^0}  \partial_{y^0} \rho(x^0,x^0;\bp)=0$ also at initial time. Here we take
\bea
\nonumber \db
F(x^0,y^0;\bp)|_{x^0=y^0=0} &\db =&\db 
\frac{1}{\omega_\bp}\left(f_\bp(0)+\frac{1}{2}\right), \nonumber\\
\db \partial_{x^0}  \partial_{y^0} F(x^0,y^0;\bp)|_{x^0=y^0=0} &\db =&\db
\omega_\vecp\left(f_\bp(0)+\frac{1}{2}\right), \nonumber\\
\db \partial_{x^0}F(x^0,y^0;\bp)|_{x^0=y^0=0} &\db =&\db  0, \quad \phi(x^0)|_{x^0=0} = 0, \quad
\partial_{x^0}\phi(x^0)|_{x^0=0} \quad
\label{eq:initial1}
\eea
with $\omega_\bp = \sqrt{\bp^2 + m^2}$ as initial conditions. Since the initial macroscopic field and its derivative are taken to be zero, this remains so at all times for the considered $Z_2$ symmetric theory with the corresponding invariance of the action under $\phi \rightarrow -\phi$. Here $f_\bp(0)$ plays the role of an initial occupation number distribution which, for instance, may be taken as
\beq \db
f_\bp(0) \, = \, {\cal N}\exp\left(-\frac{1}{2\sigma^2}
(|\bp|-|\bp_{\rm ts}|)^2\right).
\label{eq:tsunami}
\eeq
The real parameter $\sigma$ controls the width of the initial distribution and 
${\cal N}$ is a normalization constant. 
The initial condition is clearly far from thermal equilibrium and
reminiscent of two colliding high energy wave packets. More precisely, it describes a 
spatially homogeneous collection of particles which move with approximately the same momentum 
peaked around $\bp_{\rm ts}$ and $ - \bp_{\rm ts}$. Of course, other initial distributions can be considered which we will also do in the following. 

In addition, we have to state renormalization conditions in order to define the quantum field theory as for any renormalizable theory in vacuum or thermal equilibrium. From this perspective, it is rather simple to consider first a field theory in one spatial dimension. In this case, the one-loop correction (\ref{eq:sigma0F}) evaluated with a momentum regulator $\Lambda$ shows a logarithmic regulator dependence. This can be analyzed at initial time, where for the spatially homogeneous initial conditions (\ref{eq:initial1}) we find for $f_\bp(0) = 0$ that the large momentum behavior for $d=1$ is given by
\beq\db
\Sigma^{(0)}(t_0) \, = \, \frac{\lambda}{2} \int^\Lambda \frac{\rmd p}{2 \pi}\, \frac{1}{2 \omega_\bp} \, \sim \, \ln \Lambda  \, .
\eeq 
We absorb this regulator dependence in a bare mass parameter $\mu$ with the
replacement $m^2 \mapsto \mu^2 = m^2-\delta m^2$ in (\ref{eq:localself}). 
The counterterm $\delta m^2$ cancels the divergent vacuum contribution
coming from the one-loop graph. The finite part of $\delta m^2$ is fixed by
requiring that the renormalized one-loop mass parameter in vacuum 
($f_\bp(0)\equiv 0$) equals $m$ and we express all dimensionful scales
in units of $m$. Since higher loop corrections are finite for one spatial dimension as $\Lambda$ is 
sent to infinity, the renormalization procedure is straightforward in this case.\footnote{This is different for higher spatial dimensions, where a consistent renormalization will require taking into account non-Gaussian corrections. For further reading on the renormalization of 2PI effective actions, see the literature in section~\ref{sec:litQFT}.}  
\begin{figure}[t]
\centerline{\epsfig{file=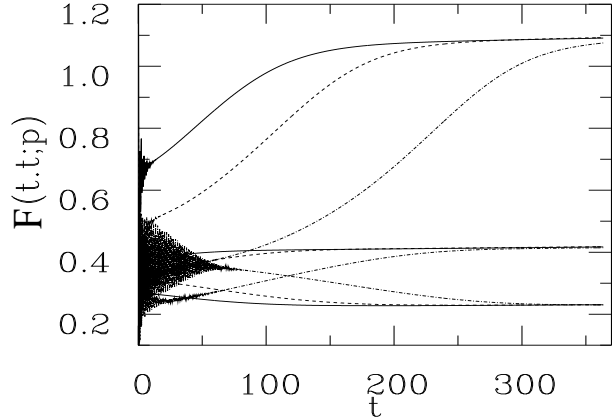,width=8.0cm,height=5.5cm}}
\caption{Evolution of the equal-time 
two-point function $F(t,t;\bp)$ with Fourier modes $|\bp|=0,3,5$ 
from the 2PI three-loop effective action. 
The evolution is shown for three very different nonequilibrium
initial conditions with the same energy density. (All in units of $m$.)} 
\label{fig:lateuni}
\end{figure}
A crucial question of nonequilibrium dynamics is 
how quickly the system {\rr \it effectively}
looses the details about the initial conditions.
Thermal equilibrium keeps no memory about 
the time history except for the values of a few conserved charges.
As a consequence, if the real scalar field theory approaches thermal equilibrium
then the late-time result would be uniquely determined by energy density.

As a first example, we show in Fig.~\ref{fig:lateuni} the
time dependence of the equal-time propagator $F(t,t;\bp)$
for three Fourier modes $|\bp|/m =0,3,5$ and three very {\rr\em different 
initial conditions with the same energy density.} The coupling is chosen for all runs to be $\lambda/m^2 = 10$. The large value suggests important corrections beyond the two-loop self-energy approximation, however, for the moment it allows us to discuss the relevant qualitative properties in a pronounced way and we will later consider the question of convergence in more detail.
For the solid line in Fig.~\ref{fig:lateuni} the initial conditions are close
to a mean field thermal solution with inverse temperature $\beta = 0.1/m$, the initial mode 
distribution for the dashed and the dashed-dotted lines deviate 
more and more substantially from a thermal equilibrium distribution. One observes that propagator modes with very different initial values but with the same momentum $|\bp|$ 
approach the same large-time value. The asymptotic behavior of the 
two-point function modes is uniquely determined by the initial 
energy density. Concerning the underlying nonequilibrium processes, we will see that this is also a very interesting observation in view of strong kinematic restrictions for scattering processes in one spacial dimension. 

The shown results are obtained from a rather simple discretization of the equations for the Fourier
components $F(t,t'; \bp)$ and $\rho(t,t'; \bp)$. They are based on a time discretization $t=n a_t$, $t'=m a_t$ with stepsize 
$a_t$ such that $F(t,t') \mapsto F(n,m)$, and  
\bea\db
 \partial_{t}^2 F(t,t') &\db \mapsto&\db \frac{1}{a_t^2}
\Big( F(n+1,m) + F(n-1,m) - 2 F(n,m) \Big),
\label{deriv} \\[0.2cm]
\db \int\limits_0^{t} dt F(t,t') &\db \mapsto&\db
a_t \Big(F(0,m)/2 + \sum\limits_{l=1}^{n-1} F(l,m) + F(n,m)/2
\Big) \label{integral}  \, ,
\eea 
where we have suppressed the momentum labels in the notation.
Above, the second derivative is replaced by a finite-difference expression
which is symmetric in $a_t \leftrightarrow -a_t$. It is obtained
from employing subsequently the lattice ``forward derivative''
$[F(n+1,m) - F(n,m)]/a_t$ and ``backward derivative''
$[F(n,m) - F(n-1,m)]/a_t$. The integral is approximated
using the trapezoidal rule with the 
average function value $[F(n,m)+F(n+1,m)]/2$ in an interval 
of length $a_t$.  The time-discretized version of 
(\ref{eq:exactrhoF}) reads then for the one-component theory
\bea\db \lefteqn{
F({\rr n+1},m;\bp) = 2 F(n,m;\bp) - F(n-1,m;\bp) } \nonumber\\[0.1cm]
&&\db - a_t^2 \left\{ \bp^2 + m^2 + \frac{\lambda}{2}
\int_{\bk} F(n,n;\bk)
\right\} F(n,m;\bp) 
\nonumber\\
&&\db - a_t^3\, \Bigg\{
\Sigma_{\rho}(n,0;\bp)\, F(0,m;\bp)/2 - \Sigma_F(n,0;\bp)\, \rho(0,m;\bp)/2
\\
&&\db \qquad +\sum\limits_{l=1}^{m-1} \Big(
\Sigma_{\rho}(n,l;\bp)\, F(l,m;\bp) - \Sigma_F(n,l;\bp)\, \rho(l,m;\bp)
\Big) \nonumber\\
&&\db \qquad +\sum\limits_{l=m}^{n-1}\, 
\Sigma_{\rho}(n,l;\bp)\, F(l,m;\bp) \Bigg\}\,  \nonumber 
\eea 
and similarly for the spectral function. These
equations are explicit in time: 
starting with $n=1$, for the time step $n+1$ one computes successively
all entries with $m=0,\ldots,n+1$ 
from known functions at earlier times. 
The above discretization leads already 
to stable numerics for small enough stepsize $a_t$, but the 
convergence properties may be easily improved with more sophisticated 
standard estimators if required. 

As for the continuum, the propagators obey 
the symmetry properties $F(n,m)=F(m,n)$ and $\rho(n,m)=-\rho(m,n)$.
Consequently, only ``half'' of the $(n,m)$--matrices have to be computed
and $\rho(n,n) \equiv 0$. Similarly, since the self-energy
$\Sigma_{\rho}$ is antisymmetric in time one can exploit that 
$\Sigma_{\rho}(n,n)$ vanishes identically. 
As initial conditions one has to specify $F(0,0;\bp)$, $F(1,0;\bp)$
and $F(1,1;\bp)$, while $\rho(0,0;\bp)$, $\rho(1,0;\bp)$ and
$\rho(1,1;\bp)$ are fixed by the equal-time commutation relations
(\ref{eq:bosecomrel}).

It is crucial for an efficient numerical 
implementation that each step forward in time does not involve 
the solution of a self-consistent or gap equation. This is 
manifest in the above discretization. The main
numerical limitation of the approach is set by the time
integrals (``memory integrals'') 
which grow with time and therefore slow down the 
numerical evaluation. Typically, the influence of early times
on the late time behavior is suppressed and can be neglected 
numerically in a controlled way. In this case, it is often
sufficient to only take into account the contributions from 
the memory integrals
for times much larger then a characteristic inverse damping 
rate. An error estimate
then involves a series of runs with increasing memory time.   

For scalars one may use for the spacial dependence a standard lattice discretization with periodic boundary conditions. For a spatial volume $V=(N_s a_s)^d$ in $d$ dimensions with lattice spacing $a_s$ 
one finds for the momenta
\beq\db
\bp^2 \,\mapsto\, \sum\limits_{i=1}^d \frac{4}{a_s^2} \sin^2 
\left(\frac{a_s p_i}{2} \right) 
\, ,
\qquad p_i=\frac{2 \pi n_i}{N_s a}  \, ,
\label{eq:latticemomenta}
\eeq
where $n_i = 0,\ldots, N_s-1$. This can be easily understood from
acting with the corresponding finite-difference expression (\ref{deriv})
for space-components: $\partial^2_x e^{-i p x} \mapsto 
e^{-i p x} [e^{ipa_s} + e^{-ipa_s} - 2]/a_s^2 =
- e^{-i p x}\, 4 \sin^2(p a_s/2)/a_s^2$. Exploiting the
lattice symmetries reduces the number of
independent lattice sites. Taken this explicitly into account becomes the more important the larger the space dimension. For instance, on the lattice there 
is only a subgroup of the rotation symmetry generated 
by the permutations of $p_x,p_y,p_z$ and the reflections $p_x
\leftrightarrow -p_x$ etc.~for $d=3$. Exploiting these
lattice symmetries reduces the number of
independent lattice sites to $(N_s+1)(N_s+3)(N_s+5)/48$.
The self-energies may be calculated in coordinate space,
where they are given by products of coordinate-space
correlation functions, and then transformed back to
momentum space. The coordinate-space correlation functions
are available by fast Fourier transformation routines. 

The lattice introduces a 
momentum cutoff $\pi / a_s$, however, the renormalized
quantities should be insensitive to cutoff variations for 
sufficiently large $\pi / a_s$.   
In order to study the infinite volume limit one has to remove finite 
size effects. This may be done by increasing the volume until
convergence of the results is observed.    
For time evolution problems
the volume which is necessary to reach the infinite volume limit 
to a given accuracy can depend on the time scale. 
This is, in particular, due to the fact that finite systems 
can show characteristic
recurrence times after which an initial effective
damping of oscillations can be reversed.

\subsubsection{Nonequilibrium evolution of the spectral function}
\label{sec:spectral}

In order to understand the nonequilibrium dynamics of the spectral function $\rho$ vs.\ the statistical two-point function $F$, it is helpful to consider for a moment the free field theory. With $\Sigma = 0$ in the nonequilibrium evolution equations (\ref{eq:exactrhoF}), the above initial conditions (\ref{eq:initial1}) lead to the plane wave solutions  
\bea \db 
F^{\rm (free)}(x^0,y^0;\bp) &\db  = &\db \frac{1}{\omega_\bp}\left(f_\bp(0)+\frac{1}{2}\right)\, \cos \left[ \omega_\bp (x^0 - y^0) \right] \, ,
\nonumber\\
\db \rho^{\rm (free)}(x^0,y^0;\bp) &\db = & \db \frac{1}{\omega_\bp}\, \sin \left[ \omega_\bp (x^0 - y^0) \right] \,  
\label{eq:freeFrho}
\eea
with frequency $\omega_\bp$. We may bring this into a more suggestive form by introducing the center coordinate $X^0 = (x^0+y^0)/2$ and the relative coordinate $s^0 = x^0 - y^0$, with respect to which a Fourier transformation to momentum
space is performed (Wigner transformation). To analyze the spectral function we may perform a Wigner transformation and write
\beq \db
i\tilde{\rho}(X^0; \omega, \bp) \, = \, \int_{-2X^0}^{2X^0}ds^0\, e^{i\omega s^0}
\rho(X^0+s^0/2, X^0-s^0/2;\bp).
\label{eqW}
\eeq
The $i$ is introduced such that $\tilde{\rho}(X^0;\omega,\bp)$ is real. Since we consider an initial-value problem with  
$x^0,y^0\geq 0\,$, the time integral over $s^0$ is bounded by its maximum value for $y^0=0$ where $s^0 = 2X^0$ and its minimum value for $x^0=0$ where $s^0 = - 2X^0$. After performing a Wigner
transformation of the free spectral function (\ref{eq:freeFrho}), we find
\beq \db
\tilde{\rho}^{\rm (free)}(X^0;\omega,\bp) \, = \,
\frac{\sin [(\omega-\omega_\bp)2X^0]}{\omega_\bp(\omega-\omega_\bp)} -
\frac{\sin [(\omega+\omega_\bp)2X^0]}{\omega_\bp(\omega+\omega_\bp)} \, . 
\label{eqW0}
\eeq 
For finite $X^0$ this spectral function shows a rapidly oscillating
behavior, while its envelope is peaked at $\omega=\pm\omega_\bp$.
In the limit $X^0\to \infty$ it reduces to 
\beq \db
\tilde{\rho}^{\rm (free)}(\omega,\bp) \, = \, 2\pi \, \sgn (\omega)\, \delta\left(\omega^2 - \omega_\bp^2\right) \, ,
\label{eq:freespectral}
\eeq
which describes the familiar form of the free spectral function in Fourier space. In general, 
the positivity condition $\mbox{sgn}(\om)\tilde{\rho}(X^0;\omega, \bp)\geq 0$ can only be shown to hold in
the special case that the initial density matrix commutes with the full
Hamiltonian, such as in thermal equilibrium. In this case the system is of
course stationary and independent of $X^0$.  As a consequence, the
interpretation of the Wigner transformed nonequilibrium spectral function as the density of
states should be taken with care. Nevertheless, as a consequence of the equal-time commutation relation 
the Wigner transform obeys the sum rule $\int d\omega/(2\pi)\, \omega\,
\tilde{\rho}(X^0;\om,\bp) = 1$ for the free as well as the interacting theory.

Going beyond the free theory, we know that the equilibrium spectral function would acquire a ``width'' in the presence of a non-zero imaginary part of the self-energy, i.e.\ $\Sigma^\rho \not = 0$.\footnote{For a vanishing $\omega$-dependence the function 
$\Gamma^{\rm (eq)}(\omega,\bp)\equiv \Sigma^{\rm (eq)}_{\rho}(\omega,\bp)/2
\omega$ plays the role of a decay rate for one-particle excited states with
momentum $\bp$.} Such a width in Fourier space is related to a damping of correlation functions in time. To obtain a qualitative understanding of this statement, we may consider the case of some spectral function which is assumed to be strictly exponentially damped, i.e.\ $\tilde{\rho}(x^0,y^0;\bp)=e^{-\gm_\bp|x^0-y^0|}E_\bp^{-1}\sin
[E_\bp(x^0-y^0)]$. The effective frequency $E_\bp$ and rate $\gamma_\bp$
are allowed to depend on the time $X^0$.  The corresponding Wigner
transform reads $\tilde{\rho}(X^0;\omega,\bp) = \tilde{\rho}_{\rm BW}(X^0;\omega,\bp) + 
\delta\tilde{\rho}(X^0;\omega,\bp)$, where $\tilde{\rho}_{\rm BW}$ denotes the 
Breit-Wigner function
\beq \db
\tilde{\rho}_{\rm BW}(X^0;\omega,\bp) =
\frac{2\omega\Gamma_\bp(X^0)}{[\omega^2-E_\bp^2(X^0)]^2
+ \omega^2\Gamma^2_\bp(X^0)}
\label{BreitWigner}
\eeq
with a width $\Gamma_\bp(X^0)=2\gamma_\vecp(X^0)$. The additional contribution
$\delta\tilde{\rho}(X^0;\om,\vecp)$ vanishes exponentially as $\exp(-\Gamma_\bp
X^0)$, such that the finite-time effect is indeed suppressed for 
$X^0 \gg 1/\Gamma_\bp$.

\begin{figure}[t]
\centerline{\epsfig{file=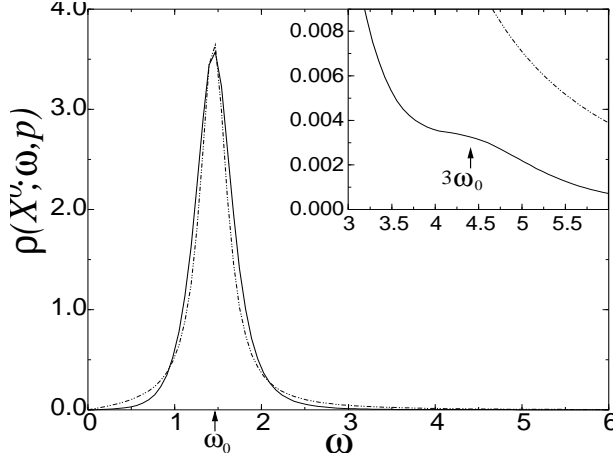,width=8.0cm,height=6.0cm}}
\caption{Wigner transform of the spectral
function as a function of $\omega$ at $X^0=35.1$ for $\bp=0$ in units of $m$. 
Also shown is a fit to a Breit-Wigner function (dotted) with 
$(\omega_{0},\Gamma_{0}) = (1.46,0.37)$. The inset shows a blow-up around 
the three-particle threshold $3\omega_{0}$. The expected bump is small but visible. Here the coupling is $\lambda/m^2=4$.}
\label{fig:spectral1}
\end{figure} 
For the numerical solution we solve the evolution equations (\ref{eq:exactrhoF}) using the above two-loop self-energy contributions. The initial conditions are given by (\ref{eq:initial1}) with the initial distribution (\ref{eq:tsunami}).
 For the plots we have used a space lattice with spacing 
$ma=0.3$ and a time lattice $a_0/a=0.25$. The system size is $mL=24$. 
In Fig.\ \ref{fig:spectral1} we display the Wigner
transform $\tilde{\rho}(X^0;\om,p)$ for the zero momentum mode for $mX^0=35.1$.  One clearly observes that the interacting theory has a
continuous spectrum described by a peaked spectral function with a nonzero
width. The inset shows a blow-up of the zero mode around the three-particle
threshold $3\omega_0/m=4.38$. The expected enhancement in the spectral function is
small but visible. 
In the figure we also present a fit to a Breit-Wigner spectral
function. While the position of the peak can be fitted
easily, the overall shape and width are only qualitatively captured.
In particular, the slope of $\rho(X^0;\om,p)$ for small $\om$ is 
quantitatively different. We also see that the Breit-Wigner fits give
a narrower spectral function (smaller width) and therefore would predict a
somewhat slower exponential relaxation in real time.

\subsection{Two-particle irreducible $1/N$ expansion} 
\label{sec:2PIN}

\subsubsection{Power counting}

The above loop or coupling expansion is restricted, a priori, to weakly coupled systems. However, even in the presence of a weak coupling the dynamics can be strongly correlated. Important examples concern systems with high occupation numbers, where the statistical correlation function $F$ grows large.  
In this section we discuss a nonperturbative 
approximation scheme for the 2PI effective action. It
classifies the contributions to the 2PI effective action
according to their scaling with powers of $1/N$, where
$N$ denotes the number of field components:
\bea\db
\Gamma_2[\phi,G] &\db\!=\!& {\db  \Gamma_2^{\rm LO}[\phi,G]} 
          \,+\, {\rr \Gamma_2^{\rm NLO}[\phi,G]} 
          \,+\, \Gamma_2^{\rm NNLO}[\phi,G]
          \,+\, \ldots \nonumber\\
 && \quad
{\db\sim N^1} \qquad\quad\,\, {\rr \sim N^0} \qquad\quad\,\,
\sim N^{-1} 
\nonumber
\eea
Each subsequent contribution $\Gamma_2^{\rm LO}$,
$\Gamma_2^{\rm NLO}$, $\Gamma_2^{\rm NNLO}$ etc.~is  
down by an additional factor of $1/N$. The importance of an expansion 
in powers of $1/N$ stems from the fact that it provides a controlled 
expansion parameter that is not based on weak couplings.
It can be applied to describe physics characterized by
nonperturbatively large fluctuations, such as encountered
near second-order phase transitions in thermal equilibrium,
or for extreme nonequilibrium phenomena such as 
parametric resonance and the subsequent emergence of strong turbulence just to mention two examples which are discussed in later chapters. For the latter cases a 2PI coupling
or loop expansion is not applicable. The method
can be applied to bosonic or fermionic theories alike if
a suitable field number parameter is available,
and we exemplify it here for the case of the scalar
$O(N)$-symmetric theory with classical action (\ref{eq:classical}).

We present a classification
scheme based on $O(N)$--invariants which parametrize the 2PI diagrams
contributing to $\Gamma[\phi,G]$. The interaction term of
the classical action in (\ref{eq:classical}) is written such that
$S[\phi]$ scales proportional to $N$. From the fields $\phi_a$ alone one
can construct only one independent invariant under $O(N)$ rotations, which
can be taken as $\tr\, \phi\phi \equiv \phi^2 = \phi_a \phi_a \sim
N$.  The minimum $\phi_0$ of the classical effective potential for this
theory is given by $\phi_0^2 = N (-6 m^2/\lambda)$ for negative
mass-squared $m^2$ and scales proportional to $N$.  Similarly, the trace
with respect to the field indices of the classical propagator $G_{0}$ is
of order~$N$.

The 2PI effective action is a singlet under $O(N)$ rotations and
parametrized by the two fields $\phi_a$ and $G_{ab}$.  To write down the
possible $O(N)$ invariants, which can be constructed from these fields, we
note that the number of $\phi$--fields has to be even in order to
construct an $O(N)$--singlet. For a compact notation we use 
$(\phi \phi )_{ab} \equiv \phi_a \phi_b$. 
All functions of $\phi$ and $G$, which are singlets under $O(N)$, can be
built from the irreducible (i.e.\ nonfactorizable in field-index space)  
invariants
\beq\rr 
\phi^2, \quad\quad \tr (G^n) \quad\quad \mbox{and}  
\quad\quad \tr (\phi \phi G^n). 
\label{oninvariants}
\eeq
We note that for given $N$ only the invariants with $n \le N$ are
irreducible --- there cannot be more independent invariants than fields. 
We will see below that for lower orders in the $1/N$ expansion and
for sufficiently large $N$ one has $n < N$. In particular, for the
next-to-leading order approximation one finds that only invariants with $n
\le 2$ appear, which makes the expansion scheme appealing from a
practical point of view. 

Since each single graph contributing to $\Gamma[\phi,G]$ is an
$O(N)$--singlet, we can express them with the help of the set of
invariants in (\ref{oninvariants}). The factors of $N$ in a given
graph have two origins: 
\bi
\item {\rr each irreducible invariant is taken to scale
proportional to $N$ since it contains exactly one trace over the field
indices,} 
\item {\rr while each vertex provides a factor of $1/N$.} 
\ei
\begin{figure}[t]
\centerline{\epsfig{file=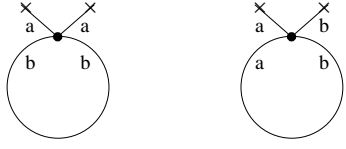,width=4.5cm}}
\caption{
Graphical representation of the $\phi$--dependent contributions for
$\Gamma_2 \equiv 0$. The crosses denote field insertions $\sim
\phi_a\phi_a$ for the left figure, which contributes at leading order, and
$\sim \phi_a\phi_b$ for the right figure contributing at next-to-leading
order.
}
\label{fig:oneloopfig}
\end{figure}
The expression (\ref{2PIaction}) for the 2PI effective
action contains, besides the classical action, the one-loop contribution
proportional to $\Tr\,\ln G^{-1} + \Tr\, G_0^{-1}(\phi) G$ and a
nonvanishing $\Gamma_2[\phi,G]$ if higher loops are taken into account.
The one-loop term contains both leading order (LO) and 
next-to-leading order (NLO) contributions.  The logarithmic
term corresponds, in absence of other terms, simply to the free field
effective action and scales proportional to the number of field
components $N$. To separate the LO and NLO contributions at the one-loop
level consider the second term $\Tr\, G_0^{-1}(\phi) G$. From the form of
the classical propagator (\ref{classprop}) one observes that it can be
decomposed into a term proportional to $\tr(G) \sim N$ and terms
$\sim (\lambda/6N) \left[ \tr(\phi\phi)\,\tr(G) + 2\,\tr(\phi\phi G) \right]$.
This can be seen as the sum of two ``2PI one-loop graphs'' with field
insertion $\sim \phi_a\phi_a$ and $\sim \phi_a\phi_b$, respectively,
as shown in Fig.~\ref{fig:oneloopfig}.
Counting the factors of $N$ coming from the traces and the prefactor, one
sees that only the first   
contributes at LO, while the second one is NLO.

According to the above rules one draws all topologically
distinct 2PI diagrams and counts the number of closed lines as well as 
the number of lines connecting two field insertions in a diagram
following the indices. For instance, the left diagram of
Fig.~\ref{fig:oneloopfig} admits one line connecting two field insertions
and one closed line. In contrast, the right figure admits only one line 
connecting two field insertions by following the indices. Therefore,
the right graph exhibits one factor of $N$ less and becomes subleading.
Similarly, for the two-loop graph below one finds:

\vspace*{0.2cm}

\centerline{\epsfig{file=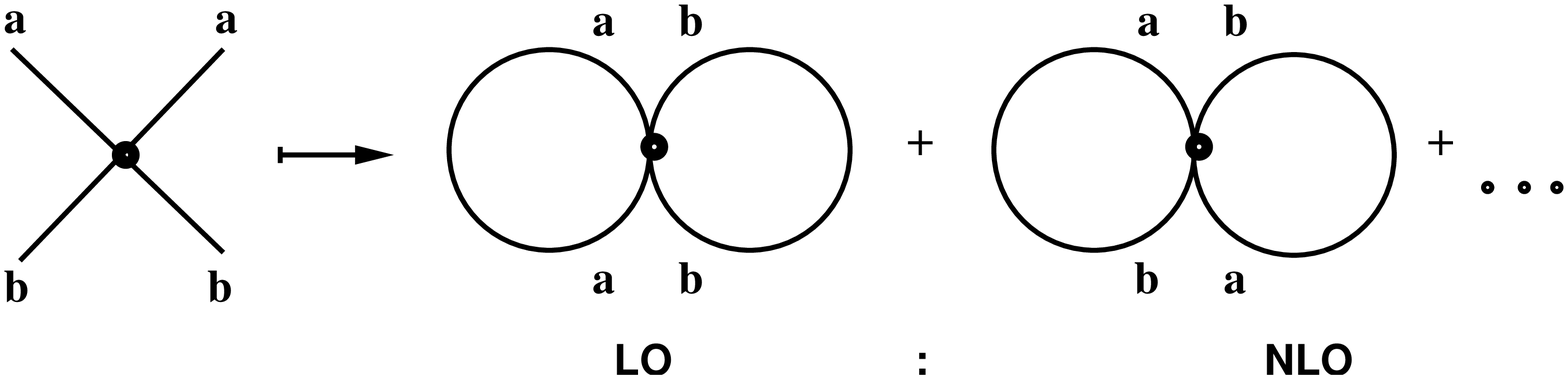,width=8.8cm}}
\hspace*{5.cm}{\db$(\tr G)^2/N \sim N$}  
\hspace*{0.7cm}{\rr$\tr G^2/N \sim N^{0}$}

\vspace*{0.2cm}

\noindent
The same can be applied to higher orders. We 
consider first the contributions to 
\mbox{$\Gamma_2[\phi=0,G] \equiv \Gamma_2[G]$},
i.e.~for a vanishing field expectation value and discuss $\phi \not = 0$
below. The LO contribution to $\Gamma_2[G]$ consists of only one 
two-loop graph, whereas to NLO there is an infinite series of
contributions which can be analytically summed:
\bea\db
\Gamma_2^{\rm LO}[G] &\db =&\db - \frac{\lambda}{4! N} 
  \int_x G_{aa}(x,x) G_{bb}(x,x) \, ,  
\label{LOcont} \\[0.2cm]
\rr \Gamma_2^{\rm NLO}[G] &\rr =&\rr  \frac{i}{2} \Tr  
\ln [\, B(G)\, ] \, , 
\label{eq:NLOcont} \\ 
\rr B(x,y;G) &\rr =&\rr \delta(x-y)
+ i \frac{\lambda}{6 N}\, G_{ab}(x,y)G_{ab}(x,y) \, .
\label{eq:Feq} 
\eea
In order to see that (\ref{eq:NLOcont}) with (\ref{eq:Feq}) 
sums the following infinite series of diagrams

\vspace*{0.2cm}

\centerline{\epsfig{file=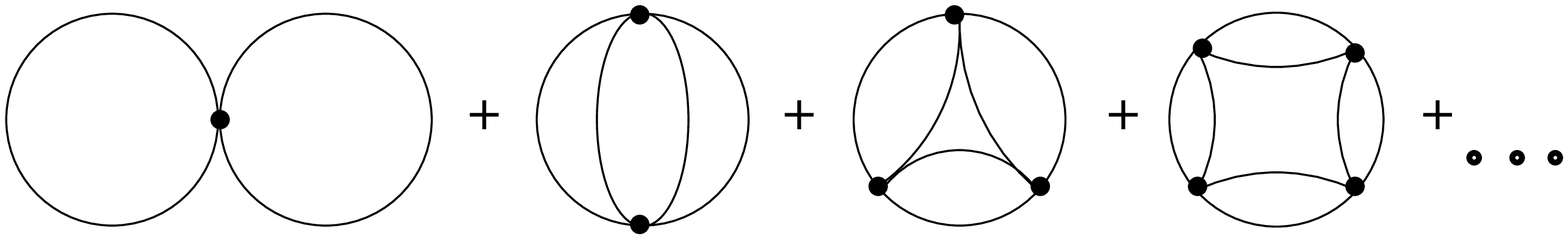,width=8.cm}}

\vspace*{0.2cm}

\noindent
one can expand:
\bea \rr
\Tr \ln [\, B(G)\, ] 
&\rr=&\rr  \int_{x} 
 \left( i \frac{\lambda}{6 N} G_{ab}(x,x)G_{ab}(x,x) \right)
\nonumber\\
&\rr -&\rr \frac{1}{2} \int_{xy}
\left( i \frac{\lambda}{6 N} G_{ab}(x,y)G_{ab}(x,y) \right)
\left( i \frac{\lambda}{6 N}\, G_{a'b'}(y,x)G_{a'b'}(y,x) \right)
\nonumber\\
&\rr +&\rr \ldots  \label{eq:logexpansion}
\eea
The first term on the RHS~of (\ref{eq:logexpansion}) 
corresponds to the two-loop graph with the index structure 
exhibiting one trace such that the contribution scales as
$\tr G^2/N \sim N^0$. One observes that each additional
contribution scales as well proportional to 
$(\tr G^2/N)^n \sim N^0$ for all $n \ge 2$. Thus
all terms contribute at the same order. 

The terms appearing in the presence of a nonvanishing field expectation
value are obtained from the effectively cubic interaction term in 
(\ref{interactionS}) for $\phi \not = 0$. One first notes that 
there is no $\phi$-dependent graph contributing at LO. To NLO there
is again an infinite series of diagrams $\sim N^0$ which can be summed:
\bea\db
\Gamma_2^{\rm LO}[\phi,G] &\db \equiv&\db   
\Gamma_2^{\rm LO}[G] \, , 
\label{eq:LObrok} \\[0.3cm]
\rr \Gamma_2^{\rm NLO}[\phi,G] &\rr =& \rr   
\Gamma_2^{\rm NLO}[\phi\equiv 0,G] + 
\frac{i\lambda}{6N} \int_{xy} I (x,y;G)\,
\phi_a(x) \, G_{ab} (x,y) \, \phi_b (y) \, ,  \label{eq:NLObrok} \\
\rr I (x,y;G) &\rr =&\rr \frac{\lambda}{6 N}\, G_{ab}(x,y) G_{ab}(x,y)
- i\, \frac{\lambda}{6 N} \int_{z}\, I (x,z;G)
 G_{ab}(z,y) G_{ab}(z,y)  \, . \nonumber\\
 \label{eq:Ifunc}
\eea
The series of terms contained in (\ref{eq:NLObrok}) with (\ref{eq:Ifunc})
corresponds to the diagrams:

\vspace*{0.2cm}
\centerline{\epsfig{file=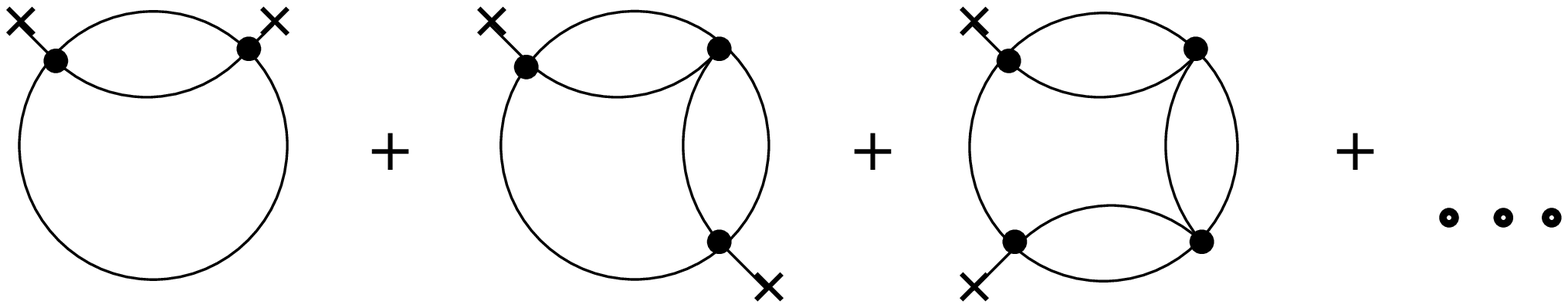,width=6.6cm}}

\vspace*{0.2cm}

\noindent
The functions $I (x,y;G)$ and the inverse 
of $B(x,y;G)$ are closely related by
\beq\db
B^{-1}(x,y;G) = \delta(x-y) - i I (x,y;G) \, ,
\label{Binverse}
\eeq
which follows from convoluting (\ref{eq:Feq}) with $B^{-1}$ and
using (\ref{eq:Ifunc}). We note that $B$ and $I$ do not depend
on $\phi$, and $\Gamma_2[\phi,G]$ is only quadratic in $\phi$ at NLO.

It is straightforward to apply the above description of an expansion of the
2PI effective action in the number of field components to boson or fermion
field theories with ``vector-like'' $N$-component fields. It is very helpful that
for the $N$-component theory discussed above one can analytically sum the infinite
series of NLO contributions analytically. The situation is 
different for theories with ``matrix'' fields, such $SU(N)$ gauge theories
relevant for QCD, where no corresponding closed expression is known. 

\subsubsection{Self-energies in the symmetric regime}
\label{nonequiN}

We consider first the case of the scalar $O(N)$ symmetric 
field theory with
a vanishing field expectation value such that 
$F_{ab}(x,y) = F(x,y) \delta_{ab}$
and $\rho_{ab}(x,y) = \rho(x,y) \delta_{ab}$. The case $\phi \not = 0$ is
treated below.
In the $1/N$--expansion of the 2PI effective action to next-to-leading
order the effective mass term $M^2(x;G)$ appearing in the 
evolution equations (\ref{eq:exactrhoF}) is given by
\beq
\db M^2(x;F) \,=\,  m^2 + \lambda \frac{N+2}{6 N}\, F(x,x) \, .
\label{eq:massNLO}
\eeq
One observes that this local self-energy 
part receives LO and NLO contributions.
In contrast, the non-local part of the self-energy (\ref{eq:sighominh})
is nonvanishing 
only at NLO: $\overline{\Sigma}(x,y;G) = - \lambda/(3 N)\, G(x,y)
I(x,y)$ and using the decomposition identities (\ref{eq:decompid})
and (\ref{eq:decompself}) one finds
\bea
{\db \Sigma_F(x,y)} &\db=&\db  - \frac{\lambda}{3 N}\, 
\Big( {\db F(x,y) I_F(x,y)} 
-\frac{1}{4} {\rr \rho(x,y) I_{\rho}(x,y)} \Big) \, ,
\label{SFFR} \\[0.2cm]
{\rr \Sigma_{\rho}(x,y)} &\db=&\db  - \frac{\lambda}{3 N}\, \Big( 
{\db F(x,y)} {\rr I_{\rho}(x,y)} 
+ {\rr \rho(x,y)} {\db I_{F}(x,y)} \Big)  \, . 
\label{SR}
\eea
Here the summation function (\ref{eq:Ifunc}) reads in terms
of its statistical and spectral components:\footnote{This follows
from using the decomposition identity for the propagator 
(\ref{eq:decompid}) and $I(x,y) = I_F(x,y) - \frac{i}{2} 
I_\rho(x,y)\, \sgn_{\C}(x^0 - y^0)$.} 
\bea
{\db I_{F}(x,y)} &\db=&\db  \Pi_F(x,y) 
- \int\limits_{t_0}^{x^0} {\rm d}z\,
{\rr I_{\rho}(x,z)} \Pi_F(z,y)  + \int\limits_{t_0}^{y^0} {\rm d}z\,  
I_F(x,z) {\rr \Pi_\rho(z,y)} ,\quad
\nonumber\\
{\rr I_{\rho}(x,y)} &\db=&\db {\rr \Pi_\rho(x,y)} 
- \int\limits_{y^0}^{x^0} {\rm d}z\,  
{\rr I_{\rho}(x,z)} {\rr \Pi_\rho(z,y)} \, ,
\label{IRFR}
\eea
where
\bea
{\db \Pi_{F}(x,y)} &\db=&\db  \frac{\lambda}{6N}\, 
\Big( {\db F_{ab}(x,y)F_{ab}(x,y)} - \frac{1}{4} {\rr \rho_{ab}(x,y)\rho_{ab}(x,y)} \Big) \, , 
\nonumber\\[0.2cm] \db
{\rr \Pi_{\rho}(x,y)} &\db=&\db \frac{\lambda}{3N}\, {\db F_{ab}(x,y)} {\rr \rho_{ab}(x,y)}  \, ,
\label{PiRFR}
\eea
using the abbreviated notation $\int_{t_1}^{t_2}
{\rm d}z \equiv \int_{t_1}^{t_2} {\rm d}z^0 
\int_{-\infty}^{\infty} {\rm d}^d z$. Here we kept in (\ref{PiRFR}) the general notation without summing over field indices, since the same expressions will be used for the case with a nonvanishing field expectation value below.
We note that $F(x,y)$ and $\rho(x,y)$ along with the other statistical and spectral components of the self-energies are real functions.

\subsubsection{Nonvanishing field expectation value} 
\label{2PINfield}

In the presence of a nonzero field expectation value $\phi_a$ the 
most general propagator can no longer be evaluated for the diagonal
configuration (\ref{eq:symmetricG}).    
For the $N$-component scalar field theory (\ref{eq:classical})
one has with $\phi^2 \equiv \phi_a\phi_a$:
\bea\db
M_{ab}^2(x;\phi,F) &\db=&\db \left( m^2 + \frac{\lambda}{6N}\, 
\left[F_{cc}(x,x)
+\phi^2(x) \right]
\right) \delta_{ab} \nonumber\\
&\db+&\db \frac{\lambda}{3N}\, \left[F_{ab}(x,x) 
+ \phi_a(x)\phi_b(x) \right] \, .
\label{Meffphi}
\eea
The self-energies 
$\Sigma^{F}_{ab}(x,y) \equiv \Sigma^{F}_{ab}(x,y;\phi,\rho,F)$ 
and $\Sigma^{\rho}_{ab}(x,y) \equiv \Sigma^{\rho}_{ab}(x,y;\phi,\rho,F)$
are obtained from (\ref{eq:NLObrok}) as
\bea \db 
\Sigma^{F}_{ab}(x,y) & \db  = & \db   - \frac{\lambda}{3 N}\Big\{
 I_F(x,y)\left[ \phi_a (x)\phi_b(y) + F_{ab}(x,y) \right] -
 \frac{1}{4} {\rr I_{\rho}(x,y)\rho_{ab}(x,y)}  
 \nonumber\\
 && \db  \qquad \quad + P_F(x,y)F_{ab}(x,y) -
 \frac{1}{4} {\rr P_{\rho}(x,y)\rho_{ab} (x,y)} \Big\},
 \label{ASFFR}\\
 \rr \Sigma^{\rho}_{ab} (x,y) & \db  = & \db   - \frac{\lambda}{3 N}\Big\{ 
 {\rr I_{\rho}(x,y)} \left[ \phi_a (x)\phi_b(y) + F_{ab} (x,y) \right] 
 + I_{F}(x,y){\rr \rho_{ab} (x,y)}  
 \nonumber\\
 && \db  \qquad \quad + {\rr P_{\rho}(x,y)} F_{ab}(x,y) + 
 P_{F}(x,y) {\rr \rho_{ab} (x,y)} \Big\}. 
 \label{ASRFR}
\eea
The functions $I_{F}(x,y) \equiv I_{F}(x,y;\rho,F)$ and 
$I_{\rho}(x,y) \equiv I_{\rho}(x,y;\rho,F)$ satisfy the
corresponding equations as for the case of a vanishing macroscopic field given above. 
The respective $\phi$-dependent summation functions 
$P_{F}(x,y) \equiv P_{F}(x,y;\phi,\rho,F)$ and 
$P_{\rho}(x,y) \equiv P_{\rho}(x,y;\phi,\rho,F)$
are given by  
\bea \db  
 P_{F} (x,y) & \db =& \db  - \frac{\lambda}{3N} \Bigg\{ H_F (x,y)
   - \int_{t_0}^{x^0} {\rm d}z\, \left[ {\rr H_{\rho}} (x,z)I_F (z,y) +
 {\rr I_{\rho} (x,z)} H_F (z,y) \right] 
 \nonumber\\
 && \db  + \int_{t_0}^{y^0} {\rm d}z\, \left[  H_F (x,z) {\rr I_{\rho} (z,y)} +
 I_F (x,z) {\rr H_{\rho} (z,y)}\right] 
 \nonumber\\
 && \db  - \int_{t_0}^{x^0} {\rm d}z\, \int_{t_0}^{y^0} {\rm d}v\,
 {\rr I_{\rho} (x,z)} H_F (z,v) {\rr I_{\rho} (v,y)}
 \nonumber\\
 && \db  + \int_{t_0}^{x^0} {\rm d}z\, \int_{t_0}^{z^0} {\rm d}v\,
 {\rr I_{\rho} (x,z) H_{\rho} (z,v)} I_F (v,y)
 \nonumber\\
 && \db  + \int_{t_0}^{y^0} {\rm d}z\, \int_{z^0}^{y^0} {\rm d}v\,
 I_F (x,z) {\rr H_{\rho} (z,v) I_{\rho} (v,y)} \Bigg\},
\label{eqB11}
\eea
\bea \rr
 P_{\rho} (x,y) & \db  = & \db  - \frac{\lambda}{3N} \Bigg\{ 
{\rr H_{\rho} (x,y)}  - \int_{y^0}^{x^0} {\rm d}z\, 
 \left[ {\rr H_{\rho} (x,z) I_{\rho} (z,y)} + 
 {\rr I_{\rho} (x,z) H_{\rho} (z,y)} \right] 
 \nonumber\\
 && \db  +  \int_{y^0}^{x^0} {\rm d}z\, \int_{y^0}^{z^0} {\rm d}v\,
 {\rr I_{\rho} (x,z) H_{\rho} (z,v) I_{\rho} (v,y)} \Bigg\},
\label{eqB12}
\eea
with 
\beq\db
H_F(x,y) \equiv -\phi_a(x) F_{ab}(x,y) \phi_b(y) \,\, , \qquad 
{\rr H_\rho(x,y) \equiv -\phi_a(x) \rho_{ab}(x,y) \phi_b(y)} \,.
\eeq
The time evolution equation for the field 
(\ref{eq:exactphi}) for the 2PI effective action
to NLO (\ref{eq:NLObrok}) is given by 
\bea\db
\lefteqn{\left(\left[\square_x + \frac{\lambda}{6N}
\phi^2(x)\right] \delta_{ab} +  M_{ab}^2(x;\phi = 0,F) 
\right) \phi_b(x)} \qquad\qquad  \nonumber\\ 
&\db\! =\!&\db \frac{\lambda}{3N} \int_{t_0}^{x^0} {\rm d}y \, 
 \left[ {\rr I_{\rho} (x,y)} F_{ab} (x,y) + 
 I_F (x,y) {\rr \rho_{ab} (x,y)} \right] \phi_b (y)
\nonumber\\
&\db\! =\!&\db - \int_{t_0}^{x^0}\! {\rm d}y \, 
 {\rr \Sigma^{\rho}_{ab}(x,y;{\db \phi = 0,F},\rho)}\, \phi_b (y)\, .
\label{eq:NLOphi}
\eea

\subsubsection{Dephasing at leading order}   
\label{sec:lofixedpoints}

For simplicity we consider spatially homogeneous field expectation 
values $\phi_a(t) = \langle \Phi_a(t,\bx) \rangle$, such that
we can use the modes $F_{ab}(t,t';\bp)$ and
$\rho_{ab}(t,t';\bp)$ after spatial Fourier transformation to describe the dynamics.

The LO contribution to the 2PI 
effective action (\ref{LOcont}) adds a time-dependent mass shift 
to the free field evolution equation. The resulting effective mass term, 
given by (\ref{Meffphi}) for $N \to \infty$, is the 
same for all Fourier modes and consequently each mode propagates 
``collisionlessly''. There are no further corrections 
since according to (\ref{eq:NLOcont}) and (\ref{eq:NLObrok}) 
the self-energies $\rr \Sigma_{F}$ and $\Sigma_{\rho}$ 
are $\Or (1/N)$ and vanish in this limit. Therefore, the evolution 
equations (\ref{eq:exactrhoF}) for this
approximation read:
\bea 
\db\left[\partial_t^2 + \bp^2 
+ M^2(t;\phi,F) 
\right]\, F_{ab}(t,t';\bp) &{\db =}& 
{\db 0}\,\, , \nonumber\\ 
{\db \left[\partial_t^2 + \bp^2 
+ M^2(t;\phi,F) \right]}\, {\rr \rho_{ab}(t,t';\bp)} 
&{\db =}& {\db 0}\,\, , 
\label{eq:LOdyn}\\
\db \left[ \partial_t^2 + \frac{\lambda}{6N}
\phi^2(t) + M^2(t;0,F) \right] \phi_b(t) 
&{\db =}& {\db 0} \,\, ,\nonumber
\eea
with
\beq\db
M^2(t;\phi,F) \, \equiv \, m^2 + \frac{\lambda}{6N}\, 
\left[\int_{\bp} F_{cc}(t,t;\bp)
+\phi^2(t) \right]  \, , 
\label{eq:LOmass}
\eeq
where $\int_{\bp} \equiv \int {\rm d}^dp/(2\pi)^d$.
In this case one observes that the evolution of $F$ and $\phi$ is  
decoupled from~$\rho$. Similar to the free 
field theory limit, at LO the spectral function does not influence 
the time evolution of the statistical propagator. 
The reason is that in this approximation the
spectrum consists only of ``quasiparticle'' modes of energy
$\omega_\bp(t) = \sqrt{\bp^2 + M^2(t)}$ with an
infinite life-time. The associated mode particle numbers are conserved 
for each momentum separately. In contrast, in the interacting quantum 
field theory beyond LO, direct scattering processes occur such that mode 
occupancies can change. We note that there are also no memory integrals appearing on the
RHS~of (\ref{eq:LOdyn}) as a consequence of this approximation. 

As an example, we consider here the dynamics for $d=1$ with the initial conditions at $t_0 = 0$:
\bea
&&\db F(0,0;\bp) \, = \, \frac{f_\bp(0)+1/2}{\sqrt{\bp^2+M^2(0)}} \quad , \quad
\partial_{t}F(t,0;\bp)|_{t=0} \, =\, 0  \,\, , 
\nonumber\\
&&\db F(0,0;\bp)\partial_{t}\partial_{t'} 
F(t,t';\bp)|_{t=t'=0} \,=\, [f_\bp(0)+1/2]^2  \, ,
\label{eq:initialtsF}\\[0.3cm]
&&\db \phi(0) \,=\, \partial_t\phi(t)|_{t=0} \,=\, 0 \, 
\label{eq:initialtsrho}
\eea
and refer to the suggested literature in section~\ref{eq:reftherm} for further studies in dimensions $d>1$ and for different initial conditions. 
Here $F_{ab}(t,t';\bp) = F(t,t';\bp) \delta_{ab}$, which is
valid for all times with these initial conditions.
The mass term $M^2(0)$ is given by the gap equation (\ref{eq:LOmass}) 
in the presence of the initial nonthermal 
particle number distribution (\ref{eq:tsunami}).

As the renormalization condition we choose the initial renormalized mass in
vacuum, $m_R\equiv M(0)|_{f(0)=0}=1$, as our dimensionful scale. In these
units the particle number is peaked around $|\bp|=p_{\rm ts}=5 m_R$ with 
a width determined by $\sigma=0.5 m_R$ and amplitude ${\cal N}=10$. 
We consider the effective coupling $\lambda/(6 m_R^2) = 1$.

\begin{figure}[t]
\epsfig{file=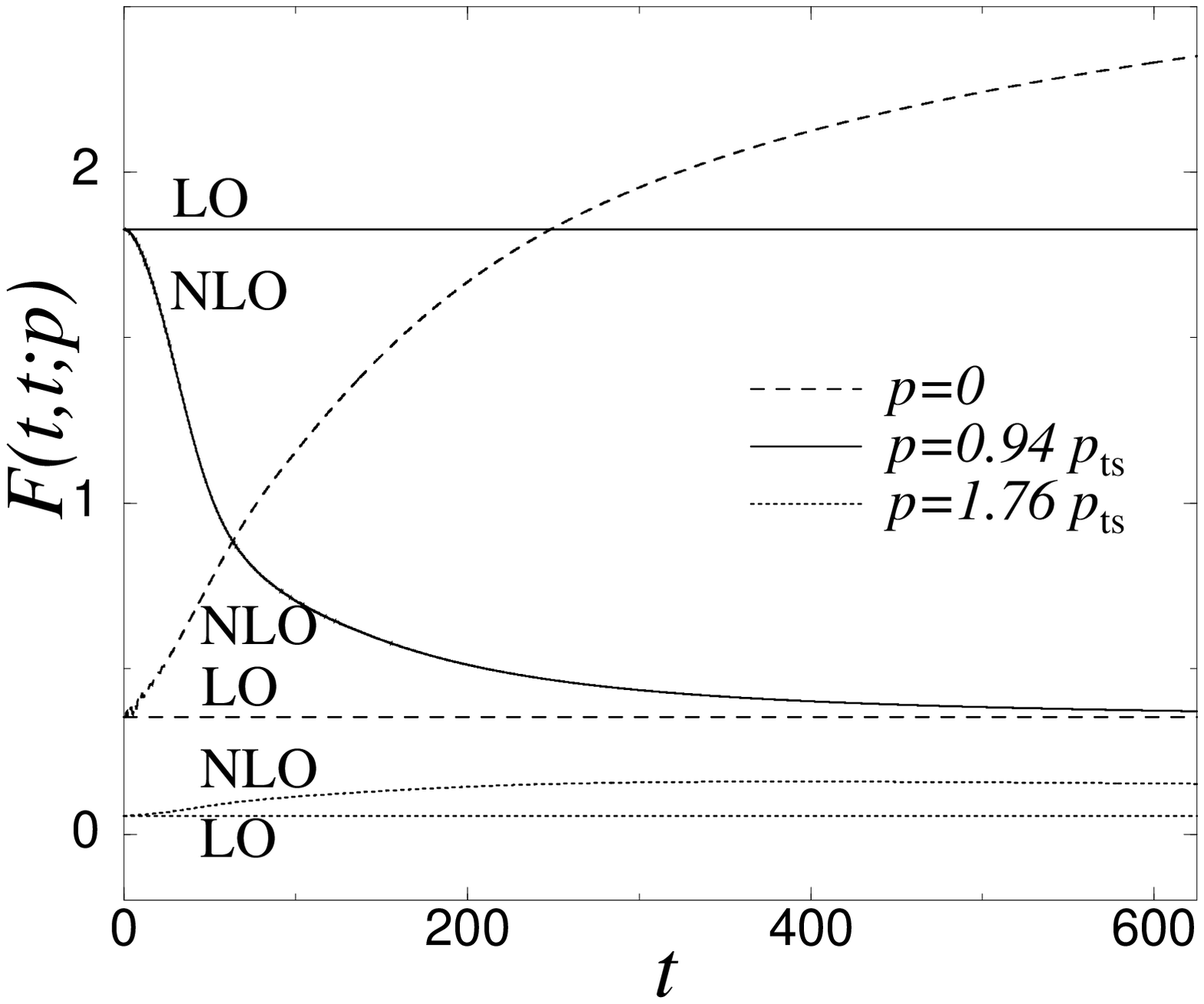,width=6.7cm}
\hspace*{0.4cm}
\epsfig{file=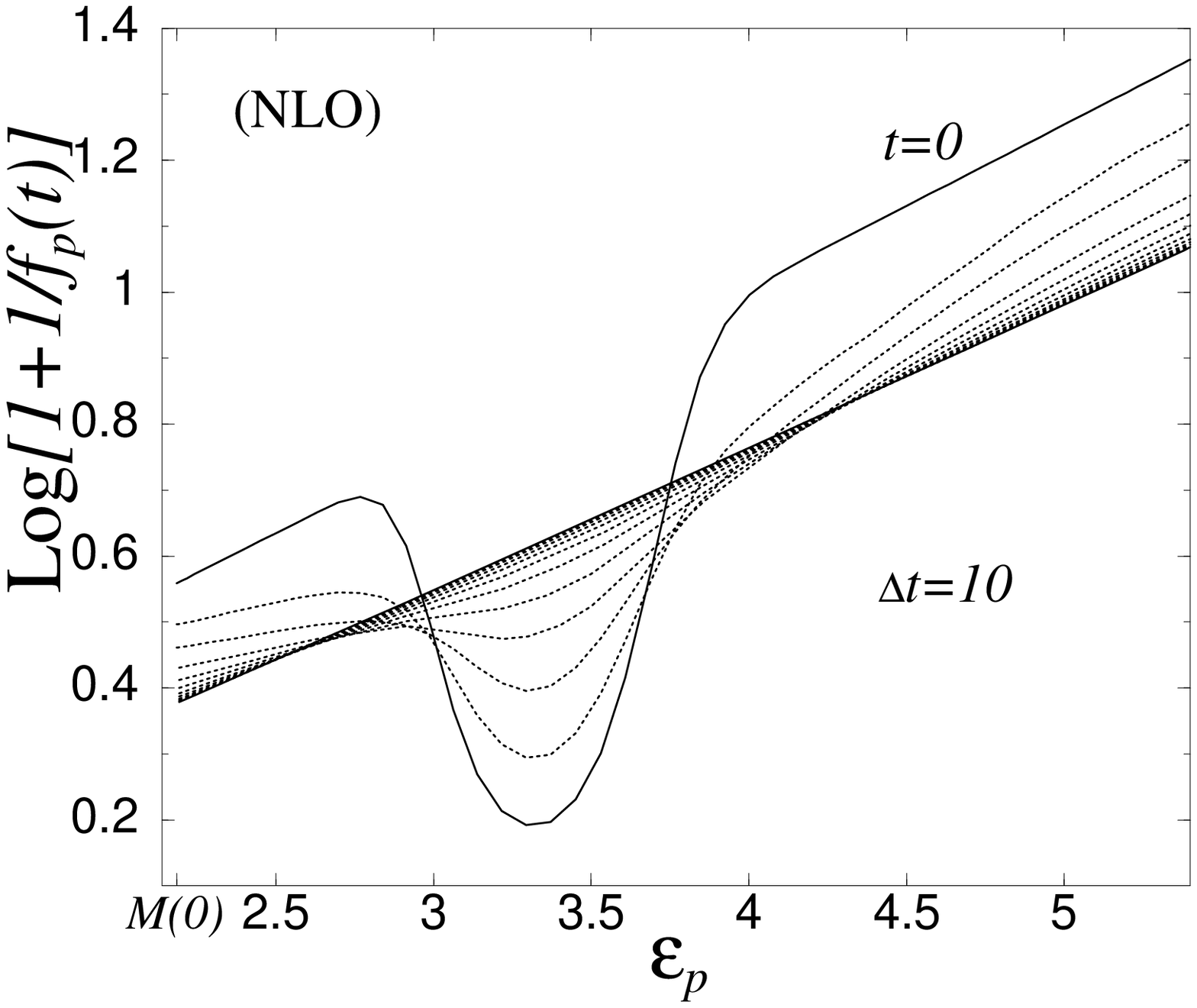,width=7.cm}
\caption{{\bf LEFT:} Comparison of the LO and NLO time dependence of the 
equal-time correlation modes
$F(t,t;\bp)$ for the initial condition 
(\ref{eq:initialtsF}). 
The importance of scattering included in the NLO approximation
is apparent: the evolution deviates from the nonthermal
LO solution and the initially peaked distribution decays,
approaching thermal equilibrium at late times. 
{\bf RIGHT:} Effective particle number distribution for a peaked initial distribution
in the presence of a thermal background. The solid line shows the
initial distribution which for low and for
high momenta follows a Bose-Einstein distribution,
i.e.\ $\ln [1+1/f_\bp(0)] \simeq \epsilon_\bp(0) / T_0$. At late times 
the nonthermal distribution
equilibrates and approaches a straight line with inverse slope
$T_{\rm eq} \, >\, T_0$.}  
\label{Figpeakphi}
\end{figure}
On the left of Fig.\ \ref{Figpeakphi} we present the time evolution of the
equal-time correlation modes $F(t,t;\bp)$
for different momenta: zero momentum, a momentum close to the 
maximally populated momentum $p_{\rm ts}$ and about twice $p_{\rm ts}$. 
One observes that the equal-time correlations at LO are strictly constant
in time. This behavior can be understood from the fact that for the 
employed initial condition 
the evolution starts at a time-translation invariant nonthermal 
solution of the LO equations. There is an infinite number of 
LO solutions which are constant in time, depending on the chosen initial condition details.
This is in sharp contrast to the well-founded expectation that 
the late-time behavior may be well described by thermal equilibrium 
physics. Scattering effects included in the NLO approximation indeed
drive the evolution  
towards thermal equilibrium, which is discussed in detail 
in section~\ref{sec:NLOtherm} below. 

\begin{figure}[t]
\epsfig{file=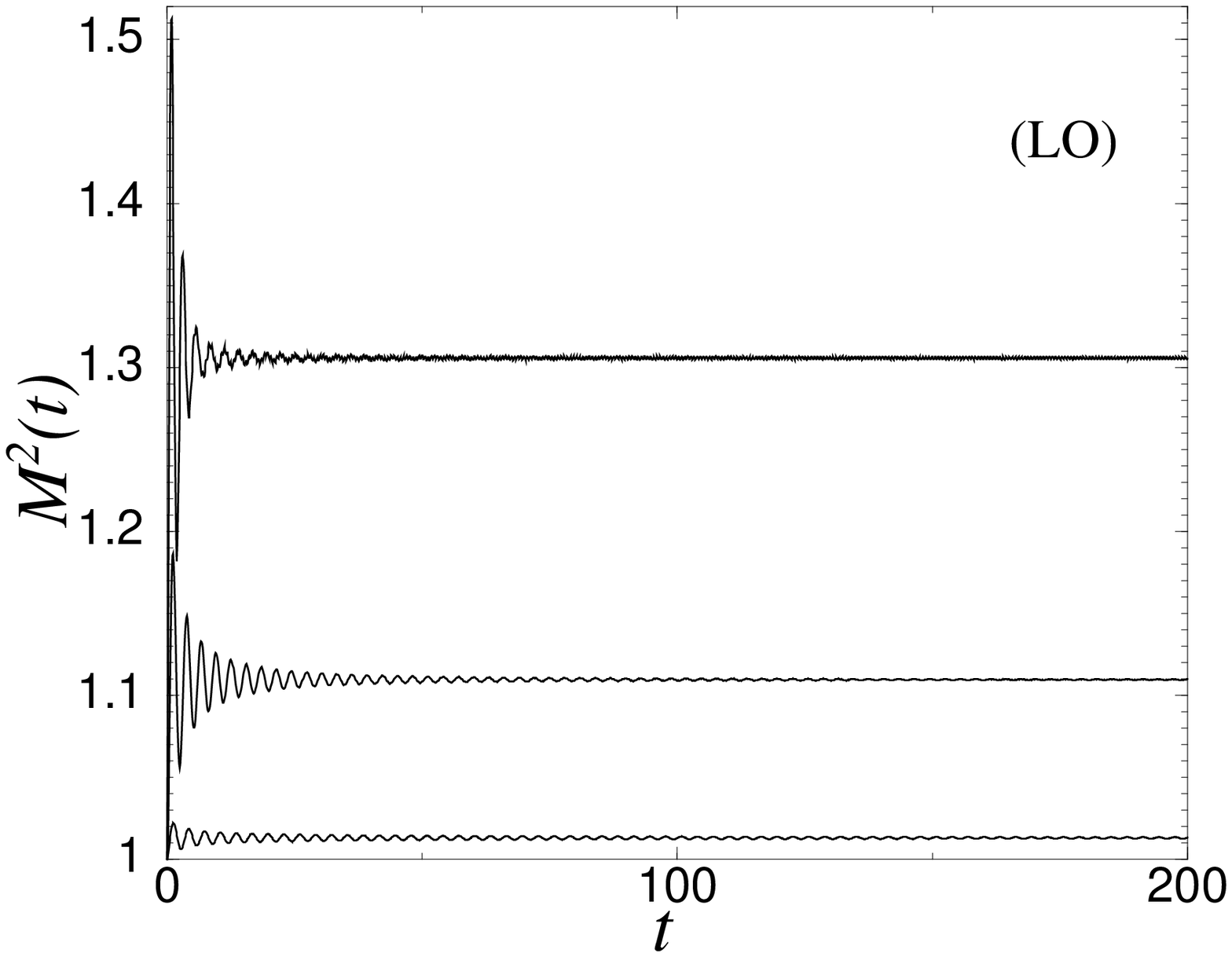,width=8.2cm}
\hspace*{-0.7cm}
\epsfig{file=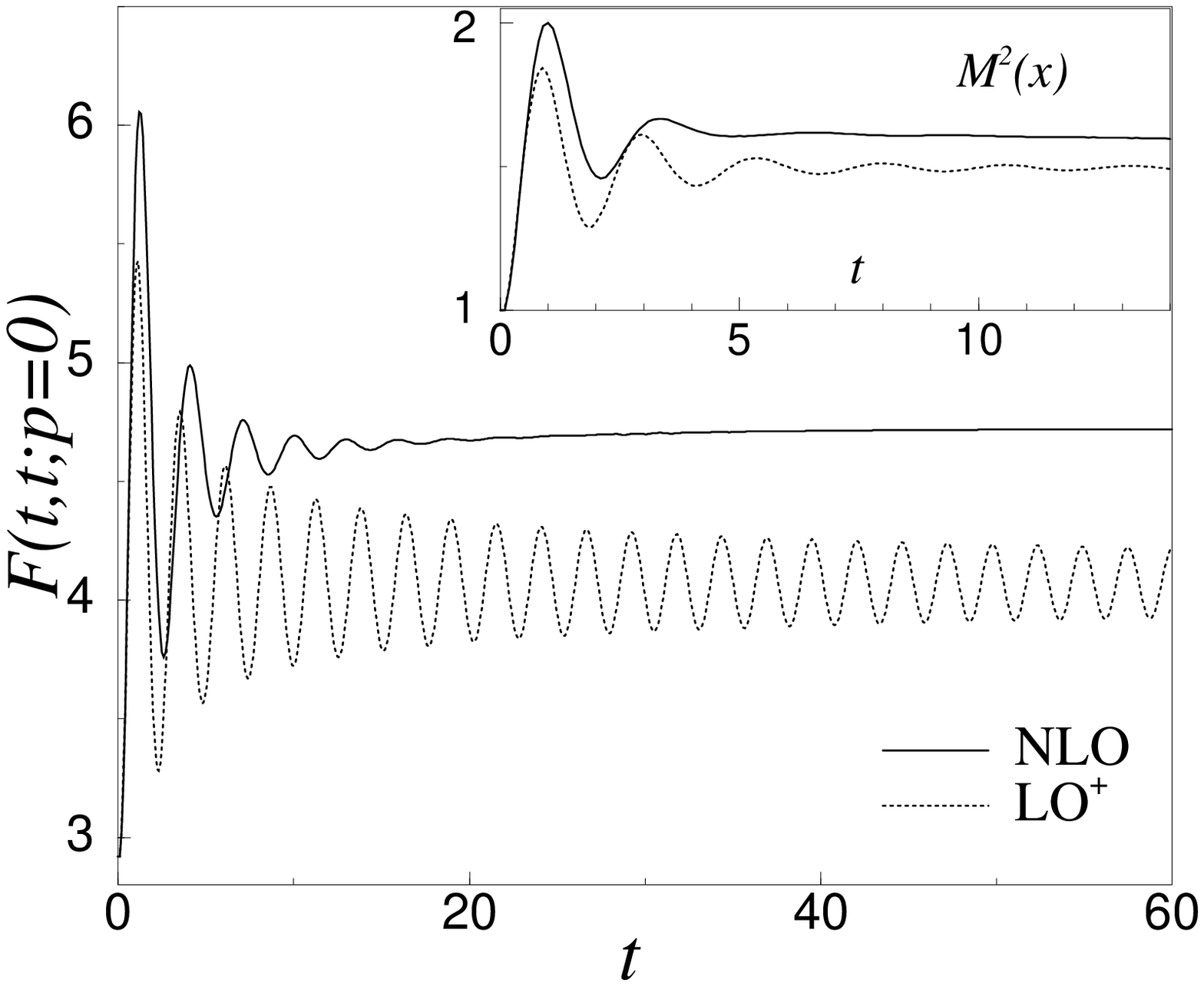,width=6.9cm}
\caption{{\bf LEFT:} Shown is the time-dependent mass term $M^2(t)$ 
in the LO approximation for three different couplings following
a ``quench''. All quantities are given in units of appropriate powers 
of $M(0)$.
{\bf RIGHT:} Time dependence of the equal-time zero-mode  $F(t,t;\bp=0)$
after a ``quench''. The inset shows the mass term $M^2(t)$, which
includes a sum over all modes. The dotted lines represent the
Hartree approximation (LO$^+$), while the solid lines give
the NLO results. The coupling is $\lambda/6N = 0.17 \, M^2(0)$ for $N=4$.}  
\label{FigLOM}
\end{figure}

A remaining question is what happens at LO if the time evolution 
does not start from a time-translation invariant nonthermal 
solution of the LO equations, as was the case for the 
initial condition employed above. 
In the left graph of Fig.~\ref{FigLOM} we plot the evolution of 
$M^2(t)$ in the LO approximation as a function of time $t$,
following a ``quench'' described by an instant drop in the
effective mass term from $2 M^2$ to $M^2$ at initial time. As a consequence, the mass term appearing in the initial conditions (\ref{eq:initialtsrho}) is not the same as the one appearing in the evolution equations (\ref{eq:LOdyn}).
The initial particle number distribution is 
$f_\bp(0)=1/(\exp[\sqrt{\bp^2+2 M^2(0)}/T_0]-1)$ with
$T_0 = 2 M(0)$ and $\phi(0) = \dot{\phi}(0) = 0$.
The sudden change in the effective mass
term drives the system out of equilibrium and one can study its
relaxation. 

In Fig.~\ref{FigLOM} we present $M^2(t)$ for three different couplings 
$\lambda=\lambda_0 \equiv 
0.5 \, M^2(0)$ (bottom), $\lambda = 10\lambda_0$ (middle)
and $\lambda = 40\lambda_0$ (top).   
All quantities are given in units of appropriate powers 
of $M(0)$. Therefore, all curves in the left graph of Fig.~\ref{FigLOM}
start at one. 
The time-dependent mass squared $M^2(t)$ shoots up in response 
to the ``quench'' and stays below the value $2 M(0)^2$ 
of the initial distribution. The amplitude of initial oscillations
is quickly reduced and the evolution is rapidly approaching constant asymptotic values.
Since there are no direct scatterings included in this approximation, the damping is due to a simple dephasing phenomenon of harmonic oscillations with a time-dependent frequency shift.
Correspondingly, the correlator $F(t,t';\bp)$, which is a function of the relative times $t-t'$ and of $t+t'$ at early times, loses its dependence on $t+t'$ asymptotically such that
$F(t,t;\bp)$ becomes a constant. 
 
To understand this in more detail, 
we compare the asymptotic values with the self-consistent
solution of the LO mass equation (\ref{eq:LOmass}) 
for constant mass squared $M_{\rm gap}^2$ and given  
particle number distribution $f_\bp(0)$:\footnote{
Here the logarithmic divergence of the one-dimensional integral 
is absorbed into the bare mass parameter $m^2$ using the same 
renormalization condition as for the dynamical evolution in
the LO approximation, i.e.\ 
$m^2+\frac{\lambda}{6} \int \frac{{\rm d}p}{2 \pi}
\left( f_\bp(0)+\frac{1}{2} \right) (\bp^2+M^2(0))^{-1/2}
= M^2(0)$.}
\beq\db
M^2_{\rm gap}\,=\, m^2 +\frac{\lambda}{6} \int \frac{{\rm d}p}{2 \pi}
\left( {\rr f_\bp(0)} + \frac{1}{2} \right) 
\frac{1}{\sqrt{\bp^2+M^2_{\rm gap}}} \, .
\label{LOgapequ}
\eeq
The result from this gap equation is 
$M^2_{\rm gap} = \{1.01,1.10,1.29\} M^2(0)$ for the three values
of $\lambda$, respectively.  
For this wide range of couplings the values are in 
good numerical agreement with the corresponding dynamical large-time 
results, which can be inferred from Fig.~\ref{FigLOM} as 
$\{1.01,1.11,1.31\}M^2(0)$. One concludes that the asymptotic behavior
at LO is well described in terms of the initial particle number distribution
$f_\bp(0)$. We emphasize that the latter is not a thermal distribution for the
late-time mass terms with values smaller than $2 M^2(0)$. 
 
The question of how strongly the LO late-time results 
deviate from thermal equilibrium depends of course on the 
details of the initial conditions. 
Typically, time- and/or momentum-averaged quantities 
are better determined by the LO approximation
than quantities characterizing a specific
momentum mode. This is related to the prethermalization of characteristic bulk quantities, and we refer to the further discussions of this topic to the literature listed in section~\ref{eq:reftherm}.

Fig.~\ref{FigLOM} shows the equal-time zero-mode $F(t,t;\bp=0)$ along
with $M^2(t)$, which includes the sum over all modes for comparison.
Here we employ a ``quench'' with a larger drop in the
effective mass term from $2.9 M^2(0)$ to $M^2(0)$. The initial 
particle number distribution is $f_\bp(0)=1/(\exp[\sqrt{\bp^2+M^2(0)}/T_0]-1)$
with $T_0=8.5 M(0)$. In the figure the dotted curves show the dynamics
obtained from an ``improved'' LO (Hartree) approximation, LO$^+$, 
that takes into account the
local part of the NLO self-energy contribution and is often employed in the literature. 
The resulting equations have the very same structure as
the LO ones, however, with the LO and NLO contribution to 
the mass term $M^2(t)$ included as given by (\ref{eq:massNLO})
below. The large-time limit of the mass term in the LO$^+$ approximation
is determined by the LO$^+$ stationary solution in complete analogy
to the above discussion. We also give in Fig.~\ref{FigLOM} the NLO results,
which are discussed below.
\begin{figure}[t]
\epsfig{file=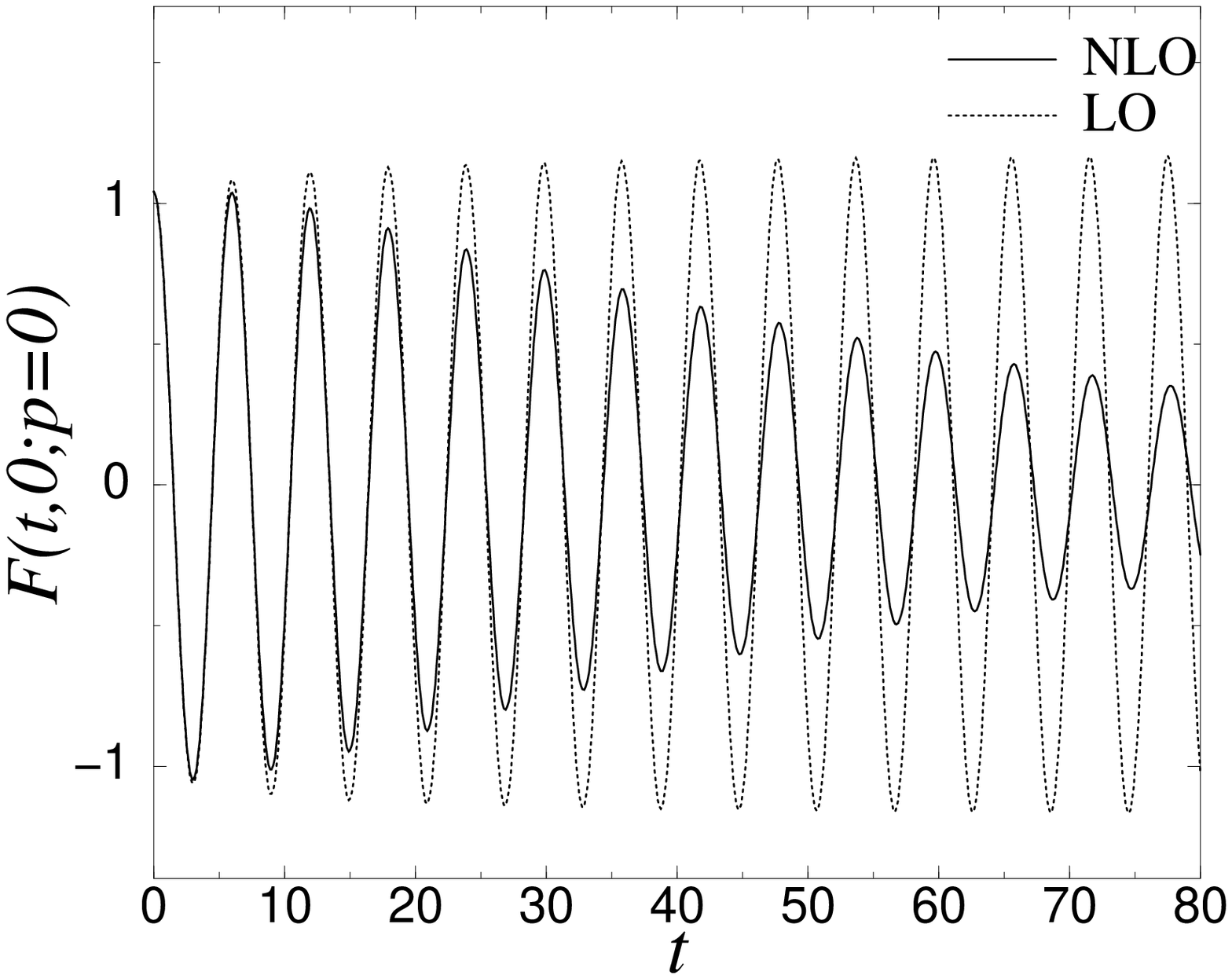,width=7.2cm}
\hspace*{0.3cm}
\epsfig{file=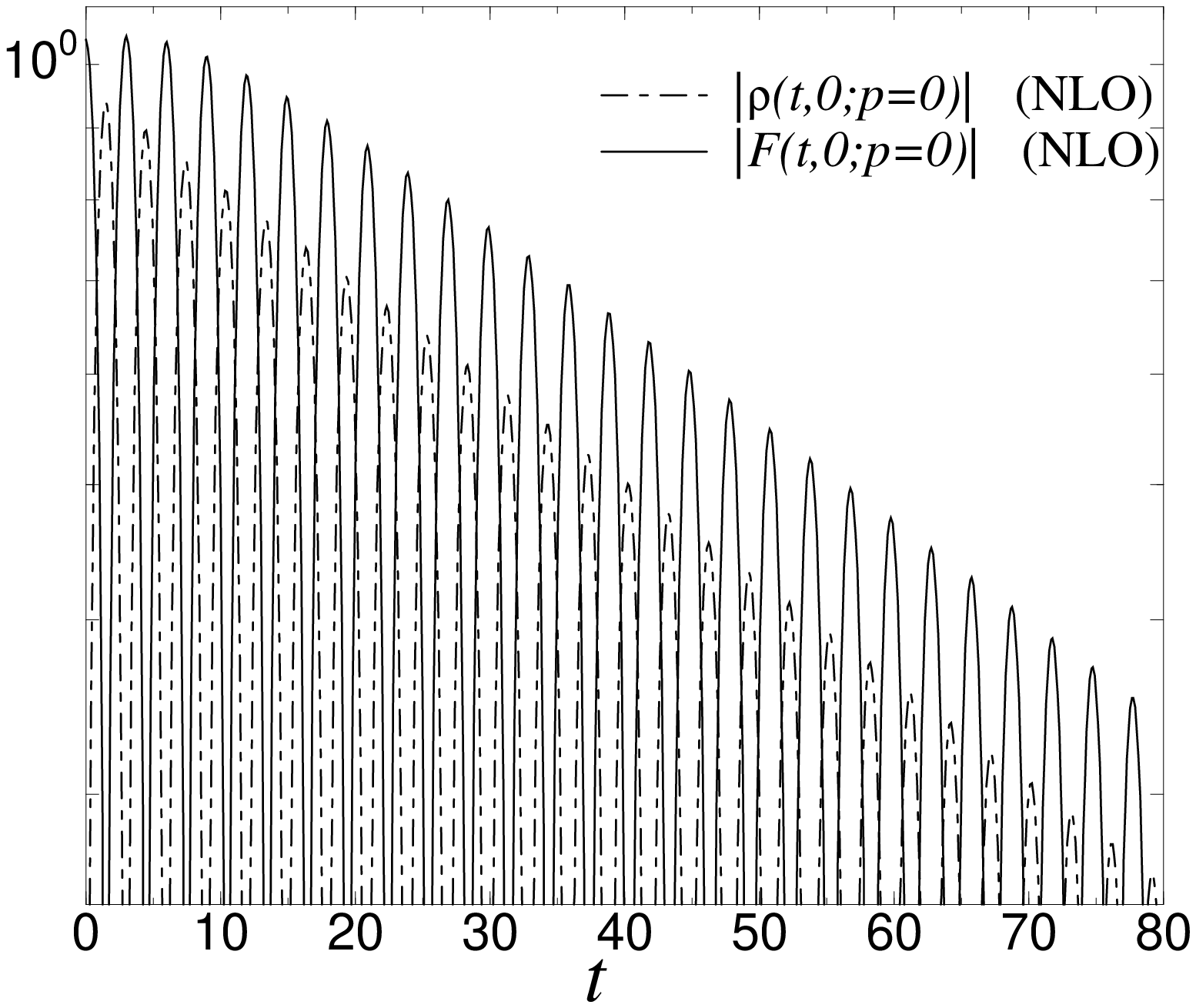,width=6.9cm}
\caption{{\bf LEFT:} Shown is the evolution of the unequal-time correlation 
\mbox{$F(t,0;\bp=0)$} after a ``quench''. Unequal-time 
correlation functions approach zero in the NLO approximation
and correlations with early times are effectively suppressed
($\lambda/6N = (5/6N\simeq 0.083) \, M^2(0)$ for $N=10$).
In contrast, there is no decay of correlations with earlier times
for the LO approximation.  
{\bf RIGHT:} The logarithmic plot of $|\rho(t,0;\bp=0)|$
and $|F(t,0;\bp=0)|$ as a function of time $t$ shows an oscillation
envelope which quickly approaches a straight line.
At NLO the correlation modes therefore approach an exponentially
damped behavior. (All in units of $M(0)$.)}  
\label{FigLogFrho}
\end{figure}

The effective loss of details about the initial conditions is 
a prerequisite for thermalization. At LO this is obstructed by
an infinite number of spurious conserved quantities (mode
particle numbers), which keep initial-time information.
An important quantity in this context is the unequal-time
two-point function $F(t,0;\bp)$, which characterizes the correlations
with the initial time. Clearly, if thermal equilibrium is
approached then these correlations should be damped.
On the left of Fig.~\ref{FigLogFrho} the dotted line shows the
unequal-time zero-mode $F(t,0;\bp=0)$ following the same ``quench''
at LO as for Fig.~\ref{FigLOM} left. One observes no decay of correlations 
with earlier times for the LO approximation. Scattering effects
entering at NLO are crucial for a sufficient effective loss of memory 
about the initial conditions, which is discussed next. 

\subsubsection{Thermalization at NLO}
\label{sec:NLOtherm}

In contrast to the LO approximation, at NLO the self-energies
$\Sigma_{F}$ and $\Sigma_{\rho} \sim {\cal O}(1/N)$ do not vanish.
For the initial conditions (\ref{eq:initialtsF}) all correlators
are diagonal in field index space and $\phi \equiv 0$ for all times.
In this case the evolution equations derived from the NLO 2PI effective
action (\ref{eq:NLOcont}) are given by (\ref{eq:exactrhoF}) with
the self-energies (\ref{eq:massNLO})--(\ref{IRFR}). 
As discussed above, the NLO evolution equations
are causal equations with characteristic 
``memory'' integrals, which integrate over the time history of the
evolution taken to start at time $t_0 = 0$. 

We consider first the same 
initial condition (\ref{eq:initialtsF}) as for the LO case 
discussed above. The result is shown on the left of Fig.~\ref{Figpeakphi}
for $N=10$. One observes that the NLO corrections quickly lead to a decay of the
initially high population of modes around $p_{\rm ts}$.
On the other hand, low momentum modes get populated such that
thermal equilibrium is approached at late times.
   
In order to make this apparent one can plot the results
in a different way. For this we note that according to
(\ref{eq:initialtsF}) the statistical propagator
corresponds to the ratio of the following
{\rr\em particle number distribution}  
\beq\db
f_{\bp}(t) + \frac{1}{2} \,=\, 
\left[ F(t,t;\bp) K(t,t;\bp) - Q^2(t,t;\bp)\right]^{1/2} \, ,
\label{eq:effpart}
\eeq
at initial time $t = 0$ and the corresponding mode energy. Here we have defined:
\beq\db
K(t,t';\bp) \,\equiv\, \partial_t \partial_{t'} F(t,t';\bp) \quad , \quad
Q(t,t';\bp) \,\equiv\, \frac{1}{2} \left[ \partial_t F(t,t';\bp)
+ \partial_{t'} F(t,t';\bp) \right] \, .
\eeq
Since we employed $Q(0,0;\bp) = 0$ for the initial conditions
(\ref{eq:initialtsF}) the {\rr\em mode energy}
is given by
\beq\db
\epsilon_{\bp}(t) \,=\,  \left(\frac{K(t,t;\bp)}{F(t,t;\bp)}\right)^{1/2} \, ,
\label{eq:effen}
\eeq
at initial time, such that $F(t,t;\bp) = [f_\bp(t) + 1/2]/\epsilon_\bp(t)$.
For illustration of the results we may use (\ref{eq:effpart}) and
(\ref{eq:effen}) for times $t > 0$ in order to {\rr\em define an
effective mode particle number and energy}, where we set 
$Q(t,t';\bp) \equiv 0$ in (\ref{eq:effpart}) for the moment. 

The behavior of the effective particle number is illustrated 
on the right of Fig.~\ref{Figpeakphi}, where we 
plot $\ln (1+1/f_{\bp}(t))$ as
a function of $\epsilon_{\bp}(t)$. 
For the corresponding plot of Fig.~\ref{Figpeakphi}
we have employed an initial peaked distribution at $p_{\rm ts}/m_R = 2.5$, 
where we added an initial ``thermal background'' distribution
with temperature \mbox{$T_0/m_R = 4$}.\footnote{The initial mass term
is $M(0)/m_R = 2.24$ and $\lambda/6N = 0.5 m_R^2$ for $N=4$.} 
Correspondingly, from the solid
line ($t=0$) in the right figure one observes 
the initial ``thermal background'' as a straight line distorted
by the nonthermal peak. 
The curves represent snapshots at equidistant time steps   
$\Delta t\, m_R = 10$. After rapid changes in 
$f_\bp(t)$ at early times the subsequent curves converge to a 
straight line to high accuracy, with inverse slope 
$T_{\rm eq}/T_0 = 1.175$. The initial high 
occupation number in a small momentum range decays quickly
with time. More and more low momentum modes get populated
and the particle distribution approaches a thermal shape.

The crucial importance of the NLO corrections for the nonequilibrium
dynamics can also be observed for other initial conditions. 
The right graph of Fig.~\ref{FigLOM} shows the equal-time zero-mode 
$F(t,t;\bp=0)$, along with $M^2(t)$ including the sum over all modes,
following a ``quench'' as described in section~\ref{sec:lofixedpoints}.
While the dynamics for vanishing self-energies $\Sigma_F$ and
$\Sigma_{\rho}$ is quickly dominated by the spurious
LO stationary solutions, this is no longer the case once the NLO self-energy 
corrections are included. 

In particular, one observes a very
efficient damping of oscillations at NLO. This becomes even more 
pronounced for unequal-time correlators as shown for $F(t,0;\bp=0)$
in the left graph of Fig.~\ref{FigLogFrho}. On the right of Fig.\ \ref{FigLogFrho} the approach to an approximately exponential behavior 
is demonstrated for $|\rho(t,0;\bp=0)|$ and $|F(t,0;\bp=0)|$ with the
same parameters as for the left figure.
The logarithmic plot shows that after a non-exponential 
period at early times the envelope of oscillations
can be well approximated by a straight line. 

The strong qualitative difference between LO and NLO appears because 
an infinite number of spurious conserved quantities is removed once 
scattering is taken into account. It should be emphasized that the
step going from LO to NLO is qualitatively very different than the 
one going from NLO to NNLO or further. In order to understand better 
what happens going from LO to NLO we consider again the
effective particle number (\ref{eq:effpart}). It is straightforward
by taking the time derivative on both sides of (\ref{eq:effpart})
to obtain an evolution equation for $f_{\bp}(t)$ with the
help of the relations (\ref{eq:exactrhoF}):
\beq
\framebox{
\begin{minipage}{13.cm} \vspace*{-0.3cm}
\bea \db
\lefteqn{
\left( f_{\bp}(t) +\frac{1}{2} \right) \partial_t f_{\bp}(t) =} \nonumber\\
&& \db \!\!\! \int_{t_0}^{t}\!\! d t'' \, \Big\{
\left[ {\rr \Sigma_{\rho}(t,t'';\bp)}F(t'',t;\bp) 
- {\rr \Sigma_{F}(t,t'';\bp)}
\rho(t'',t;\bp) \right] \partial_{t}F(t,t';\bp)|_{t=t'}
\nonumber\\ 
&&\,\qquad \db 
- \left[ {\rr \Sigma_{\rho}(t,t'';\bp)} \partial_{t} F(t'',t;\bp) 
- {\rr \Sigma_{F}(t,t'';\bp)} \partial_{t} \rho(t'',t;\bp) \right] 
F(t,t;\bp) \Big\} \nonumber
\eea
\end{minipage}}
\label{eq:exactn}
\eeq
Here $t_0$ denotes the initial time which was set to zero in
(\ref{eq:exactrhoF}) without loss of generality. 
Since $\rr \Sigma_F \sim \Or (1/N)$ as well as $\rr \Sigma_{\rho}$,
one directly observes that at LO, i.e.~for $N \to \infty$, the
particle number for each momentum mode is strictly conserved:
{\rr\em $\partial_t f_{\bp}(t) \equiv 0$ at LO.} 
Stated differently, (\ref{eq:effpart}) just 
specifies the infinite number of additional constants of motion 
which appear at LO. In contrast, once corrections beyond LO 
are taken into account then (\ref{eq:effpart}) no longer
represent conserved quantities.

\subsection{Transport equations}

We have seen in the previous sections that after a transient rapid evolution, which is characterized by damped oscillations of correlation functions with frequency determined by the renormalized mass for zero spatial momentum, the subsequent evolution of equal-time correlation functions such as $F(t,t;\bp)$ becomes comparably smooth. Therefore, after the oscillations become effectively damped out, an approximate description taking into account only low orders in an expansion in derivatives with respect to the center coordinates $\sim (t+t')$ is expected to be suitable. This will be used in the following to obtain a set of equations, which form the basis of kinetic theory and which will be used to describe the transport of conserved quantities in section~\ref{sec:transport}.

\subsubsection{Gradient expansion}
\label{sec:gradient}

We first note that the spectral function (\ref{eq:comrho}) is directly related to the retarded propagator, $G_R$, or the advanced one, $G_A$, by
\beq\db
G_R(x,y)=\Theta(x^0-y^0)\rho(x,y)\, , \quad
G_A(x,y)=-\Theta(y^0-x^0)\rho(x,y)
\label{eq:GRGA}\, .
\eeq
Similarly, the retarded and advanced self-energies are 
\beq\db
\Sigma_R(x,y)=\Theta(x^0-y^0)\Sigma_\rho(x,y)\, , \quad
\Sigma_A(x,y)=-\Theta(y^0-x^0)\Sigma_\rho(x,y)
\label{eq:SRSA}\, .
\eeq
Here we consider the case of a vanishing field expectation value, $\phi(x) = 0$.
With the help of the above notation,
interchanging $x$ and $y$ in the evolution equation
(\ref{eq:exactrhoF}) for $F(x,y)$ and subtraction one obtains
\begin{eqnarray} \db
\lefteqn{\Big( \square_x - \square_y + M^2 \left( x \right) 
- M^2 \left( y \right) \Big) F \left( x, y \right)}  
\nonumber \\
  &\db = &\db \int {\rm d}^{d+1}{z}\,  \theta \left( z^0 \right) 
\Big( F \left( x, z \right) \Sigma_A \left( z, y \right) 
+ G_R \left( x, z \right) \Sigma_F \left( z, y \right) 
\nonumber \\ 
  &   &\db {} - \Sigma_R \left( x , z \right) F \left( z, y \right) 
- \Sigma_F \left( x, z \right) G_A \left( z, y \right) \Big) 
\label{eq:Fint} \, .
\end{eqnarray}
The same procedure yields for the spectral function
\begin{eqnarray} \db
\lefteqn{\Big( \square_x - \square_y + M^2 \left( x \right) 
- M^2 \left( y \right) \Big) \rho \left( x, y \right)} 
\nonumber \\
  &\db = &\db \int {\rm d}^{d+1}{z}\, 
\Big( G_R \left( x, z \right) \Sigma_{\rho} \left( z, y \right) 
+ \rho \left( x, z \right) \Sigma_A \left( z, y \right) 
\nonumber \\ 
  &   &\db {} - \Sigma_{\rho} \left( x , z \right) G_A \left( z, y \right) 
- \Sigma_R \left( x, z \right) \rho \left( z, y \right) \Big) 
\label{eq:rhoint} \,.
\end{eqnarray}
So far, the equations (\ref{eq:Fint}) and (\ref{eq:rhoint}) 
are fully equivalent to the exact equations (\ref{eq:exactrhoF}).

Transport equations are
obtained by prescribing $F$, $\rho$ and derivatives 
at a {\em finite} time using the equations with $t_0 \to -\infty$ 
as an approximate description. Furthermore, one 
employs a gradient expansion to (\ref{eq:Fint}) and
(\ref{eq:rhoint}).  In practice, this expansion is carried out to low order in the number of derivatives with respect to the center coordinates
\beq\db
X^{\mu} \equiv \frac{x^{\mu} + y^{\mu}}{2} 
\label{eq:center}
\eeq
and powers of the relative coordinates
\beq\db
s^{\mu} \equiv x^{\mu} - y^{\mu} \, .
\label{eq:relative}
\eeq
Even for finite $X^0$ one assumes that the 
relative-time coordinate $s^0$ ranges from $-\infty$ to $\infty$
in order to achieve a convenient description in Wigner space, i.e.\ in Fourier 
space with respect to the relative coordinates (\ref{eq:relative}). 
This requires a loss of information about the details of the initial state, which enters in the derivation as an assumption, and limits the use of the approximate equations to not too early times.
\begin{figure}[t]
\centerline{
\epsfig{file=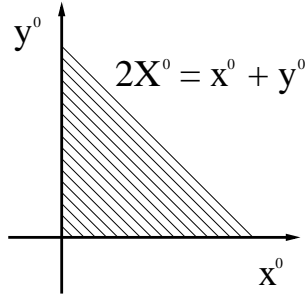,width=4.cm}
}
\caption{\small 
For given finite $2X^0 = x^0 + y^0$ the relative coordinate 
$s^0 = x^0 - y^0$ has a finite range from $-2 X^0$ to $2 X^0$.
\label{fig:srange}
}
\end{figure}

For the description in Wigner space 
we introduce the Fourier transforms with respect to
the relative coordinates, such as
\bea\db
F(X,\omega,\bp)&\db =&\db \int_{-2X^0}^{2X^0}\! {\rm d}s^0\, e^{i\omega s^0} 
\int_{-\infty}^{\infty}\! {\rm d}^d s\, e^{-i \bp\, \bs} F\left(X+\frac{s}{2},X-\frac{s}{2}\right)
\nonumber\\
&\db =&\db 2 \int_{0}^{2X^0} \!{\rm d}s^0\, \cos(\omega s^0)  
\int_{-\infty}^{\infty}\! {\rm d}^d s\, e^{-i \bp\, \bs} F\left(X+\frac{s}{2},X-\frac{s}{2}\right)
\label{eq:Frho}, \qquad \quad
\eea
using the symmetry property 
$F(x,y) = F(y,x)$ for the second line.
We emphasize that the time integral
over $s^0$ is bounded by $\pm 2 X^0$. The time
evolution equations are initialized at time
$x^0 = y^0 = 0$ such that $x^0 \ge 0$ and $y^0 \ge 0$. 
According to (\ref{eq:relative}),
the minimum value of the relative coordinate $s^0$ is then given by
$-y^0 = - 2X^0$ for $x^0 = 0$ while its maximum value is $x^0 = 2X^0$ 
for $y^0 =0$ as illustrated in Fig.~\ref{fig:srange}. Similarly, we define 
\bea\db
\tilde{\rho}(X,\omega,\bp)&\db =&\db
-i\int_{-2X^0}^{2X^0}\! {\rm d}s^0\, e^{i\omega s^0} 
\int_{-\infty}^{\infty}\! {\rm d}^d s\, e^{-i \bp\, \bs}
\rho\left(X+\frac{s}{2},X-\frac{s}{2}\right)
\nonumber\\
&\db =&\db 2 \int_{0}^{2X^0} \! {\rm d}s^0\, 
\sin(\omega s^0) \int_{-\infty}^{\infty}\! {\rm d}^d s\, e^{-i \bp\, \bs}
\rho\left(X+\frac{s}{2},X-\frac{s}{2}\right), \qquad 
\label{eq:Wigrho}
\eea
where a factor of $i$ is included in the definition 
to have $\tilde{\rho}(X,\omega,\bp)$ real and we 
have used $\rho(x,y) = - \rho(y,x)$. 
The equivalent transformations
are done to obtain the self-energies $\Sigma_F(X,\omega,\bp)$ and
$\tilde{\Sigma}_\rho(X,\omega,\bp)$.

In order to exploit the convenient properties of a
Fourier transform it is a standard procedure to extend
the limits of the integrals
over the relative time coordinate
in (\ref{eq:Frho}) and (\ref{eq:Wigrho}) 
to $\pm\infty$. For instance, using the chain rule 
\beq \db
\square_x - \square_y = 2 \frac{\partial}{\partial s_{\mu}} 
\frac{\partial}{\partial X^{\mu}} 
\eeq 
and the gradient expansion of the mass terms in (\ref{eq:Fint})
to\footnote{The expression is actually correct to NNLO
since the first correction is 
${\cal O}\left(\left(s^{\mu}\partial_{X^{\mu}}\right)^3\right)$.}
NLO:
\beq \db
M^2 \left( X + \frac{s}{2} \right) - M^2 \left( X - \frac{s}{2} \right) 
\,\simeq\, s^{\mu}\, \frac{\partial M^2 ( X )}{\partial X^{\mu}}  
\,,
\eeq
then the LHS of (\ref{eq:Fint}) becomes in Wigner space
\bea \db
\int_{-\infty}^{\infty} {\rm d}^{d+1} s\, e^{i p_\mu s^\mu} 
\left[ 2 \frac{\partial}{\partial s_{\mu}} \frac{\partial}{\partial X^{\mu}} 
+ s^{\mu} \frac{\partial M^2 ( X )}{\partial X^{\mu}}  
\right] 
F \left( X+\frac{s}{2},X-\frac{s}{2} \right) &&
\nonumber\\
\db   =  - i \left[ 2 p^{\mu} \frac{\partial}{\partial X^{\mu}} 
+  \frac{\partial M^2 ( X )}{\partial X^{\mu}}  
\frac{\partial}{\partial p_{\mu}} \right] F \left( X, p \right)\, ,&& 
\eea
with $p^0 \equiv \omega$. 

Similarly, one transforms the
RHS of (\ref{eq:Fint}) and (\ref{eq:rhoint}) using
the above prescription. To derive this expression, we consider for functions $f(x,y)$ and $g(x,y)$ the integral $\int \rmd^{d+1} z f(x,z) g(z,y)$. For
\begin{equation} \db
f(x,z) \, \equiv \, f\left( X_{xz} + \frac{s_{xz}}{2}, X_{xz} - \frac{s_{xz}}{2} \right)
\end{equation}
we introduce the Fourier transform with respect to the relative coordinate:
\begin{equation}
\db f\left( X_{xz}, p \right) \, = \, \int \rmd^{d+1} s_{xz} \exp \left\{ i p_\mu s^\mu_{xy} \right\}
f\left( X_{xz} + \frac{s_{xz}}{2}, X_{xz} - \frac{s_{xz}}{2} \right) \, .
\label{eq:Fourierf}
\end{equation}
This quantity may also be written as
\begin{equation}
\db f\left( X_{xz}, p \right) \, \equiv \, f\left( X_{xy} + \frac{s_{zy}}{2}, p \right)
\, = \, \exp \left\{ \frac{s^\mu_{zy}}{2} \frac{\partial}{\partial X^\mu_{xy}} \right\} f(X_{xy},p) \,,
\label{eq:taylorf}
\end{equation}
where we Taylor expanded to obtain the last equality. Similar steps for $g(z,y)$ lead to
\begin{eqnarray}
\db g\left( X_{zy}, p \right) & \db = &\db \int \rmd^{d+1} s_{zy} \exp \left\{ i p_\mu s^\mu_{zy} \right\}
g\left( X_{zy} + \frac{s_{zy}}{2}, X_{zy} - \frac{s_{zy}}{2} \right)
\nonumber\\
& \db = &\db g\left( X_{xy} - \frac{s_{xz}}{2}, p \right)
\, = \, \exp \left\{ -\frac{s^\mu_{xz}}{2} \frac{\partial}{\partial X^\mu_{xy}} \right\} g(X_{xy},p) \, .
\end{eqnarray}
Inverting (\ref{eq:Fourierf}) gives with (\ref{eq:taylorf})  
\begin{eqnarray}
\db f(x,z) &\db = & \db \int \frac{\rmd^{d+1} p}{(2\pi)^{d+1}} \exp\left\{ -i p_\mu s^\mu_{xz} \right\} \exp \left\{ \frac{s^\mu_{zy}}{2} \frac{\partial}{\partial X^\mu_{xy}} \right\} f(X_{xy},p) \, ,
\nonumber\\
\db g(z,y) &\db = & \db \int \frac{\rmd^{d+1} p^\prime}{(2\pi)^{d+1}} \exp\left\{ -i p_\mu^\prime s^\mu_{zy} \right\} \exp \left\{
-\frac{s^\mu_{xz}}{2} \frac{\partial}{\partial X^\mu_{xy}} \right\} g(X_{xy},p^\prime) \,
\end{eqnarray}
using also the equivalent expressions for $g(z,y)$. With this we can write employing a partial integration:
\begin{eqnarray}
&&\db \!\!\!\!\!\!\!\!\!\!\!\! \int \rmd^{d+1} s_{xy}\, \exp \left\{ i p_\mu s^\mu_{xy} \right\} \int \rmd^{d+1} z f(x,z)\, g(z,y)
\nonumber\\
&&\db = \, \exp \left\{ \frac{i}{2} \left( \frac{\partial}{\partial p_\mu} \frac{\partial}{\partial{X^{\mu\,\prime}_{xy}}}
- \frac{\partial}{\partial{p_\mu^\prime}} \frac{\partial}{\partial{X^\mu_{xy}}} \right) \right\}
f(X_{xy},p) g(X_{xy}^\prime,p^\prime)|_{X_{xy} = X_{xy}^\prime, p = p^\prime} 
\nonumber \\
&&\db \simeq \, f(X_{xy},p) g(X_{xy},p)
+ \frac{i}{2} \left\{ f(X_{xy},p); g(X_{xy},p)
\right\}_{PB} \, ,\qquad
\end{eqnarray}
where the Poisson bracket reads
\begin{equation}
\db \left\{ f(X,p); g(X,p)
\right\}_{PB} \, = \, \frac{\partial f(X,p)}{\partial p_\mu} \frac{\partial g(X,p)}{\partial X^\mu} - \frac{\partial f(X,p)}{\partial X^\mu} \frac{\partial g(X,p)}{\partial p_\mu} \, .
\end{equation}

The lowest-order equations are obtained by 
neglecting ${\cal O}\left(\partial_{X^{\mu}} 
\partial_{p_{\mu}}\right)$
and higher contributions in the gradient expansion. To this
order the transport equations then read:
\bea\rr
2 p^{\mu} \frac{\partial F ( X, p)}{\partial X^{\mu}} 
&\rr =&\rr
\tilde{\Sigma}_{\rho} \left( X, p \right) 
F \left( X, p \right) 
- 
\Sigma_F \left( X, p \right)
\tilde{\rho} \left( X, p \right) 
\, ,
\label{eq:LOgradF}\\[0.2cm]
\rr 2 p^{\mu} \frac{\partial \tilde{\rho} ( X, p)}{\partial X^{\mu}} 
&\rr =& \rr 0 \, .
\label{eq:LOgradrho}
\eea
We note that the compact form of the gradient expanded equation
to lowest order is very similar to the exact equation for the statistical
function (\ref{eq:exactrhoF}). The main technical difference is that 
there are no integrals over the time history. 
Furthermore, to this
order in the expansion the evolution equation for the
spectral function $\tilde\rho(X,p)$ becomes trivial.
Accordingly, this approximation describes changes in the
occupation number while neglecting its impact on the
spectrum and vice versa. Moreover, one observes that in thermal equilibrium the fluctuation-dissipation relations (\ref{eq:flucdissbose}) and (\ref{eq:sigmadecom}) yield $\partial_X F ( X, p) = 0$.

\subsubsection{Vertex-resummed kinetic equation}
\label{sec:vertexres}

In the following we consider the lowest-order expressions (\ref{eq:LOgradF}) and (\ref{eq:LOgradrho}) of the gradient expansion, where the self-energies are obtained from the 2PI $1/N$ expansion to NLO. It is convenient to introduce a suitable ``occupation number distribution'' $f(X,p)$. For the real scalar field theory one has
\begin{equation}\db
F(X,-p) \, = \, F(X,p) \quad , \quad \tilde{\rho}(X,-p) \, = \, - \tilde{\rho}(X,p) \, .
\label{eq:symFrho}
\end{equation}
Without loss of generality we can write 
\begin{equation} \db
F(X,p) \, = \, \left( f(X,p) + \frac{1}{2} \right) \tilde{\rho}(X,p) \, ,
\label{eq:defn}
\end{equation}
which defines the function $f(X,p)$ for any given $F(X,p)$ and $\tilde{\rho}(X,p)$. We emphasize that without additional assumptions (\ref{eq:defn}) does not represent a fluctuation-dissipation relation (\ref{eq:flucdissbose}), which holds only if $f(X,p)$ is replaced by a thermal distribution function. We will not assume this in the following and keep $f(X,p)$ general at this stage. In particular,
(\ref{eq:symFrho}) then implies the identity
\begin{equation} \db
f(X,-p) \, = \, - \left( f(X,p) + 1 \right) \, .
\end{equation}
For spatially homogeneous ensembles (\ref{eq:LOgradrho}) implies a constant $\tilde{\rho}(p)$ that does not depend on time. In contrast, the statistical function $F(t,p)$ to this order can depend on time $t \equiv X^0$ and using (\ref{eq:defn}) we can write
\begin{equation} \db
\int_0^\infty \frac{{\mathrm d} p^0}{2\pi}\, 2 p^0 \frac{\partial}{\partial t} F(t,p) \, = \, 
\int_0^\infty \frac{{\mathrm d} p^0}{2\pi}\, 2  p^0 \tilde{\rho}(p)\, \frac{\partial f(t,p)}{\partial t} \, = \, C[f](t,\bp) \, .
\label{eq:genev} 
\end{equation}
Here $C[f]$ describes the effects of interactions to lowest order in the gradient expansion. 

It is instructive to consider for a moment a free spectral function given by
\begin{equation} \db
\tilde{\rho}^{(0)}(p) \, = \, 2 \pi\, {\mathrm{sgn}}(p^0) \, \delta\!\left((p^0)^2-\omega_\bp^2\right)
\label{eq:freerho}
\end{equation}
for a relativistic scalar field theory with particle energy $\omega_\bp$. Choosing a free field theory type spectral function with zero ``width'' leads to the characterization of the dynamics in terms of a ``gas'' of particles, where
\begin{equation}\db
f(t,\bp) \, = \, f(t,p^0=\omega_\bp,\bp) \, = \, \int_0^\infty \frac{{\mathrm d} p^0}{2\pi}\, 2  p^0 \tilde{\rho}^{(0)}(p)\, f(t,p) \, 
\end{equation}
denotes the ``on-shell'' number distribution. In the following we will consider the resummed $1/N$ expansion of the 2PI effective action to next-to-leading order. If the spectral function $\tilde{\rho}(p)$ for such an approximation is taken, it is not of the free field form (\ref{eq:freerho}) but acquires a non-zero ``width'' as pointed out in section~\ref{sec:2PI}. We may nevertheless use the same procedure to define for spatially homogeneous systems an effective number distribution  
\begin{equation} \db
f(t,\bp) \, \equiv \, \int_0^\infty \frac{{\mathrm d} p^0}{2\pi}\, 2  p^0 \tilde{\rho}(p)\, f(t,p) \, ,
\label{eq:neff}
\end{equation}
which depends on time and spatial momenta only. This will allow us to formulate the quantum field theory in a way reminiscent of a Boltzmann equation. The time evolution of that distribution is, according to (\ref{eq:LOgradF}), given by 
\begin{eqnarray}\db
\frac{\partial f(t,\bp)}{\partial t} &\db = & \db C[f](t,\bp) 
\, = \, \int_0^\infty \frac{{\mathrm d} p^0}{2\pi}\, \left[ \tilde{\Sigma}_{\rho} ( t, p) F ( t, p) - \Sigma_F ( t, p ) \tilde{\rho}( p ) \right] . \qquad
\label{eq:neffchange}
\end{eqnarray}
For the following it is useful to note that the last expression can be rewritten with
\begin{eqnarray} \db
\tilde{\Sigma}_{\rho} ( t, p) F ( t, p) - \Sigma_F ( t, p ) \tilde{\rho}( p )
&\db \equiv &\db \left[ \Sigma_F(t,p) +\frac{1}{2} \tilde{\Sigma}_\rho(t,p) \right]
\left[ F(t,p) - \frac{1}{2} \tilde{\rho}(p) \right] 
\nonumber\\
&\db - &\db \left[ \Sigma_F(t,p) -\frac{1}{2} \tilde{\Sigma}_\rho(t,p) \right]
\left[ F(t,p) +\frac{1}{2} \tilde{\rho}(p) \right] \,.
\nonumber\\
\end{eqnarray} 
With (\ref{eq:defn}) one also has
\begin{eqnarray} \db
F(t,p) - \frac{1}{2} \tilde{\rho}(p) &\db = & \db f(t,p)\, \tilde{\rho}(p) \, ,
\nonumber\\
\db F(t,p) +\frac{1}{2} \tilde{\rho}(p) &\db = & \db \left[ f(t,p) + 1 \right] \tilde{\rho}(p) \, , 
\end{eqnarray} 
which will allow us to conveniently express everything in terms of factors of $f(t,p)$ or, including ``Bose enhancement'', $[ f(t,p) + 1 ]$ and $\tilde{\rho}(p)$.  

In the following, we consider the collision term $C[f]$ at next-to-leading order in the $1/N$ expansion of the 2PI effective action as described in section \ref{sec:2PIN}. To this end, we introduce retarded and advanced quantities as $I_R(x,y)=\Theta(x^0-y^0)I_\rho(x,y)$, $I_A(x,y)=-\Theta(y^0-x^0)I_\rho(x,y)$ and equivalently for $\Pi_R(x,y)$ and $\Pi_A(x,y)$ using the expressions (\ref{IRFR}) and (\ref{PiRFR}) with $t_0 \to - \infty$. Then we Fourier transform with respect to the relative coordinates to obtain the lowest-order gradient expansion result. To ease the notation, we suppress the dependence on the global central coordinate and only write the momentum dependencies, with $\int_q \equiv \int \rmd^{d+1} q/(2\pi)^{d+1}$. For the retarded and advanced summation functions one finds from (\ref{IRFR}):
\begin{equation}
\db	I_R(t,p) = \frac{\Pi_R(t,p)}{1+\Pi_R(t,p)} \quad , \quad I_A(t,p) = \frac{\Pi_A(t,p)}{1+\Pi_A(t,p)} \, .
\end{equation}
Proceeding in the same way also for $I_F$, one obtains with $I_\rho = I_R - I_A$:
\begin{equation}
\db	I_\rho(t,p)= v_{\rm{eff}}(t,p)\, \Pi_\rho(t,p) \quad , \quad
	I_F(t,p)= v_{\rm{eff}}(t,p)\, \Pi_F(t,p) \, . 
\end{equation}
The vertex correction $v_{\rm{eff}}(t,p)$ sums the infinite chain of ``ring'' diagrams appearing at NLO in the $1/N$ expansion and is given by
\begin{equation} \db
v_{\rm{eff}} (t,p) \, = \, \frac{1}{| 1 + \Pi_{R} (t,p)|^2} ~,
\label{EffectiveCoupling}
\end{equation}
with $| 1 + \Pi_{R}|^2\equiv (1+\Pi_R)(1+\Pi_A)$ and the one-loop retarded self-energy
\begin{equation}
\db \Pi_{R} (t,p) \, = \, \frac{\lambda}{3} \int_{q} F (t,p-q) G_{R} (q) ~.
\label{OneLoopRA}
\end{equation}
At next-to-leading order in the 2PI $1/N$ expansion, the above linear combinations of self-energies are then given by
\begin{eqnarray}
&&\hspace{-10pt} \db \Sigma_{F} (t,p) \pm \frac{1}{2} \tilde{\Sigma}_{\rho} (t,p) \, = \,- \frac{\lambda^2}{18 N} \int_{q l} v_{\rm{eff}} (t,p-q) \nonumber\\
&& \db \times\: \left[ F(t,p-q-l) \pm \frac{1}{2} \tilde{\rho}(p-q-l)\right] \left[ F(t,q) \pm \frac{1}{2} \tilde{\rho}(q) \right] \left[ F(t,l) \pm \frac{1}{2} \tilde{\rho}(l) \right] \, , \qquad\quad
\end{eqnarray}
which has the structure of a two-loop self-energy, however, with a time and momentum dependent ``effective coupling'' entering via $v_{\rm{eff}}(t,p)$. 
In terms of $f(t,p)$ the above self-energy combinations read
\begin{eqnarray}
&&\hspace{-10pt} \db \Sigma_{F} (t,p) + \frac{1}{2} \tilde{\Sigma}_{\rho} (t,p) \, =\, -  \frac{\lambda^2}{18 N} \int_{ql} v_{\rm{eff}} (t,p-q) \nonumber\\
&& \db \times\: \left[ f(t,p-q-l) + 1 \right] \tilde{\rho}(p-q-l)\, \left[ f(t,q) + 1 \right] \tilde{\rho}(q)\, \left[ f(t,l) + 1 \right] \tilde{\rho}(l) , \qquad \\
&&\hspace{-10pt} \db \Sigma_{F} (t,p) - \frac{1}{2} \tilde{\Sigma}_{\rho} (t,p) \, = \, - \frac{\lambda^2}{18 N} \int_{ql} v_{\rm{eff}} (t,p-q) \nonumber\\
&& \db \times\: f(t,p-q-l) \tilde{\rho}(p-q-l)\, f(t,q) \tilde{\rho}(q) \, f(t,l) \tilde{\rho}(l) ~.
\end{eqnarray}
Putting everything together one obtains 
\begin{eqnarray}
&& \hspace{-20pt} \db \left( \tilde{\Sigma}_{\rho} F - \Sigma_{F} \tilde{\rho} \right) (t,p) \, = \, - \frac{\lambda^2}{18 N} \int_{qlr} (2 \pi)^{d+1} \delta(p-q-l-r) \nonumber\\
&& \db \times\: v_{\rm{eff}} (t,p-q) \bigg\{ \left[ f(t,q) + 1 \right] \left[ f(t,l) +1\right] \left[ f(t,r) + 1\right] f(t,p) \nonumber\\
&& \db -\: f(t,q) f(t,l) f(t,r) \left[ f(t,p) + 1  \right] \bigg\} \, \tilde{\rho}(q) \tilde{\rho}(l) \tilde{\rho}(r) \tilde{\rho}(p) ~.  
\label{Stationarity2}
\end{eqnarray} 
To bring this expression into a form which can be directly compared to kinetic or Boltzmann descriptions, we map onto positive frequencies. For this we split the frequency integrals $\int_{-\infty}^{\infty} \rmd q^0 \ldots = \int_{-\infty}^0 \rmd q^0 \ldots + \int_0^{\infty} \rmd q^0 \ldots$ with $q^0 \rightarrow - q^0$ for the negative frequency part and employ $f(t,-q^0,\bq) = - [f(t,q^0,\bq) +1]$. 

Collecting the $2^{3} = 8$ contributions from the different orthants in frequency space one finds
a ``collision integral'' (\ref{eq:neffchange}) that is accurate to next-to-leading order in the large-$N$ expansion, including processes to all orders in the coupling constant. We obtain with $f_p \equiv f(t,p)$ suppressing the global time dependence:
\begin{eqnarray} \db
C^{{\mathrm{NLO}}}[f](\bp) \!\!\! & \db = & \rr \!\! \int\! {\mathrm{d}}\Omega^{2\leftrightarrow 2}[f](p,l,q,r)\,  \left[ 
(f_p+1) (f_l+1) f_q f_r - f_p f_l (f_q+1) (f_r+1) \right] 
\nonumber\\
&\db +& \db \!\! \int\! {\mathrm{d}}\Omega^{1\leftrightarrow 3}_{(a)}[f](p,l,q,r)\,  \left[ 
(f_p+1) (f_l+1) (f_q+1) f_r - f_p f_l f_q (f_r+1) \right]
\nonumber\\
&\db +& \db \!\! \int \! {\mathrm{d}}\Omega^{1\leftrightarrow 3}_{(b)}[f](p,l,q,r)\,  \left[ 
(f_p+1) f_l f_q f_r - f_p (f_l+1) (f_q+1) (f_r+1) \right]
\nonumber\\
&\db +& \db \!\! \int \! {\mathrm{d}}\Omega^{0\leftrightarrow 4}[f](p,l,q,r)\,  \left[ 
(f_p+1) (f_l+1) (f_q+1) (f_r+1) - f_p f_l f_q f_r \right] .
\nonumber\\
\label{eq:CNLO}
\end{eqnarray} 
Here 
\begin{eqnarray} \rr 
\int {\mathrm{d}}\Omega^{2\leftrightarrow 2}[f](p,l,q,r) &\rr  = &\rr \frac{\lambda^2}{18 N} \int_0^\infty \frac{{\mathrm{d}}p^0{\mathrm{d}}l^0{\mathrm{d}}q^0{\mathrm{d}}r^0}{(2\pi)^{4-(d+1)}} \int_{\bl \bq \br} \delta(p+l-q-r)
\nonumber\\ 
&\rr \times& \rr \tilde{\rho}_p \tilde{\rho}_l \tilde{\rho}_q \tilde{\rho}_r \left[ v_{\mathrm{eff}}(p+l) + v_{\mathrm{eff}}(p-q) + v_{\mathrm{eff}}(p-r) \right] \, ,
\nonumber\\
\db \int {\mathrm{d}}\Omega^{1\leftrightarrow 3}_{(a)}[f](p,l,q,r) & \db = & \db \frac{\lambda^2}{18 N} \int_0^\infty \frac{{\mathrm{d}}p^0{\mathrm{d}}l^0{\mathrm{d}}q^0{\mathrm{d}}r^0}{(2\pi)^{4-(d+1)}} \int_{\bl \bq \br} \,  \delta(p+l+q-r) 
\nonumber\\
&\db \times& \db \tilde{\rho}_p \tilde{\rho}_l \tilde{\rho}_q \tilde{\rho}_r \left[ v_{\mathrm{eff}}(p+l) + v_{\mathrm{eff}}(p+q) + v_{\mathrm{eff}}(p-r) \right] \, ,
\nonumber\\
\db \int {\mathrm{d}}\Omega^{1\leftrightarrow 3}_{(b)}[f](p,l,q,r) & \db = & \db \frac{\lambda^2}{18 N} \int_0^\infty \frac{{\mathrm{d}}p^0{\mathrm{d}}l^0{\mathrm{d}}q^0{\mathrm{d}}r^0}{(2\pi)^{4-(d+1)}} \int_{\bl \bq \br} \,  \delta(p-l-q-r) 
\nonumber\\
&\db \times& \db \tilde{\rho}_p \tilde{\rho}_l \tilde{\rho}_q \tilde{\rho}_r \, v_{\mathrm{eff}}(p-l) \, , 
\nonumber\\ 
\db \int {\mathrm{d}}\Omega^{0 \leftrightarrow 4}[f](p,l,q,r) & \db = & \db \frac{\lambda^2}{18 N} \int_0^\infty \frac{{\mathrm{d}}p^0{\mathrm{d}}l^0{\mathrm{d}}q^0{\mathrm{d}}r^0}{(2\pi)^{4-(d+1)}} \int_{\bl \bq \br} \,  \delta(p+l+q+r)
\nonumber\\
&\db \times& \db \tilde{\rho}_p \tilde{\rho}_l \tilde{\rho}_q \tilde{\rho}_r \, v_{\mathrm{eff}}(p+l) \, 
\label{eq:omegaNLO}
\end{eqnarray}
with $\int_\bq \equiv \int \rmd^dq/(2\pi)^d$. We emphasize that the above expressions still contain the integrations over frequencies and spectral functions. No quasi-particle assumptions using a free-field form of the spectral function has been employed yet and the only approximations are the $1/N$ expansion to NLO and the gradient expansion underlying (\ref{eq:genev}). 

One observes that for sufficiently large $p$, for which $\Pi^{R}(p) \ll 1$, the vertex function (\ref{EffectiveCoupling}) approaches one.  In this case the $2 \leftrightarrow 2$ contribution of the first line in (\ref{eq:CNLO}) is reminiscent of the two-to-two scattering process in a Boltzmann equation. The main difference is that the latter assumes a $\delta$-like spectral function such that all momenta are on shell. Therefore, in the perturbative expression (\ref{eq:2to2}) off-shell processes involving the decay of one into three particles or corresponding $3 \to 1$ annihilation processes or even $0 \leftrightarrow 4$ processes are absent. They can occur in principle at NLO in the 2PI $1/N$ expansion, which leads to the different terms contributing to the RHS of (\ref{eq:CNLO}), but they are typically small. At sufficiently high momenta these off-shell contributions should be suppressed along with all quantum-statistical corrections such that the spectral function approaches a $\delta$-like behavior. In this case one recovers the standard Boltzmann equation for elastic two-to-two scattering. In the infrared $v_{\mathrm{eff}}(p)$ may have a nontrivial momentum dependence, which incorporates important vertex corrections for the $2 \leftrightarrow 2$ scattering term that are discussed, in particular, in section~\ref{sec:transport}.

\subsection{Bibliography}
\label{eq:reftherm}

\begin{itemize}
\item Fig.~\ref{fig:lateuni} is taken from J.~Berges and J.~Cox, {\it Thermalization of Quantum Fields from Time-Reversal Invariant 
Evolution Equations}, Phys.\ Lett.\ {\bf B517} (2001) 
369, where thermalization in relativistic quantum field theory has been shown for $1+1$ dimensions. Similar studies in $2+1$ dimensions include 
S.~Juchem, W.~Cassing and C.~Greiner, {\it Quantum dynamics and thermalization for out-of-equilibrium phi**4-theory,}
Phys.\ Rev.\ D {\bf 69} (2004) 025006. For thermalization in $3+1$ dimensions including fermions and bosons see J.~Berges, S.~Borsanyi and J.~Serreau,
  {\it Thermalization of fermionic quantum fields,}
  Nucl.\ Phys.\  B {\bf 660} (2003) 51. For a related study of thermalization in the context of ultracold quantum gases, see J.~Berges and T.~Gasenzer,
  {\it Quantum versus classical statistical dynamics of an ultracold Bose gas},
  Phys.\ Rev.\ A {\bf 76} (2007) 033604. 
\item Fig.~\ref{fig:spectral1} is taken from G.~Aarts and J.~Berges,
  {\it Nonequilibrium time evolution of the spectral function in quantum field
  theory,} Phys.\ Rev.\  D {\bf 64} (2001) 105010.
\item Figs.~\ref{Figpeakphi} -- \ref{FigLogFrho} are taken from J.~Berges,
  {\it Controlled nonperturbative dynamics of quantum fields out of
  equilibrium,} Nucl.\ Phys.\  A {\bf 699} (2002) 847, which compares thermalization dynamics from the $1/N$ expansion at LO and NLO. The above presentation is also based on G.~Aarts, D.~Ahrensmeier, R.\ Baier, J.~Berges and J.~Serreau, {\it Far-from-equilibrium dynamics with broken symmetries
from the 1/N expansion of the 2PI effective action}, Phys.\ Rev.\ {\bf D66} (2002) 045008.
\item  The phenomenon of prethermalization has been pointed out in J.~Berges, S.~Borsanyi and C.~Wetterich, {\it Prethermalization}, Phys.\ Rev.\ Lett.\  {\bf 93} (2004) 142002. Applications to condensed matter systems include M. Moeckel, S. Kehrein, {\it Interaction Quench in the Hubbard model}, Phys.\ Rev.\ Lett.\ {\bf 100} (2008) 175702, and for experimental investigations with ultracold atoms see M.~Gring et al., {\it Relaxation and Prethermalization in an Isolated Quantum System}, Science {\bf 337} (2012) 6100.
\item The presentation of section~\ref{sec:gradient} follows J.~Berges and S.~Borsanyi,
  {\it Range of validity of transport equations},
  Phys.\ Rev.\ D {\bf 74} (2006) 045022. Section~\ref{sec:vertexres} about the vertex-resummed kinetic theory is based on J.~Berges and D.~Sexty, {\it Strong versus weak wave-turbulence in relativistic field theory},
  Phys.\ Rev.\ D {\bf 83} (2011) 085004. For a discussion of transport in quantum field theory, see also~S.~Juchem, W.~Cassing and C.~Greiner, {\it Nonequilibrium quantum field dynamics and off-shell transport for phi**4 theory in (2+1)-dimensions}, Nucl.\ Phys.\ A {\bf 743} (2004) 92. For a corresponding nonrelativistic study, see A.~Branschadel and T.~Gasenzer,
  {\it 2PI nonequilibrium versus transport equations for an ultracold Bose gas},
  J.\ Phys.\ B {\bf 41} (2008) 135302.
\end{itemize}

%% file: ch_clstatLH.tex
\section{Classical aspects of nonequilibrium quantum fields}
\label{sec:clstat}

It is an important question to what extent nonequilibrium quantum field
theory can be approximated by classical-statistical field theory.  
It is a frequently employed strategy in the literature to consider
nonequilibrium classical dynamics instead of quantum dynamics since the
former can be simulated numerically up to controlled statistical errors.
Classical-statistical field theory
indeed gives important insights when the number of bosonic field quanta per mode 
is sufficiently large such that quantum fluctuations 
are suppressed compared to statistical fluctuations. 
We will derive below a sufficient condition for the
validity of classical approximations to nonequilibrium quantum
dynamics. The description in terms of spectral and statistical
correlation functions as introduced in section~\ref{sec:specstat}
is particularly suitable for comparisons 
since these correlation functions possess well-defined
classical counterparts.

Classical Rayleigh-Jeans divergences and the lack of genuine quantum 
effects --- such as the approach to quantum thermal equilibrium
characterized by Bose-Einstein statistics --- limit
the use of classical-statistical field theory. To find out its
use and its limitations we perform below direct comparisons
of nonequilibrium classical and quantum evolutions for same initial 
conditions. One finds
that classical methods can give an accurate description of
quantum dynamics for the case of large enough characteristic occupation numbers.
This typically limits their application to not too late times, before the approach to quantum thermal
equilibrium sets in.  

Classical methods have been extensively used in the past
to rule out ``candidates'' for approximation schemes applied 
to nonequilibrium quantum field theory. Approximations that 
fail to describe classical nonequilibrium dynamics should be
typically discarded also for the quantum case. If the dynamics is
formulated in terms of correlation functions then approximation
schemes for the quantum evolution can be straightforwardly implemented
as well for the respective classical theory. For instance,
the 2PI $1/N$-expansion introduced in section~\ref{sec:2PIN}
can be equally well implemented in
the classical as in the quantum theory. Therefore, in the
classical-statistical approach one can compare NLO results 
with results including 
all orders in $1/N$. This gives a rigorous answer to
the question of what happens at NNLO or beyond in this case.
In particular, for increasing occupation numbers per mode the classical and
the quantum evolution can be shown to approach each other 
if the same initial conditions are applied and for not too late times.
For sufficiently high particle number densities one
can therefore strictly verify how rapidly the $1/N$ series
converges for far-from-equilibrium dynamics. 

Though we restrict ourselves here to scalar fields, it is important to note that crucial applications arise also by coupling classical-statistical bosonic fields, including lattice gauge fields, to fermions. For instance, this is done to describe the nonequilibrium real-time evolution in quantum electrodynamics for ultrastrong laser fields or quantum chromodynamics related to collision experiments of heavy nuclei. Fermions are never classical in the sense that they cannot be strongly occupied because of the Pauli exclusion principle. However, since the fermions occur quadratically in the respective actions, their functional integral can be treated on the lattice without loosing their genuine quantum nature. We refer to the reference list in section~\ref{sec:litCl} for further reading on these extensions of the classical-statistical lattice simulation approach.

\subsection{Nonequilibrium quantum field theory revisited}

\subsubsection{Functional integral}

In order to discuss the different origins of quantum and of classical-statistical
fluctuations, it is convenient to rewrite the nonequilibrium generating functional
of section \ref{sec:functionalintegral}. We 
start by employing the notation introduced in section~\ref{sec:nonequgenfunc}, where the superscripts `$+$' and `$-$'
indicate that the fields are taken on the
forward branch (${\cal C}^+$)  starting at $t_0$ and backward (${\cal C}^-$) along the
closed time path, respectively. To be specific, we consider again 
the $N$-component scalar field theory for which the classical action
(\ref{eq:classical}) can be written as
\begin{eqnarray}
&& \!\! \!\! \!\! \!\! \!\! \!\! \db S[\varphi^+,\varphi^-] 
\nonumber\\
&\db = & \db \int_{x,t_0}
	\Bigg\{\frac{1}{2}\partial^\mu\varphi^+_a(x)\partial_\mu\varphi^+_a(x)  
	-\frac{m^2}{2}\varphi^+_a(x)\varphi^+_a(x) 
	-\frac{\lambda}{4!N}\left(\varphi^+_a(x)\varphi^+_a(x)\right)^2
\nonumber\\
&\db - & \db \frac{1}{2}\partial^\mu\varphi^-_a(x)\partial_\mu\varphi^-_a(x)  
	+\frac{m^2}{2}\varphi^-_a(x)\varphi^-_a(x) 
	+\frac{\lambda}{4!N}\left(\varphi^-_a(x)\varphi^-_a(x)\right)^2	
 \Bigg\}  \, .
\label{eq:classicalpm}
\end{eqnarray}
Here the minus sign in front of the `$-$' terms accounts
for the reversed time integration of the closed time contour and $\int_{x,t_0} \equiv \int_{t_0} \rmd x^0 \int \rmd^d x$.
The corresponding generating functional for correlation functions then reads in this notation
\begin{eqnarray}
\label{eq:NEgenFuncZphipm}
 &&  \!\! \!\! \!\! \!\! \!\! \!\! \db Z[J^+,J^-,R^{++},R^{+-},R^{-+},R^{--};\varrho_0] \, = \,  
  \int [{\rm d}\varphi_0^+][{\rm d}\varphi_0^-]\,\,
  \varrho_0\left[\varphi_0^+,\varphi_0^-\right] 
     \nonumber\\
     &\db \times & \db \int\limits_{\varphi_0^{+}}^{\varphi_0^{-}} \mathscr{D}^\prime\varphi^+ \, \mathscr{D}^\prime\varphi^-
    \exp i \Bigg\{ S'[\varphi^+,\varphi^-]
   + \int_{x}\,\left( \varphi^+_a(x), \varphi^-_a(x)\right)\left(\!\begin{array}{c} J^+_a(x) \\ - J^-_a(x)
  \end{array}\!\right)
  \nonumber\\
  &\db +&\db  \frac{1}{2}\int_{xy}\!\! \left( \varphi^+_a(x),\varphi^-_a(x)
  \right)\left(\!\begin{array}{cc} R^{++}_{ab}(x,y) & -R^{+-}_{ab}(x,y)\\
  -R^{-+}_{ab}(x,y) & R^{--}_{ab}(x,y) \end{array}\!\right)
  \left(\!\begin{array}{c} \varphi^+_b(y) \\ \varphi^-_b(y)
  \end{array}\!\right)
\Bigg\}.
\nonumber\\
\end{eqnarray}
Again the superscripts `$+$' and `$-$' indicate that the respective time arguments of the sources are taken on the forward or backward branch of the
closed time path, and the prime on the measure means that the integration over the fields at
initial time $t_0$ is excluded.

In order to simplify the comparison with the classical-statistical field theory, a standard linear transformation $A$ of the fields is introduced as
\begin{equation}
\db  \left(\begin{array}{c}
  \varphi \\
  {\tilde\varphi}
  \end{array}\right)
  \,\equiv\, A\,\left(\begin{array}{c}
  \varphi^+ \\
  \varphi^-
  \end{array}\right),
\label{eq:RTransf}
\end{equation}
where
\begin{equation}
\db  A \,
  = \, \left(\begin{array}{rr} \frac{1}{2} & \frac{1}{2} \\
                             1           & -1
                             \end{array}\right), \qquad
  A^{-1}
  \, = \, \left(\begin{array}{rr} 1 & \frac{1}{2} \\
                             1 & -\frac{1}{2}
                             \end{array}\right)
\label{eq:RMatrix}
\end{equation}
such that $\varphi = (\varphi^+ + \varphi^-)/2$ and ${\tilde
\varphi} = \varphi^+ - \varphi^-$, or $\varphi^\pm =
\varphi \pm {\tilde \varphi}/2$, respectively. To avoid a proliferation of
symbols we have used here $\varphi$, which agrees with the
defining field in (\ref{eq:classical}) only for $\varphi^+ =
\varphi^-$. Since this will be the case for expectation values
in the absence of sources, where physical observables are
obtained, and since there is no danger of confusion in the
following we keep this notation.

\begin{figure}[!t]
\centering
\begin{picture}(20,42) (295,-164)
    \SetWidth{0.5}
    \SetColor{Black}
\Text(242,-135)[lb]{$\varphi $}
\Text(273,-135)[lb]{$\varphi $}
\Text(242,-164)[lb]{${\tilde \varphi}$}
\Text(273,-164)[lb]{$\varphi $}
    \Text(333,-135)[lb]{${\tilde \varphi}$}
    \Text(363,-135)[lb]{${\tilde \varphi}$}
    \Text(333,-164)[lb]{$\varphi $}
    \Text(363,-164)[lb]{${\tilde \varphi}$}
\SetWidth{1.0}
    \Line(250,-135)(270,-155)
    \Line(260,-145)(270,-135)
    \DashLine(260,-145)(250,-155){2}
\SetWidth{1.0}
    \DashLine(340,-135)(350,-145){2}
    \DashLine(350,-146)(360,-135){2}
    \Line(350,-145)(340,-155)
    \DashLine(360,-155)(350,-145){2}
\end{picture}
\caption{Classical (left) and quantum vertex (right) in the scalar field theory.}
\label{fig:clvsqm}
\end{figure}
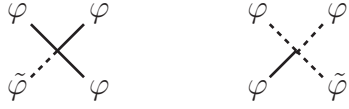

Correspondingly, we write for the source terms
\begin{eqnarray}
\db  \left(\begin{array}{c} J \\ \tilde{J} \end{array}\right)
  &\db \equiv& \db A\,\left(\begin{array}{c} J^+ \\ J^-
  \end{array}\right) ,
\label{eq:JTransf} \\
\db \left(\begin{array}{rr}
  R^F & R^{\mathrm R} \\
  R^{\mathrm A} & R^{\tilde F}
  \end{array}\right)
  &\db \equiv& \db A
  \left(\begin{array}{rr}
  R^{++}  & R^{+-} \\
  R^{-+}  & R^{--}
  \end{array}\right)A^{T} .
\label{eq:KMatrix}
\end{eqnarray}
Inserting these definitions into the functional integral
(\ref{eq:NEgenFuncZphipm}) and using that (\ref{eq:JTransf}) and
(\ref{eq:KMatrix}) can be equivalently written as
\begin{eqnarray}
\db  \left(\begin{array}{c}
  \tilde{J} \\ J
  \end{array}
  \right)
  &\db \equiv& \db \left(A^{-1}\right)^T
  \left(\begin{array}{c}
  J^+ \\ - J^-
  \end{array}\right),
\label{eq:JTransf2}\\
\db \left(\begin{array}{rr}
  R^{\tilde F}     & R^{\mathrm A} \\
  R^{\mathrm R}      & R^F
  \end{array}\right)
  &\db \equiv&\db \left(A^{-1}\right)^T
  \left(\begin{array}{rr}
  R^{++}  & -R^{+-} \\
  -R^{-+} & R^{--}
  \end{array}\right)A^{-1}\qquad
\label{eq:KMatrix2}
\end{eqnarray}
one finds:
\begin{eqnarray}
  && \!\!\!\!\!\!\db Z[J,\tilde{J},R^F,R^\mathrm{R},R^\mathrm{A},R^{\tilde
  F};\varrho_0] \, = \, \int [{\rm d}\varphi_0][{\rm d}{\tilde\varphi}_0]\,
       \varrho_0\left[\varphi_0+{\tilde\varphi}_0/2,
                   \varphi_0-{\tilde\varphi}_0/2\right] \nonumber\\
  &&\db \times
     \int\limits_{\varphi_0,{\tilde\varphi}_0}
        \mathscr{D}^\prime\varphi\, \mathscr{D}^\prime{\tilde\varphi}
     \,\exp i \Bigg\{S[\varphi,{\tilde\varphi}]
     + \int_{x,t_0}\,\left( \varphi_a(x), {\tilde\varphi}_a(x)\right)
               \left(\begin{array}{c}{\tilde J}_a(x) \\ J_a(x)\end{array}\right)
  \nonumber\\
  &&\db +\, \frac{1}{2}\int_{xy,t_0}\,
  \left( \varphi_a(x),{\tilde\varphi}_a(x) \right)
  \left(\begin{array}{cc}
  R^{\tilde F}_{ab}(x,y)         & R^\mathrm{A}_{ab}(x,y)\\[1pt]
  R^\mathrm{R}_{ab}(x,y)         & R^{F}_{ab}(x,y)
  \end{array}\right)
  \left(\begin{array}{c} \varphi_b(y) \\ {\tilde\varphi}_b(y) \end{array}\right)
  \Bigg\}.
\label{eq:ZQuantPhiXi}
\end{eqnarray}
Here $S[\varphi,{\tilde\varphi}] = S_0[\varphi,{\tilde\varphi}] +
S_{\rm int}[\varphi,{\tilde\varphi}]$ consists of the action for
the free field theory
\begin{equation}
\db  S_0[\varphi,{\tilde\varphi}]
  \, = \, \db \int_{x,t_0}
	\left[ \partial^\mu{\tilde \varphi}_a(x)\partial_\mu\varphi_a(x)  
	- m^2 \tilde{\varphi}_a(x)\varphi_a(x) \right]
  \label{eq:freeS}
\end{equation}
and the interaction part
\begin{equation}
{\db  S_{\rm int}[\varphi,{\tilde\varphi}]
  = -\frac{\lambda}{6 N}\int_{x,t_0} {\tilde\varphi}_a(x)\varphi_a(x)\varphi_b(x)\varphi_b(x)}
  {\rr  -\frac{\lambda}{24 N}\int_{x,t_0} {\tilde\varphi}_a(x){\tilde\varphi}_a(x)
                         {\tilde\varphi}_b(x)\varphi_b(x)}.
\label{eq:SqPhiXi}
\end{equation}
The two types of vertices appearing in the interaction part (\ref{eq:SqPhiXi}) are illustrated in Fig.~\ref{fig:clvsqm}. To understand their role for the dynamics, it is important to note that one can also write down a functional integral for the corresponding nonequilibrium classical-statistical field theory. This is done in section~\ref{sec:FIntCl}, where we will observe that the generating functionals for correlation
functions are very similar in the quantum and the classical-statistical theory. A crucial difference is that the quantum
theory is characterized by an additional vertex: The interaction term $\sim \tilde{\varphi}^3$, appearing with (\ref{eq:SqPhiXi}) in the functional integral of the quantum theory, does not occur for the classical theory.  

\subsubsection{Connected one- and two-point functions}
\label{sec:QCorrFcts}
For later use, we define in the following correlation functions in the above basis. From the generating functional for connected correlation functions, $W = -i\, {\rm ln} Z$, we define the macroscopic field $\phi_a$ and ${\tilde\phi}_a$ by
\begin{equation}
\db \frac{\delta W}{\delta {\tilde J}_a(x)} \, = \, \phi_a(x) , \quad
  \frac{\delta W}{\delta J_a(x)} \, = \, {\tilde\phi}_a(x)\, .
\label{eq:Defphiphitilde}
\end{equation}
The connected statistical correlation function $F_{ab}(x,y)$, the
retarded/advanced propagators $G^\mathrm{R/A}_{ab}(x,y)$, and
the ``anomalous'' propagator ${\tilde F}_{ab}(x,y)$ are defined by
\begin{eqnarray}
\db  \frac{\delta W}{\delta R^{\tilde F}_{ab}(x,y)}
  &\db =&\db
  \frac{1}{2}\left( \phi_a(x)\phi_b(y) + F_{ab}(x,y) \right) \, ,
  \nonumber\\
  \db \frac{\delta W}{\delta R^\mathrm{A}_{ab}(x,y)}
  &\db =&\db \frac{1}{2}\left(\phi_a(x){\tilde\phi}_b(y)-iG^\mathrm{R}_{ab}(x,y) \right) \, ,
  \nonumber\\
  \db \frac{\delta W}{\delta R^\mathrm{R}_{ab}(x,y)}
  &\db =&\db \frac{1}{2}\left({\tilde\phi}_a(x)\phi_b(y)-iG^\mathrm{A}_{ab}(x,y) \right) \, ,
  \nonumber\\
  \db \frac{\delta W}{\delta R^{F}_{ab}(x,y)}
  &\db =&\db  \frac{1}{2}\left({\tilde \phi}_a(x) {\tilde \phi}_b(y)
                     + {\tilde F}_{ab}(x,y)\right) \, .
  \label{eq:propagators}
\end{eqnarray}
Equivalently, we can define the same connected two-point correlation functions by the second functional derivatives
\begin{eqnarray} \db
\frac{\delta^{2} W}{\delta \tilde{J}_{a} (x) \delta J_{b} (y)} \, = \, G^\mathrm{R}_{a b} (x,y) & \db ,&\db  
\frac{\delta^{2} W}{\delta J_{a} (x) \delta \tilde{J}_{b} (y)} \,=\, G^\mathrm{A}_{a b} (x,y) ~, 
\\
\db \frac{\delta^{2} W}{\delta \tilde{J}_{a} (x) \delta \tilde{J}_{b} (y)} \, = \, i F_{a b} (x,y) &\db ,&\db 
\frac{\delta^{2} W}{\delta J_{a} (x) \delta J_{b} (y)} \,=\, i \tilde F_{a b} (x,y)~.
\label{eq:connectedtwopoint}
\end{eqnarray}
We note that the propagators satisfy the symmetry properties
$G_{a b}^{\mathrm A} (x,y) = G^{\mathrm R}_{b a} (y,x)$,
$F_{a b} (x,y) = F_{b a} (y,x)$ and ${\tilde F}_{a b} (x,y) = {\tilde F}_{b a} (y,x)$.
These properties follow directly from the definition of the propagators in terms of the second functional derivatives with respect to $J$ and ${\tilde J}$.
The spectral function $\rho$ is given by the difference of the retarded and advanced propagators
$\rho_{a b} (x,y) = G^{\mathrm R}_{a b} (x,y) - G_{a b}^{\mathrm A} (x,y)$,
and $G^{\mathrm R}_{a b} (x,y) = \rho_{a b} (x,y) \theta (x^{0}-y^{0})$.

It is important to note that the anomalous propagator ${\tilde F}$ vanishes in the limit where the external sources are set to zero. This is a consequence of the algebraic identity (\ref{eq:AlgebraicIdentity}), since ${\tilde F}= G^{++} + G^{--} - G^{-+} - G^{+-}$, as one may readily check by changing the basis: 
\begin{eqnarray}
\!\!\!\! &\!\! & \db \left(\begin{array}{cc}
    F &
    -i G^\mathrm{R} \\
    -i G^\mathrm{A} &
    {\tilde F} \end{array}\right)
    \, = \, A
   \left(\begin{array}{cc}
   G^{++} & G^{+-}\\
   G^{-+} & G^{--}
   \end{array}\right)
   A^{T}
\nonumber\\
 \!\!\!\!  &\!\! &\db =\,    \left(\begin{array}{cc}
   \left[G^{++}+G^{--}+G^{+-}+G^{-+}\right]/4 & \left[G^{++}-G^{--}+G^{-+}-G^{+-}\right]/2\\
   \left[G^{++}-G^{--}-G^{-+}+G^{+-}\right]/2 & G^{++}+G^{--}-G^{-+}-G^{+-}
   \end{array}\right)  
   , \quad \qquad
   \label{eq:GR}
\end{eqnarray}
More generally, in the absence of sources, we have 
\begin{eqnarray}
\db \frac{\delta W}{\delta J_{a} (x)} \Big|_{J,{\tilde J} , R^{R,A,F,{\tilde F}} = 0} &\db =&\db {\tilde \phi}_{a} (x) = 0 ~,
\\ \db
 \frac{\delta^{2} W}{\delta J_{a} (x) \,\delta J_{b} (y)} \Big|_{J,{\tilde J} ,R^{R,A,F,{\tilde F}} = 0} &\db =& \db i {\tilde F}_{a b} (x,y) = 0 ~,
\end{eqnarray} 
and, correspondingly, arbitrary functional derivatives of the generating functional with respect to $J$ vanish in the absence of sources. 

We finally mention that the inverse of the two-point function matrix (\ref{eq:GR})
in the absence of sources, where ${\tilde F}=0$, reads
\begin{equation}
\db   \left(\begin{array}{cc}
   0            &  i(G^\mathrm{A})^{-1} \\
   i(G^\mathrm{R})^{-1} &  (G^\mathrm{R})^{-1} \cdot F \cdot (G^\mathrm{A})^{-1}
   \end{array}\right) \, .
   \label{eq:inverse}
\end{equation}
Here we employed the compact matrix notation
\begin{equation}
\db  \left[(G^\mathrm{R})^{-1} \cdot F \cdot (G^\mathrm{A})^{-1}
  \right]_{ab}(x,y)
  \,\equiv\, \int_{z w}
  (G^\mathrm{R})^{-1}_{ac}(x,z)
  F_{cd}(z,w)
  (G^\mathrm{A})^{-1}_{db}(w,y) .
  \label{eq:Xinv}
\end{equation}
For vanishing sources, the exact inverse two-point function
(\ref{eq:inverse}) can then be written
as
\begin{equation}
\label{eq:DSeqn}
   \db \left(\begin{array}{cc}
   0            & i(G^\mathrm{A})^{-1} \\
   i(G^\mathrm{R})^{-1} & (G^\mathrm{R})^{-1} \cdot F \cdot (G^\mathrm{A})^{-1}
   \end{array}\right)
   \ =\  \left(\begin{array}{cc}
   0        & G_{0}^{-1} \\
   G_{0}^{-1}    & 0
   \end{array}\right)
   -  \left(\begin{array}{cc}
    0           & -i\Sigma^\mathrm{A}\\
    -i \Sigma^\mathrm{R} & \Sigma^F
    \end{array}\right),
\end{equation}
where the retarded,
advanced and statistical self-energies are obtained as 
\begin{equation}
\db \left(\begin{array}{cc}
    0 &
    -i \Sigma^\mathrm{A} \\
    -i \Sigma^\mathrm{R} &
    \Sigma^F \end{array}\right)
    \, = \, \left(A^{-1}\right)^T
   \left(\begin{array}{cc}
   \Sigma^{++} & -\Sigma^{+-}\\
   -\Sigma^{-+} & \Sigma^{--}
   \end{array}\right)
   A^{-1} 
   \label{eq:Sigmarotated}
\end{equation}
from the self-energies written in the `$\pm$' basis. To the
retarded/advanced self-energy $\Sigma^\mathrm{R/A}$ and the
statistical self-energy $\Sigma^F$ contribute only graphs with
propagator lines associated to $G^\mathrm{R,A}$ and $F$, which can
be obtained from closed two-particle irreducible graphs by opening
one propagator line.

\subsubsection{2PI effective action}
\label{app:2PIEA}

The 2PI effective action written in terms of the rotated variables becomes a functional of the field expectation values $\phi,{\tilde \phi}$, and the propagators $G^{\mathrm R},G^{\mathrm A},F,{\tilde F}$, 
\begin{eqnarray}
\db  \Gamma
  &\db =& \db \ W - \int_x \left(\phi_a(x) {\tilde J}_a(x)
     + {\tilde\phi}_a(x) J_a(x) \right)
  \nonumber\\
  &&\db -\ \frac{1}{2} \int_{xy} \Big\{ R^{\tilde F}_{ab}(x,y)
    \Big(\phi_a(x)\phi_b(y) + F_{ab}(x,y) \Big)
  \nonumber\\
  &&\db \qquad +\ R^\mathrm{A}_{ab}(x,y)
  \left(\phi_a(x){\tilde \phi}_b(y) - iG^\mathrm{R}_{ab}(x,y) \right)
  \nonumber\\
  &&\db \qquad +\ R^\mathrm{R}_{ab}(x,y)
  \left({\tilde \phi}_a(x)\phi_b(y) -iG^\mathrm{A}_{ab}(x,y)\right)
  \nonumber\\
  &&\db \qquad +\ R^{F}_{ab}(x,y)
  \left( {\tilde \phi}_a(x){\tilde \phi}_b(y) + {\tilde F}_{ab}(x,y) \right)
  \Big\} .
  \label{eq:effact}
\end{eqnarray}
From (\ref{eq:effact}) one observes the equations of motion for the fields
\begin{eqnarray}
  \db \frac{\delta \Gamma}{\delta \phi_a(x)}
  &\db =&\db - {\tilde J}_a(x) -\int_y
      \Big( R^{\tilde F}_{ab}(x,y) \phi_b(y)
  \nonumber\\
  &&\db +\ \frac{1}{2} R^\mathrm{A}_{ab}(x,y) {\tilde \phi}_b(y)
    +  \frac{1}{2}{\tilde \phi}_b(y) R^\mathrm{R}_{ba}(y,x) \Big),
  \\
  \db \frac{\delta \Gamma}{\delta \tilde{\phi}_a(x)}
  &\db =&\db - J_a(x)
     -\int_y \Big( R^{F}_{ab}(x,y) \tilde{\phi}_b(y)
  \nonumber\\
  &&\db + \frac{1}{2} R^\mathrm{R}_{ab}(x,y) \phi_b(y)
    +  \frac{1}{2} \phi_b(y) R^\mathrm{A}_{ba}(y,x) \Big),
    \label{eq:fields}
\end{eqnarray}
as well as for the two-point functions
\begin{eqnarray}
&&\db  \frac{\delta \Gamma}{\delta F_{ab}(x,y)}
  \, = \, - \frac{1}{2}R^{\tilde F}_{ab}(x,y) \, , \qquad
  i\frac{\delta \Gamma}{\delta G^\mathrm{R}_{ab}(x,y)}
  \, = \, - \frac{1}{2}R^\mathrm{A}_{ab}(x,y)\, ,
\nonumber  \\
&&\db  i\frac{\delta \Gamma}{\delta G^\mathrm{A}_{ab}(x,y)}
  \, = \, - \frac{1}{2}R^\mathrm{R}_{ab}(x,y) \, , \qquad
  \frac{\delta \Gamma}{\delta \tilde{F}_{ab}(x,y)}
  \, = \, - \frac{1}{2}R^{F}_{ab}(x,y)\, .
\end{eqnarray}

In the presence of a non-vanishing field value, $\phi\not = 0$ but 
${\tilde \phi} = 0$,
the interaction vertices are obtained from $(\ref{eq:SqPhiXi})$ by
shifting in $S[\varphi, {\tilde \varphi}]$ the field $\varphi \to
\phi + \varphi$, and collecting all cubic and quartic terms in the
fluctuating fields $\varphi$ and ${\tilde\varphi}$, i.e.\
\begin{eqnarray}
\db  S_{\rm int}[\varphi,{\tilde \varphi};\phi]
  &\db =& \db
      -\frac{\lambda}{6 N}\int_x
  {\tilde\varphi}_a(x)        \varphi_a(x)        \varphi_b(x)\varphi_b(x)
   {\rr -\frac{\lambda}{24 N}\int_x
  {\tilde\varphi}_a(x){\tilde\varphi}_a(x){\tilde\varphi}_b(x)\varphi_b(x)}
  \nonumber\\
  && \db -   \frac{\lambda}{3 N}       \int_x
  {\tilde\varphi}_a(x)        \varphi_a(x)        \varphi_b(x)   \phi_b(x)
  -\frac{\lambda}{6 N}\int_x
  {\tilde\varphi}_a(x)           \phi_a(x)        \varphi_b(x)\varphi_b(x)
  \nonumber\\
  && \rr -\frac{\lambda}{24 N}\int_x
  {\tilde\varphi}_a(x){\tilde\varphi}_a(x){\tilde\varphi}_b(x)   \phi_b(x).
\label{eq:SqPhiXiVev}
\end{eqnarray}
The quadratic terms in the fluctuating fields are taken into
account in the classical inverse propagator 
corresponding to a field
dependent $iG_0^{-1}(x,y;\phi)$ given by (\ref{classprop}).
For the classical-statistical field theory only those terms 
of (\ref{eq:SqPhiXiVev}) appear, which are linear in ${\tilde \varphi}$. This is shown in the following.

\subsection{Functional integral for the classical-statistical theory}
\label{sec:FIntCl}

The classical field equation of motion 
for the $N$-component scalar field $\varphi_a(x)$ with
action (\ref{eq:classical}) is given by
\bea \db
\left[-\square_x -m^2 - \frac{\lambda}{6N} 
\varphi_b(x)\varphi_b(x)\right]\varphi_a(x) \, =\, 0 \, ,
\label{eqmotion} 
\eea
Its solution, $\varphi_a^\cl(x)$, requires the specification of the initial conditions
$\varphi_a^\cl(t_0,\bx) = \varphi_{0,a}(\bx)$ and
$\pi_a^\cl(t_0,\bx) = \pi_{0,a}(\bx)$, with $\pi_{a}^\cl(x) = \partial_{x^0}\varphi_a^\cl(x)$ for the considered scalar field theory. We define the {\rr\it macroscopic or average classical field} by
\begin{equation}\db
\phi_{a}^\cl (x) \, = \, \langle \varphi_a(x)\rangle_\cl \, = \, 
\int \left[\rmd \pi_0 \right] \left[\rmd \varphi_0 \right] W^\cl[\varphi_0,\pi_0]\, \varphi_a^\cl(x) \, .
\label{eq:phicl}
\end{equation}
Here $W^\cl[\varphi_0,\pi_0]$ denotes the normalized
probability functional at initial time. The measure indicates integration
over classical phase-space,
\beq\db
\int \left[\rmd \pi_0 \right] \left[\rmd \varphi_0 \right]  \, = \, \int
{\prod\limits_{a=1}^N\prod\limits_{\bx}} 
\, \rmd \pi_{0,a}(\bx)\, {\rm d}\varphi_{0,a}(\bx) \, ,
\eeq
and the theory may be defined on a spatial lattice.
Similarly, the connected {\rr\it classical-statistical propagator} $F_{ab}^\cl (x,y)$
is defined by
\beq\db
F_{ab}^\cl (x,y) + \phi_{a}^\cl (x) \phi_{b}^\cl (y) 
\, = \, \langle \varphi_a(x)\varphi_b(y)\rangle_\cl \equiv 
\int \left[\rmd \pi_0 \right] \left[\rmd \varphi_0 \right] W^\cl[\varphi_0,\pi_0]
\varphi_a^\cl(x)\varphi_b^\cl(y) \, .
\label{eqFcl}
\eeq
The classical equivalent of the quantum spectral function  
is obtained by replacing
$-i$ times the commutator with the Poisson bracket:\footnote{
Recall that the Poisson bracket with respect to the initial fields is
\beq
\{ A(x), B(y) \}_{\rm PB} 
= \sum_{a=1}^N\int_\bz \left[ 
 \frac{\delta A(x)}{\delta\varphi_{0,a}(\bz)} 
 \frac{\delta B(y)}{\delta\pi_{0,a}(\bz)} 
-\frac{\delta A(x)}{\delta\pi_{0,a}(\bz)} 
 \frac{\delta B(y)}{\delta\varphi_{0,a}(\bz)}\right] \, .
\eeq} 
\beq \db
 \rho_{ab}^\cl(x,y) \, = \, -\langle\,
\{\varphi_a(x),\varphi_b(y)\}_{\rm PB}\, \rangle_\cl \, .
\label{eq:classspec}
\eeq
As a consequence, one finds the {\rr\it equal-time relations for the
classical spectral function}
\beq
\db \rho_{ab}^\cl(x,y)|_{x^0=y^0} \, = \, 0,\;\;\;\;
 \partial_{x^0}\rho_{ab}^\cl(x,y)|_{x^0=y^0} \, = \, \delta_{ab} \delta(\bx-\by) \, .
\label{eq:cleqtime}
\eeq
Though their origin is different, we note that they are 
in complete correspondence with the respective quantum relations  
(\ref{eq:bosecomrel}). 

In general, classical-statistical correlation functions are obtained as phase-space
averages over trajectories given by solutions of the classical
field equation (\ref{eqmotion}). Such averages, for an
arbitrary functional of the field $f[\varphi_a]$, are defined as
\begin{equation}\db
  \langle f[\varphi_a]\rangle_\mathrm{cl}
  = \int[\mathrm{d}\varphi_0][\mathrm{d}\pi_0] \, 
     W^\cl[\varphi_0,\pi_0]
    \, f[\varphi_a^\mathrm{cl}].
  \label{eq:classaverage}
\end{equation}
This expression will be the starting point for constructing a functional integral for the classical-statistical field theory similar
to the expression (\ref{eq:ZQuantPhiXi}) for the quantum theory.

We define
\begin{equation}
\db S^\mathrm{cl}[\varphi,{\tilde \varphi}] \, \equiv \, 
S_0[\varphi,{\tilde \varphi}] + S_{\rm
int}^\mathrm{cl}[\varphi,{\tilde \varphi}] \, .
\end{equation}
The free part, $S_0[\varphi,{\tilde \varphi}]$, is given by (\ref{eq:freeS}). Integrating by parts, we have to take care of the finite initial-time boundary values ${\tilde \varphi}_{0,a}({\bf x}) = {\tilde \varphi}(t_0,{\bf x})$ and $\pi_{0,a}({\bf x}) = \partial_{x^0}\varphi(x)|_{x^0=t_0}$ and we can write  
\begin{equation}
\db  S_0[\varphi,{\tilde\varphi}]
  \, = \, \db - \int_\bx \pi_{0,a}(\bx)\, {\tilde \varphi}_{0,a}(\bx) + \int_{x,t_0}
	{\tilde \varphi}_a(x)\left( - \square_x  
	- m^2 \right) \varphi_a(x) \, .
  \label{eq:freeScl}
\end{equation}
The interaction part, $S_{\rm int}^\mathrm{cl}[\varphi,{\tilde \varphi}]$, reads 
\begin{equation}
\db  S_{\rm int}^\mathrm{cl}[\varphi,{\tilde \varphi}]
  \, = \, -\frac{\lambda}{6N}\int_x {\tilde\varphi}_a(x)\varphi_a(x)\varphi_b(x)\varphi_b(x) \,.
  \label{eq:SqPhiXicl}
\end{equation}
This interaction part differs from (\ref{eq:SqPhiXi}) in that it contains fewer
vertices. The absence of vertices beyond those which are linear in
${\tilde\varphi}$ turns out to be a crucial difference between a
classical-statistical and a quantum field theory as is shown
in the following. From the definition of $S^\mathrm{cl}[\varphi,{\tilde \varphi}]$,
the classical equation of motion (\ref{eqmotion}) for the field $\varphi_a(x)$ can be obtained as
\begin{equation}
\db \frac{\delta S^\mathrm{cl}[\varphi,{\tilde \varphi}]}
       {\delta {\tilde \varphi}_a(x)} \, = \, \left(-\square_x -m^2 - \frac{\lambda}{6N} 
\varphi_b(x)\varphi_b(x)\right)\varphi_a(x) \, = \, 
       0 \, .
\end{equation}
We now rewrite this equation of motion 
as a $\delta$-constraint in a functional
integral using the Fourier transform representation
\begin{eqnarray}
  &&\db 
  \delta \left[ \frac{\delta S^\mathrm{cl}[\varphi,{\tilde \varphi}]}
       {\delta {\tilde \varphi}} \right]
\nonumber\\
 &&\db = \int \mathscr{D} {\tilde\varphi}\,
    \exp\left\{i \int_{x,t_0} {\tilde \varphi}_a(x) \left(-\square_x -m^2 - \frac{\lambda}{6N} 
\varphi_b(x)\varphi_b(x)\right)\varphi_a(x) \right\}
  \nonumber\\
  &&\db = \int \mathscr{D} {\tilde\varphi}\,
    \exp\left\{i S^\mathrm{cl}[\varphi,{\tilde \varphi}]+{i}
    \int_{\mathbf{x}}\pi_{0,a}({\bf x})\tilde\varphi_{0,a}({\bf x})\right\}
\label{eq:deltaconstraint}\, ,
\end{eqnarray}
where we have used (\ref{eq:freeScl}) for the last equality.

In order to complete the construction of the functional integral for classical-statistical correlation functions, we note that any functional $f[\varphi^\mathrm{cl}]$ of the classical field solution can be written as
\begin{eqnarray}
\db   f[\varphi^\mathrm{cl}]
  &\db = &\db \int\limits_{\varphi_0} \mathscr{D}^\prime \varphi \, f[\varphi]\,
      \delta\left[\varphi - \varphi^\mathrm{cl}\right]
  \nonumber\\
  &\db = &\db \int\limits_{\varphi_0} \mathscr{D}^\prime \varphi\, f[\varphi]\,
      \delta\left[\frac{\delta S^\mathrm{cl}[\varphi,{\tilde\varphi}]}
                       {\delta {\tilde \varphi}}\right] {\cal J}[\varphi]
  \nonumber\\
  &\db =& \db \int\limits_{\varphi_0}
     \mathscr{D}^\prime\varphi\, \mathscr{D}{\tilde \varphi}\,
      f[\varphi]\, \exp\Big\{i S^\mathrm{cl}[\varphi,{\tilde \varphi}]
      + {i}\int_{\mathbf{x}}\pi_{0,a}(\bx)\tilde\varphi_{0,a}(\bx)\Big\}\,
      {\cal J}[\varphi]. \quad
  \label{eq:functionalasPI}
\end{eqnarray}
The prime on the measure again signifies that no integration over the initial time is implied, since this is fixed by the initial condition. What we have employed in the second equality of (\ref{eq:functionalasPI}) is the generalization of
$\delta\left(g(x)\right) = \sum_i \delta(x-x_i)/|g'(x_i)|$ for zeros $x_i$ of an ordinary function $g(x)$ to functionals.
The Jacobian reads
\begin{equation}
\db  {\cal J}[\varphi]
  = \left|{\rm det}\left(
  \frac{\delta^2 S^\mathrm{cl}[\varphi,{\tilde \varphi}]}
       {\delta \varphi \delta {\tilde\varphi}}\right)\right|.
  \label{eq:Jacobean}
\end{equation}
Here it turns out that the Jacobian plays the role of an irrelevant normalization constant.
Putting things together, we can write 
\begin{eqnarray}\db
  \langle f[\varphi_a]\rangle_\mathrm{cl}
  &\db = & \db \int[\mathrm{d}\varphi_0][\mathrm{d}\pi_0] \, 
     W^\cl[\varphi_0,\pi_0] 
   \label{eq:classaverage2}\\  
   &\db \times &\db  \int\limits_{\varphi_0}
     \mathscr{D}^\prime\varphi\, \mathscr{D}{\tilde \varphi}\,
      f[\varphi]\, \exp\Big\{i S^\mathrm{cl}[\varphi,{\tilde \varphi}]+i\int_{\mathbf{x}}\pi_{0,a}(\bx)\tilde\varphi_{0,a}(\bx)\Big\}\,
      {\cal J}[\varphi].
  \nonumber
\end{eqnarray} 
The initial conditions may also be specified by a classical density matrix $\varrho_0^\cl\left[\varphi_0+{\tilde \varphi}_0/2,\varphi_0-{\tilde \varphi}_0/2\right]$, which is
characterized by the Fourier transform of the phase-space probability distribution $W^\cl[\varphi_0,\pi_0]$:
\begin{equation}
\db \varrho_0^\cl
    \left[\varphi_0+{\tilde \varphi}_0/2,\varphi_0-{\tilde \varphi}_0/2\right]  
    \, =\, \int[\mathrm{d}\pi_0]
    W^\cl[\varphi_0,\pi_0]\exp\left\{i\int_{\mathbf{x}}
                    \pi_{0,a}(\bx)\,\tilde\varphi_{0,a}(\bx)\right\}.
  \label{eq:FTrhoW}
\end{equation}
Adding also sources, we obtain the generating functional for classical-statistical correlation functions:
\begin{eqnarray}
  && \!\!\!\!\!\!\db Z^\cl[J,\tilde{J},R^F,R^\mathrm{R},R^\mathrm{A},R^{\tilde
  F};\varrho_0] \, = \, \int [{\rm d}\varphi_0][{\rm d}{\tilde\varphi}_0]\,
       \varrho_0^\cl\left[\varphi_0+{\tilde\varphi}_0/2,
                   \varphi_0-{\tilde\varphi}_0/2\right] \nonumber\\
  &&\db \times
     \int\limits_{\varphi_0,{\tilde\varphi}_0}
        \mathscr{D}^\prime\varphi\, \mathscr{D}^\prime{\tilde\varphi}
     \,\exp i \Bigg\{S^\cl[\varphi,{\tilde\varphi}]
     + \int_{x,t_0}\,\left( \varphi_a(x), {\tilde\varphi}_a(x)\right)
               \left(\begin{array}{c}{\tilde J}_a(x) \\ J_a(x)\end{array}\right)
  \nonumber\\
  &&\db +\, \frac{1}{2}\int_{xy,t_0}\,
  \left( \varphi_a(x),{\tilde\varphi}_a(x) \right)
  \left(\begin{array}{cc}
  R^{\tilde F}_{ab}(x,y)         & R^\mathrm{A}_{ab}(x,y)\\[1pt]
  R^\mathrm{R}_{ab}(x,y)         & R^{F}_{ab}(x,y)
  \end{array}\right)
  \left(\begin{array}{c} \varphi_b(y) \\ {\tilde\varphi}_b(y) \end{array}\right)
  \Bigg\}\, {\cal J}[\varphi]. \qquad
\label{eq:ZClPhiXi}
\end{eqnarray}
Comparing with the quantum generating functional in
(\ref{eq:ZQuantPhiXi}), and using that the Jacobian ${\cal
J}[\varphi]$ plays the role of an irrelevant normalization constant, 
we find that the generating functionals for correlation functions are very similar in the quantum and the classical-statistical theory. 
The main difference is that the quantum theory is characterized by more vertices. 

In particular, all definitions for correlation functions given in section~\ref{sec:QCorrFcts} for the quantum theory apply as well for the respective classical correlators. For instance, from (\ref{eq:connectedtwopoint}) we infer for the classical spectral function in the absence of sources:
\begin{eqnarray}
  && \!\!\!\!\!\!\db \rho^\cl_{ab}(x,y) \, = \,  i\int [{\rm d}\varphi_0][{\rm d}{\tilde\varphi}_0]\,
       \varrho_0^\cl\left[\varphi_0+{\tilde\varphi}_0/2,
                   \varphi_0-{\tilde\varphi}_0/2\right] \nonumber\\
  &&\db \times
     \int\limits_{\varphi_0,{\tilde\varphi}_0} \!\!\!\!
        \mathscr{D}^\prime\varphi\, \mathscr{D}^\prime{\tilde\varphi} \, 
				\big[\varphi_a(x)\tilde\varphi_b(y)-\tilde\varphi_a(x)\varphi_b(y)\big]
     \, e^{i S^\cl[\varphi,{\tilde\varphi}]} \, .
	\label{eq:rhoppt}	
\end{eqnarray}
In order to understand the equivalence with the definition as a phase-space average of the Poisson bracket in (\ref{eq:classspec}), we may consider the retarded case $x^0 < y^0$ first and take $x^0=t_0$. With the help of (\ref{eq:FTrhoW}) we can then write  
\begin{eqnarray}
\!\! &&\!  \db - i\int [{\rm d}\varphi_0][{\rm d}{\tilde\varphi}_0]\,
       \varrho_0^\cl\left[\varphi_0+{\tilde\varphi}_0/2,
                   \varphi_0-{\tilde\varphi}_0/2\right] \!\!
									\int\limits_{\varphi_0,{\tilde\varphi}_0} \!\!\!\!
        \mathscr{D}^\prime\varphi\, \mathscr{D}^\prime{\tilde\varphi} \,
				\tilde\varphi_a(x)\varphi_b(y)\, e^{i S^\cl[\varphi,{\tilde \varphi}]} \nonumber\\	
\!\! &\db =&\!  \db -i \int[\mathrm{d}\varphi_0][\mathrm{d}\pi_0]  
     W^\cl[\varphi_0,\pi_0] \!\!
		\int\limits_{\varphi_0,{\tilde\varphi}_0} \!\!\!\! \mathscr{D}^\prime\varphi\, \mathscr{D}{\tilde \varphi}\,
      \tilde\varphi_a(x)\varphi_b(y)\, e^{i S^\mathrm{cl}[\varphi,{\tilde \varphi}]+ i\int_{\mathbf{x}}\pi_{0,a}\tilde\varphi_{0,a}}{\cal J}[\varphi]\nonumber\\
\!\! &\db =&\! 	\db -\int[\mathrm{d}\varphi_0][\mathrm{d}\pi_0]  
     W^\cl[\varphi_0,\pi_0] \frac{\delta}{\delta\pi_{0,a}(\mathbf{x})} \!
		\int\limits_{\varphi_0,{\tilde\varphi}_0} \!\!\!\! \mathscr{D}^\prime\varphi\, \mathscr{D}{\tilde \varphi}\,
      \varphi_b(y)\, e^{i S^\mathrm{cl}[\varphi,{\tilde \varphi}]+ i\int_{\mathbf{x}}\pi_{0,a}\tilde\varphi_{0,a}}{\cal J}[\varphi]\nonumber\\		
\!\! &\db =&\!\db -\int [{\rm d}\varphi_0][\mathrm{d}\pi_0]  
     W[\varphi_0,\pi_0]
     \frac{\delta\varphi^\mathrm{cl}_b(y)}
          {\delta\pi_{0,a}(\mathbf{x})}
					\nonumber\\
 \!\! &\db =&\!\db - \int [{\rm d}\varphi_0][\mathrm{d}\pi_0]  
     W[\varphi_0,\pi_0]\int_\mathbf{z}
     \frac{\delta\varphi^\mathrm{cl}_a(t_0,\mathbf{x})}
          {\delta\varphi_{0,c}(\mathbf{z})}
     \frac{\delta\varphi^\mathrm{cl}_b(y)}
          {\delta\pi_{0,c}(\mathbf{z})}\, , 	\nonumber
\end{eqnarray}
where for the last line we have employed (\ref{eq:functionalasPI}) and summation over repeated field indices is implied. Doing the corresponding steps for the retarded case $x^0 > y^0$ taking $y^0 = t_0$ shows the equivalence.

\subsection{Classicality condition}
\label{sec:classicality}

Since the generating functionals for
correlation functions are very similar in the classical and the
quantum theory, the same techniques can be used
to derive time evolution equations of classical correlation
functions. In particular, the classical dynamic 
equations have the very same form (\ref{eq:exactrhoF}) and (\ref{eq:exactphi}) as their quantum analogues with the replacements $\phi_a(x)\to \phi_a^\cl$, $F_{ab}(x,y) \to F_{ab}^\cl(x,y)$, and $\rho_{ab}(x,y) \to \rho_{ab}^\cl(x,y)$. The corresponding classical-statistical self-energies $\Sigma_{ab}^{F,\cl}(x,y)$ and $\Sigma_{ab}^{\rho,\cl}(x,y)$ have in general the same diagrammatic contributions but are lacking certain terms due to the reduced number of vertices.

In order to compare classical and quantum corrections to self-energies, we note that
the classical-statistical generating functional
(\ref{eq:ZClPhiXi}) exhibits an important reparametrization
property: If the fluctuating fields are rescaled according to
\begin{equation}
\db  \varphi_a(x) \to \varphi_a^\prime(x) 
  =  \sqrt{\lambda}\,\varphi_a(x)\quad ,\quad
  \tilde{\varphi}_a(x) \to \tilde{\varphi}^\prime_a(x)
  = \frac{1}{\sqrt{\lambda}}\,{\tilde{\varphi}_a(x)} \label{eq:rescaling}
\end{equation}
then the coupling $\lambda$ drops out of $S^\mathrm{cl}[\varphi,{\tilde
\varphi}]=S_0[\varphi,{\tilde \varphi}]+S_{\rm
int}^\mathrm{cl}[\varphi,{\tilde \varphi}]$ defined in
(\ref{eq:freeS}) and (\ref{eq:SqPhiXicl}). The free part
$S_0[\varphi,{\tilde \varphi}]$ remains unchanged and the
interaction part becomes
\begin{equation}
 \db S_{\rm int}^\mathrm{cl}[\varphi^\prime,{\tilde \varphi}^\prime]
  = -\frac{1}{6 N}\int_x {\tilde\varphi}^\prime_a(x)\varphi^\prime_a(x)\varphi^\prime_b(x)\varphi^\prime_b(x).
  \label{eq:SqPhiXiclRescaled}
\end{equation}
Moreover, the functional measure in (\ref{eq:ZClPhiXi}) is
invariant under the rescaling (\ref{eq:rescaling}), and the
sources can be redefined accordingly. Therefore, the classical
statistical generating functional becomes independent of $\lambda$,
except for the coupling dependence entering the probability
distribution fixing the initial conditions. Accordingly, the
coupling does not enter the classical dynamic equations for
correlation functions. All the $\lambda$-dependence enters the initial
conditions which are required to solve the dynamic equations.

In contrast to the classical case, this reparametrization property
is absent for the quantum theory: After the rescaling
(\ref{eq:rescaling}) one is left with
$S[\varphi^\prime,\tilde{\varphi}^\prime]$ whose coupling
dependence is given by the interaction part
\begin{equation}
{\db  S_{\rm int}[\varphi^\prime,\tilde{\varphi}^\prime]
  = -\frac{1}{6 N}\int_x {\tilde\varphi}^\prime_a(x)\varphi^\prime_a(x)\varphi^\prime_b(x)\varphi^\prime_b(x)}
  {\rr  -\frac{\lambda^2}{24 N}\int_x {\tilde\varphi}^\prime_a(x){\tilde\varphi}^\prime_a(x)
                         {\tilde\varphi}^\prime_b(x)\varphi^\prime_b(x)},
\label{eq:SqPhiXiRescaled}
\end{equation}
according to (\ref{eq:SqPhiXi}). Comparing to
(\ref{eq:SqPhiXiclRescaled}) one observes that the quantum
vertex, which is absent in the classical-statistical theory,
encodes all the $\lambda$-dependence of the dynamics.

The comparison of quantum versus classical dynamics becomes
particularly transparent using the above rescaling. The rescaled
macroscopic field and statistical correlation function are given
by
\begin{equation}
\db \phi_a^\prime(x) = \sqrt{\lambda} \phi_a(x) \,, \quad F_{ab}^\prime(x,y)
= \lambda F_{ab}(x,y)\,, \label{eq:phiFRescaled}
\end{equation}
while the spectral function $\rho_{ab}(x,y)$ remains unchanged
according to (\ref{eq:rhoppt}).
Similarly, we define for the statistical self-energy $\Sigma^{F\prime}_{ab}(x,y) = \lambda \Sigma^F_{ab}(x,y)$.
For instance, for the two-loop self-energies for vanishing macroscopic field given in (\ref{eq:2loopresF}) and (\ref{eq:2loopresr}) we obtain
\bea 
\db \Sigma_F^\prime(x,y) &\db =&\db
-\frac{N+2}{18N^2}\,F^\prime(x,y)\left[F^{\prime 2}(x,y) 
{\rr -\frac{3}{4} \lambda^2 \rho^2(x,y)}\right] \, ,
\label{eq:2loopresF}\\
\db\Sigma_{\rho}(x,y) &\db =&\db
-\frac{N+2}{6N^2}\,\rho(x,y)\left[F^{\prime 2}(x,y) 
{\rr -\frac{1}{12} \lambda^2 \rho^2(x,y)}\right] \, .
\label{eq:2loopresr}
\eea
The corresponding classical-statistical self-energies $\Sigma_F^{\prime\cl}(x,y)$ and $\Sigma_\rho^{\cl}(x,y)$ are given by the same expressions by dropping the $\lambda$-dependent terms in (\ref{eq:2loopresF}) and (\ref{eq:2loopresr}), which are proportional to $\rho^2$. More precisely, one observes that the quantum evolution equations would be accurately described by the classical ones if the {\rr\it classicality condition} 
\begin{equation}
\mbox{\framebox{\rr 
$\displaystyle  \,\, F^2(x,y) \gg \rho^2(x,y) \,\,$}}
\label{eq:classicality}
\end{equation} 
in terms of the non-rescaled correlation functions holds. A similar analysis of the loop-corrections in the presence of a nonzero macroscopic field given in section~\ref{sec:sigmaloops}, or for the NLO $1/N$ expansion of section~\ref{2PINfield} yields the same condition. 

However, the requirement to fulfill (\ref{eq:classicality}) for all space-time arguments is too restrictive and it can be typically only achieved for a limited range of time and relevant momenta. One expects that the classical description
becomes a reliable approximation for the quantum theory if
the number of field quanta in each mode is sufficiently high. 
The classicality condition (\ref{eq:classicality}) entails
the justification of this expectation. In order to illustrate the 
condition in terms of a more intuitive picture of occupation numbers,
we employ the free-field theory type form of the spectral function and
statistical propagator with mode frequency $\omega_\bp$ as given in (\ref{eq:freeFrho}).
From this one obtains the following estimates for the
time-averaged correlators:
\begin{equation} \db
\overline{F^2}(t,t';\bp) \equiv
\frac{\omega_{\bp}}{2 \pi} \int_{t-2 \pi/\omega_{\bp}}^{t}  
\!\!\!\!\!\!\!\!\!\!\!\!\!\! \rmd t' \,
F^2(t,t';\bp) \,\rightarrow\, 
\frac{(f_{\bp}(t)+1/2)^2}{2 \omega_{\bp}^2(t)} , \quad  
\overline{\rho^2}(t,t';\bp) \, \rightarrow \,
\db \frac{1}{2 \omega_{\bp}^2(t)}  .
\end{equation}
Inserting these estimates in (\ref{eq:classicality}) for equal momenta 
yields 
\beq
{\db\left[f_{\bp}(t)+\frac{1}{2} \right]^2 \, \gg \, 1}
\qquad {\rm or} \qquad {\db f_{\bp}(t) \, \gg \, \frac{1}{2} } \, .
\label{eq:estimatencl}
\eeq
If the dominant momentum modes have occupancies much larger than the quantum-half, then a classical-statistical description can be an accurate approximation of the quantum dynamics. It is important to note that this is in general {\it\rr not} the case in thermal equilibrium, where at temperature $T$ the typical momenta $p \sim T$ have an occupancy of order one. Consequently, the late-time approach to thermal equilibrium in quantum theories is beyond the range of applicability of classical-statistical approximations. Important examples, where a classical-statistical description is accurate, include continuous thermal phase transitions, since the relevant momenta for scaling behavior have $p \ll T$ with occupancy $T/p \gg 1$. Other examples are nonequilibrium instabilities, which yield high occupation numbers of characteristic modes, such as described in section~\ref{sec:instab}, or wave turbulence and nonthermal scaling phenomena such as described in section~\ref{sec:transport}.

\subsection{Precision tests of quantum versus classical-statistical dynamics}

When the nonequilibrium quantum dynamics of a highly occupied system can be accurately mapped onto a classical-statistical field theory evolution, one can solve it without further approximations using lattice simulation techniques. This mapping is valid as long as the classicality condition (\ref{eq:classicality}) is fulfilled for typical momenta. The lattice field theory is then defined on a spatial grid with spacing $a_s$ and side length $N_s a_s$ in a box with periodic boundary conditions, as explained around equation (\ref{eq:latticemomenta}) at the end of section~\ref{sec:solneqeq}. Classical-statistical simulations consist of numerically solving the classical field equations of motion and Monte Carlo sampling of initial conditions according to (\ref{eq:classaverage}). Thus, observables are obtained by averaging over the different classical trajectories that arise from the different initial field configurations.  

For the relativistic second-order differential equation of the scalar field (\ref{eqmotion}), the numerical integrations on a $d$-dimensional grid can be efficiently done using a standard leapfrog algorithm. For the nonrelativistic first-order equation (\ref{eq:gpe}) a conventional split-step method may be applied. These classical equations are then supplemented by suitable quantum initial conditions. For instance, typical Gaussian initial conditions represented by spatially homogeneous ensembles for a relativistic field equation as explained in section~\ref{sec:initialconditions} can be specified by a macroscopic field $\phi(t_0) = \phi_0$, its derivative $\dot{\phi}(t_0) = \dot{\phi}_0$ and the distribution function $f_\bp(t_0)$ as 
\begin{eqnarray}
\db \varphi(t_0,\bx) = \phi_0 + \int \frac{\rmd^d p}{(2\pi)^d}  \sqrt{\frac{f_\bp(t_0) + 1/2}{\omega_\bp(t_0)}}\, c_{\bp}\, e^{i\bp \bx}\;,
 \label{mat:class-field-init}
\end{eqnarray}
with the initial frequency $\omega_\bp(t_0) = \sqrt{\bp^2 + M^2}$. For Gaussian initial conditions, the coefficients $c_{\boldsymbol{p}}$ have to satisfy the relations
\begin{eqnarray}
\db \langle c_{\bp}c^{*}_{\bp'} \rangle_\cl = (2 \pi)^d \delta(\bp - \bp')\quad , \quad \langle c_{\bp}c_{\bp'} \rangle_\cl = \langle c^{*}_{\bp}c^{*}_{\bp'} \rangle_\cl = 0\,.
 \label{mat:gaussian-numbers-relations}
\end{eqnarray}
They can be realized by taking $c_{\bp}$ as Gaussian random numbers multiplied by complex random phase factors. This can be formulated as $c_{\bp} = A(\bp)~e^{i 2 \pi \alpha(\bp)}$ with a Gaussian distributed amplitude $A(\bp)$ and uniformly distributed phase $\alpha(\bp)$ between \mbox{$0$ and $1$}. The random numbers $c_{\bp}$ have additionally to satisfy $c^*_{-\bp} = c_{\bp}$ to ensure that $\varphi(t_0,\bx)$ is real-valued. 

The conjugate momentum field $\pi(t_0,\bx) = \dot{\varphi}(t_0,\bx)$ is initialized in a very similar way:
\begin{eqnarray}
\db \pi(t_0,\bx) = \dot{\phi}_0 + \int \frac{\rmd^d p}{(2\pi)^d}  \sqrt{\left(f_\bp(t_0) + 1/2\right) \omega_\bp(t_0)}\, \tilde{c}_{\bp}\, e^{i\bp \bx}\;,
 \label{mat:class-momfield-init}
\end{eqnarray}
with complex Gaussian random numbers $\tilde{c}_{\bp}$ satisfying the same relations as $c_{\bp}$ such as (\ref{mat:gaussian-numbers-relations}). Since $c_{\bp}$ and $\tilde{c}_{\bp}$ are independent random numbers, one also has $\langle c_{\bp}\tilde{c}_{\bp'} \rangle_\cl = 0$. Therefore, $\langle \varphi(t_0,\bx) \pi(t_0,\by) + \pi(t_0,\bx) \varphi(t_0,\by) \rangle_\cl$ vanishes automatically at initial time $t=t_0$ for the initial conditions discussed here. For instance, the macroscopic field or the statistical correlation function at times larger than $t_0$ are then given by the averages (\ref{eq:phicl}) or (\ref{eqFcl}),
and equivalently for higher statistical $n$-point correlation functions. Since $\rho^{\rm cl}(t,t',\bx-\bx')$ is given by the Poisson bracket (\ref{eq:classspec}), spectral properties cannot be obtained from simple products of classical fields. However, in contrast to solving the equations (\ref{eq:exactrhoF}) for correlation functions, knowledge of the spectral function is not a prerequisite for the computation of correlation functions from classical-statistical simulations.\\  

In the following, we apply these simulation techniques to the scalar $N$-component field theory.
Since the nonequilibrium evolution of classical-statistical
correlation functions can be obtained numerically
up to controlled statistical errors, the results include all orders 
in $1/N$. Consequently, they can be used for a precision test of approximation schemes such as the 2PI $1/N$
expansion implemented in classical-statistical field theory.
We emphasize that this compares two very different calculational
procedures: the results from the simulation involve thousands of 
individual runs from which the correlators are constructed, while the 
corresponding results employing the 2PI $1/N$ expansion involve
only a single run solving directly the evolution equation for the
correlators. The accuracy of the simulations manifests itself 
also in the fact that the time-reversal invariant dynamics can 
be explicitly reversed in practice for not too late times.
We then compare to the corresponding NLO results of the quantum 
theory and check the classicality condition of section~\ref{sec:classicality}.

Here, we consider a system in $d=1$ that is invariant under space translations and work
in momentum space. Similar calculations have been performed also for $d=3$ and for theories including fermions, which we refer to in the literature section~\ref{sec:litCl}. At $t_0=0$, we choose the same initial conditions for the classical and the corresponding quantum theory: A Gaussian initial state with zero macroscopic field and a statistical propagator $F(0,0;p) = [f_p(0)+1/2]/\om_p(0)$, where the initial $f_p(0)$ represents a peaked distribution around the momentum $p = p_{\rm ts}$ as in section~\ref{sec:solneqeq}. The initial mode energy is given by $\om_p(0) = \sqrt{p^2+M^2}$, where $M$ is the one-loop renormalized mass in the presence of the
nonequilibrium medium, determined from the corresponding one-loop gap equation as in~(\ref{LOgapequ}). 
As a renormalization condition we choose the one-loop renormalized mass in
vacuum $m_R\equiv M|_{f_p(0)=0}$ as our scale. The results shown below are obtained using a fixed coupling constant 
$\lambda/m_R^2=30$. 

\begin{figure}[t]
 \centerline{
\epsfig{file=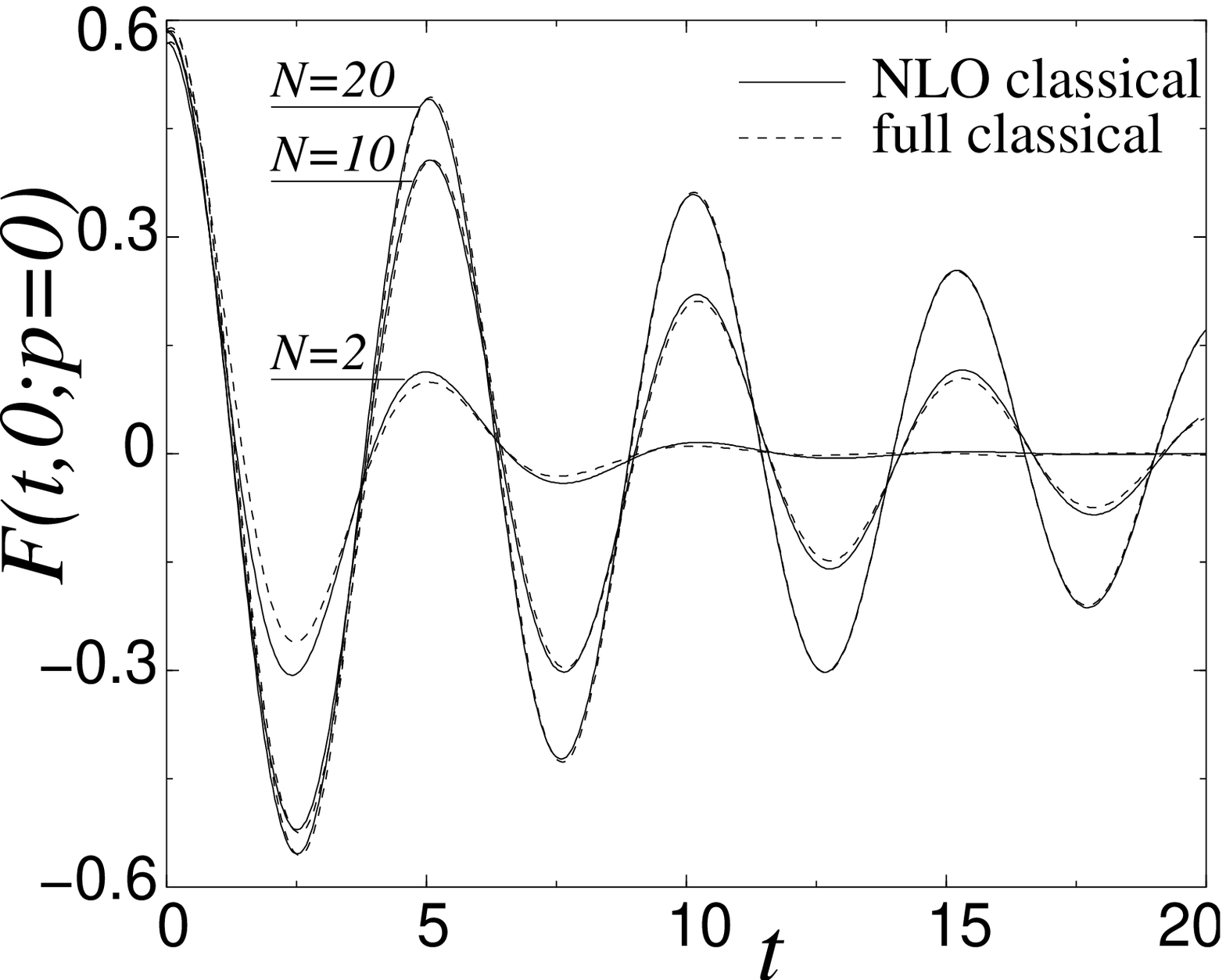,width=7.2cm}
 \hspace{.cm}
\epsfig{file=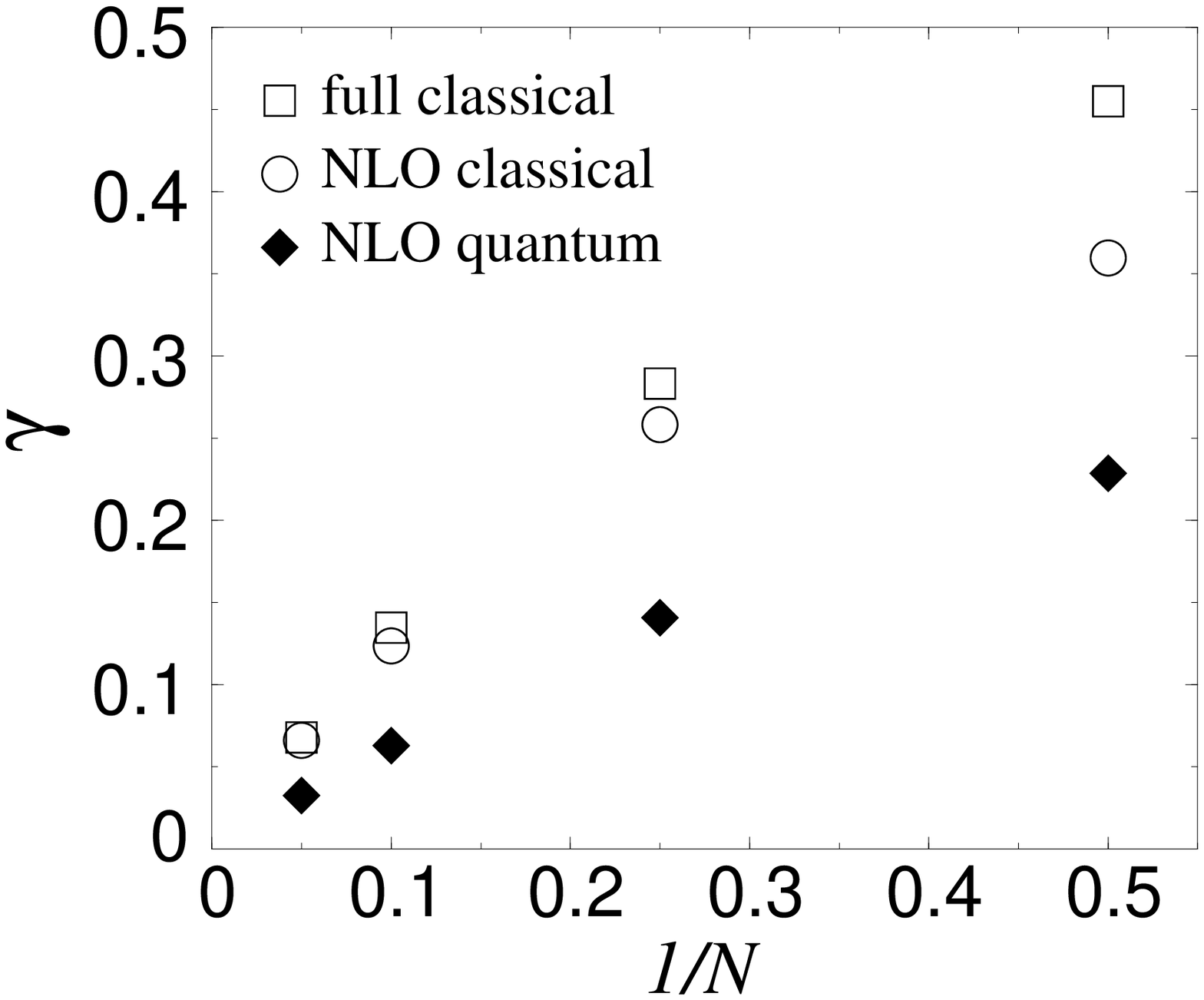,width=7.2cm}}
 \caption{\label{fig:classearly} {\bf Left:} Unequal-time two-point 
function $F(t,0;p=0)$ at zero
momentum for $N=2,10,20$. The solid lines show
results from the NLO classical evolution and the dashed lines from the
full classical-statistical simulation. 
One observes a convergence of 
classical NLO and full results already for moderate values of $N$.
 {\bf Right:} Damping rate extracted 
 from $F(t,0;p=0)$ as a function of $1/N$. 
 Open symbols represent NLO and full classical evolution.
 The quantum NLO results are shown with black symbols for comparison.
 The initial conditions are characterized by low occupation 
 numbers so that quantum effects become sizeable. One observes
 that in the quantum theory
 the damping rate is reduced compared to the classical theory. 
 (All in units of $m_R$.)}
\end{figure}

On the left of Fig.~\ref{fig:classearly} the classical-statistical propagator 
$F(t,0;p=0)$ is presented for three values of $N$. All other parameters
are kept constant. The figure compares the time evolution using the 2PI
$1/N$ expansion to NLO and the classical-statistical simulation.\footnote{The results presented below have been obtained from sampling
50000-80000 independent initial conditions to approximate the exact
evolution of correlation functions. In general, far fewer configurations are required for $d > 1$ because of self-averaging.}  One observes that the
approximate time evolution of the correlation function shows a rather good
agreement with the full result even for small values of $N$. For $N=20$ the exact and NLO evolution can hardly be
distinguished. A very sensitive quantity for comparisons is the damping rate
$\gamma$, which is obtained from an exponential fit to the envelope of
$F(t,0;p=0)$.  The systematic convergence of the NLO and the Monte Carlo
result as a function of $1/N$ can be observed from the right graph of
Fig.~\ref{fig:classearly}. The
accuracy of the description of far-from-equilibrium processes
within the classical-statistical NLO approximation of the 2PI effective action is remarkable.

The right graph of Fig.~\ref{fig:classearly} 
also shows the damping rate from the quantum
evolution, using the same initial conditions and parameters. One observes
that the damping in the quantum theory differs and, in particular, is
reduced compared to the classical result. The effective loss of details 
about the initial conditions takes more time for the quantum system
than for the corresponding classical one. In the limit $N\to \infty$
damping of the unequal-time correlation function goes to zero
since the nonlocal part of the self-energies
vanishes identically at LO large-$N$ and scattering is absent. In
this limit there is no difference between evolution in a quantum and
classical-statistical field theory for same initial conditions. 

\begin{figure}[t]
 \centerline{
\epsfig{file=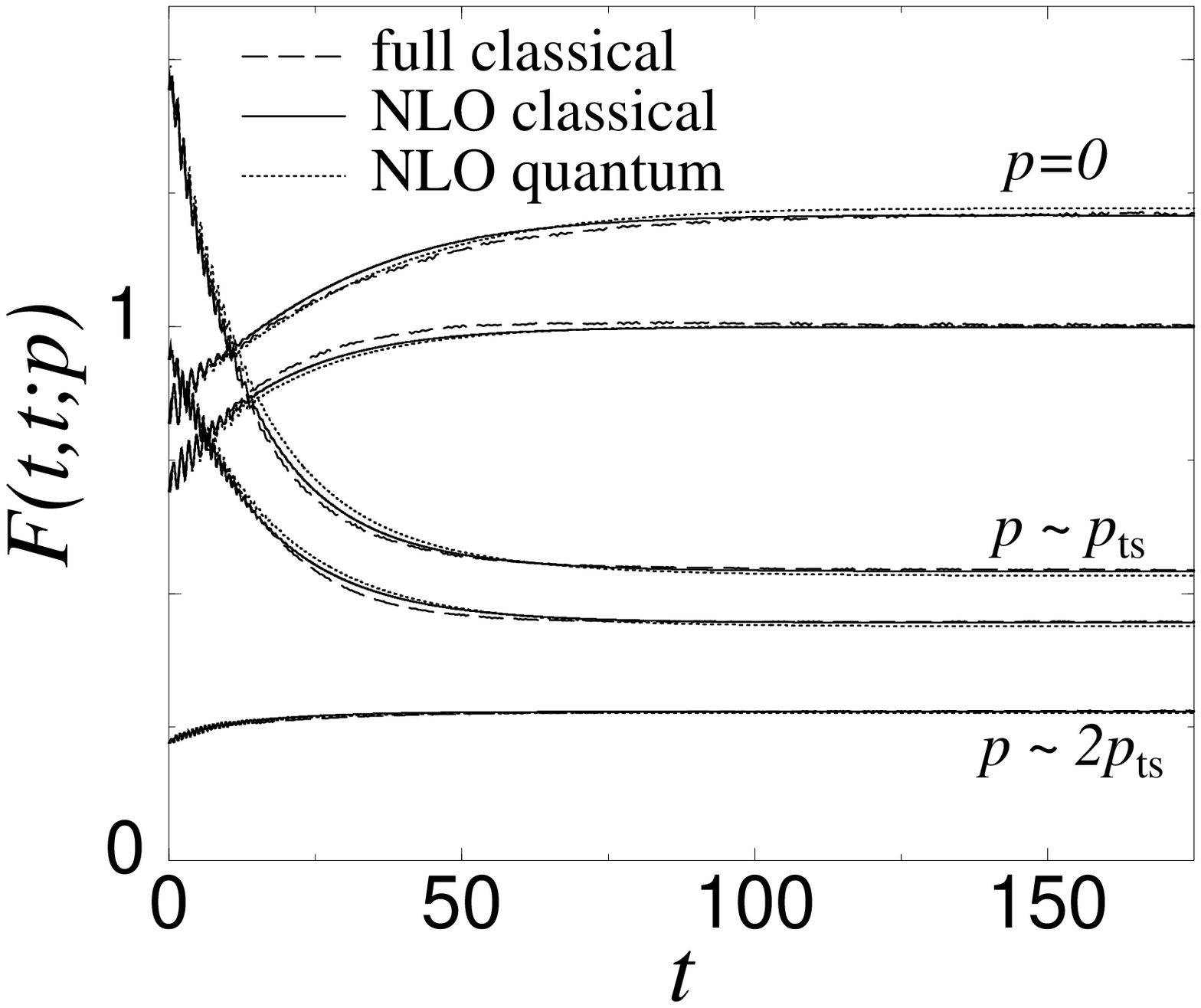,width=7.2cm}
 \hspace{.1cm}
\epsfig{file=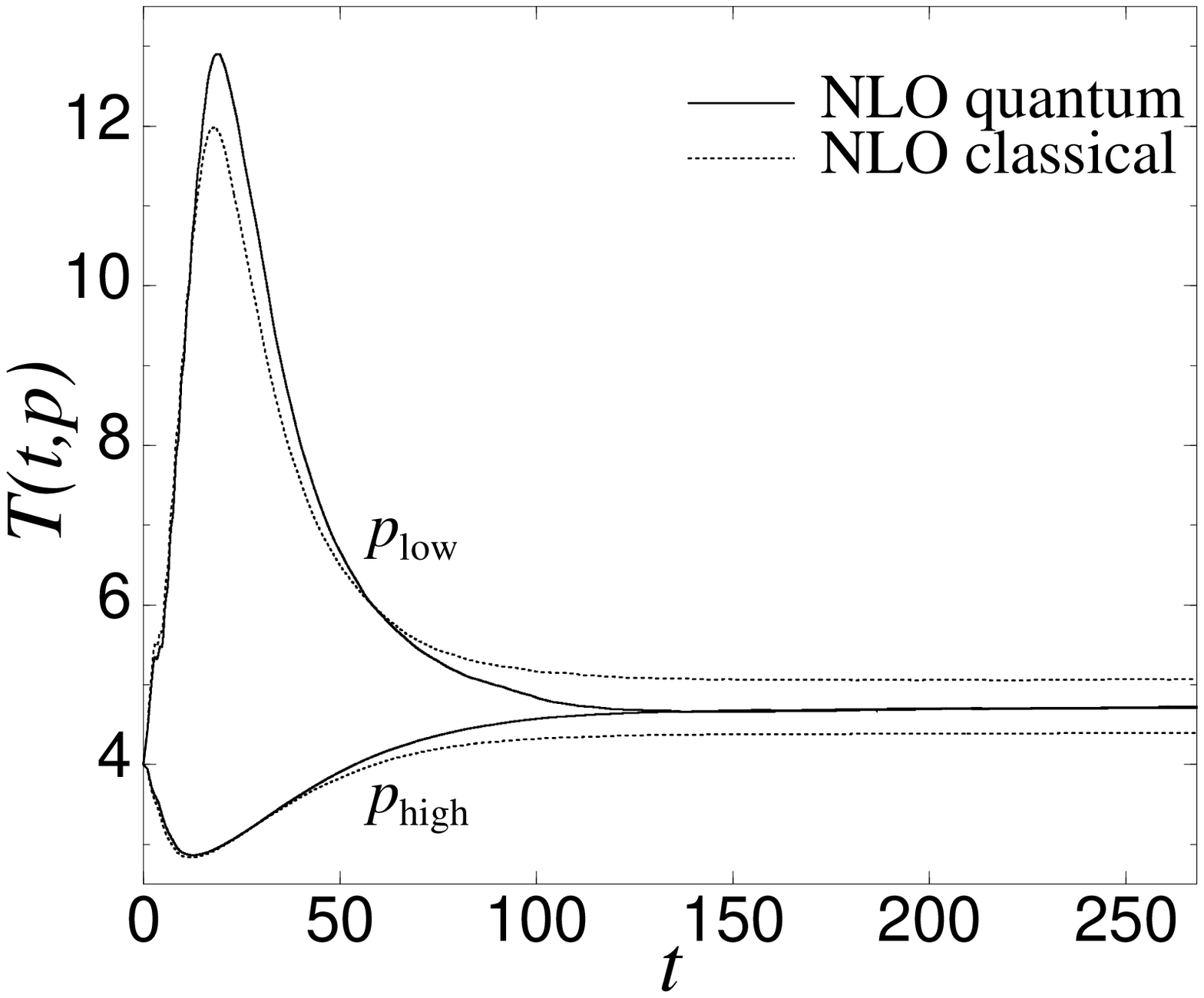,width=6.9cm}}
 \caption{\label{fig:classlate} {\bf Left:} Nonequilibrium evolution of the 
equal-time two-point function $F(t,t;p)$ for $N=10$ for various momenta $p$. 
One observes a good agreement between the full simulation (dashed) and
the NLO classical result (full). The quantum evolution is shown with
dotted lines. The integrated initial particle density is six times as high as
in Fig.\ \ref{fig:classearly}. 
 {\bf Right:} A very sensitive quantity to 
study deviations is the time dependent inverse slope 
$T(t,p)$ defined in the text. When a Bose-Einstein distributed occupation number is approached, all 
modes get equal $T(t,p)=T_{\rm eq}$, as can be observed to high 
accuracy for the quantum evolution. For classical 
thermal equilibrium the defined inverse slope remains momentum dependent.}
\end{figure}

For finite $N$ scattering is present and quantum and classical
evolution differ in general. However, as discussed in 
section~\ref{sec:classicality}, the classical field approximation
may be expected to become a reliable description for the quantum theory if
the number of field quanta in each field mode is sufficiently high. 
We observe that increasing the initial particle number density leads
to a convergence of quantum and classical time evolution at not too late
times. In Fig.~\ref{fig:classlate} (left) the time evolution of 
the equal-time
correlation function $F(t,t;p)$ is shown for several momenta $p$ and $N=10$.
Here the initial integrated particle density $\int dp/(2\pi) f_p(0)/M=1.2$ 
is six times as high as in Fig.~\ref{fig:classearly}. 
At $p = 2 p_{\rm ts}$ one finds  
$f_{2 p_{\rm ts}}(0) \simeq 0.35$ and a slightly larger value 
at this momentum of about $\simeq 0.5$ at later times shown. 
For these evolutions the classicality condition (\ref{eq:estimatencl}) is therefore 
approximately fulfilled up to momenta $p \simeq 2 p_{\rm ts}$, and
one indeed observes from the left of Fig.~\ref{fig:classlate} 
a rather good agreement of quantum and classical evolution 
in this range. For an estimate of 
the NLO truncation error we also give the full Monte Carlo
result for $N=10$ showing a quantitative agreement 
with the classical NLO evolution during these times.

From the left of Fig.~\ref{fig:classlate} one observes that the initially 
highly occupied modes ``decay'' as time proceeds and the
low momentum modes become more and more populated. At late times
the classical theory and the quantum theory approach their 
respective equilibrium distributions. Since classical and quantum 
thermal equilibrium are distinct, the classical and quantum time evolutions
have to deviate at sufficiently late times.
Figure~\ref{fig:classlate} shows the time dependent inverse slope 
parameter 
\beq\db
T(t,p) \equiv - f_p(t) \mbox{$[f_p(t)+1]$} 
\left(\frac{{d} f_p}{{d} \epsilon_p}\right)^{-1}  \, ,
\label{alsoinvslope}
\eeq
which has been introduced in section~\ref{sec:NLOtherm} to
study the approach to a Bose-Einstein distribution.\footnote{Note
that $\,{d} {\rm Log}(f_p^{-1}(t) + 1)/{d} \epsilon_p(t) = T^{-1}(t,p)$.} 
It employs the effective particle number $f_p(t)$ defined in
(\ref{eq:effpart}) and mode energy $\epsilon_p(t)$ given by
(\ref{eq:effen}).
Initially one observes a very different behavior of $T(t,p)$ for the low
and high momentum modes, indicating that the system is far from
equilibrium. The quantum evolution approaches
a Bose-Einstein distributed occupation number with a momentum-independent inverse slope 
$T_{\rm eq}=4.7\, m_R$ to very good accuracy. 
In contrast, in the classical theory the slope parameter remains momentum
dependent since the classical dynamics does, of course, not reach a 
Bose-Einstein distribution.

To see this in more detail we note that 
for a Bose-Einstein distribution, $f_\beta(\ep_p) =
1/[\exp(\epsilon_p/T_{\rm eq})-1]$, the inverse slope (\ref{alsoinvslope})
is independent of the mode energies and
equal to the temperature $T_{\rm eq}$.  During the nonequilibrium evolution
effective thermalization can therefore be observed if $T(t,p)$ 
becomes time and momentum independent, $T(t,p) \to T_{\rm eq}$. This is 
indeed seen on the left of Fig.~\ref{fig:classlate} for the quantum 
system. If the system is approaching classical equilibrium 
at some temperature $T_\cl$ and is not too strongly coupled, the following behavior 
is expected.  From the definition (\ref{eq:effpart}) of $f_p(t)$ in
terms of two-point functions, we expect to find approximately
\beq\db
T(t,p)  \to  T_\cl\left( 1- \ep_p^2/T_\cl^2\right) \, ,
\label{eqinvsl}
\eeq
i.e.~a remaining momentum dependence with $T(t,p) <
T(t,p')$ if $\ep_p>\ep_p'$. Indeed, this is
what one observes for the classical field theory 
result in Fig.~\ref{fig:classlate}.

For a classical theory a very simple test 
for effective equilibration
is available. An exact criterion can be obtained 
from the classical counterpart of the
``KMS'' condition for thermal equilibrium discussed in 
section~\ref{sec:detourthermal}. In coordinate space the 
classical equilibrium ``KMS'' 
condition reads
\beq\db
\frac{1}{T_\cl} \frac{\partial}{\partial x^0}F^{\rm (eq)}_\cl(x-y) = 
-\rho^{\rm (eq)}_\cl(x-y),
\label{eqclKMS}
\eeq
and in momentum space 
\beq\db
F^{\rm (eq)}_\cl(k) = -if_\cl(k^0)\rho^{\rm (eq)}_\cl(k), \;\;\;\;\;\;\;\;
f_\cl(k^0) = \frac{T_\cl}{k^0}.
\eeq
Differentiating equation
(\ref{eqclKMS}) with respect to $y^0$ at $x^0=y^0=t$ gives
\beq\db
\frac{1}{T_\cl}\frac{\partial}{\partial y^0}\frac{\partial}{\partial x^0}
F^{\rm (eq)}_\cl(x-y)\Big|_{x^0=y^0=t} 
= -\frac{\partial}{\partial y^0}\rho^{\rm (eq)}_\cl(x-y)\Big|_{x^0=y^0=t}. 
\eeq
Combining this KMS relation with the equal-time condition
(\ref{eq:cleqtime}) for the spectral function leads to
\beq\db
\partial_t\partial_{t'}F_\cl^{\rm (eq)}(t,t';\bx-\by)|_{t=t'} 
= T_\cl\delta(\bx-\by).
\eeq
In terms of the classical conjugate momentum fields 
$\pi_a(x) \equiv \partial_{x^0}\varphi_a(x)$ 
this represents the well-known equilibrium relation
\mbox{$\langle\pi_a(t,\bx)\pi_b(t,\by)\rangle_\cl^{\rm (eq)} =
T_\cl\delta(\bx-\by)\delta_{ab}$}.
Out of equilibrium one can define an effective classical mode temperature
\beq\db
T_\cl(t,p) = \partial_t\partial_{t'}F_\cl(t,t';p)|_{t=t'} \, .
\eeq
Effective classical equilibration is observed if $T_\cl(t,p)$
becomes time and momentum independent, $T_\cl(t,p)\to T_\cl$. This is indeed the case for the results presented at 
sufficiently late times for given lattice regularization.

\begin{figure}[t]
\centerline{\epsfig{file=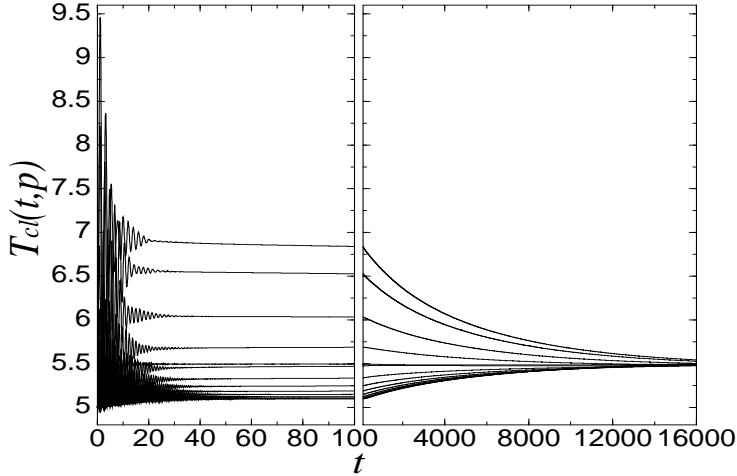,height=6.5cm,width=9.5cm}}
\caption{
Nonequilibrium time evolution in $1+1$ dimensions from the
three-loop approximation of the classical-statistical 2PI effective
action for one scalar field, $N=1$. Shown is the effective mode 
temperature $T_\cl(t,p)$. One observes the approach to classical  
equilibrium, $T_\cl(t,p) \to T_{\rm cl}$, at 
sufficiently late times for given lattice regularization.
}
\label{fixedpointclassical}
\end{figure}
Apart from the 2PI $1/N$ expansion,
the late-time behavior can also be studied from the 2PI loop expansion. 
For the quantum theory this has been demonstrated in 
section~\ref{sec:solneqeq} using two-loop self-energy corrections.
We show that this can also be employed for the classical-statistical theory 
and use the corresponding loop approximation in the following to demonstrate the above 
statements about classical equilibration.

In Fig.~\ref{fixedpointclassical} the nonequilibrium evolution of the
classical mode temperature $T_\cl(t,p)$ is shown 
for various momentum modes in the
$(1+1)$-dimensional classical scalar field theory for $N=1$.
The equations are solved by a lattice discretization with spatial
lattice spacing $m_R a_s=0.4$, time step $a_t/a_s=0.2$, and $N_s=24$ sites for $\lambda/m_R^2=1$. 
For the initial
ensemble we take $F_\cl(0,0;p) = T_0/(p^2+m_R^2)$ and 
$\partial_t\partial_{t'}F_\cl(t,t';p)|_{t=t'=0} = T_0$ with
$T_0/m_R=5$. We have observed that at sufficiently late times the
contributions from early times to the dynamics are effectively suppressed.
This fact has been
employed in Fig.\ \ref{fixedpointclassical} to reach the very late times.
One sees that at sufficiently late times the system relaxes towards
classical equilibrium with a final temperature $T_{\rm cl}/m_R \approx 5.5$.  
Though the typical 
classical equilibration times observed are substantially
larger than the times required to approach thermal equilibrium
in the respective quantum theory, this is not a regularization independent statement for the classical theory. In contrast to the
quantum theory, statements about equilibration times are sensitive to the employed lattice 
regularization for the classical theory because of the
Rayleigh-Jeans divergence. 

\pagebreak

\subsection{Bibliography}
\label{sec:litCl}
\begin{itemize}
\item Figs.~\ref{fig:classearly}--\ref{fixedpointclassical} are taken from
G.~Aarts and J.~Berges, {\it Classical aspects of quantum fields far from equilibrium}, Phys.\ Rev.\ Lett.\ {\bf 88}
(2002) 0416039 and J.~Berges, {\it Nonequilibrium quantum fields and the classical field theory limit}, Nucl.\ Phys.\ A {\bf 702} (2002) 351.
\item The above presentation follows also to a large extent J.~Berges and T.~Gasenzer,
  {\it Quantum versus classical-statistical dynamics of an ultracold Bose gas,}
  Phys.\ Rev.\ A {\bf 76} (2007) 033604. See also S.~Jeon, {\it The Boltzmann equation in classical and quantum field theory}, Phys.\ Rev.\ C {\bf 72} (2005) 014907.
\item Many of these ideas can be traced back to the original work of P.\ C.\ Martin, E.\ D.\ Siggia, and H.\ A.\ Rose, {\it Statistical Dynamics of Classical Systems}, Phys.\ Rev.\ A {\bf 8} (1973) 423.
\item Classical-statistical field theory for the relativistic reheating problem was pointed out in 
  D.~T.~Son, {\it Classical preheating and decoherence}, hep-ph/9601377; 
  S.~Y.~Khlebnikov and I.~I.~Tkachev,
  {\it Classical decay of inflaton},
  Phys.\ Rev.\ Lett.\  {\bf 77} (1996) 219.
\item Classical-statistical field theory studies related to
approximation schemes in QFT can be found in:
{\it Exact and truncated dynamics in nonequilibrium field theory},
G.~Aarts, G.~F.~Bonini and C.~Wetterich,
Phys.\ Rev.\ D {\bf 63} (2001) 025012.
{\it On thermalization in classical scalar field theory},
G.~Aarts, G.~F.~Bonini, C.~Wetterich,
Nucl.\ Phys.\  {\bf B587} (2000) 403. 
{\it Classical limit of time-dependent quantum field theory: A  
Schwinger-Dyson approach}, F.~Cooper, A.~Khare and H.~Rose,
Phys.\ Lett.\ B {\bf 515} (2001) 463.
A.~Arrizabalaga, J.~Smit and A.~Tranberg, {\it Tachyonic preheating using 2PI - 1/N dynamics and the classical
approximation}, JHEP {\bf 0410} (2004) 017.
\item Classical-statistical simulations including fermions have been pioneered in G.~Aarts and J.~Smit,
  {\it Real time dynamics with fermions on a lattice},
  Nucl.\ Phys.\ B {\bf 555} (1999) 355 for $d=1$, and became possible for $d=3$ in J.~Berges, D.~Gelfand and J.~Pruschke,
  {\it Quantum theory of fermion production after inflation},
  Phys.\ Rev.\ Lett.\  {\bf 107} (2011) 061301.
  For a review about real-time lattice QED and QCD see V.~Kasper, F.~Hebenstreit and J.~Berges,
  {\it Fermion production from real-time lattice gauge theory in the classical-statistical regime},
  Phys.\ Rev.\ D {\bf 90} (2014) 025016.
\item For diagrammatics in classical field theory see:
{\it Classical approximation for time-dependent quantum field
theory:  Diagrammatic analysis for hot scalar fields},
G.~Aarts and J.~Smit,
Nucl.\ Phys.\ {\bf B511} (1998) 451.
\end{itemize}

%% file: ch_instabLH.tex
\section{Nonequilibrium instabilities}
\label{sec:instab}

\subsection{Parametric resonance} 
\label{sec:parametricresonance}

In classical mechanics 
parametric resonance is the phenomenon of resonant amplification of 
the amplitude of an oscillator having a time-dependent periodic frequency.
In the context of quantum field theory a similar phenomenon describes 
the amplification of quantum fluctuations, which can be interpreted as 
particle production. It provides an important building block for our 
understanding of the (pre)heating of the early universe at the end of an 
inflationary period, and may also be operative 
at certain stages in relativistic heavy-ion 
collision experiments. The example of parametric resonance can lead to nonperturbative phenomena such as strong turbulence and Bose condensation far from equilibrium, even in the presence of arbitrarily small couplings, which will be discussed in section~\ref{sec:transport}. Here we will consider the phenomenon
as a ``paradigm'' for far-from-equilibrium dynamics following nonequilibrium instabilities. Much of the nonlinear physics turns out to be universal and thus independent of the details of the underlying mechanism that triggers the instability.

We recall first the classical mechanics example of 
resonant amplitude growth 
for an oscillator with time-dependent periodic frequency. A physical 
realization of this situation is a pendulum with a periodically 
changing length as displayed:  

\vspace*{0.2cm}

\centerline{
\epsfig{file=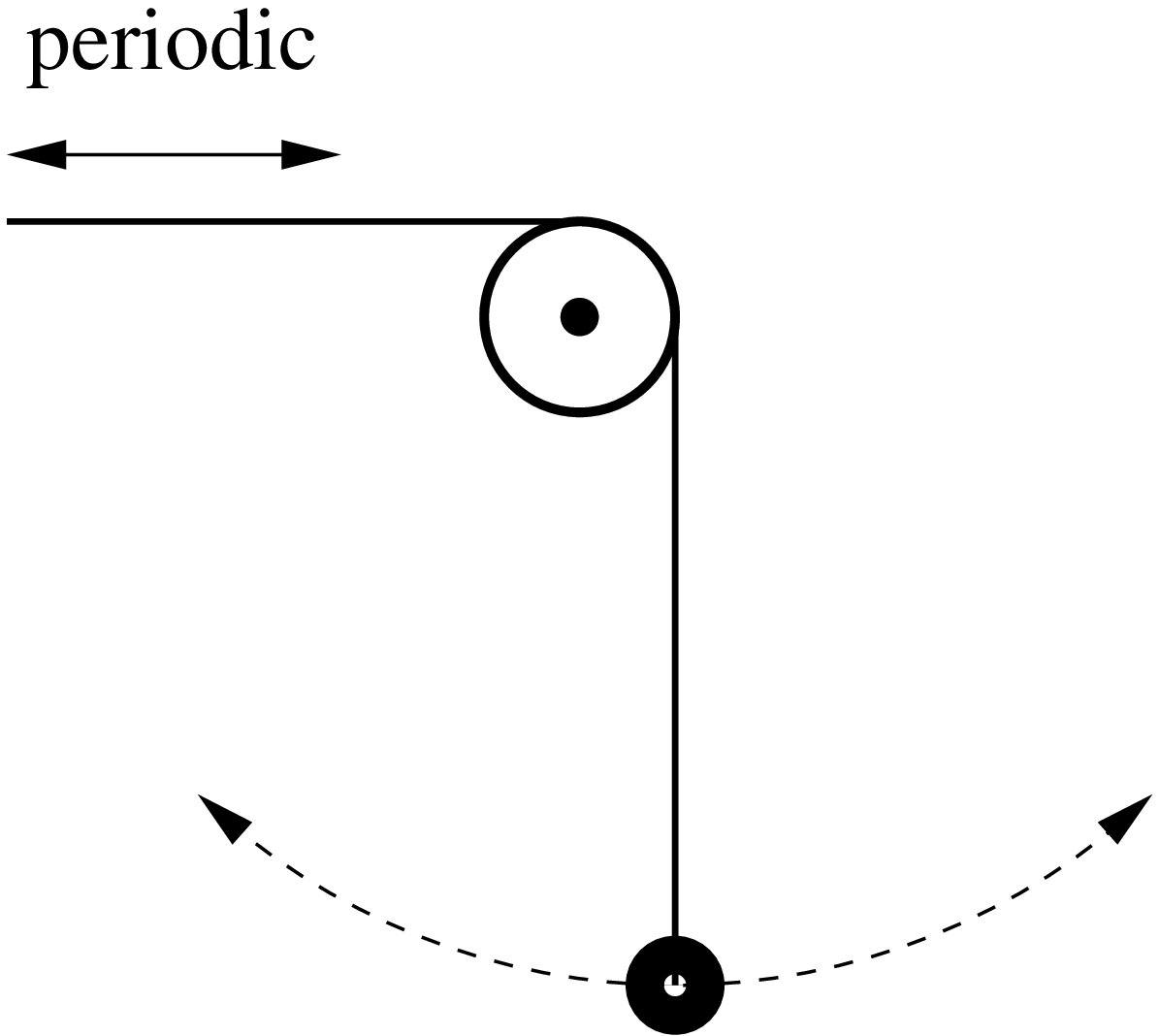,width=4.cm}
}

\vspace*{0.2cm}

\noindent
In the linear regime, the amplitude $y(t)$ is described by the second-order differential equation 
\begin{equation}
\db \ddot{y}(t) + \omega^2(t)\, y(t) \, = \, 0
\end{equation}
with periodic $\omega(t+ \Delta T)=\omega(t)$ of period $\Delta T$.
Since the equation is invariant under $t \to t + \Delta T$  
one expects periodic solutions 
\begin{equation}
\db y(t+\Delta T) \, = \, A\, y(t) \,.
\end{equation}
Writing the time-independent amplitude as
\begin{equation}
\db A \, = \, e^{\alpha \, \Delta T} \, 
\end{equation}
in terms of the so-called Floquet index $\alpha$, these solutions can be expressed as
\begin{equation}
\db y(t) \, = \, e^{\alpha t}\, \Pi(t)  
\label{eq:exponential}
\end{equation}
with periodic $\Pi(t+ \Delta T)=\Pi(t)$. This can be directly verified since
\begin{eqnarray}
\db y(t+\Delta T) 
&\db = &\db e^{\alpha (t + \Delta T)}\, \Pi(t+\Delta T)
\, = \,  A \,  \underbrace{e^{\alpha t} \, \Pi(t)} \, .
\\
&&\db \hspace*{5cm} y(t) \nonumber
\end{eqnarray}
With $y(t)$ also $y(-t)$ is a solution and one concludes from (\ref{eq:exponential}) that for real Floquet index $\alpha \neq 0$ there is an instability characterized by an exponential growth. 

In contrast to this mechanics example, in closed systems described by quantum field theory there will be no external periodic source. Below we will see that a large coherent 
field amplitude coupled to its own quantum fluctuations
can trigger the phenomenon of parametric resonance. 
Mathematically, however, important aspects are very similar 
to the above classical example for sufficiently early times. There is even a precise mapping: The mechanical oscillator amplitude $y$ plays the role of the statistical 
two-point function $F$ in quantum field theory, and the periodic
$\omega^2(t)$ plays the role of an effective mass
term $M^2(\phi(t))$ whose time dependence is induced by an
oscillating macroscopic field $\phi(t)$. Simple linear
approximations to the problem turn out to be mathematically equivalent to the
above mechanics example. Accordingly, well-known Lam{\'e}--type
solutions of the mechanics problem will also play a role 
in the quantum field theory study. Substantial deviations
do, however, quickly set in with important non-linear effects.

\subsubsection{Linearized classical analysis}

We consider the $O(N)$ symmetric scalar field theory with classical action (\ref{eq:classical}). Since the phenomenon of parametric resonance is essentially classical, for the purpose of comparison with the quantum treatment we start with a linearized classical analysis. The $N$-component field is written as
\begin{equation}
\db \phi_a(x) \, = \, \phi \, \delta_{a1} + \delta \phi_a(x) \, , 
\end{equation}
where we will consider a spatially homogeneous but time-dependent ``background field'', i.e.\ $\phi = \phi(x^0)$. The classical equations of motion are obtained from the stationarity of the action, and in the following we will consider them in linear approximation in the ``fluctuations'' $\delta \phi_a(x)$. These equation will also be seen to correspond to the evolution equations in the corresponding quantum field theory in the limit where all nonlinear or loop corrections are neglected.

We denote the fluctuations in the ``longitudinal'' field direction as $\delta \phi_\parallel(x) \equiv \delta \phi_1(x)$ and in the ``transverse'' direction as
$\delta \phi_\perp(x) \equiv \delta \phi_{a > 1}(x)$. The square of the fields appearing in the action (\ref{eq:classical}) then reads
\begin{equation}
\db \phi_a(x)\phi_a(x) \, = \, \phi^2 + 2 \phi\, \delta\phi_\parallel(x)
+ \delta\phi_\parallel(x)^2 + (N-1) \delta\phi_\perp(x)^2 \, ,
\label{eq:quad}
\end{equation}
whereas the quartic interaction term involves
\begin{equation}
\db \left[\phi_a(x)\phi_a(x)\right]^2 \, = \, \phi^4 + 4 \phi^3\, \delta\phi_\parallel(x)
+ 2 \phi^2 \left[ 3\, \delta\phi_\parallel(x)^2 + (N-1)  \delta\phi_\perp(x)^2 \right] + \mathcal{O}\left( \delta \phi^3 \right) \, .
\label{eq:quart}
\end{equation}
Here we expanded the quartic term neglecting cubic and quartic powers of the fluctuations, since this is sufficient to obtain the evolution equations to linear order.   

The classical field equation for the longitudinal component is given by
the stationarity condition $\delta S[\phi+ \delta \phi]/\delta \, \delta \phi_\parallel(x) = 0$. Using (\ref{eq:quad}) and (\ref{eq:quart}) one observes that  to lowest order in an expansion in powers of $\delta \phi_\parallel(x)$ it leads to the background field equation
\begin{equation}
\db \left(\frac{\partial^2}{\partial x_0^2} + m^2 + \frac{\lambda}{6N}\, \phi^2(x_0) \right) \phi(x_0) \, = \, 0 \, .
\label{eq:fieldexp1}
\end{equation}
To next-to-leading order in powers of $\delta \phi_\parallel(x)$ the same stationarity condition gives
\begin{equation}
\db
\left( \square_x + m^2 + \frac{\lambda}{2N}\, \phi^2(x_0) \right) \delta \phi_\parallel(x)
\, = \, 0 \, . 
\label{eq:fieldexp2}
\end{equation}
For the transverse components one finds from $\delta S[\phi+ \delta \phi]/\delta \, \delta \phi_\perp(x) = 0$ the equation
\begin{equation}
\db
\left( \square_x + m^2 + \frac{\lambda}{6N}\, \phi^2(x_0) \right) \delta \phi_\perp(x)
\, = \, 0 \, . 
\label{eq:fieldexp3}
\end{equation}

In the following we will choose $m=0$ as motivated in section \ref{sec:Earlyuniverseinflation} and denote $t \equiv x_0$. It is convenient to introduce the rescaled background field 
\begin{equation}
\db \sigma(t) \, = \, \sqrt{\frac{\lambda}{6N}} \, \phi(t) \, .
\end{equation}
We will consider the relevant case of a parametrically large initial field amplitude, $\phi(t=0) \sim \sqrt{6N/\lambda}$ with $\lambda \ll 1$, such that the initial $\sigma_0 \equiv \sigma(t=0)$ is of order one. In terms of the rescaled field the evolution equation (\ref{eq:fieldexp1}) reads
\begin{equation}
\db \ddot{\sigma}(t) + \sigma^3(t) \, = \, 0 \, .
\label{eq:sigma}
\end{equation}
The solution of this equation can be given in terms of Jacobi elliptic functions. For an initial amplitude $\sigma_0$ and derivative $\dot{\sigma}(t=0) = 0$ one has
\begin{equation}
\db \sigma(t) \, = \, \sigma_0 \, \mathrm{cn}(\sigma_0 t) \, . 
\end{equation}
Here the Jacobi cosine $\mathrm{cn}(z)\equiv \mathrm{cn}(z;n=1/2)$ is a doubly periodic function in $z$ with periods $4 K_n$ and $4 i K_{1-n}$, where 
$K_n = \int_0^{\pi/2} \rmd \Theta/\sqrt{1-n \sin^2 \Theta}$ is the complete elliptic integral of the first kind. We denote the characteristic frequency of the oscillations of $\sigma(t)$ as
\begin{equation}
\db \omega_0 \, = \, \frac{\pi \sigma_0}{2 K_{1/2}} \, \simeq \, 0.847\, \sigma_0 \, ,
\label{eq:chfreq}
\end{equation}
where $K_{1/2} \simeq 1.854$. Since the function will enter quadratically the equations for fluctuations, it will also be important that it is a quasi-periodic function with period $\mathrm{cn}(z+2 K_{1/2}) = - \mathrm{cn}(z)$. Averaged over one period, the square of the field amplitude gives $\,\overline{\sigma^2}= \int_0^{2 K_{1/2}} \rmd t\,
\sigma^2(t)/(2 K_{1/2}) \simeq 0.457\, \sigma_0^2$.

In order to study the fluctuations, we will consider the products
\begin{equation}
\db F_\parallel(x,y) \, = \, \delta \phi_\parallel(x) \, \delta \phi_\parallel(y)
\quad , \quad
F_\perp(x,y) \, = \, \delta \phi_\perp(x) \, \delta \phi_\perp(y) \, ,
\end{equation}
which will be the relevant quantities to compare to in the quantum treatment below. For spatially homogeneous systems they only depend on the relative spatial coordinates, $F_{\parallel,\perp}(x,y)= F_{\parallel,\perp}(x^0,y^0; \bx - \by)$.
According to (\ref{eq:fieldexp2}) and (\ref{eq:fieldexp3}) their spatial Fourier transforms, $F_{\parallel,\perp}(t,t';\bp)=\int \rmd^d s \exp(-i \bp {\bf s}) F_{\parallel,\perp}(t,t';{\bf s})$, obey in this linearized classical approach
\begin{eqnarray}
\db
\left[ \partial_t^2 + \bp^2 + {\rr 3 \sigma_0^2 \, \mathrm{cn}^2(\sigma_0 t)} \right]F_{\parallel}(t,t';\bp) 
&\db =& \db 0 \, ,\\[0.1cm]
\db \left[ \partial_t^2 + \bp^2 
+ {\rr \sigma_0^2 \, \mathrm{cn}^2(\sigma_0 t)} \right]F_{\perp}(t,t';\bp) 
&\db =& \db 0 \, .
\label{eq:Fppcl}
\end{eqnarray}
We will consider initial conditions characterizing a pure-state initial density matrix, where we take $\partial_{t}F_{\parallel,\perp}(t,0;\bp)|_{t=0} = 0$ and
$\partial_{t}  \partial_{t'} F_{\parallel,\perp}(t,t';\bp)|_{t=t'=0} \equiv
F^{-1}_{\parallel,\perp}(0,0;\bp)/4 = \omega_{\parallel,\perp}(\bp)/2$ in accordance with the results of section (\ref{sec:initialconditions}). The frequency of the longitudinal modes is $\omega_\parallel(\bp) = \sqrt{\bp^2 + 3 \sigma_0^2}$, and $\omega_\perp(\bp) = \sqrt{\bp^2 + \sigma_0^2}$ for the transverse modes.

Of course, in these linear equations the two-point functions can be factorized
as products of momentum-dependent single-time functions $f_{\parallel,\perp}(t;\bp)$ with
\begin{equation} \db 
 F_{\parallel,\perp}(t,t';\bp) \, = \,  \frac{1}{2}  \left[f_{\parallel,\perp}(t;\bp)f_{\parallel,\perp}^*(t';\bp)+f_{\parallel,\perp}^*(t;\bp)f_{\parallel,\perp}(t';\bp)\right] \, , 
\end{equation}
where $f_{\parallel,\perp}^*(t;\bp)$ denotes the complex conjugate of $f_{\parallel,\perp}(t;\bp)$. In terms of these so-called 
mode functions the equations of motion read, e.g., for the
transverse modes
\bea 
\db \left[ \partial_t^2 + \bp^2 + {\rr \sigma_0^2 \, \mathrm{cn}^2(\sigma_0 t)} \right]
f_\perp(t;\bp) \,=\, 0 \, .
\label{eq:lametype}
\eea
Up to an overall arbitrary phase,
the above initial conditions for the two-point functions translate into
\begin{equation}  
\db f_\perp(0,\bp)\, = \, 1/\sqrt{2\,\omega_\perp(\bp)} \quad, \quad  
\partial_t f_\perp(t,\bp)|_{t=0} \, = \, -i\, \sqrt{\frac{\omega_\perp(\bp)}{2}}
\end{equation} 
and equivalently for $f_\parallel$. For given momentum $\bp$, equation (\ref{eq:lametype}) is mathematically 
equivalent to the classical mechanics oscillator described
above. Before we summarize the analytical solution of the Lam{\'e}--type equation (\ref{eq:lametype}), we discuss some general properties. In order to be able to observe a significant resonance, the frequency $\omega_0$ of the field $\sigma(t)$ and the characteristic frequency of the fluctuations $f_{\parallel,\perp}(t;\bp)$ should be similar. Therefore, for transverse modes the dominant resonances should fulfill the approximate condition
\begin{equation}
\db \omega_\perp(t,\bp) \,\equiv \, \sqrt{\bp^2 + \sigma^2(t)} \, \simeq \, \sqrt{\bp^2 + \sigma^2_0/2} \,\, \stackrel{!}{\sim} \,\, \omega_0 \, ,
\end{equation}  
where we replaced the field square by its approximate time average.
Since $\omega_0 \sim \sigma_0$, this condition could be met for not too high momenta $\bp^2 \lesssim \sigma_0^2/2$. In contrast, for longitudinal modes the same analysis leads to the approximate condition $\sqrt{\bp^2 + 3 \sigma^2_0/2} \stackrel{!}{\sim} \omega_0$ which even for vanishing momenta is not fulfilled such that only subdominant resonances may be expected. The results of this qualitative analysis are indeed validated by the analytic solutions of the Lam{\'e}--type equation which we turn to next.      

In the following, we concentrate on the dominant transverse modes to analyze the instability dynamics at early times. We choose $U_p(t)$ and $U_p(-t)$ as an independent set of solutions of (\ref{eq:lametype}) such that $f_\perp(t;\bp)$ is a linear combination of those. Here we will be interested in the unstable modes which show exponential amplification and state the corresponding growth rates. Similar to the above classical oscillator example, we consider a Floquet analysis of the periodic solution
\begin{equation}
\db U_p(t + \Delta T) \, = \, e^{\alpha_p \, \Delta T}\, U_p(t) \, ,
\end{equation}
where the time-independent index $ \alpha_p$ is a function of spatial momentum. Correspondingly, we can write
\begin{equation}
\db U_p(t) \, = \, e^{\alpha_p \, t}\, \Pi_p(t) 
\end{equation}
with the periodic function $\Pi_p(t+\Delta T) = \Pi_p(t)$ of period $\Delta T = 2 K_{1/2} \sigma_0$. Again, this can be directly verified with $U_p(t+\Delta T) = \exp\{ \alpha_p (t+\Delta T)\}\Pi(t+\Delta T) = \exp\{ \alpha_p \Delta T\} U_p(t)$. The solutions of Lam{\'e} equations can be found in textbooks on differential equations and we only give here the relevant properties of the Floquet index $\alpha_p$ for (\ref{eq:lametype}). In accordance with our qualitative discussion above, resonant amplification of modes can be found for not too large momenta. More precisely, for $\bp^2 \ge \sigma_0^2/2$ purely oscillating solutions are obtained. In contrast, for $0 < \bp^2 < \sigma_0^2/2$ the Floquet index  
\begin{equation}
\db \alpha_p \, = \, i\, \omega_0 - \gamma_p
\end{equation}
has a real part $\gamma_p$ and an imaginary part $\omega_0$, given by the characteristic frequency (\ref{eq:chfreq}). The real part implies exponential solutions with a growth or decay rate for either $U_p(t)$ or $U_p(-t)$ given by 
\begin{equation}
\db \gamma_p \, = \, \sigma_0 \, Z\left(\mathrm{cn}^{-1}\left(\sqrt{\frac{2 \bp^2}{\sigma_0^2}}\right) \right) \,.
\label{eq:rate1}
\end{equation}
Here, the Jacobi zeta function $Z(x)$ has the series expansion
\begin{eqnarray}
\db Z(x) &\db = & \db \frac{2\pi}{K_{1/2}} \sum\limits_{n=1}^{\infty} \frac{e^{- n \pi}}{1-e^{- 2 n \pi}}\, \sin\left( \frac{n \pi\, x}{K_{1/2}} \right)
\nonumber\\
&\db \simeq & \db \frac{2\pi}{K_{1/2}}\, e^{-\pi} \, \sin\left( \frac{\pi\, x}{K_{1/2}} \right) \, ,
\end{eqnarray}
where the latter approximation gives an excellent description for our purposes.
Taking into account that $\mathrm{cn}^{-1}(0) = K_{1/2}$ and $\mathrm{cn}^{-1}(1) = 0$, we note that the rate $\gamma_p$ vanishes at the boundaries $\bp^2 = 0$ and $\bp^2 = \sigma_0^2/2$ of the instability band.

In order to get a further analytic understanding we may use the somewhat crude approximation
\begin{equation}
\db \mathrm{cn}^{-1}(x) \, \simeq \, \frac{2 K_{1/2}}{\pi}\, \mathrm{arccos}(x) \, ,
\end{equation}
which in our case will be good at the few percent level, and the identity
\begin{equation}
\db \sin \left( 2\, \mathrm{arccos} \left( \sqrt{x}\, \right) \right) \, = \, 2 \sqrt{2 x (1-x)} \, .
\end{equation}
Then the rate (\ref{eq:rate1}) can be expressed as 
\begin{equation}
\db \gamma_p \, \simeq \, \frac{4\pi\sigma_0}{K_{1/2}}\, e^{-\pi}\, \sqrt{\frac{2\bp^2}{\sigma_0^2}\left( 1 - \frac{2\bp^2}{\sigma_0^2} \right)}
\end{equation}
and the maximally amplified growth rate $\gamma_0 \equiv \gamma(\bp^2= \bp_0^2)$ occurs for $\bp_0^2 = \sigma_0^2/4$ with
\begin{equation}
\db \gamma_0 \, \simeq \, \frac{2\pi\sigma_0}{K_{1/2}}\, e^{-\pi}\, \simeq \, 0.146\, \sigma_0 \, . 
\end{equation}
Therefore, there is a separation of scales between the characteristic frequency of oscillations, $\omega_0$, and the characteristic growth rate, $\gamma_0$:
\begin{equation}
\db \gamma_0 \, \ll \, \omega_0 \,.
\end{equation}
When analyzing the growth in the nonlinear regime below, we will often exploit this separation of scales by considering suitable time averages to smooth out the rapid oscillations.

\subsection{Nonlinear regime: secondary instabilities}

\subsubsection{Dynamical power counting}

Above we obtained the set of evolution equations (\ref{eq:sigma}) and (\ref{eq:Fppcl}) for the field $\phi(t) = \sigma(t) \, \sqrt{6N/\lambda}$ and the fluctuations $F_{\parallel,\perp}$ from linearized classical dynamics. According to section \ref{sec:exactevoleq}, the very same set of equations can also be obtained starting from the quantum evolution of the rescaled field expectation value and the statistical two-point functions by neglecting all loop corrections: Setting $\Sigma^F_{\parallel,\perp}$, $\Sigma^\rho_{\parallel,\perp}$ and $\Sigma^{(0)}_{\parallel,\perp}$ to zero in the field equation (\ref{eq:exactphi}) and the equations for the two-point functions (\ref{eq:exactrhoF}) leads precisely to (\ref{eq:sigma}) and (\ref{eq:Fppcl}). We note that the evolution equation for the spectral function $\rho_{\parallel,\perp}$ decouples from the rest in this linear approximation. It is important to point out that this linearized classical approximation is only valid at early times for $\lambda \ll 1$. For strong coupling loop corrections cannot be neglected in this case.

In turn, we may use the quantum evolution equations (\ref{eq:exactphi}) and (\ref{eq:exactrhoF}) to determine the range of validity of the above classical approach. Parametrically the initial statistical propagators are of order one, i.e.\
\begin{equation}
\db F_{\parallel,\perp}(0,0;\bp) \, \sim \, {\mathcal O}(N^0 \lambda^0) \, ,
\end{equation} 
and the initial field is
\begin{equation}
\db \phi(0) \, \sim {\mathcal O}\left(N^{1/2} \lambda^{-1/2}\right) \, .
\end{equation}
As a consequence, all loop corrections are initially suppressed by powers of the coupling constant $\lambda \ll 1$ and the classical approximation represents an accurate description of the early quantum dynamics. However,
parametric resonance leads to an exponential amplification of modes, where the dominant growth behavior is parametrically given by
\begin{equation}
\db F_\perp(t,t';\bp) \, \sim \, e^{\gamma_p(t+t')} \, .
\end{equation}
Accordingly, the maximally amplified mode $F_{\perp}(t,t;\bp_0)$ grows with rate $2 \gamma_0$. As a consequence of the exponential amplification
of the statistical propagator, there is a characteristic time when $F_{\perp}(t,t;\bp_0)$ turns out to be no longer parametrically of order one. This is the time where the linearized classical approximation breaks down. To make further analytical progress, the nonlinear regime is most efficiently described using the evolution equations of section \ref{sec:exactevoleq}. This allows us to study what happens when loop corrections start to become relevant. Indeed, we will find that there is a well separated hierarchy of times, where different loop integrals start to become of order one at corresponding separate times. At some point an infinite number of diagrams with an arbitrary number of loops becomes of order one and no power counting based on a small coupling can be performed. We will address this later stage in a subsequent section on nonthermal fixed points using nonperturbative large-$N$ techniques.

A general loop contribution to the evolution equations contains powers of $\lambda$, the field $\phi$, the propagators $F_{\parallel,\perp}$ and the spectral functions $\rho_{\parallel,\perp}$. Here it is important to note that the ``weight'' of the spectral functions in loop integrals remains parametrically of order one at all times as encoded in the equal-time commutation relations (\ref{eq:bosecomrel}).
It is also important to take into account the fact that transverse fluctuations ($F_{\perp}$) exhibit the dominant growth in the linear regime. Consequently, contributions containing more transverse propagators ($F_{\perp}$) can become important earlier than those diagrams containing longitudinal propagators ($F_{\parallel}$) instead. For instance, an expression containing powers $\lambda^n F_{\perp}^m \phi^{2l}$ with integers $n,m$ and $l$ may be expected to give sizable corrections to the linearized evolution equations once $F_{\perp} \sim 1/\lambda^{(n-l)/m}$ for typical momenta. Here $n$ yields the suppression factor from the coupling constant, whereas $m$ introduces the enhancement due to large fluctuations for typical momenta and $l$ due to a large macroscopic field. The power counting can become more involved as time proceeds, and it is remarkable that one can indeed identify a sequence of characteristic time scales with corresponding growth rates.

\begin{figure}[t]
\centerline{
\epsfig{file=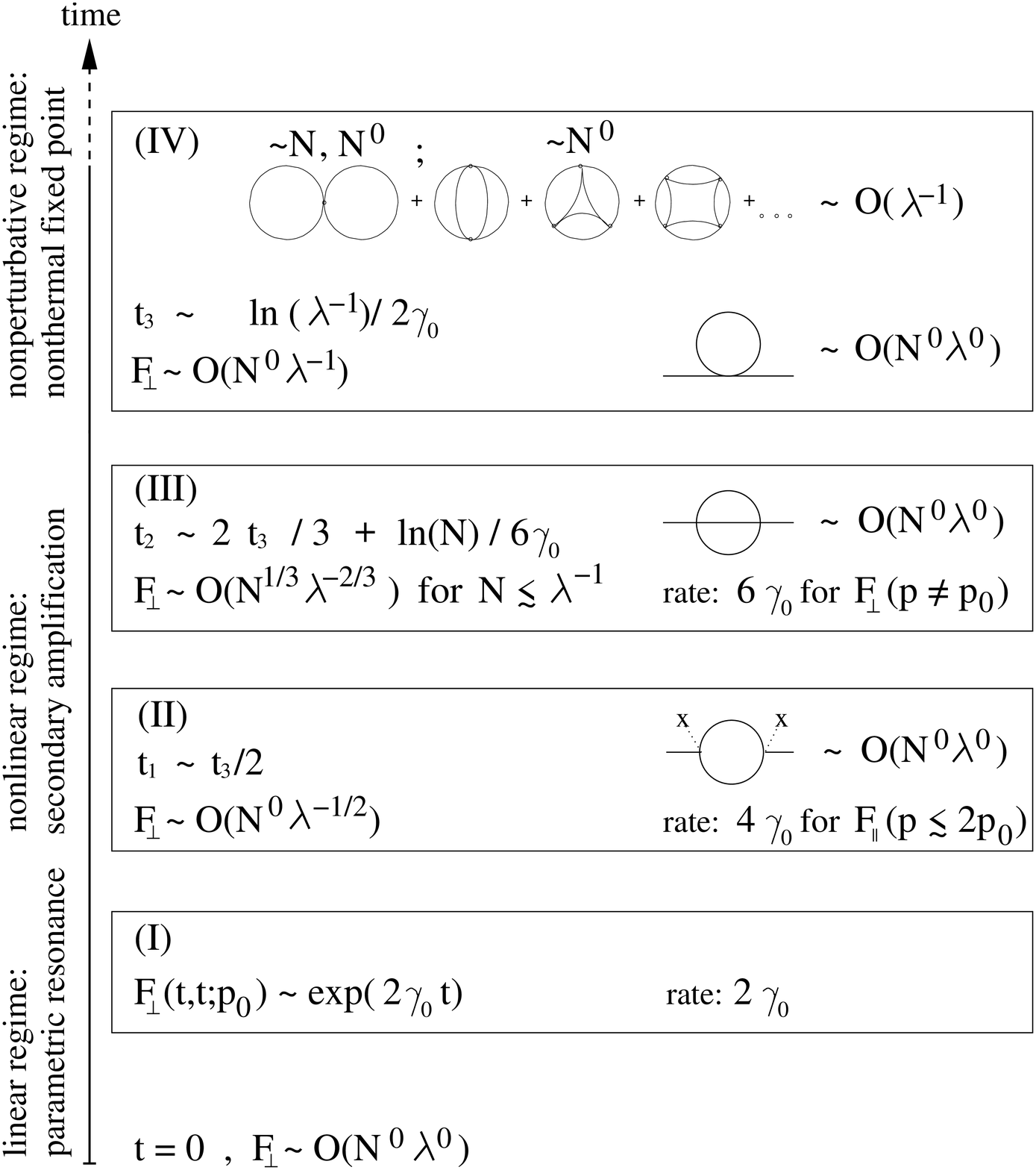,width=11.cm}}
\caption{Schematic overview of the characteristic time scales and the
respective relevant diagrammatic contributions.}
\label{fig:timescheme}
\end{figure}
Before we consider the calculations in more detail, the situation 
is schematically summarized in Fig.~\ref{fig:timescheme}.
The time when $F_{\perp}(t,t;\bp_0) 
\sim \Or (N^0 \lambda^{-1/2})$ is denoted by $t = t_1$.
At this time the one-loop diagram with two field insertions
indicated by crosses as depicted in Fig.~\ref{fig:timescheme}
will give a contribution of order one to the evolution equation for 
$F_{\parallel}(t,t;\bp)$. For instance, the two powers of the coupling coming 
from the vertices of that diagram are canceled by the field amplitudes and by propagator lines associated
to the amplified $F_{\perp}(t,t';\bp_0)$. Similarly, at the time
$t = t_2$ the maximally amplified transverse propagator 
mode has grown to $F_{\perp}(t,t';\bp_0) 
\sim \mathcal{O} (N^{1/3}\lambda^{-2/3})$. As a consequence, the 
``setting sun'' diagram in Fig.~\ref{fig:timescheme} becomes
of order one and is therefore of the same order as the classical
contributions. Though the loop corrections become of order one
later than the initial time, they induce amplification rates
that are multiples of the rate $2\gamma_0$ which lead to
a very rapid growth of modes in a wide momentum range. 
Finally, when the fluctuations have grown 
nonperturbatively large with $F_{\perp}(t,t';\bp_0) 
\sim \mathcal{O} (N^0\, \lambda^{-1})$
any loop correction will no longer be suppressed by
powers of the small coupling $\lambda$. In this case
the nonperturbative $1/N$ expansion of the 2PI effective action, which will be discussed later, becomes of crucial
importance for a quantitative description of the
dynamics.

\subsubsection{Nonlinear amplification}

The $\mathcal{O} (\lambda^0)$ approximation for longitudinal modes breaks down at 
the time 
\beq
\qquad t\simeq t' ={\rr t_1}\,\,:
\hspace*{0.3cm} {\rr F_{\perp}(t,t';\bp_0) 
\,\sim\, \mathcal{O} (N^0 \lambda^{-1/2})}  \, .
\eeq
This can be derived from the ${\mathcal O} (\lambda)$ evolution equations
to which one-loop self-energies contribute, which diagrammatically
are given by
\epsfig{file=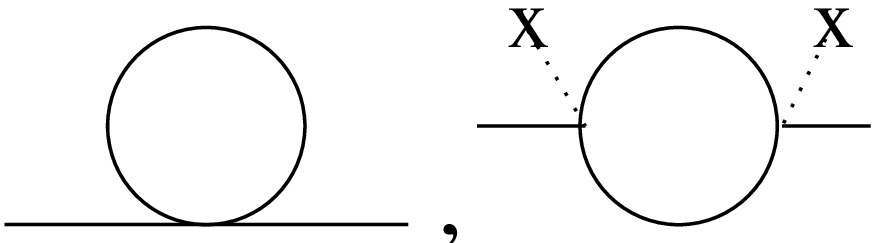,width=2.cm}.
The approximate evolution equation reads
\bea\db
\lefteqn{
\Big(\partial_t^2 + {\bp}^2 +  M^2(t) + 3 \sigma^2(t) \Big) 
F_\parallel(t,t';{\bp}) \simeq   } \nonumber \\
&&\db  { \frac{2 \lambda (N-1)}{3N} \sigma(t) \Bigg\{
\int_0^t {\rm d} t'' \sigma(t'')
\Pi^\rho_\perp(t,t'';{\bp}) F_\parallel(t'',t';{\bp})} \nonumber\\ 
&&\db { -\frac{1}{2}\int_0^{t'} {\rm d} t'' \sigma(t'')
\Pi^F_\perp(t,t'';{\bp})\rho_\parallel(t'',t';{\bp})\Bigg\}} \,\equiv\, RHS  \, .
\label{eq:orderlambda}
\eea
where
\bea\db 
 M^2(t) &\db =&\db  m^2 + \frac{\lambda}{6N} 
 \Big[3 T_\parallel(t) + (N-1) T_\perp(t)\Big] \, , \\
\label{eq:massscale}
\db  T_{\parallel,\perp}(t) &\db =&\db 
 \int^{\Lambda} \frac{{\rm d}^3 p}{(2\pi)^3} F_{\parallel,\perp}(t,t;\bp) \, .
\label{eq:tadpoles}
\eea
The ``tadpole'' contributions  
$T_{\parallel}$ and $T_{\perp}$ from the longitudinal and transverse
propagator components, respectively, are regularized by some \mbox{$\Lambda \gg p_0$} whose precise value is irrelevant as long as the integral is dominated by momenta much smaller than the cutoff. Here we have abbreviated $\Pi^X_\perp(t,t'';{\bp}) 
= \int {\rm d}^3 q/(2 \pi)^3\, F_\perp(t,t'';{\bp}-\bq)\, X_\perp(t,t'';\bq)$
with \mbox{$X = \{F, \rho\}$}, and we used that
$F_{\perp}^2 \gg \rho_{\perp}^2$.\footnote{This corresponds to the classicality condition of section~\ref{sec:classicality}.} One observes that indeed for $F_{\perp} 
\sim \mathcal{O} (N^0 \lambda^{-1/2})$ the 
RHS of (\ref{eq:orderlambda}) becomes $\sim \mathcal{O} (1)$ 
and cannot be neglected.

\begin{figure}[t]
\centerline{
\epsfig{file=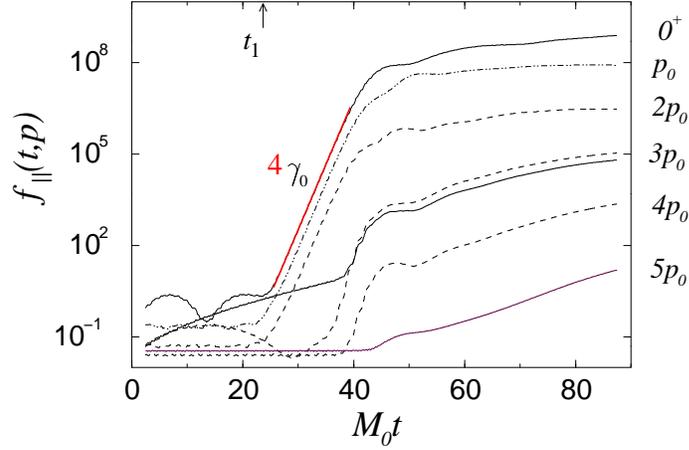,width=9.cm}}  
\caption{Effective particle number for the 
 longitudinal modes as a function of time for various momenta. For $t \gtrsim t_1$ the nonlinear corrections trigger an exponential growth with 
  rate $4\gamma_0$ for $p \lesssim 2p_0$. 
  The thick line corresponds to a mode in the parametric resonance band, 
  and the long-dashed line for a similar one outside the band. The resonant 
  amplification is quickly dominated by source-induced amplification.}
  \label{fig:number_lg}
\end{figure}

In order to make analytical progress, one has to evaluate the ``memory 
integrals'' in the above equation. This is dramatically simplified
by the fact that the integral is at this stage approximately local in time,
since the exponential growth lets the latest-time contributions 
dominate the integral. For the approximate evaluation of the memory integrals we
consider 
\beq
\db \int_0^t {\rm d} t'' \;\;\longrightarrow\;\;  
\int_{t - c/\omega_0}^t {\rm d} t'' 
\eeq
with $c \sim \mathcal{O}(1)$. As long as we are only interested in exponential growth rates and characteristic time scales, the value for the constant $c$ will turn out to be irrelevant to ``leading-log'' accuracy. We then perform a Taylor expansion around the latest time $t$ $(t')$,
\bea\db
\rho_{\parallel,\perp}(t,t'';\bp) &\db \simeq& \db
\partial_{t''}\rho_{\parallel,\perp}(t,t'';\bp)|_{t=t''}
(t'' - t) \equiv (t - t'') \, , \nn
\db F_{\parallel,\perp}(t,t'';\bp) &\db \simeq& \db F_{\parallel,\perp}(t,t;\bp) \, ,
\eea
where we have used the equal-time commutation relations (\ref{eq:bosecomrel}).
With these approximations 
the RHS of (\ref{eq:orderlambda}) can be evaluated as:
\bea \db
RHS
&\db\simeq&\db
{\db \lambda}\, \sigma^2(t)\,  
\frac{c^2}{\omega_0^2} \frac{(N-1)}{3N}\, {\db T_{\perp}(t)} \, 
{\db F_{\parallel}(t,t';{\bp})}\,
\!\!\!\!\!\qquad {\db\rm (mass\; term)}
\label{eq:massesta}
\\
&\rr +&\rr {\rr \lambda}\, 
\sigma(t)\sigma(t')\, \frac{c^2}{\omega_0^2}\,\frac{(N-1)}{6N}
{\rr \Pi^F_\perp(t,t';{\bp})} \!\!\!\!\!\qquad\;\; {\rr\rm (source\; term)} 
\label{eq:sourceasta}
\eea
The first term is a contribution to the effective mass, whereas the second 
term represents a source. Note that both the ``tadpole'' mass term 
and the above correction to this mass are of the same order in 
$\lambda$, however, with opposite sign.
To evaluate the momentum integrals,
we use a saddle point approximation around the dominant
$p\simeq p_0$, valid for $t,t'\gg \gamma_0^{-1}$, 
with $F_\perp(t,t',\bp) \simeq F_\perp(t,t',\bp_0) 
\exp [-|\gamma_0^{\prime\prime}|(t+t')(p-p_0)^2/2]$.\footnote{Here
$\gamma(p) \simeq \gamma_0 + \gamma_0^{\prime\prime} (p-p_0)^2/2$
with $\gamma_0^{\prime\prime} < 0$.}
From this one obtains for the above mass term:
\beq\db
 \lambda\, T_{\perp}(t) \,\simeq\, 
\lambda\, \frac{p_0^2\,F_{\perp} (t,t;\bp_0)}{2 
(\pi^3 |\gamma_0^{\prime\prime}| t)^{1/2}} \, .
\label{eq:mest}
\eeq
The result can be used to obtain an estimate at what time $t$
this loop correction becomes an important contribution to the
evolution equation. Note that 
to this order in $\lambda$ it is correct to use 
$F_{\perp} (t,t';\bp_0) \sim e^{\gamma_0 (t+t')}$ on the
RHS of (\ref{eq:mest}). The condition 
\mbox{$\lambda\, T_{\perp}(t=t_3)$} $\sim \mathcal{O}(1)$
can then be written for $\lambda \ll 1$ as:   
\beq
\mbox{\framebox{\db$\displaystyle t_3 \simeq 
\frac{1}{2\gamma_0}\, \ln \lambda^{-1}$}}
\label{eq:tnonpert}
\eeq
The same saddle point approximation can be performed to
evaluate the source term (\ref{eq:sourceasta}):
\beq \rr
\lambda\, \Pi^F_\perp(t,t';0) \,\simeq \, \lambda\,
\frac{p_0^2\,F_{\perp}^2(t,t';\bp_0) }{4 (\pi^3 |\gamma_0^{\prime\prime}| 
(t+t'))^{1/2}} \, .
\eeq
Here we only wrote the source term for $\bp = 0$ where it has its maximum, 
although it affects all modes with $p \lesssim 2p_0$. Again this can
be used to estimate the time $t = t_1$ at which 
$\lambda\, \Pi^F_{\perp} \sim \mathcal{O}(1)$:
\beq
\mbox{\framebox{\rr $\displaystyle t_1 \simeq \frac{1}{2}\,
t_3$}}
\eeq
One arrives at the important conclusion that the source term 
(\ref{eq:sourceasta}) becomes {\rr\em earlier} of order one than
the mass term (\ref{eq:massesta}): For $t_1 \lesssim t 
\lesssim t_3$ the source term dominates the dynamics! 
Using these estimates in
(\ref{eq:orderlambda}) one finds that the longitudinal modes
with $0 \lesssim p \lesssim 2p_0$ get amplified with twice the 
rate $2\gamma_0$:
\beq\rr 
F_{\parallel}(t,t;\bp) \, \sim\,  \lambda\, F_{\perp}^2(t,t;\bp_0) \sim 
\lambda\, e^{4\gamma_0 t} \, . 
\label{eq:Fpaamp}
\eeq
Though the non-linear contributions start later, they grow
twice as fast. The analytical estimates for $t_1$ 
and rates agree well with the numerical solution\footnote{The plots are obtained for the $N$-component field theory at NLO in the $1/N$ expansion of the 2PI effective action with nonzero mass parameter $m$. The latter is not relevant for our purposes.} of the evolution equations without memory expansion as 
shown in Fig.~\ref{fig:number_lg}, where we plot the effective particle number distribution
\begin{equation}
\db f_{\parallel,\perp}(t,\bp) \, = \, \sqrt{F_{\parallel,\perp}(t,t;\bp) \partial_t\partial_{t'} F_{\parallel,\perp}(t,t';\bp)}|_{t=t'}-\frac{1}{2} \, 
\end{equation}
and $M_0^2 \equiv M^2(t=0)$.
\begin{figure}[t]
\centerline{
\epsfig{file=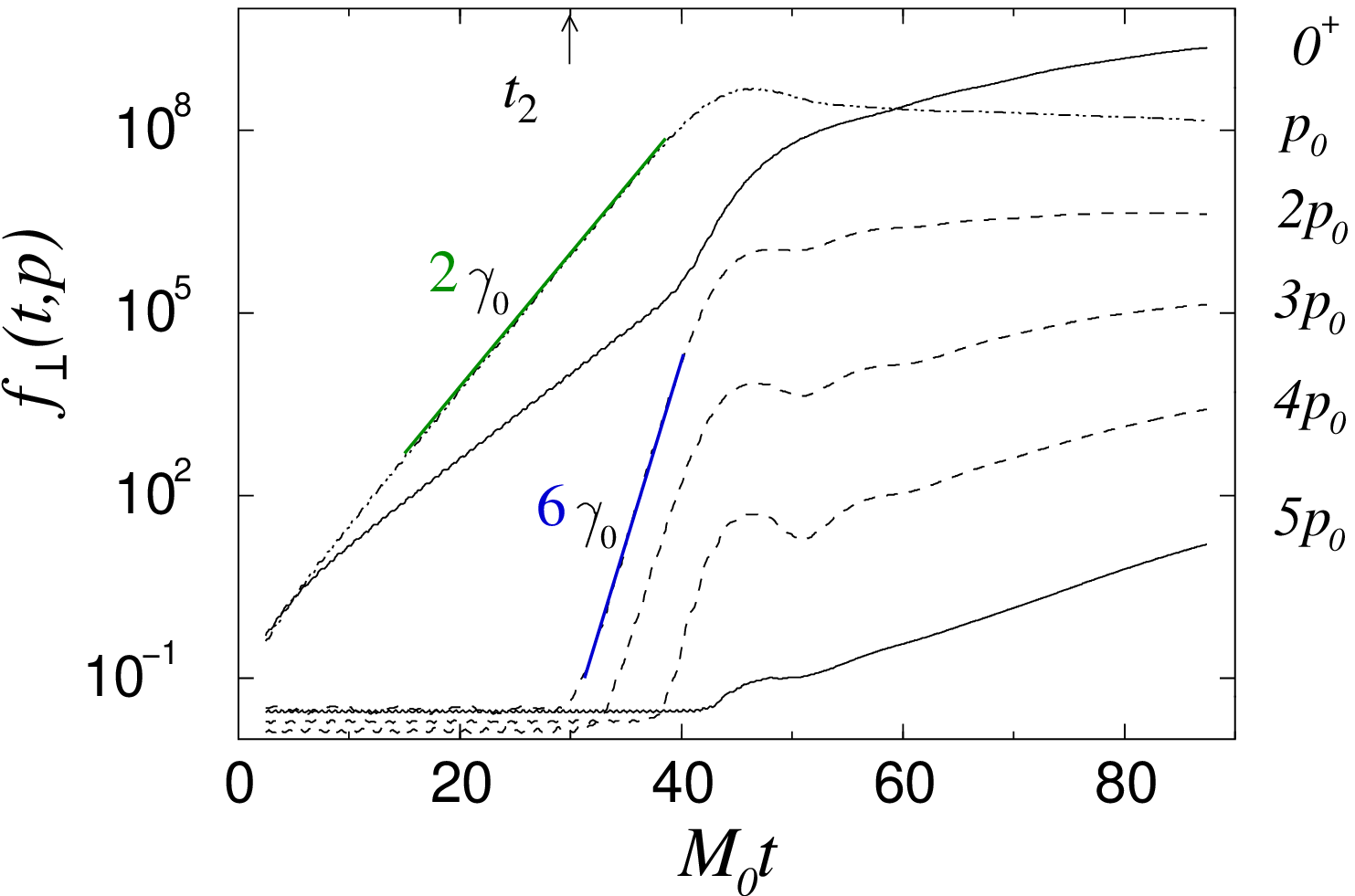,width=9.cm}} 
\caption{Effective particle number for the 
 transverse modes as a function of time for various momenta 
 $p \le 5 p_0$. At early times, modes with $p \simeq p_0$ are  
 exponentially amplified with a rate $2 \gamma_0$. Due to 
 nonlinearities, one observes subsequently an enhanced growth
 with rate~$6\gamma_0$ for a broad momentum range.}
 \label{fig:number_tr}
\end{figure}

A similar analysis can be made for the transverse modes. 
Beyond the Lam{\'e}--type ${\mathcal O}(\lambda^0)$ description, the 
evolution equation for $F_{\perp}$ receives contributions from the 
feed-back of the longitudinal modes at ${\mathcal O} (\lambda)$ 
as well as from 
the amplified transverse modes at ${\mathcal O} (\lambda^2)$. 
They represent {\rr\em source terms} in the evolution equation
for $F_{\perp}(t,t';\bp)$ which are both 
parametrically of the form $\sim \lambda^2 F_\perp^3/N$ as is depicted
below: 

\vspace*{0.3cm}

\parbox{8.6cm}{
\hspace*{0.3cm}\parbox{3cm}{
\centerline{
\epsfig{file=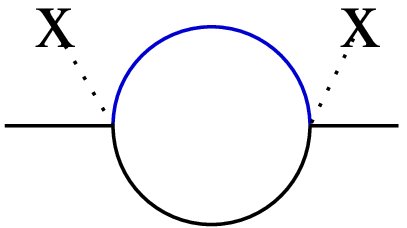,width=2.cm}
}}
{\rr $\displaystyle \sim \frac{\lambda}{N} {\db F_{\parallel}} F_{\perp}
\,\,\stackrel{\mbox{\footnotesize \db cf.~(\ref{eq:Fpaamp})}}{\sim}\,\,
 \frac{\lambda^2}{N} F_{\perp}^3$}

\hspace*{0.3cm}\parbox{3cm}{
\centerline{
\epsfig{file=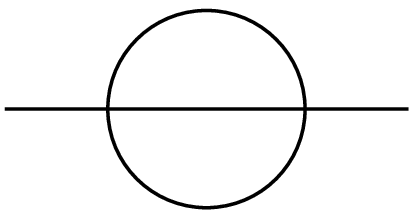,width=2.cm}
}}
{\rr $\displaystyle \sim \frac{\lambda^2}{N} F_{\perp}^3$}
}
\parbox{3.5cm}{$\Bigg\}$ \, \, 
$\displaystyle\rr \sim \frac{\lambda^2}{N}
\, e^{6 \gamma_0 t}$}

\vspace*{0.3cm}

\noindent
Following along the lines of the above paragraph and using
(\ref{eq:tnonpert}) this leads to 
the characteristic time $t =  t_2$ at which
these source terms become of order one:
\beq
\mbox{\framebox{\rr$\displaystyle t_2 \simeq \frac{2}{3}\, 
t_3 + \frac{\ln N}{6 \gamma_0}$}} 
\label{eq:tcollect}
\eeq
For $t\simeq t' \simeq  t_2$ the dominant 
transverse mode has grown to 
\beq\rr
F_{\perp}(t,t';\bp_0) 
\,\sim\, \mathcal{O} (N^{1/3}\lambda^{-2/3})  \, .
\eeq
Correspondingly, for $t_2 \lesssim t \lesssim t_3$
one finds a large growth rate $\sim 6 \gamma_0$ 
in a momentum range $0 \lesssim p \lesssim 3 p_0$, in 
agreement with the numerical results
shown in Fig.\ \ref{fig:number_tr}. In this time range
the longitudinal modes exhibit an enhanced amplification as well 
(cf.~Fig.\ \ref{fig:number_lg}). It is important to realize that the phenomenon
of source-induced amplification repeats itself: the newly amplified modes, 
together with the primarily amplified ones, act as a source for other 
modes, and so on. In this way, even higher growth rates of multiples of 
$\gamma_0$ can be observed
and the amplification rapidly propagates towards
higher momentum modes.
We note that this regime is present if $t_2\le t_3$, i.e.\ as long as $N \lesssim \lambda^{-1}$.

Around $t  \lesssim t_3$ is the earliest 
time when sizeable corrections to the maximally amplified mode $F_\perp(t,t;\bp_0)$ and to the field appear. Around $t_3$ there are corrections of order one which come from diagrams with an arbitrary number of loops. As a consequence, the dynamics is no longer characterized in terms of the small expansion parameter $\lambda$. This is the nonperturbative regime, where the $1/N$ expansion may be used to describe also the subsequent evolution governed by a nonthermal fixed point as is explained in section~\ref{sec:transport}. 
 
\begin{figure}[t]
\centerline{\includegraphics[scale=0.4,angle=-90]{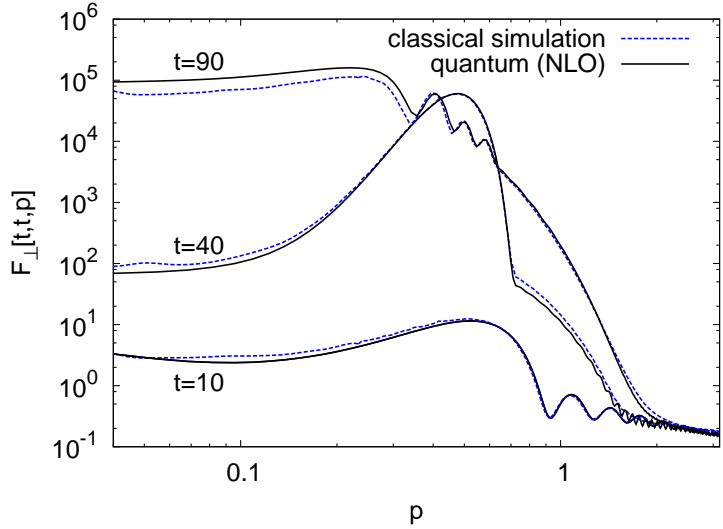}}
\caption{\label{fig:comp} Two-point function $F_{\perp}(t,t;{\bf p})$ as a function of momentum $|{\bf p}|$ for three different times. The quantum evolution (solid) from the 2PI $1/N$ expansion to NLO and the full classical-statistical simulation using Monte Carlo sampling (dashed) agree remarkably well.}
\end{figure}

The accuracy of the above description based on the 2PI effective action can be tested using numerical simulations on a space-time lattice. This exploits the fact that classical-statistical field theory descriptions of the dynamics have an overlapping range of validity with the underlying quantum field theory, as described in section~\ref{sec:clstat}. As an example,
Fig.~\ref{fig:comp} shows $F_{\perp}(t,t,{\bf p})$ for times $t=10$, $t=40$ in the nonlinear regime, and $t=90$ in the nonperturbative regime. The solid lines give the results for the quantum evolution for $\lambda = 0.01$ and $N=4$ by solving the NLO equations from the 2PI effective action of section~\ref{2PINfield} in $d=3$ numerically. For comparison, the dashed lines give the same quantity as obtained from simulations of the corresponding classical-statistical field theory on a lattice with same initial conditions. The level of agreement between the different results is remarkable. Quantum fluctuations are expected to be suppressed if the classicality condition $F^2 \gg \rho^2$ of section~\ref{sec:classicality} is fulfilled, which is very well realized for typical $F \sim 1/\lambda$ with $\rho$ being order unity. In turn, contributions beyond NLO in the $1/N$ expansion seem to play an inferior role even for the nonperturbative regime in this case.

\subsection{Bibliography}

\begin{itemize} 
\item The presentation for the scalar field theory is based on J.~Berges and J.~Serreau, {\it Parametric resonance in quantum field theory,}
Phys.\ Rev.\ Lett.~{\bf 91} (2003) 111601, where the 2PI $1/N$ expansion to next-to-leading order has been used to describe the relevant nonlinear phenomena. Fig.~\ref{fig:number_lg} and \ref{fig:number_tr} are taken from that reference.
Parts of the linearized analysis are based on  
D.~Boyanovsky, H.~J.~de Vega, R.~Holman and J.~F.~Salgado, {\it Analytic and numerical study of preheating dynamics,}
Phys.\ Rev.\ D {\bf 54} (1996) 7570, where a leading-order large-$N$ approximation 
is employed. Fig.~\ref{fig:comp} is taken from J.~Berges, A.~Rothkopf and J.~Schmidt,
  {\it Non-thermal fixed points: Effective weak-coupling for strongly correlated systems far from equilibrium,}  Phys.\ Rev.\ Lett.\  {\bf 101} (2008) 041603.
\item The preheating phenomenon was first developed in L.~Kofman, A.~D.~Linde and A.~A.~Starobinsky, {\it Reheating after inflation}, Phys.\ Rev.\ Lett.\  {\bf 73} (1994) 3195. Its description in terms of classical-statistical simulations has been worked out in S.~Yu.~Khlebnikov and I.~I.~Tkachev, {\it Classical Decay of the Inflaton}, Phys.\ Rev.\ Lett.\ {\bf 77} (1996) 219.	
\item For a review about the inclusion of fermions, see J.~Berges, D.~Gelfand and D.~Sexty,
  {\it Amplified Fermion Production from Overpopulated Bose Fields},
  Phys.\ Rev.\ D {\bf 89} (2014) 025001.
\end{itemize}

\pagebreak

%% file: ch_transport_fixedpointsLH.tex
\section{Nonthermal fixed points and turbulence}
\label{sec:transport}

From the example of parametric resonance in section~\ref{sec:instab} we have seen that once the exponential growth of fluctuations stops, the systems slows down considerably. At this stage there is an infinite number of loop corrections of order unity. The coupling ``drops out'' of the problem and no power counting in terms of a small coupling parameter can be given. In this section we want to study the physics of this slow evolution for the examples of relativistic $N$-component scalar field theory as well as nonrelativistic (Gross-Pitaevskii) field theory. They serve here as a paradigm to investigate universal behavior of isolated many-body systems far from equilibrium, which is relevant for a wide range of applications from high-energy particle physics to ultracold quantum gases as pointed out in section~\ref{sec:intro}. The universality is based on the existence of {\it\rr nonthermal fixed points}, which represent nonequilibrium attractor solutions with self-similar scaling behavior. The corresponding dynamic universality classes turn out to be remarkably large, encompassing both relativistic as well as nonrelativistic quantum and classical systems.  

We start with a standard introduction to the phenomenon of {\it\rr wave turbulence} in section~\ref{eq:waveturbulence}, where perturbative kinetic theory can describe the stationary transport of energy from a cascade towards higher momenta. Perturbative kinetic theory has also been employed in the literature to describe infrared phenomena such as Bose condensation, however, such an approach neglects important vertex corrections since the large occupancies at low momenta lead to strongly nonlinear dynamics. Here we apply the vertex-resummed kinetic theory based on the 2PI $1/N$ expansion to NLO, which is described in section~\ref{sec:vertexres}. This allows us to gain analytic understanding of the formation of a {\it\rr dual cascade} and the phenomenon of {\it\rr far-from-equilibrium Bose condensation}. 

However, isolated systems out of equilibrium have no external driving forces that could realize stationary transport solutions. Instead, the transport in isolated systems is described in terms of the more general notion of a {\it\rr self-similar} evolution, where the physics is described in terms of universal scaling exponents and scaling functions as outlined in section~\ref{sec:intro}. We can use the vertex-resummed kinetic theory for an analytic description of the phenomena, which we present in 
section~\ref{sec:fixedpointsselfsimilarity}. Because of the typical high occupancies, this can be compared to results from classical-statistical lattice simulations using the methods of section~\ref{sec:clstat}.

\subsection{Stationary transport of conserved charges}
\label{eq:waveturbulence}

While many aspects of stationary transport of conserved charges associated to turbulence have long reached textbook level, there is still rather little known about turbulent behavior in nonperturbative regimes of quantum field theories. Here the strong interest is also driven by related questions concerning the dynamics of relativistic heavy-ion collisions. In this section we consider stationary solutions of transport equations with a net flux of energy and particles across momentum scales, thus violating detailed balance. We start with standard perturbative results relevant for high momenta in section~\ref{eq:weakwaveturbulence} and consider the nonperturbative infrared regime in section~\ref{sec:nonpert}.

\subsubsection{Perturbative regime: Weak wave turbulence}
\label{eq:weakwaveturbulence}

In perturbative kinetic theory, when two particles scatter into two particles, the time evolution of the distribution function $f_\bp(t)$ for a spatially homogeneous system is given by
\begin{equation}
\db \frac{\partial f_\bp(t)}{\partial t} = C^{2\leftrightarrow 2}\left[f\right](t,\bp) \, .
\label{eq:kin22}
\end{equation}
The collision integral is of the form
\begin{equation} \db
C^{2\leftrightarrow 2}[f](\bp) \, =  \, \int {\mathrm{d}}\Omega^{2\leftrightarrow 2}(\bp,\bl,\bq,\br)
\left[(f_\bp+1) (f_\bl+1) f_\bq f_\br - f_\bp f_\bl (f_\bq+1) (f_\br+1)\right] \, ,
\label{eq:2to2}
\end{equation}
where we suppress the global time dependence to ease the notation. The details of the model enter $\int {\mathrm{d}}\Omega^{2\leftrightarrow 2}(\bp,\bl,\bq,\br)$, which for the example of the relativistic $N$-component scalar field theory with quartic $\lambda/(4!N)$-interaction reads: 
\begin{eqnarray} \db
\int {\mathrm{d}}\Omega^{2\leftrightarrow 2}(\bp,\bl,\bq,\br) & \db = & \db \lambda^2\, \frac{N+2}{6 N^2} \int_{\bl \bq \br} \!\! (2\pi)^{d+1}\, \delta(\bp+\bl-\bq-\br) 
\nonumber\\
&\db \times& \db \delta(\omega_\bp + \omega_\bl - \omega_\bq - \omega_\br)\, \frac{1}{2\omega_\bp 2\omega_\bl 2\omega_\bq 2\omega_\br} \, 
\label{eq:measurepert}
\end{eqnarray}
with $\omega_\bp = \sqrt{\bp^2 + m^2}$ and $\int_\bp \equiv \int \rmd^dp/(2\pi)^d$. This kinetic equation can be obtained from quantum field theory using the two-loop self-energies (\ref{eqsigmaF3}) and (\ref{eqsigmarho3}) for the lowest-order gradient expansion along the lines of section~\ref{sec:vertexres}, and taking the on-shell spectral function (\ref{eq:freespectral}) of the free theory.
The expression (\ref{eq:2to2}) with (\ref{eq:measurepert}) for the collision integral represents a standard Boltzmann equation for a gas of relativistic particles. 

Clearly, this approximation cannot be used if the occupation numbers per mode become too large such that higher loop-corrections become sizeable. Parametrically, for a weak coupling $\lambda$ a necessary condition for its validity is $f_\bp \ll 1/\lambda$ since otherwise any loop-order contributes significantly as explained in detail in section~\ref{sec:instab}. On the other hand, the phenomenon
of weak wave turbulence is expected for not too small occupation numbers per mode as is explained in the following. For the corresponding regime $1 \ll f_\bp \ll 1/\lambda$ one may use the above Boltzmann equation, which approximately becomes 
\begin{equation} \db 
\frac{\partial f_\bp(t)}{\partial t} \, \simeq \, \int {\mathrm{d}}\Omega^{2\leftrightarrow 2}(\bp,\bl,\bq,\br) \left[ 
(f_\bp + f_\bl) f_\bq f_\br - f_\bp f_\bl (f_\bq + f_\br)\right] \, .
\label{eq:2to2cl}
\end{equation}
For number conserving $2 \leftrightarrow 2$ scatterings, apart from the energy density $\epsilon$ also the total particle number density $n$ is conserved, which are given by
\begin{equation} \db
\epsilon \, = \, \int_\bp \omega_\bp f_\bp \quad , \quad n \, = \, \int_\bp f_\bp \, .
\label{eq:epsntot}
\end{equation}
The fact that they are conserved 
may be described by a continuity equation in momentum space, such as 
\begin{equation} \db
\frac{\partial}{\partial t}\left( \omega_\bp f_\bp \right) + \nabla_\bp \cdot {\bf j}_\bp \, = \, 0
\label{eq:continuity}
\end{equation}
for energy conservation. Similarly, particle number conservation is described by formally replacing $\omega_\bp \to 1$ in the above equation and a corresponding 
substitution of the flux density. For the isotropic situation we can consider the energy flux $A(k)$ through a momentum sphere of radius $k$. Then only the radial component of the flux density ${\bf j}_\bp$ is nonvanishing and
\begin{equation} \db 
\int_\bp^k \nabla_\bp \cdot {\bf j}_\bp \, = \, \int_{\partial k} {\bf j}_\bp \cdot {\mathrm d}{\mathbf{A}}_\bp \, \equiv \, (2\pi)^d A(k)\, .
\end{equation}
Since in this approximation $\omega_\bp$ is constant in time, we can write with the help of (\ref{eq:continuity}) and the kinetic equation (\ref{eq:kin22}): 
\begin{equation} \db
A(k) \, = \, - \frac{1}{2^d \pi^{d/2} \Gamma(d/2+1)}\, \int^k \! {\mathrm d}p \, |\bp|^{d-1} \omega_\bp C^{2\leftrightarrow 2}(\bp) \, .
\label{eq:flux}
\end{equation}
Stationary wave turbulence is characterized by a {\it\rr scale-independent flux} $A(k)$, for which the respective integral does not depend on the integration limit $k$. To this end, we consider scaling solutions 
\begin{equation} \db
f_{\bp} \, = \, s^{\kappa} f_{s\bp} \quad , \quad 
\omega_{\bp} \, = \, s^{-1}\, \omega_{s\bp} \, ,
\label{eq:scalingprop}
\end{equation}
with occupation number exponent $\kappa$ and assuming a linear dispersion relation relevant for momenta $|\bp| \gg m$.
Since the physics is scale invariant, we can choose $s = 1/|\bp|$ such that $f_\bp = |\bp|^{-\kappa}\, f_1$ and $\omega_\bp = |\bp|\, \omega_1$.

Using these scaling properties, one obtains for the collision integral (\ref{eq:measurepert}) and (\ref{eq:2to2cl}) of the theory with quartic self-interaction:
\begin{equation} \db
C^{2\leftrightarrow 2}(\bp) \, = \, s^{-\mu_4}\, C^{2\leftrightarrow 2}(s\bp) \, , 
\end{equation}
where the scaling exponent is given by 
\begin{equation} \db
\mu_4 \, = \, (3d-4)-(d+1) - 3\kappa\, = \, 2d - 5 - 3\kappa\, .
\label{eq:Del4}
\end{equation}
The first term in brackets comes from the scaling of the measure, the second from energy-momentum conservation for two-to-two scattering and the third from the three factors of the distribution function appearing in (\ref{eq:2to2cl}). Apart from the 4-vertex interaction considered, it will be relevant to investigate also scattering in the presence of a non-vanishing field expectation value such that an effective 3-vertex appears according to section~\ref{sec:effectiveactions}. 

To keep the discussion more general, we may write for the scaling behavior of a generic collision term 
\begin{equation}
\db C(\bp) = |\bp|^{\mu_l}\, C({\mathbf 1}) 
\end{equation}
in terms of the scaling exponent $\mu_l$ for $l$-vertex scattering processes. In the general case of an $l$-vertex one obtains along these
lines
\begin{equation} \db
\mu_l = (l-2) d - (l+1) - (l-1) \kappa\, .
\label{eq:mul}
\end{equation}
For the scaling properties of the energy flux we can then write
\begin{eqnarray} \db
A(k) & \db = & \db - \frac{1}{2^d \pi^{d/2} \Gamma(d/2+1)}\, \int^k \! {\mathrm d}p\, |\bp|^{d+\mu_l} \, \omega_1  C({\mathbf 1}) . \quad 
\label{eq:intA}
\end{eqnarray}
If the exponent in the integrand is nonvanishing, the integral gives
\begin{equation} \db
A(k) \, \sim \,  \frac{k^{d+1+\mu_l}}{d+1+ \mu_l}\, \omega_1 C({\mathbf 1}) \, .
\end{equation} 
Thus, scale invariance may be obtained for
\begin{equation} \db
d+1+\mu_l \, = \, 0 \, .
\end{equation}
This gives the scaling exponent for the perturbative 
\begin{equation} 
\!\!\!\!\!\!\mbox{\it relativistic energy cascade:}\quad {\rr \kappa \, = \, d - \frac{l}{l-1} }\, .
\label{eq:ekappaweak}
\end{equation}
One observes that stationary turbulence requires in this case the existence of the limit
\begin{equation} \db
\lim_{d+1+\mu_l \to 0} \,\, \frac{C({\mathbf 1})}{d+1+\mu_l} \, = \, {\mathrm{const}} \, \neq \, 0 \, ,
\end{equation}
such that the collision integral must have a corresponding zero of
first degree. Similarly, starting from the continuity equation for
particle number one can study stationary turbulence associated to
particle number conservation. This leads to the perturbative 
\begin{equation} 
\!\!\!\!\!\!\mbox{\it relativistic particle cascade:}\quad {\db \kappa\, = \, d - \frac{l+1}{l-1}} \, .
\end{equation}
Accordingly, for quartic self-interactions we get $\kappa = d - 4/3$ for the energy cascade and $\kappa = d - 5/3$ for the particle cascade. In the presence of a coherent field, when a 3-vertex can become relevant, the associated scaling exponents are $\kappa = d - 3/2$ for the energy cascade and $\kappa = d - 2$ for the particle cascade.

\subsubsection{Nonperturbative regime: Strong turbulence}
\label{sec:nonpert}

The above perturbative description is expected to become invalid at sufficiently low momenta. In particular, the occupation numbers $f_\bp \sim |\bp|^{-\kappa}$ for $\kappa > 0$ would grow nonperturbatively large in the infrared such that the approximation (\ref{eq:2to2}) becomes questionable. This concerns, for instance, the relevant case of $d=3$ for the description of the early universe dynamics outlined in section~\ref{sec:Earlyuniverseinflation}. To understand where the picture of weak wave turbulence breaks down and to compute the properties of the infrared regime, we have to consider nonperturbative approximations. For this purpose, we employ the vertex-resummed kinetic theory of section~\ref{sec:vertexres} based on the expansion of the 2PI effective action in the number of field components to NLO. 

For a relativistic theory with dispersion $\omega_\bp = \sqrt{\bp^2 + m^2}$, the scaling assumption (\ref{eq:scalingprop}) should be valid for sufficiently high momenta $|\bp| \gg m$ such that the dispersion is approximately linear with $\omega_\bp \sim |\bp|$. However, this is more involved at low momenta if a mass gap exists. An effective mass gap is typically expected because of medium effects even if the mass parameter in the Lagrangian is set to zero. In that case, the infrared modes behave effectively nonrelativistic as explained in section~\ref{eq:nonrelqft}. In the following, we will analyze the two cases of a relativistic theory without mass gap and a nonrelativistic theory separately for comparison. 
 
Following similar lines as in section~\ref{eq:weakwaveturbulence}, we first look for relativistic scaling solutions. To be able to cope with occupancies of order $\sim 1/\lambda$, we replace the perturbative collision term (\ref{eq:2to2cl}) by the vertex-resummed expression (\ref{eq:CNLO}), which can be approximated for $f_p \gg 1$ accordingly by 
\begin{eqnarray} \db
C^{{\mathrm{NLO}}}[f](\bp) & \db \simeq & \db \!\!\int {\mathrm{d}}\Omega^{{\mathrm{NLO}}}[f](p,l,q,r)  \left[ 
(f_p + f_l) f_q f_r - f_p f_l (f_q + f_r) \right] , \quad
\label{eq:CNLOcl}
\end{eqnarray}
with
\begin{eqnarray} \db
\int {\mathrm{d}}\Omega^{{\mathrm{NLO}}}(p,l,q,r) &\db  \simeq &\db \frac{\lambda^2}{18 N} \int_0^\infty \frac{{\mathrm{d}}p^0{\mathrm{d}}l^0{\mathrm{d}}q^0{\mathrm{d}}r^0}{(2\pi)^{4-(d+1)}} \int_{\bl \bq \br} \delta(p+l-q-r)
\nonumber\\ 
&\db \times& \db \tilde{\rho}_p \tilde{\rho}_l \tilde{\rho}_q \tilde{\rho}_r \left[ {\rr v_{\mathrm{eff}}(p+l)} + {\rr v_{\mathrm{eff}}(p-q)} + {\rr v_{\mathrm{eff}}(p-r)} \right] . \quad
\label{eq:NLOomega}
\end{eqnarray}
We emphasize that this still involves integration over frequencies as well as spatial momenta since no free-field form for the spectral function $\tilde{\rho}_p \equiv \tilde{\rho}(p^0,\bp)$ is used so far, and $f_p \equiv f(p^0,\bp)$. A crucial difference to the perturbative kinetic equation is the appearance of the vertex function $v_{\mathrm{eff}}(p^0,\bp)$ given by (\ref{EffectiveCoupling}), which encodes the emergence of a momentum-dependent effective coupling from the NLO corrections of the $1/N$ expansion. It should be emphasized that $v_{\mathrm{eff}}$ depends itself on the distribution function via the retarded self-energy (\ref{OneLoopRA}). By writing down (\ref{eq:CNLOcl}), we neglected off-shell processes included in (\ref{eq:CNLO}). However, their contributions are expected to be small for the scaling behavior considered and we discard them in the following.

In principle, nonperturbative scaling phenomena may involve an anomalous scaling exponent for $\tilde{\rho}(p^0,\bp)$. Using isotropy we write 
\begin{equation}\db
\tilde{\rho}(p^0,\bp) \, = \, s^{2-\eta}\, \tilde{\rho}(s^z p^0,s \bp) \, , 
\label{eq:scalingrho}
\end{equation}
with a nonequilibrium ``anomalous dimension'' $\eta$. The dynamical scaling exponent $z$ appears since only spatial momenta are related by rotational symmetry and frequencies may scale differently because of the presence of medium effects. Accordingly, the scaling behavior of the statistical correlation function
\begin{equation}\db
F(p^0,\bp) \, = \, s^{2+\kappa_{\mathrm s}}\, F(s^z p^0,s \bp)
\label{eq:scalingF}  
\end{equation}
will be described using a scaling exponent $\kappa_{\mathrm s}$. This translates with the definition (\ref{eq:defn}) for $f_p \gg 1$ into
\begin{equation}\db
f(p^0,\bp) \, = \, s^{\kappa_{\mathrm s} + \eta}\, f(s^z p^0, s\bp) \, .
\label{eq:scalingn}
\end{equation}
Using these definitions, one can infer the scaling behavior of $v_{\mathrm{eff}}(p^0,\bp)$, which is given in terms of the ``one-loop'' retarded self-energy $\Pi_{R}(p^0,\bp)$ according to (\ref{EffectiveCoupling}). From (\ref{OneLoopRA}) follows
\begin{equation}\db
\Pi_{R}(p^0,\bp) \, = \, s^\Delta\, \Pi_{R}(s^z p^0, s\bp)
\end{equation}
with  
\begin{equation}\db
\Delta = 4 - d -z + \kappa_{\mathrm s} - \eta .
\label{eq:delta}
\end{equation}
If $\Delta > 0$ one finds from (\ref{EffectiveCoupling}) the infrared scaling behavior
\begin{equation}\rr
v_{\mathrm{eff}}(p^0,\bp) \, = \, s^{-2\Delta}\, v_{\mathrm{eff}}(s^z p^0, s\bp) \, . 
\end{equation}
(For $\Delta \leq 0$ the effective coupling would become trivial with $v_{\mathrm{eff}} \simeq 1$, on which we comment below.) 
Employing these scaling properties, (\ref{eq:NLOomega}) gives
\begin{equation}\db
\int {\mathrm{d}}\Omega^{{\mathrm{NLO}}}(p,l,q,r) \, = \, s^{-2\kappa_{\mathrm s}-z-2\eta} \int {\mathrm{d}}\Omega^{{\mathrm{NLO}}}(s^z p^0,s^z l^0,s^z q^0,s^z r^0;s\bp,s\bl,s\bq,s\br) \, .
\end{equation}

Following the procedure of section~\ref{eq:waveturbulence}, for any conserved quantity we can compute the flux through a momentum sphere $k$. Stationary turbulence solutions then require that the respective integral does not depend on $k$. Energy conservation, expressed in terms of the effective particle number distribution (\ref{eq:neff}), corresponds to the fact that
\begin{equation}
\db \epsilon = \int_0^\infty \frac{\rmd p^0}{2\pi} \int \frac{\rmd^d p}{(2\pi)^d}\,  2 (p^0)^2 \tilde{\rho}_p f_p
\label{eq:energynlo}
\end{equation} 
is a constant of motion in this description. Also the effective particle number density 
\begin{equation}
\db n = \int_0^\infty \frac{\rmd p^0}{2\pi} \int \frac{\rmd^d p}{(2\pi)^d}\,  2 p^0 \tilde{\rho}_p f_p
\end{equation}
is constant for the collision integral (\ref{eq:CNLOcl}). Similar to (\ref{eq:flux}), the flux for this effective particle number reads 
\begin{equation}\db
A(k) \, = \, - \frac{1}{2^d \pi^{d/2} \Gamma(d/2+1)}\, \int^k \! {\mathrm d}p\, |\bp|^{d-1}\, C^{{\mathrm{NLO}}}(\bp) \, .
\label{eq:Aeff}
\end{equation}
The momentum integral can be evaluated along similar lines as before using the above scaling properties 
such that 
\begin{equation}\db
A(k) \, \sim \,  \frac{k^{d-\kappa_{\mathrm s}+z-\eta}}{d-\kappa_{\mathrm s}+z-\eta}\, C^{{\mathrm{NLO}}}({\mathbf 1}) \, .
\end{equation} 
Therefore, scale invariance may be obtained for the  
\begin{equation} 
\!\!\!\!\!\!\mbox{\it particle cascade:} \quad {\rr \kappa_{\mathrm s} \, = \, d + z - \eta } 
\label{eq:kappasn}
\end{equation}
in the nonperturbative low-momentum regime.
Similarly, for the scaling solution associated to energy conservation one finds, taking into account the additional power of $p^0$ in the integrand of (\ref{eq:energynlo}), the exponent for the
\begin{equation} 
\!\!\!\!\!\!\mbox{\it energy cascade:} \quad {\db \kappa_{\mathrm s} \, = \, d + 2z - \eta }\, .
\label{eq:kappase}
\end{equation}
With these solutions, we can now reconsider the above assumption that $\Delta > 0$ by plugging (\ref{eq:kappasn}) or (\ref{eq:kappase}) into (\ref{eq:delta}). Indeed, it is fulfilled under the sufficient condition that $\eta < 2$. Taking into account that the anomalous dimension for scalar field theory is expected to be small for not too low dimension, with $\eta$ maybe of the order of a few percent for $d=3$, and employing a relativistic $z \simeq 1$, we find $\kappa_{\mathrm s} \simeq d+1$ for the particle cascade and $\kappa_{\mathrm s} \simeq d +2$ for the energy cascade solution. 

The scaling of the effective occupation number distribution $f(\bp)$, which depends only on spatial momentum, can finally be obtained from the definition (\ref{eq:neff}). This we write as
\begin{equation}
\db  f(\bp) + \frac{1}{2} = \int_0^\infty \frac{{\mathrm d} p^0}{2\pi}\, 2  p^0 \tilde{\rho}(p^0,\bp)\, \left[f(p^0,\bp) + \frac{1}{2}\right] =  \int_0^\infty \frac{{\mathrm d} p^0}{2\pi}\, 2  p^0 F(p^0,\bp) , \quad
\label{eq:frel}
\end{equation}
using that $\int_0^\infty \rmd p^0/(2\pi) p^0 \tilde{\rho}(p^0,\bp) = 1/2$ from the commutation relation (\ref{eq:bosecomrel}) in Fourier space. Of course, the scaling behavior (\ref{eq:scalingF}) for $F(p^0,\bp)$ may only be observed in a momentum regime with sufficiently high occupancies $f(\bp)$ as implied in the above derivation. Then (\ref{eq:frel}) yields:  
\begin{equation}
{\db f(\bp)} {\db = s^{\kappa_{\mathrm s} + 2 - 2z} f(s\bp)} {\db \sim}
\left\{ \begin{array}{ll}
{\db |\bp|^{-(d+2-z-\eta)}} & \mbox{\it relativistic particle cascade}\\[0.1cm]
{\db |\bp|^{-(d+2-\eta)}} & \mbox{\it relativistic energy cascade}
\end{array} \right.
\, 
\label{eq:relfscal}
\end{equation}
These estimates show that vertex corrections can lead to a strongly modified infrared scaling behavior as compared to the perturbative treatment of section~\ref{eq:weakwaveturbulence}.   

Anticipating that there is a mass gap in the relativistic theory, and because of the very interesting applications such as to the physics of ultracold atoms, we would like to compare this to the nonrelativistic case. The nonrelativistic limit is outlined in section (\ref{eq:nonrelqft}) for the complex Gross-Pitaevskii field theory and we present here the relevant changes as compared to the relativistic case. To this end, we consider a nonrelativistic $N$-component complex scalar field theory and perform again the $1/N$ expansion to NLO. Since already the relativistic scaling exponents (\ref{eq:relfscal}) indicate no explicit dependence on $N$ at NLO -- it can only enter indirectly via $\eta$ and $z$ -- this seems to be a very good starting point to understand also the single complex field case of Gross-Pitaevskii.   

To proceed with the analytic estimate, we first note that for the nonrelativistic theory (\ref{eq:Zmunr}) with quartic $(g/2) (\chi^* \chi)^2$ interaction the RHS of the evolution equation (\ref{eq:evol_eq_F_nonrel}) for $F^{({\rm nr})}$ has the same functional dependence on $\rho^{({\rm nr})}$ and $F^{({\rm nr})}$ as its relativistic counterpart. Therefore, proceeding in the same way as for the relativistic theory, with the corresponding scaling ansatz (\ref{eq:scalingrho}) for $\rho^{({\rm nr})}$ and (\ref{eq:scalingF}) for $F^{({\rm nr})}$, leads to the very same solutions (\ref{eq:kappasn}) and (\ref{eq:kappase}). A crucial difference arises when the occupation number distribution $f_{\rm nr}(\bp)$ is determined. Using the nonrelativistic definition (\ref{eq:n_def_stat_nonrel}) for the distribution function, we have with the notation (\ref{eq:indexnotation}) in the absence of a condensate:
\begin{equation}
\db f_{\rm nr}(\bp) + \frac{1}{2} =  \int_0^\infty \frac{{\mathrm d} p^0}{2\pi}\, F_{aa}^{({\rm nr})}(p^0,\bp) \equiv \frac{1}{2} \int \rmd^3x e^{-i {\mathbf p} {\mathbf x}} \langle \chi(t,\mathbf x)\chi^*(t,0) + \chi(t,0)\chi^*(t,\mathbf x)\rangle. \quad
\label{eq:nonfrel}
\end{equation} 
Comparison with the relativistic case shows that a difference in the scaling behavior is caused by the additional factor of $\sim p^0$ in the integrand of (\ref{eq:frel}). Therefore, we find that 
\begin{equation}
{\rr f_{\rm nr}(\bp) } {\db = s^{\kappa_{\mathrm s} + 2 - z} f_{\rm nr}(s\bp) } {\rr \sim}
\left\{ \begin{array}{ll}
{\rr |\bp|^{-(d + 2 - \eta)}} & \mbox{\it nonrel.\ particle cascade}\\[0.1cm]
{\db |\bp|^{-(d + 2 + z - \eta)}} & \mbox{\it nonrel.\ energy cascade}
\end{array} \right.
\, \label{eq:nonrelfscal}
\end{equation}
scales with one ``$z$'' difference than the relativistic solution (\ref{eq:relfscal}). This can have important consequences, such as the fact that the scaling exponent for the nonrelativistic particle cascade is now independent of the dynamic exponent $z$ describing the dispersion $\omega_\bp \sim |\bp|^z$. As a consequence, the same particle cascade scaling solution is found also in the presence of a condensate, where for the Gross-Pitaevskii theory the approximate (Bogoliubov) dispersion is given by 
\begin{equation}
\db \omega_\bp = \sqrt{\frac{\bp^2}{2m} \left(\frac{\bp^2}{2m} + 2g|\chi_0|^2\right)}  \, .  
\end{equation}
At large momenta, or in the absence of a condensate, one recovers $\omega_{\bp}=\bp^2/(2m)$, while for low momenta one has $\omega_\bp \sim |\bp|$. The particular importance of the particle cascade and the associated phenomenon of Bose condensation far-from-equilibrium, as outlined in the introductory section~\ref{sec:intro}, will be addressed in more detail next.

\subsection{Nonthermal fixed points}
\label{sec:fixedpointsselfsimilarity}

\subsubsection{Self-similarity}

In contrast to the stationary turbulence described above, we now turn to time evolution. In general, isolated systems out of equilibrium cannot realize stationary transport solutions. Instead, we have to consider the more general notion of a {\it\rr self-similar}
evolution, where the dynamics is described in terms of {\it\rr time-independent scaling functions} and {\it\rr scaling exponents}. 

A self-similar evolution of the distribution function $f(t,\bp)$ for a spatially homogeneous and isotropic system is characterized as
\begin{equation}
\db f(t,\bp)=s^{\alpha/\beta} f(s^{-1/\beta}t,s \bp) \, 
\label{eq:gen_selfsim_ansatz_pert}
\end{equation}
with the real scaling exponents $\alpha$ and $\beta$. Again, all quantities are considered to be dimensionless by use of some suitable momentum scale. Choosing $s^{-1/\beta}t = 1$ we recover (\ref{eq:selfsim}), i.e.\
\begin{equation}
\rr f(t,\bp)=t^\alpha\,f_S(t^\beta \bp),
\label{eq:selfsim_ansatz_pert}
\end{equation}
where the time-independent scaling function $f_S(t^\beta \bp)\equiv f(1,t^\beta \bp)$ denotes the {\it\rr fixed point distribution}. 
This scaling form represents an enormous reduction of the possible dependence of the dynamics on variations in time and momenta, since $t^{-\alpha}f(t,\bp)$ depends on the product $t^\beta |\bp|$ instead of separately depending on time and momenta. Therefore, an essential part of the time evolution is encoded in the momentum dependence of the fixed point distribution $f_S(\bp)$.
Moreover, the values for $\alpha$ and $\beta$ determine the rate and direction of transport processes, since a given characteristic momentum scale $K(t_1)=K_1$ evolves as $K(t)=K_1(t/t_1)^{-\beta}$ with amplitude $f(t,K(t))\sim t^\alpha$. This aspect is also further discussed in the introductory section~\ref{sec:intro}. 

For the self-similar distribution (\ref{eq:selfsim_ansatz_pert}), the scaling behavior of a generic collision integral is then given by 
\begin{equation}
\db C[f](t,\bp) = s^{-\tilde{\mu}}\, C[f](s^{-1/\beta}t,s \bp) = t^{-\beta \tilde{\mu}}\, C[f_S](1,t^\beta \bp) \, ,
\label{eq:selfsimC}
\end{equation}
where $\tilde{\mu}$ is a function of scaling exponents similar to the discussion above. Substituting this scaling form into the kinetic equation leads to the time-independent {\it\rr fixed point equation} for $f_S(\bp)$,
\begin{equation}
\rr \left[ \alpha+\beta\,\bp \cdot\mathbf{\nabla}_{\bp} \right] f_S(\bp) = C[f_S](1,\bp) \, ,
\end{equation}
and the {\it\rr scaling relation}
\begin{equation}
\rr \alpha - 1 = - \beta \tilde{\mu} \, .
\label{eq:scalingrelab}
\end{equation}
This follows from comparing the LHS of the kinetic equation,
\begin{equation}
\db \frac{\partial}{\partial t}\left[ t^\alpha\, f_S(t^\beta \bp) \right]
	= t^{\alpha-1}\left[ \alpha+\beta\,\bq\cdot\nabla_\bq \right] f_S(\bq) \big|_{\bq=t^\beta \bp} \, ,
\label{eq:lhs_pert}
\end{equation}
to its RHS given by (\ref{eq:selfsimC}).

Further relations can be obtained by either imposing energy conservation or particle number conservation if applicable.
For constant  
\begin{equation}
\db n = \int \frac{\rmd^dp}{(2\pi)^d}\, f(t,\bp) = t^{\alpha - \beta d} \int \frac{\rmd^dq}{(2\pi)^d}\, f_S(\bq)
\end{equation}
one obtains the relation for 
\begin{equation} 
\!\!\!\!\!\!\mbox{\it particle conservation:} \quad  {\db \alpha = \beta d}.
\label{eq:pcrel}
\end{equation}
Similarly, one obtains from
\begin{equation} 
\!\!\!\!\!\!\mbox{\it energy conservation:} \quad  {\db \alpha = \beta (d+z)}.
\label{eq:ecrel}
\end{equation}
One observes that there is no single scaling solution conserving both energy and particle number. As outlined already in the introductory section~\ref{sec:intro}, in this case a {\it\rr dual cascade} is expected to emerge such that in a given inertial range of momenta only one conservation law governs the scaling behavior.\\ 

\noindent  
{\bf\emph{Perturbative scaling behavior:}} We start again with a perturbative analysis of the relativistic theory, which is relevant at momenta above a possible mass gap such that $\omega_\bp = s^{-1} \omega_{s\bp}$, and afterwards we consider the highly nonlinear infrared regime. For the self-similar distribution (\ref{eq:selfsim_ansatz_pert}), the scaling behavior of a generic collision integral for $l$-vertex scattering is then given by 
\begin{equation}
\db C(t,\bp) = s^{(l-1)\alpha/\beta + l+1 - (l-2) d}\, C(s^{-1/\beta}t,s \bp) \, .
\end{equation}
The reasoning is the same as for (\ref{eq:mul}), with the only difference that now the distribution function scales according to (\ref{eq:gen_selfsim_ansatz_pert}) instead of (\ref{eq:scalingprop}). Consequently, the factor $(l-1)\alpha/\beta$ appears, replacing the term $(l-1)\kappa$ of (\ref{eq:mul}). Using the scaling relation (\ref{eq:scalingrelab}) and (\ref{eq:pcrel}) from particle conservation gives the perturbative solution for 
\begin{equation} 
\mbox{\it relativistic particle transport:}\quad {\db \alpha = - \frac{d}{l+1} \quad , \quad \beta = -\frac{1}{l+1} }\, .
\label{eq:nab}
\end{equation}
Similarly, one finds the perturbative solution for 
\begin{equation} 
\mbox{\it relativistic energy transport:}\quad {\db \alpha = - \frac{d+1}{2l-1} \quad , \quad \beta = -\frac{1}{2l-1} }\, .
\label{eq:eab}
\end{equation}
For instance, for quartic self-interactions the perturbative energy transport is characterized by $\alpha = -(d+1)/7$, and $\beta=-1/7$, where the latter is independent of the dimensionality of space $d$. Likewise, for the 3-vertex in the presence of a coherent field one has for the energy transport $\alpha = -(d+1)/5$ and $\beta=-1/5$. The latter is used for $d=3$ to describe the direct energy cascade of the inflaton dynamics in section~\ref{sec:Earlyuniverseinflation}.\\   

\noindent  
{\bf\emph{Infrared regime:}} In order to go beyond the perturbative estimates, which are not applicable in the infrared, we again consider the $1/N$ expansion to NLO. At the lowest-order gradient expansion employed, the spectral function is time-independent
for a homogeneous system according to (\ref{eq:LOgradrho}). Consequently, its scaling properties are still described by (\ref{eq:scalingrho}) also for the self-similar evolution. However, for $f(t,p^0,\bp)$ we have to take into account the scaling with time:
\begin{equation}
\db f(t,p^0,\bp) =s^{\alpha_{\mathrm s}/\beta}\,f(s^{-1/\beta}t,s^zp^0,s \bp) \, ,
\end{equation}
which increases the number of scaling exponents as compared to (\ref{eq:scalingn}) for the more restrictive case of stationary turbulence. With these definitions we can analyze the scaling of the collision integral (\ref{eq:CNLOcl}) and (\ref{eq:NLOomega}), which gives
\begin{equation}
\db C^{{\mathrm{NLO}}}(t,\bp) = s^{\alpha_{\mathrm s}/\beta - z}\, C^{{\mathrm{NLO}}}(s^{-1/\beta}t,s\bp)
= t^{\alpha_{\mathrm s} - \beta z}\, C^{{\mathrm{NLO}}}(1,t^\beta \bp)
\label{eq:CNLOselfsim}
\end{equation} 
following the corresponding steps done in section~\ref{sec:nonpert}. Using the scaling relation (\ref{eq:scalingrelab}) we infer
\begin{equation}
\db \alpha = \alpha_{\mathrm s} -\beta z + 1 \, .
\label{eq:NLOscaling}
\end{equation}
In particular, with the same reasoning as before we note that (\ref{eq:CNLOselfsim}) applies as well to the corresponding nonrelativistic collision integral.

Again, the crucial difference emerges when the occupation number distributions $f(t,\bp)$ for the relativistic and $f_{\rm nr}(t,\bp)$ for the nonrelativistic theories are determined. For the former, we have according to (\ref{eq:frel})
\begin{eqnarray}
\db  f(t,\bp) &\db =& \db s^{\alpha_{\mathrm s}/\beta +2 - 2z -\eta} \int_0^\infty \frac{{\mathrm d} p^0 s^z}{2\pi}\, 2  (s^z p^0) \tilde{\rho}(s^z p^0,s\bp)\, f(s^{-1/\beta}t,s^z p^0,s\bp) \nonumber\\
&\db \equiv&\db s^{\alpha_{\mathrm s}/\beta +2 - 2z -\eta}  f(s^{-1/\beta}t,s\bp)
\, = \, t^{\alpha_{\mathrm s} + \beta(2 - 2z - \eta)} f_S(t^\beta \bp)
\end{eqnarray}
such that relativistically $\alpha = \alpha_{\mathrm s} + \beta(2 - 2z - \eta)$ by comparison to (\ref{eq:selfsim_ansatz_pert}). Together with (\ref{eq:NLOscaling}) one finds in the absence of a mass gap for 
\begin{equation} 
\mbox{\it relativistic transport:}\quad {\db \beta =  \frac{1}{2-z-\eta} } \quad \mbox{\it of}\,\,  
\left\{ \begin{array}{ll}
\mbox{\it particles:} & \db \alpha = d/(2-z-\eta)\\[0.1cm]
\mbox{\it energy:} & \db \alpha = (d+z)/(2-z-\eta)
\end{array} \right.
\label{eq:reltrans}
\end{equation}
Remarkably, the scaling exponent $\beta$ is obtained without using in addition energy or particle conservation, whereas the different solutions for $\alpha$ arise from imposing (\ref{eq:pcrel}) or (\ref{eq:ecrel}), respectively.

Similarly, the nonrelativistic case yields according to the definition (\ref{eq:nonfrel}): 
\begin{equation}
\db f_{\rm nr}(t,\bp) = t^{\alpha_{\mathrm s} + \beta(2 - z - \eta)} f_{S,{\rm nr}}(t^\beta \bp) \, .
\end{equation}
Therefore, $\alpha = \alpha_{\mathrm s} + \beta(2 - z - \eta)$ by comparing to (\ref{eq:selfsim_ansatz_pert}), which scales again with one ``$z$'' difference than in the relativistic case. Correspondingly, one finds for 
\begin{equation} 
\mbox{\it nonrel.\ transport:}\quad {\rr \beta =  \frac{1}{2-\eta}} \, \quad \mbox{\it of} \,\,
\left\{ \begin{array}{ll}
\mbox{\it particles:} & \rr \alpha = d/(2-\eta)\\[0.1cm]
\mbox{\it energy:} & \db \alpha = (d+z)/(2-\eta)
\end{array} \right.
\label{eq:nonreltrans}
\end{equation}
The nonrelativistic particle transport solution with taking $\eta \rightarrow 0$ is referred to in (\ref{eq:exponentsIPC}) for the description of the inverse cascade in the nonequilibrium evolution of an ultracold Bose gas in section~\ref{sec:Bosegas}.

\subsubsection{Inverse particle cascade and condensate formation}
\label{sec:condensationnonrel}

In order to check the above analytic estimates, one may use the fact that the nonequilibrium quantum dynamics of the highly occupied system can be accurately mapped onto a classical-statistical field theory evolution as described in section~\ref{sec:clstat}. This mapping is valid for the system considered, as long as $f_{\rm nr}(t,\bp) \gg1$ for typical momenta $\bp$ according to (\ref{eq:estimatencl}). We employ initial conditions with high occupation numbers as motivated in section~\ref{sec:intro}, where it is also described that the nonrelativistic and the relativistic theories show the same universal behavior for low momenta. 

In the following, we consider the nonrelativistic Bose gas described by (\ref{eq:Zmunr}) in the dilute regime ($\zeta \ll 1$). The initial distribution function at $t=0$ is taken as
\begin{equation}
\db f_{\rm nr}(0,\bp) \sim \frac{1}{\zeta} \,\Theta(Q-|\bp|)\,,
 \label{eq:fluctuationIC-nonrel}
\end{equation}
which describes overoccupation up to the characteristic momentum $Q$. The initial condensate fraction is taken to be zero, i.e.~$|\psi_0|^2(t=0) = 0$ in the defining equation (\ref{eq:n_def_stat_nonrel}), with an initial $f_{\rm nr}(0,\bp) = 50 / (2 m g Q) \,\Theta(Q-|\bp|)$. We always plot dimensionless quantities obtained by the rescalings $f_{\rm nr}(t,\bp)\rightarrow f_{\rm nr}(t,\bp)\,2mgQ$, $t\rightarrow t\,Q^2/(2m)$ and $\bp\rightarrow \bp/Q$. This reflects the classical-statistical nature of the dynamics in the highly occupied regime, which has the important consequence that if we measure time in units of $2m/Q^2$ and momentum in units of $Q$ then the combination $f_{\rm nr}(t,\bp)\,2mgQ$ does not depend on the values of $m$, $g$ and $Q$.

\begin{figure}[t]
\centerline{
\epsfig{file=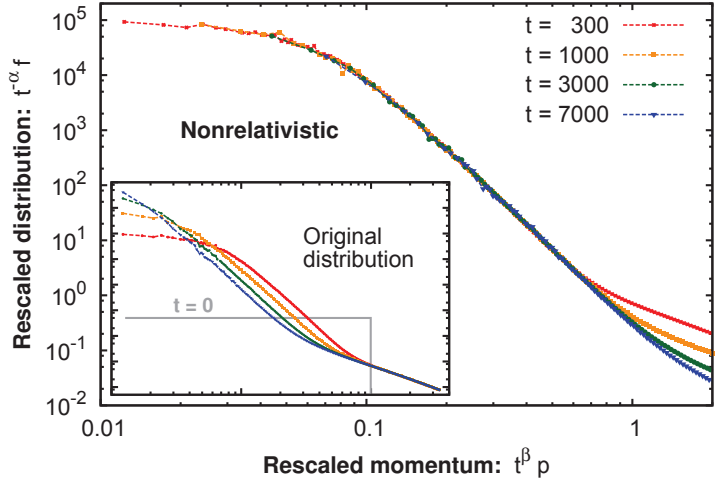,width=10.cm}}
\caption{\label{fig:ab_nonrel_1} Rescaled distribution function of the nonrelativistic theory as a function of the rescaled momentum for different times. The inset shows the original distribution without rescaling.}
\end{figure}

The initial mode occupancies (\ref{eq:fluctuationIC-nonrel}) are expected to get quickly redistributed at the beginning of the nonequilibrium evolution and then a slower behavior sets in. The latter reflects the dynamics near the nonthermal fixed point, where universality can be observed. We concentrate on the low-momentum part of the distribution and analyze its infrared scaling properties. 

Fig.~\ref{fig:ab_nonrel_1} shows the results for the rescaled distribution $(t/t_{\rm ref})^{-\alpha}f_{\rm nr}(t,\bp)$ as a function of $(t/t_{\rm ref})^\beta |\bp|$, where the reference time $t_{\rm ref} Q^2/(2m) = 300$ after which self-similarity is well developed is the earliest time shown. In contrast, the inset gives the curves at different times together with the initial distribution without rescaling for comparison. With the appropriate choice of the infrared scaling exponents $\alpha$ and $\beta$, all the curves at different times lie remarkably well on top of each other after rescaling. This is a striking manifestation of the self-similar dynamics (\ref{eq:selfsim}) near the nonthermal fixed point. The numerical estimates for the scaling exponents obtained are
\begin{equation}
\db	\alpha=\,1.66 \pm 0.12\,,\qquad \beta=\,0.55 \pm 0.03\, .
\label{eq:ab_results_nonrel}
\end{equation}
Comparing these values to the analytic estimates for the particle cascade in (\ref{eq:exponentsIPC}) or (\ref{eq:nonreltrans}), one observes that the numerical results (\ref{eq:ab_results_nonrel}) agree rather well with the NLO approximation for a vanishing anomalous dimension $\eta$. Furthermore, the simulation results confirm that $\alpha = 3\beta$ to very good accuracy as expected for $d=3$ from number conservation (\ref{eq:pcrel}) in the infrared scaling regime. 

The positive values for the exponents imply that particles are transported towards low momenta, which has important consequences.
For the initial conditions (\ref{eq:fluctuationIC-nonrel}), there is no condensate present at $t=0$. However, the inverse particle cascade continuously populates the zero-mode, which will lead to the formation of a condensate far from equilibrium.

\begin{figure}[t]
\centerline{
\epsfig{file=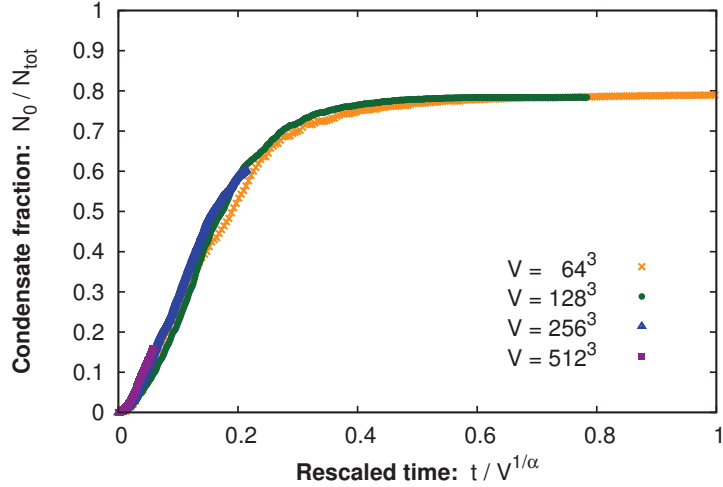,width=10.cm}}
\caption{\label{fig:conden_lin_nonrel} Evolution of the condensate fraction for the nonrelativistic Bose gas for different volumes $V$. The different curves become approximately volume independent after rescaling of time by $V^{-1/\alpha}$, in agreement with (\ref{eq:condens-time-estimate}).}
\end{figure}

Of course, the time needed to fill the entire volume with a single condensate increases with the volume $V$ in which the simulations are done. Using that the parametrically slow power-law dynamics dominates the time until condensation is completed, we can use the scaling exponent $\alpha$ describing the growth in the infrared to estimate this condensation time. Taking the value of the zero-momentum correlator $V^{-1} F^{({\rm nr})}(t,t,\bp=0)$ at the initial time $t_0$ of the self-similar regime as $V^{-1}f_{\rm nr}(t_0,0)$ and its final value at the time $t_f$ as $|\psi_0|^2(t_f)$, we can estimate from $V^{-1} F^{({\rm nr})}(t,t,\bp=0) \sim t^\alpha$ the condensation time as 
\begin{equation}
\db t_f \,\simeq\,  t_0 \, \left(\frac{|\psi_0|^2(t_f)}{f_{\rm nr}(t_0,0)}\right)^{1/\alpha} \, V^{1/\alpha} \, .
\label{eq:condens-time-estimate}
\end{equation}
The time-dependent condensate fraction at zero momentum is given by 
\begin{equation}
\db \frac{N_0(t)}{N_{\rm total}}= \frac{|\psi_0|^2(t)}{\int \rmd^3p/(2\pi)^3 f_{\rm nr}(t,p) + |\psi_0|^2(t)} \,.
\label{eq:condfrac} 
\end{equation}
Accordingly, we may define the condensate fraction for the case of finite volumes as $N_0/N_{\rm total} \rightarrow V^{-1} F^{({\rm nr})}(t,t,\bp=0)/F^{({\rm nr})}(t,t,{\mathbf x}=0)$, using that $N_{\rm total} = F^{({\rm nr})}(t,t,{\mathbf x}=0)$. In Fig.~\ref{fig:conden_lin_nonrel} we show the evolution of the condensate fraction for different volumes and rescale the time axis by $V^{-1/\alpha}$. Indeed, as predicted by (\ref{eq:condens-time-estimate}), the different curves are approximately volume independent. One finds that the condensate fraction saturates at $N_0/N_{\rm total} \simeq 0.8$.

\subsection{Bibliography}

\begin{itemize}
\item Figures~\ref{fig:ab_nonrel_1}, \ref{fig:conden_lin_nonrel} and much of the discussion about self-similar evolution are taken from A.~Pi\~neiro Orioli, K.~Boguslavski and J.~Berges, {\it Universal self-similar dynamics of relativistic and nonrelativistic field theories near nonthermal fixed points}, http://arxiv.org/abs/1503.02498. The presentation of the perturbative scaling behavior follows mostly the review by R.~Micha and I.~I.~Tkachev,
{\it Turbulent thermalization}, Phys.\ Rev.\ D {\bf 70} (2004) 043538.
\item The discussion on strong turbulence in relativistic theories follows partly J.~Berges, A.~Rothkopf and J.~Schmidt,
  {\it Non-thermal fixed points: Effective weak-coupling for strongly correlated systems far from equilibrium,}  Phys.\ Rev.\ Lett.\  {\bf 101} (2008) 041603, and J.~Berges and D.~Sexty,
  {\it Strong versus weak wave-turbulence in relativistic field theory},
  Phys.\ Rev.\ D {\bf 83} (2011) 085004. The presentation of nonrelativistic aspects follows partly C.~Scheppach, J.~Berges and T.~Gasenzer,
  {\it Matter Wave Turbulence: Beyond Kinetic Scaling,}  Phys.\ Rev.\ A {\bf 81} (2010) 033611, and B.~Nowak, J.~Schole, D.~Sexty and T.~Gasenzer,
  {\it Nonthermal fixed points, vortex statistics, and superfluid turbulence in an ultracold Bose gas,}  Phys.\ Rev.\ A {\bf 85} (2012) 043627.
\item The discussion of Bose condensation out of equilibrium follows to some extent J.~Berges and D.~Sexty,
  {\it Bose condensation far from equilibrium,}  Phys.\ Rev.\ Lett.\  {\bf 108} (2012) 161601, and B.~Nowak, J.~Schole and T.~Gasenzer,
  {\it Universal dynamics on the way to thermalization},
  New J.\ Phys.\  {\bf 16} (2014) 9,  093052.	
\item For a presentation of nonthermal fixed points in the language of the renormalization group, see J.~Berges and G.~Hoffmeister,
  {\it Nonthermal fixed points and the functional renormalization group,}  Nucl.\ Phys.\ B {\bf 813} (2009) 383, and J.~Berges and D.~Mesterhazy,
  {\it Introduction to the nonequilibrium functional renormalization group,}  Nucl.\ Phys.\ Proc.\ Suppl.\  {\bf 228} (2012) 37.  
\item Nonthermal fixed points in the presence of fermions are discussed in
J.~Berges, D.~Gelfand and J.~Pruschke,
  {\it Quantum theory of fermion production after inflation,}  Phys.\ Rev.\ Lett.\  {\bf 107} (2011) 061301.    
\end{itemize}